\documentclass[final,3p,times,twocolumn]{elsarticle}
\pdfoutput=1   % visible for top 5 lines; arxiv tweak

\usepackage{amsmath}  %%%added by myself

\usepackage{graphicx}
\usepackage{color}
\usepackage{amsmath}
\usepackage{amsfonts}
\usepackage{mathrsfs}
\usepackage{sidecap}
\usepackage{subcaption}

\usepackage[utf8]{inputenc}
\usepackage[english]{babel}
\usepackage{graphicx}
\graphicspath{{images/}{../images/}}
\usepackage{tabularx}
 
\usepackage{cuted}
\usepackage{flushend}

\usepackage{graphics}
\usepackage{caption}
\usepackage[colorlinks]{hyperref}
\usepackage{amsfonts} 
\usepackage{subfiles}
\usepackage{blindtext}
\usepackage{subfiles}
\usepackage{blindtext}

\numberwithin{equation}{section}

\journal{Physics Reports}
 
\begin{document}
\begin{frontmatter}

\title{Network Resilience}

\author[First]{Xueming Liu} 
\author[Second_0,Second_1]{Daqing Li} 
\author[Third]{Manqing Ma}
\author[Fourth]{Boleslaw K. Szymanski}
\author[Fifth]{H Eugene Stanley}
\author[Sixth]{Jianxi Gao}

\address[First]{Key Laboratory of Image Information Processing and Intelligent Control, School of Artificial Intelligence and Automation, Huazhong University of Science and Technology, Wuhan 430074, Hubei, China}

\address[Second_0]{School of Reliability and Systems Engineering, Beihang University, Beijing 100191, China}

\address[Second_1]{College of Safety Science and Engineering, Civil Aviation University of China, Tianjin 300300, China}

\address[Third]{Department of Computer Science, Rensselaer Polytechnic Institute, Troy, NY 12180; Network Science and Technology Center, Rensselaer Polytechnic Institute, Troy, NY 12180}  

\address[Fourth]{Department of Computer Science, Rensselaer Polytechnic Institute, Troy, NY 12180; Network Science and Technology Center, Rensselaer Polytechnic Institute, Troy, NY 12180}   

\address[Fifth]{Center for Polymer Studies, Department of Physics, Boston University, Boston, MA 02215;}

\address[Sixth]{Department of Computer Science, Rensselaer Polytechnic Institute, Troy, NY 12180; Network Science and Technology Center, Rensselaer Polytechnic Institute, Troy, NY 12180 (e-mail: gaoj8@rpi.edu)}

\begin{abstract}
Many systems on our planet shift abruptly and irreversibly from the desired state to an undesired state when forced across a ``tipping point''. Some examples are mass extinctions within ecosystems, cascading failures in infrastructure systems, and changes in human and animal social networks. The ability to avoid such regime shifts or to recover quickly from such a non-resilient state demonstrates a system's resilience; system resilience is a quality that enables a system to adjust its activities to retain its basic functionality when errors and failures occur. 
In the past 50 years, attention has been paid almost exclusively to low-dimensional systems; scholars have focused on the calibration of the resilience functions of such systems and the identification of indicators of early warning signals based on two to three connected components. 
In recent years, taking advantage of network theory and the availability of lavish real datasets, network scientists have begun to explore real-world complex networked multidimensional systems, as well as their resilience functions and early warning indicators. This report presents a comprehensive review of resilience functions and regime shifts in complex systems in domains such as ecology, biology, society, and infrastructure. The research approach includes empirical observations, experimental studies, mathematical modeling, and theoretical analysis. We also review the definitions of some ambiguous terms, including robustness, resilience, and stability.

\end{abstract}


%\maketitle
%\date{\today}

%\newpage

\end{frontmatter}

\tableofcontents

%\linenumbers

\section{Introduction}\label{Introduction}

\subsection{Resilience is essential}
Despite their widespread consequences for human health \cite{venegas2005self}, the economy \cite{perrings1998resilience} and the environment \cite{may1977thresholds}, events leading to loss of resilience---from cascading failures in infrastructure systems \cite{lyapunov1992general} to mass extinctions in ecological networks~\cite{gao2016universal} and cell fate induction in biological systems \cite{chang2011systematic}---are rarely predictable and often irreversible. The cost of resilience loss is sometimes unaffordable: the outbreak of the COVID-19 pandemic caused over 2.65 million deaths worldwide as of March 15, 2021, and continues to kill more people and further shut down economies. In seven East African countries, swarms of desert locusts have been wreaking havoc as they descend on crops and pasturelands, devouring everything in a matter of hours \cite{devi2020locust}. Moreover, a recent bushfire in Australia burned through 10 million hectares of land, with its raging fires displacing thousands of people and killing countless animals \cite{Resilienceshift}. In August 2017, Hurricane Harvey caused 107 confirmed deaths and inflicted \$125 billion in damage in the US.

Although resilience is a fundamental property of many systems in different fields, each field adds its own unique perspective to these complex systems to achieve resilience. A rich body of work across disciplines has led to many different results and opinions on the concept of resilience, each motivated by the needs of the particular system in question. Examples are the resilience of smart grids~\cite{chen2012smart} and intelligent transportation systems~\cite{isaac2010security} in engineering, the regime shift of coral reefs~\cite{mumby2007thresholds} and tropical forests~\cite{van2012response} in ecology, the population collapse of planktonic organisms due to light irradiance~\cite{veraart2012recovery} and budding yeast due to dilution~\cite{dai2013slower} in biology, social behavior change~\cite{centola2018experimental} in socioeconomic systems \cite{prakash2012threshold}, and the resilience of health systems~\cite{legido2020high} and supply chains~\cite{ivanov2020viability} to the COVID-19 pandemic.

Globalization and technological revolutions are constantly changing our planet. Today, we have a worldwide exchange of people, goods, money, information, and ideas, which has produced many new opportunities, services and benefits for humanity. Simultaneously, however, the underlying networks have created pathways along which dangerous and damaging events can spread rapidly and globally \cite{helbing2013globally}. Consequently, the occurrence of resilience loss in one system may increase the likelihood of that in another or simply correlate at distant places \cite{rocha2018cascading}. When a system undergoes a regime shift, it moves from one set of self-reinforcing processes and structures to another. Changes in a key variable (for example, the temperature in coral reefs) often make a system more susceptible to shifting regimes when exposed to shock events (such as hurricanes) or external driver actions (such as fishing). For example, Australian bushfires and locust swarms are linked to oscillations in the Indian Ocean dipole, which is one aspect of the growth of global climate change \cite{steffen2011australian}. How a system responds to shock events and external disturbances is defined as its resilience, which characterizes its ability to adjust its activity to retain its basic functionality in the face of internal disturbances or external changes \cite{gao2016universal}.

\subsection{Multidisciplinary feature}
\begin{figure}[!ht]
\centering
\includegraphics[width=0.95\linewidth]{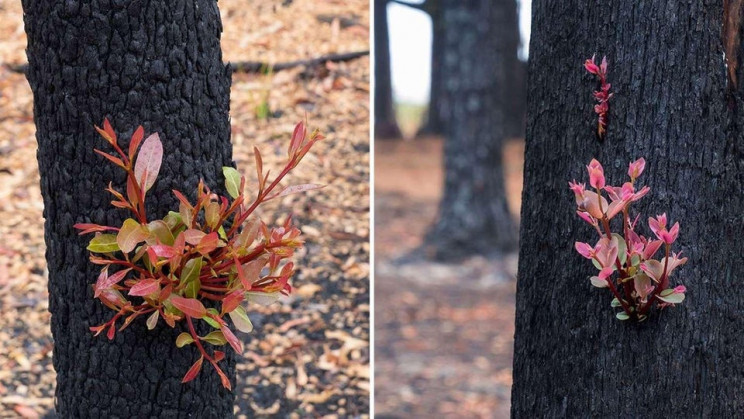} % Figure 1
\caption{
Regrowth during the 2019-2020 Australia bushfire.\\
The figure is from \cite{AustraliaBushfires}.
}
\label{AlstraliaFireResilience}
\end{figure}

As stated in a {\em Science} article~\cite{rohr2014structural}, current studies have either looked into local stability or used numerical simulations \cite{ives2007stability} but have not yet found a unified theory \cite{allesina2012stability}. Therefore, the current challenge is to develop a general framework for unifying the implications of network topology and dynamics, namely, a unified theory for network resilience. Each field has its own focus on specific problems, but the concepts, methods, and algorithms from one discipline can be helpful in another~\cite{motiian2017few}. Disciplinary progress accelerates resilience research in all fields~\cite{chen2015net2net} by integrating knowledge and expertise and forming novel frameworks with which to catalyze scientific discovery and innovation. For example, unlike traditional system resilience \cite{mumby2007thresholds}, engineered systems are designed to be able to recover from disasters, so their resilience is usually defined as the speed at which the system bounces back from the degradation of its functions~\cite{dong2019robust}. Accordingly, borrowing system resilience concepts---regime shift, adaptation, and transformation---can enhance the resilience of the engineered system. For example, like the Phoenix from ancient Greek mythology, which was cyclically reborn from its own ashes, the resilience of our natural systems demonstrates the exceptional renewal of forests through such an adaptive cycle \cite{holling2002resilience}. In the past billions of years, many natural systems have co-adapted to environmental changes and become increasingly resilient to failures. As shown in Fig. \ref{AlstraliaFireResilience}, the regeneration of the Australian forests has occurred, despite fires still blazing there. Damaged by this same Australian bushfire, the water, electricity network, communication network, transport and supply chains have been hit hard, and their slow recovery speed highlights their lack of resiliency \cite{Resilienceshift}. This comparison promotes the concept of how to learn from nature to improve the resilience of infrastructural systems. Therefore, this review aims to promote the research workforce, algorithms, methodologies, and innovations from science to engineering and offer avenues through which to obtain a better understanding of resilience and its enhancement across disciplines.

\subsection{Three resilience frameworks}
Due to the diversity and complexity of the system's response to perturbations, the concept of resilience is multifaceted \cite{bhamra2011resilience}. There are more than 70 definitions of resilience in the scientific literature \cite{fisher2015more}. Some of them generally span multiple disciplines, while others are proposed for specific systems. For instance, Holling \cite{holling1973resilience} used the notion of ``resilience'' to characterize the degree to which a system can endure perturbations without collapsing or being carried into some new and qualitatively different state.
Haimes \cite{haimes2009complex} pointed out two essential elements of resilience: the ability to withstand the presence of errors and that to recover to the original stable state after a perturbation. Bruneau et al. \cite{bruneau2003framework} conceptualized seismic resilience as the ability of both physical and social systems to withstand earthquake-generated forces and demands and cope with earthquake impacts through situation assessment, rapid response, and active recovery strategies \cite{ahn2015reflections}. This review summarizes three frameworks for resilience: system resilience, engineering resilience, and adaptive cycle resilience. (1) System resilience (or ecological resilience) focuses on the magnitude of the change or perturbation that a system can endure without shifting to another stable state \cite{holling1973resilience}. In ecology, the distribution of tree cover in Africa, Australia, and South America reveals strong evidence of the existence of three distinct attractors: forest, savanna, and a treeless state. The empirical results indicate that precipitation universally determines the basins of attraction and resilience states \cite{hirota2011global}. In biology, laboratory populations of the budding yeast Saccharomyces cerevisiae demonstrate a bifurcation diagram experimentally, captured by the fluctuations in population density in response to the size and duration near the tipping point \cite{dai2012generic,dai2013slower}. (2) Engineering resilience is defined by the recovery rate or time \cite{holling1996engineering}. For instance, a 2003 electricity blackout \cite{buldyrev2010catastrophic} affected much of Italy and led chaos: 110 trains were cancelled, with 30,000 passengers stranded on trains in the railway network, and all flights in Italy were cancelled. However, after several hours, electricity was restored gradually in most places, and people's daily lives returned to normal.
(3) Adaptive resilience characterizes the capacity of socioecological systems to adapt or transform in response to unfamiliar, unexpected and extreme shocks \cite{holling2002resilience}. Most definitions attempt to achieve a balance among these three types of resilience. Some helpful examples include aquatic algal blooms, commodity crop markets, and cities such as ancient Rome, Jerusalem, or San Francisco, which were repeatedly attacked or damaged and then rebuilt.

\subsection{Low-dimensional system resilience}
\begin{SCfigure*}
\centering
\includegraphics[width=0.667\textwidth]{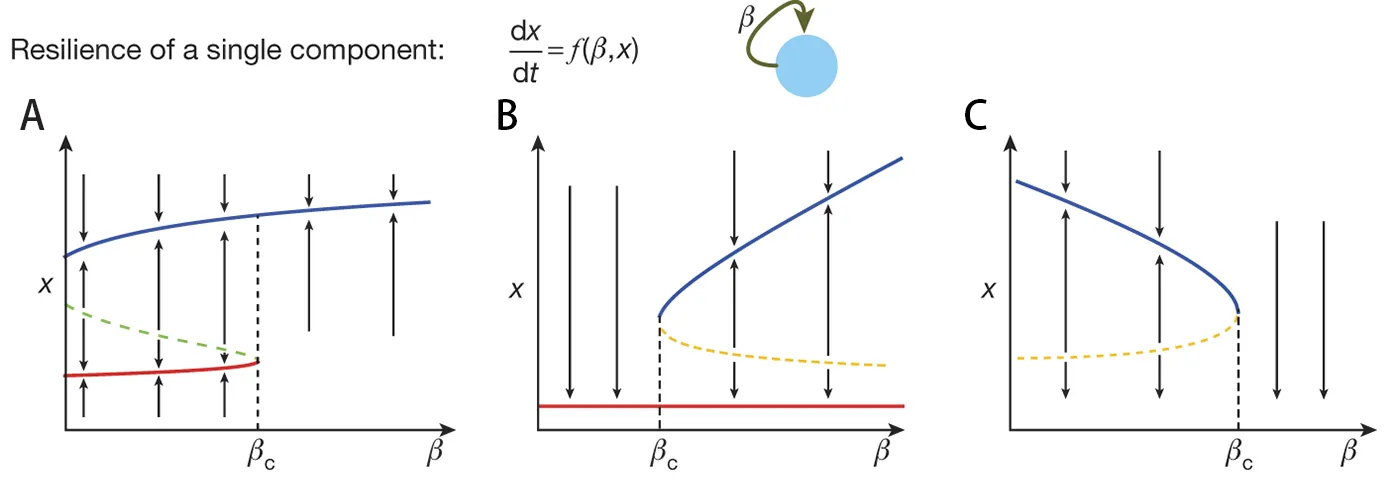} % Figure 2
\caption{
Resilience of a one-dimensional system. A, The system exhibits a single stable fixed point for $\beta>\beta_c$ (blue) and two (or more) stable fixed points otherwise. B, Resilience function with a first-order transition from the desired (blue) state to an undesired (red) state. C, Resilience function with a stable solution for $\beta<\beta_c$ and no solution above $\beta_c$. \\
The figure is from \cite{gao2016universal}.
}
\label{OneDimensionalResilience}
\end{SCfigure*}

During the past decades, resilience has been increasingly employed throughout the science and engineering fields, which makes it a multidisciplinary concept \cite{bhamra2011resilience}. Due to the often unknown intrinsic dynamics in large-scale systems and the limitations of analytic tools, most studies have concentrated on low-dimensional systems or time-series data analysis without modeling\cite{scheffer2001catastrophic, scheffer2009early, dai2013slower}. In studies with dynamical models, resilience behavior is usually captured by a one-dimensional (or low-dimensional) nonlinear dynamic equation, ${\rm d}x/{\rm d}t=f(\beta, x)$, where the functional form of $f(\beta, x)$ shows the system dynamics, and parameter $\beta$ represents the environmental conditions \cite{may1977thresholds}. The rule of the stability of motion \cite{lyapunov1992general} requires that $f(\beta, x_0)=0$ and ${\partial f }/{\partial x}|_{x=x_0}<0$; thus, one can obtain resilience function $x(\beta)$, which represents the possible states of the system as a function of $\beta$. As shown in Fig. \ref{OneDimensionalResilience}, at some critical point, $\beta_c$, the resilience function may undergo a bifurcation or become nonanalytic, indicating that the system loses its resilience by experiencing a sudden transition to a different, often undesirable, fixed point of equation.

\subsection{High-dimensional system resilience}
Extracting the complex system resilience function requires an accurate wiring network of the system and a description of the nonlinear dynamics that govern the complex interactions between components. The emergence of network science has provided a powerful and general tool with which to characterize the structure of large-scale complex real-world systems \cite{barabasi1999emergence}, such as the Internet \cite{cohen2010complex}, power grids \cite{zhou2019bayesian}, genome-scale gene regulatory networks \cite{teichmann2004gene} and metabolic networks \cite{liu2014detection}. Their nontrivial topologies have been uncovered and characterized over the past two decades \cite{liu2016control}. In addition, the accumulation of massive data and the rapid development of computational methods make it possible to identify and predict the exact forms of dynamic models directly from empirical data \cite{wang2016data, barzel2015constructing}. These two prerequisites make it possible for us to develop tools for analyzing the resilience of high-dimensional systems, i.e., network resilience.

However, such a one-dimensional approach rules out its application to many natural and physical systems that are usually multidimensional. One solution is to lift (or embed) the nonlinear dynamics into a higher-dimensional space, where its evolution is approximately linear \cite{korda2018linear}. Such lifting can be achieved by changing the focus from the ``dynamics of states'' to the ``dynamics of observables'' \cite{budivsic2012applied}. A set of scalar observables can measure a system's state, and there are different choices for these observations. If one can find a collection of observables with dynamics that appear to be governed by a linear evolution law, then resilience in nonlinear dynamical systems can be determined entirely by the spectrum of the evolution operator \cite{williams2015data}. The two primary candidates for the study of dynamical systems via operators are the Koopman operator \cite{koopman1931hamiltonian} and the Perron-Frobenius operator \cite{kohda1994explicit}, which complement one another under appropriate function space. The Perron-Frobenius operator represents a picture of the ``dynamics of densities'', in which it is difficult to compute invariant densities, while the Koopman operator presents the ``dynamics of observables'', which are more suitable for physical experiments. Thus, the Koopman operator has been widely used for analyzing system dynamics, as it is a linear operator that generalizes linear mode analysis from linear systems to nonlinear systems \cite{budivsic2012applied, mauroy2013isostables}. Such a linearizing method may cause the system to simplify by shedding its nonlinear nature, but this requires computing the eigenfunctions of the Koopman operator \cite{mauroy2013isostables}, which limits its applications because doing so is extremely time-consuming for large-scale networks \cite{korda2018linear}.

Recently, Gao et al. \cite{gao2016universal} developed a set of analytical tools for identifying the natural control and state parameters of a large-scale complex system, helping derive effective one-dimensional dynamics \cite{tu2017collapse} that accurately predict the critical point (tipping point) at which the system loses its ability to recover. The proposed analytical framework allows us to systematically separate the roles of the system dynamics and topology. Although the critical point depends only on the system's intrinsic dynamics, three topological characteristics---density, heterogeneity, and symmetry---can enhance or diminish a system's resilience. The dimension-reduction method \cite{gao2016universal} has been applied to a class of bipartite mutualistic networked systems in ecology, arriving at a two-dimensional system that captures the essential mutualistic interactions in the original large-scale system \cite{jiang2018predicting}. Using the dominant eigenvalues and eigenvectors of the network adjacency matrix, Laurence et al. reduced its dimensions and increased its accuracy \cite{laurence2019spectral}. This method is able to predict the multiple activation of modular networks as well as the critical points of random networks with arbitrary degree distributions.

\subsection{Review topics}

Based on the theoretical tools for both low-dimensional and large-scale networks above and the advanced data analysis techniques \cite{gao2017complex}, studies on the resilience in real networks from various fields, ranging from the natural to humanmade world, have been carried out. The goal of this article is to review the advances in resilience in real networks. To achieve this, we discuss a series of topics that are essential to the understanding of network resilience, such as alternative stable states (bistability) \cite{beisner2003alternative}, regime shifts \cite{ghaffarizadeh2014multistable}, and early warning signals \cite{bauch2016early}. We illustrate these topics according to their application scenarios in ecology, biology, and social and infrastructural systems.
\begin{itemize}
\item Ecological network resilience: in ecology, species extinction/co-existence is a critical issue that has attracted much attention. Many studies have been carried out to predict the thresholds and tipping points at which certain species become extinct. These studies have focused on the analytical solutions of low-dimensional systems or the numerical simulations of high-dimensional systems. Recently, there have been additional studies carried out on the analytical prediction of resilience in high-dimensional systems. \\
\item Biological network resilience: in biology, many living systems exhibit drastic state shifts in response to small changes in environmental parameters, leading to disease or apoptosis. We review recent studies on predicting tipping points and discovering early warning signals in organisms, which can help prevent or delay the onset of disease. \\
\item Social network resilience: the behavior of social groups of humans and social animals are likely to exhibit tipping points. We first distinguish between cultural and survival resilience. Then, we summarize the concepts behind tipping points and show that they arise in both types of resilience. We also show instances in which they are likely to occur in human and animal societies. The study of tipping points may open up new inquiry lines in behavioral ecology and generate novel questions, methods, and approaches to human and animal behavior. \\
\item Infrastructural network resilience: most infrastructural network systems are considered recoverable due to effective human interventions. The traditional concept of ``engineering resilience'' measures the time it takes for a system to recover or the relative change in its recovery back to equilibrium after a disturbance. Very recently, a few groups have studied the system resilience and prediction of tipping points in engineering systems. For example, certain traffic congestion may be avoided in transportation systems if effective early warnings can be provided.
\end{itemize}

Resilience problems are ubiquitous, with broad applications to numerous disciplines in addition to the four domains discussed above, such as environmental science, computer science, management science, economics, political science, business administration, and phycology \cite{fraccascia2018resilience}. For instance, the extensively studied concept of psychological resilience has multiple definitions, with the adversary and positive adaptation being two core concepts, which characterize the ability to mentally or emotionally cope with a crisis \cite{fletcher2013psychological}.
For business and economic systems, {\em resilience} is defined as the capacity to survive, adapt and grow in the face of change and uncertainty related to disturbances, regardless whether they are caused by resource stresses, societal stresses, or acute event stresses \cite{kupers2014turbulence}. Furthermore, topics related to resilience, such as alternative stable states, also exist in other disciplines. For example, in materials science, most solid matter types have a single stable solid state for a particular set of conditions. Nevertheless, Yang et al. \cite{yang2019hybrid} described a material composed of a polymer impregnated with a supercooled salt solution, termed sal-gel, which assumes two distinct but stable and reversible solid states under the same conditions for a range of temperatures. In addition to the various definitions of resilience itself, there are conceptual overlaps between resilience and other concepts, such as robustness and stability, which we discuss in this review to demonstrate their differences. The breakthroughs reviewed here can advance readers' understanding of the complex systems surrounding them and enable them to design more resilient infrastructure or social systems.

\section{Essential topics in network resilience}\label{Robustness}

The word ``resilience'' is derived from the Latin terms \textit{resiliere} or \textit{resilio}, meaning ``bounce'' or ``rebound'', respectively 
\cite{capano2017resilience}. The action of ``bouncing back'' characterizes the basic meaning of resilience, which is a dynamical property that requires a shift in a system's core activities. In other words, network resilience is a concept that describes a networked system's ability to retain its essential functions, defined in terms of its network dynamics within a particular region, and there is no requirement that it remain at a specific fixed point, which is distinct from the concept of stability. ``Resilience'' was initially used in material science to describe the resistance of materials to physical shocks \cite{winson193227} and has been widely used to characterize an individual's ability to cope with adversity, trauma, or other sources of stress \cite{murphy1974coping, urruty2016stability}.

In 1973, Holling et al. \cite{holling1973resilience} defined resilience as a measure of the persistence of systems and their ability to absorb change and disturbance, and it has since become a popular concept used in ecology. Later, resilience was adopted as a generic concept and used to describe the response to changes in systems in other fields such as biology \cite{meredith2018applying}, social sciences \cite{parsons2017social}, and engineering \cite{hollnagel2006resilience}. In most of these applications, resistance to perturbations, measured as the recovery rate of the system after the occurrence of perturbations, is considered a critical aspect of resilience \cite{holling1996engineering}. Based on the adaptive cycle that includes four phases---growth, consolidation, release, and reorganization \cite{ holling2004complex}---a more general meaning of resilience was proposed \cite{fisher2015more}, i.e., the capacity of socioecological systems to adapt and transform in response to unfamiliar, unexpected, and extreme shocks.

In general, system resilience determines whether the system can tolerate a significant perturbation without shifting to another stable state; it measures the ability of the system to absorb changes in state variables, driving variables, and parameters and still persist \cite{standish2014resilience}. 
As discussed in Chapter \ref{Introduction}, traditional mathematical treatment of resilience used from ecology to engineering approximates the behavior of a complex system with a one-dimensional (1D) nonlinear dynamic equation $dx/dt =f(\beta,x)$, and the critical transitions related to resilience can be captured by the solution of this function.

Whereas, real-world systems are composed of numerous components connected via a complex set of weighted, often directed, interactions and controlled by not one microscopic parameter but by a large family of parameters, such as the weights of all interactions. Hence, instead of a 1D function $f(\beta, x)$, characterized by a single parameter $\beta$, their state should be described by a network of coupled nonlinear equations that capture the interactions between the system's many components and account for the complex interplay between the system's dynamics and changes in the underlying network topology. Therefore, the resulting resilience function is a multi-dimensional manifold over the complex parameter space characterizing the system. Thus, to extend the concept of 1D system resilience to network resilience, we should understand and review:
\begin{itemize}
    \item[(1)] The multi-dimensional equations on networks generalized $f(\beta, x)$ for the 1D equation (Sec. \ref{1mnd});
    \item[(2)] The resilience phenomenon and possible critical transitions of high dimensional systems (Sec. \ref{1rpids});
    \item[(3)] A set of novel analytical tools that is general for various resilience functions of networks (Sec. \ref{1atfenr}).
\end{itemize}	
The significant difference between 1D system resilience and network resilience is the underlying network topology of the system, which promotes us to also review the differences among related network concepts in the literature, such as stability and robustness.

\subsection{Modeling network dynamics}\label{1mnd}
Network dynamics describe how entities evolve, for example, how gene expression levels in gene regulatory networks change \cite{nelson2004oscillations}, how species abundances evolve during a period \cite{volkov2003neutral}, or how many individuals are newly affected by or have recovered \cite{kitsak2010identification} from a particular disease. The modeling of such processes requires the determination of equation forms and parameters, a task that is not easy to achieve due to the complexity and unknown mechanisms within these processes. Thus, compared to extensive studies on reconstructing networks' static structures \cite{newman2003structure}, there is much less research on modeling the dynamics of large-scale networks.
In the following, we review the dynamic models of several large-scale systems.

\begin{figure}
\centering
\includegraphics[width=0.48\textwidth]{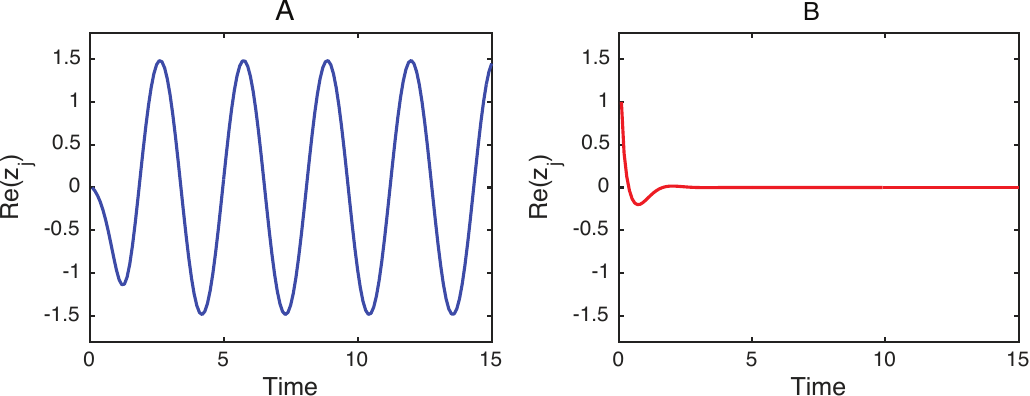}  % figure 3
\caption{
(A) Active and (B) inactive dynamics in isolated oscillators.
Periodic oscillation occurs in the active oscillator, while the inactive oscillator becomes quiescent after transient damping oscillation.
Here, we set control parameter $\alpha_j =2$ for the active oscillator and $\alpha_j =-2$ for the inactive oscillator.
The inherent frequency $\Omega_j$ is 2.
}
\label{SL_Osillator}
\end{figure}

Systems may exhibit various nonlinear dynamics \cite{pan2020phase, ma2018randomly} such as oscillation, spreading and bifurcation. These dynamic models involve different equation forms and various parameters.
For example, the networked Stuart-Landau (SL) oscillator system \cite{tanaka2015dynamical} can be described by the following coupled ordinary differential equations (ODEs):
\begin{equation}\label{Oscillators}
\frac{dz_j}{dt}=(\alpha_j+i\Omega_j-|z_j|^2)z_j+\sigma \sum_{k=1}^Nc_{kj}(z_k-z_j),
\end{equation}
where $z_j$ and $\Omega_j$ are the complex amplitude and the inherent frequency of the $j$th SL oscillator, respectively,
and $\alpha_j$ is its control parameter.
The parameters $\sigma$ and $c_{ij}$ describe the interactions between the $j$th and other SL oscillators.
Figure \ref{SL_Osillator} shows the (a) active and (b) inactive dynamics in an isolated oscillator.

The dynamics of gene regulatory networks can be modeled as Michaelis-Menten equations \cite{alon2006introduction}:
\begin{equation}\label{MM}
\frac{dx_i}{dt}=-Bx_i^{a}+\sum_{j=1}^NA_{ij}\frac{x_j^h}{1+x_j^h},
\end{equation}
where $B$ is the rate constant, $a$ is the level of self-regulation, and the Hill coefficient $h$ describes the level of interactions between genes.
The spreading process can be modeled as a susceptible infected susceptible (SIS) model \cite{boguna2013nature}:
\begin{equation}\label{SIS}
\frac{dx_i}{dt}=-Bx_i+\sum_{j=1}^NA_{ij}R(1-x_i)x_j.
\end{equation}
All these dynamic equations can be generalized according to the following equation:
\begin{equation}\label{GeneralDynamics}
\frac{dx_i}{dt}=W(x_i(t))+\sum_{j=1}^NA_{ij}Q(x_i, x_j),
\end{equation},
where the first term, $W(x_i(t))$, describes the self-dynamics of $x_i$, and the second term captures the interactions between component $i$ and its neighbors.
Barzel et al. \cite{barzel2013universality} developed a self-consistent theory of dynamical perturbations in an attempt to uncover the universal characteristics of a broad range of dynamical processes. By analyzing the macroscopically accessible distributions of certain dynamical measures---correlation ($G$), impact ($I$), stability ($S$), propagation and cascade dynamics---the system's universality class could be determined, even without knowledge of the analytical formulation of the system's dynamics. Duan et al. \cite{duan2019universal} extended the approach to interdependent networks and tested their approach by modeling epidemic spreading, birth-death processes, and biochemical and regulatory dynamics.

Despite the difficulties associated with uncovering the complex and nonlinear mechanisms of network dynamics, the accumulations of massive data and the rapid development of computational methods make it possible to identify and predict dynamic models directly from empirical data. Takens \cite{takens1981detecting} showed that the underlying dynamical system could be faithfully reconstructed from time series under fairly general conditions, establishing a one-to-one correspondence between the reconstructed and existing but unknown dynamical systems \cite{wang2016data}.
Wang et al. \cite{wang2011predicting} developed a framework based on compressive sensing that could predict the exact forms of both system equations and parameter functions.

Recently, Barzel et al. \cite{barzel2015constructing} developed a method for inferring the microscopic dynamics of a complex system from observations of its response to external perturbations, enabling the construction of nonlinear pairwise dynamics that are guaranteed to recover the observed behavior.
Consider a complex networked system with $N$ components whose states $x_i(t)$ and $(i=1,...N)$ are governed by the following ODEs:
\begin{equation}\label{NnetDynamics}
\frac{dx_i}{dt}=M_0(x_i(t))+\sum_{j=1}^{N}A_{ij}M(x_i(t), x_j(t)),
\end{equation}
where $M_0(x_i(t))$ describes the $i$th component's self-dynamics, $M(x_i(t), x_j(t))$ captures the impact of neighbor $j$ on the state of $i$, and $A_{ij}$ is the adjacency matrix.
Factoring the interaction term as $M(x_i(t), x_j(t))=M_1(x_i(t))M_2(x_j(t))$, the system's dynamics are uniquely characterized by three independent equations:
\begin{equation}\label{ThreeEquations}
\textbf{m}=(M_0(x), M_1(x), M_2(x)),
\end{equation}
which is a point in the model space $\mathbb{M}$.
For systems of unknown microscopic dynamics, the challenge is to infer the appropriate model by identifying $M_0(x)$, $M_1(x)$, and $M_2(x)$, which accurately describe the system's observable behavior $\mathcal{X}$.
Define a subspace $\mathbb{M}(\mathcal{X})\in \mathbb{M}$ comprising all models \textbf{m} that can be validated against $\mathcal{X}$.
Barzel et al. \cite{barzel2015constructing} proposed a method to link the observed system response to the leading terms of \textbf{m}. 
The method defines the exact boundaries of $\mathbb{M}(\mathcal{X})$ rather than providing a specific model 
\textbf{m},
thereby providing the most general class of dynamics that can be used to describe the observed responses captured by two quantities, the transient response $\mathcal{T}$ and the asymptotic response $\mathcal{G}$.
By applying both numerical data on the gene regulatory dynamic model and empirical data on cell biology and human activity, the effective dynamic model can predict the system's behaviors and provide crucial insights into its inner workings.

In summary, resilience is defined based on transitions between states, which are usually characterized by network dynamical equations. The prediction of resilience in large-scale networks has for a long time been restricted by unclear internal network dynamics. Although the discovery of the dynamics of complex networks remains a major challenge, using the methods reviewed above to infer network dynamics, it is possible to model the dynamics of large-scale networks and thereby to construct a fundamental framework for network resilience analysis.

\subsection{Resilience phenomena in dynamical systems}\label{1rpids}

\begin{SCfigure*}
 \centering
 \includegraphics[width=0.667\textwidth]{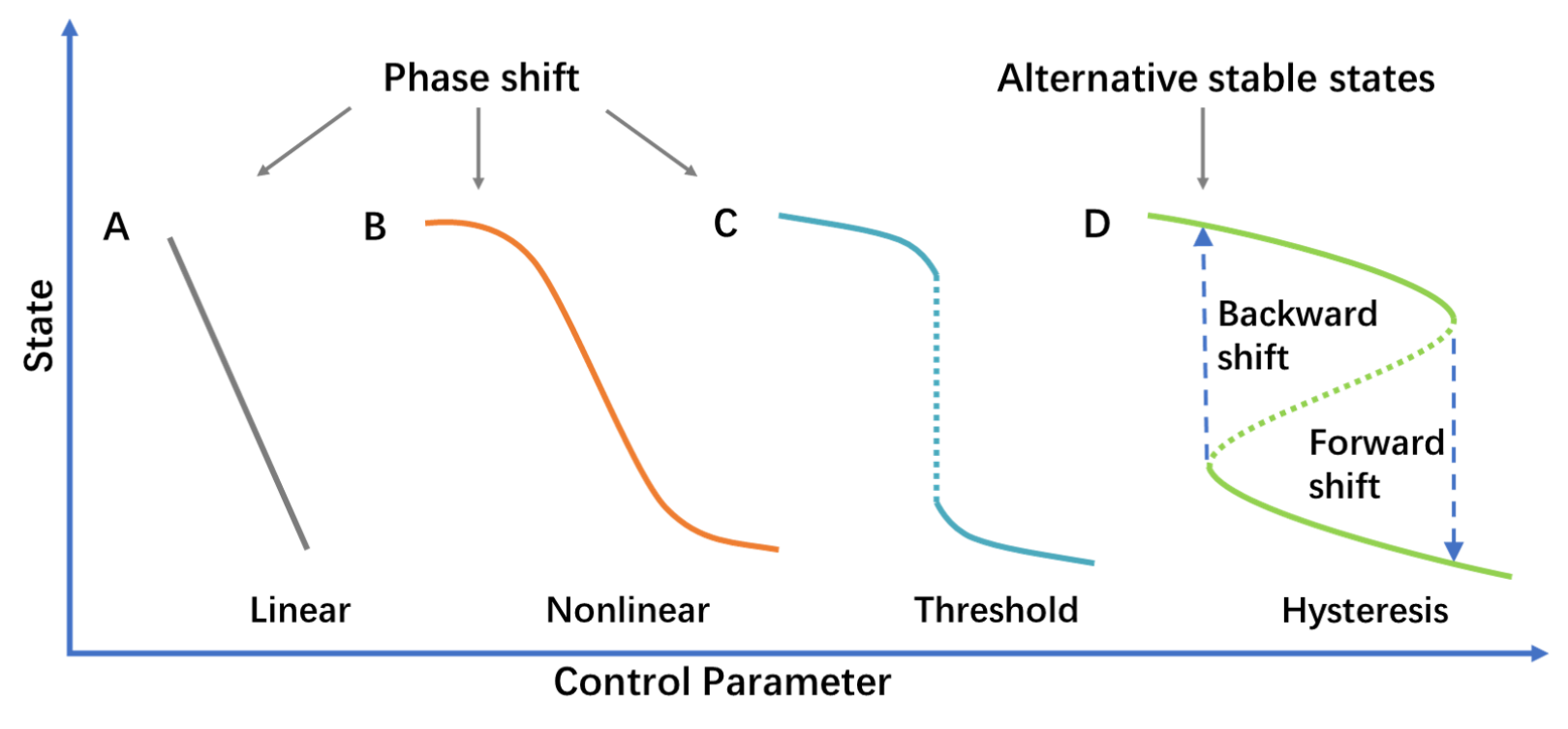} % figure 4
\caption{ Schematic showing different types of state transitions in system states.
(A) The state in a system responds linearly to changes in conditions;
(B) A continuous and gradual ``phase shift'' from ``upper'' to ``lower'' mutually exclusive states as the background environmental parameter increases;
(C) A limiting case of sudden regime shift between two mutually exclusive states;
(D) A sudden catastrophic shift between multiple stable states in a hysteretic system with different thresholds for forward and backward shifts. \\
\textit{Source:} The figure has been modified from \cite{moffett2015multiple}.
}
\label{FourTypesShifts}
\end{SCfigure*}

The state of a dynamical system may change as the external conditions change, and the response is usually nontrivial. When external conditions change gradually with time \cite{tilman2001forecasting}, the states of some highly resilient systems may respond linearly, as shown in Fig. \ref{FourTypesShifts}A. For example, in a coral reef system, there is a simple linear relationship between coral and fishing \cite{mumby2013evidence}.
A more common and complex relationship between the state of a system and conditions is nonlinear:
a continuous and gradual phase shift from upper to lower mutually exclusive states as the background environmental parameter increases (Fig. \ref{FourTypesShifts}B) \cite{moffett2015multiple}. In both the linear and smooth nonlinear cases, there is only one equilibrium state, and the system can return to the former state if the control parameters return to the previous level \cite{mumby2013evidence}.
Some systems may be inert over a specific range of conditions but may respond abruptly when the control parameter approaches a threshold, exhibiting an unexpected shift between two mutually exclusive states (Fig. \ref{FourTypesShifts}C). These three types of state transitions are called phase shifts or regime shifts. They are characterized by dominant populations of an ecological community responding smoothly and continuously along an environmental gradient until a threshold is reached, at which point the community shifts to a new dominant or suite of dominant populations. In any given environment, there is at most one stable state \cite{dudgeon2010phase}, except at the tipping point, where the system shows an abrupt shift and may, in principle, have more than one attractor.

A different crucial situation arises when the system response curve is folded backwards, creating alternative stable states (or multiple stable states) for a specific range of parameters \cite{scheffer2001catastrophic}; the parameters may describe an external condition rather than an interactive part of the system, or the change in the condition may be very slow relative to the rate of change in the system. In such a case, the system has two alternative stable states that are separated by an unstable equilibrium that marks the border between the basin of attraction of the states, which is shown as the green dashed line in Fig. \ref{FourTypesShifts}D. Over a specific range of conditions, the upper and lower states need not be mutually exclusive but may coexist. As the condition or the control parameter changes, an abrupt shift between states occurs at two critical points: (1) the system shifts from an upper to a lower state, and (2) the system shifts from a lower to an upper state. These two abrupt shifts usually differ because it is difficult for a system in a lower state to return to its upper state.

\begin{figure}[!ht]
\centering
 \includegraphics[width=0.95\linewidth]{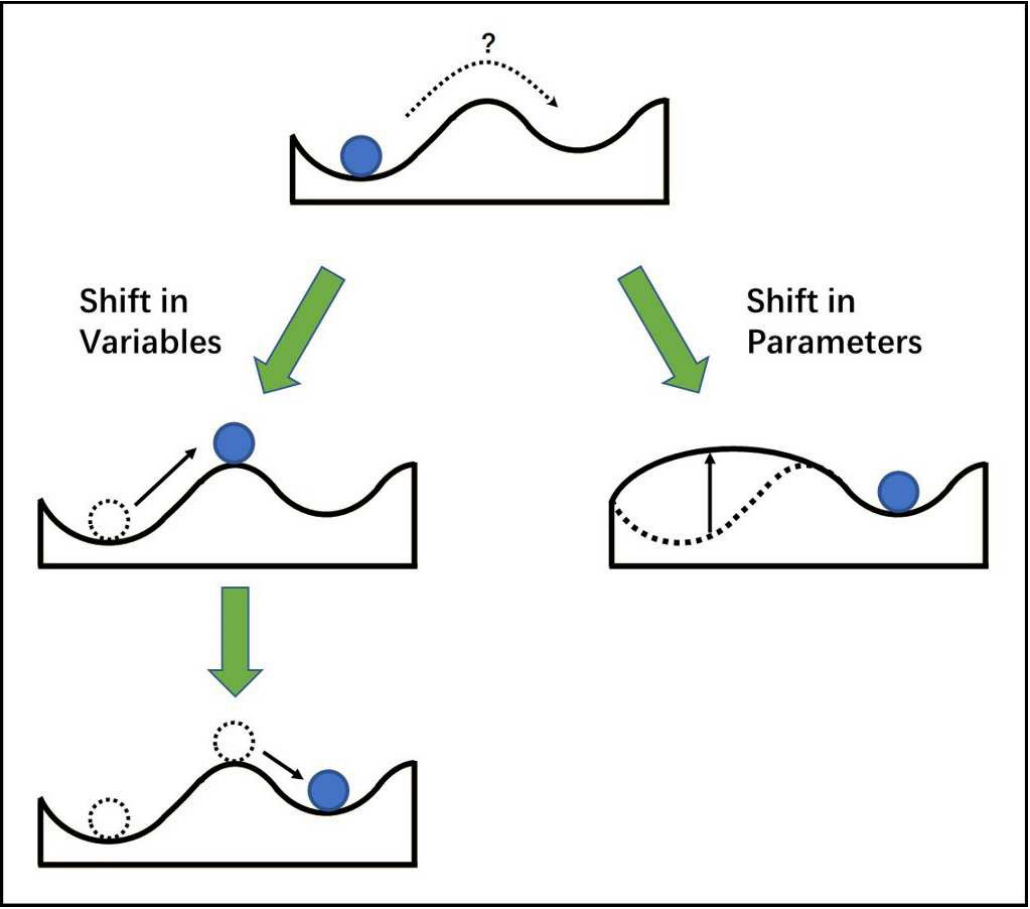} % figure 5
\caption{Two-dimensional ball-in-cup diagrams showing (left) the way in which a shift in state variables causes the ball to move and (right) the way in which a shift in parameters causes the landscape itself to change, resulting in the movement of the ball. \\
\textit{Source:} The figure has been reproduced from \cite{beisner2003alternative}.
}
\label{Basin_Ball_scheme}
\end{figure}

\noindent
\textit{\textbf{Alternative stable states.}}
Alternative stable states in a system can be illustrated by the ball-in-cup analogy \cite{beisner2003alternative} outlined in Fig. \ref{Basin_Ball_scheme}. In that analogy, all conceivable states of the system are represented by a surface on which the system's actual state (for example, the abundance of all populations or the expression level of a gene) is represented as a ball residing on it.
The ball's movement can be anticipated from the nature of the surface: in the absence of external intervention, the ball always rolls downhill.
In the most straightforward representation of alternative stable states, the surface has two basins, and the ball resides on one of them. Valleys or dips on the surface represent domains of attraction for a state. The question then becomes the following: How does the ball move from one basin to the other? There are two ways in which this can occur: the ball can be moved (Fig. \ref{Basin_Ball_scheme}, left), or the landscape upon which it sits can be altered (Fig. \ref{Basin_Ball_scheme}, right). The first of these ways requires substantial perturbation to the state variables (for example, population densities). The latter requires a change in the parameters governing interactions within the dynamical system, such as birth rates, death rates, carrying capacity and migration in ecosystems or individual activities in social systems; all of these can be changed by human interventions or natural disasters.

Although we use the terms stable states and equilibria, ecosystems are never stable in the sense that they do not change, and slow trends and fluctuations are always occurring. Therefore, Scheffer et al. \cite{scheffer2003catastrophic} suggested that the words ``regimes' and ``attractors' are more appropriate for showing these dynamics.
Since the terms multiple stable states and ``alternative stable states'' have been used extensively in the literature, we also use these terms while keeping in mind that they refer to a kind of relative dynamic balance rather than to a stable state excluding dynamics.

\noindent
\textit{\textbf{Abrupt phase shifts and tipping points.}}
When the external situation or control parameter goes beyond a tipping point, catastrophic shifts occur, leading to disastrous consequences. Once this happens, the results may be largely irreversible. For example, some cloud forests were established under a wetter rainfall regime thousands of years previously, and the moisture they require is now supplied by the condensation of water from clouds intercepted by the canopy. If the trees are cut, then this water input ceases, and the resulting conditions can be too dry to permit the recovery of the forest \cite{folke2004regime}. Thus, the accurate prediction of abrupt shifts and tipping points is among the essential topics related to network resilience.

Both regime shifts and multiple stable states are underlying mechanisms that can explain the catastrophic shifts in complex systems. Abrupt shifts with alternative stable states are usually related to bifurcation \cite{crandall1971bifurcation} and catastrophe theory \cite{zeeman1979catastrophe}, as shown in Fig. \ref{FourTypesShifts}(D), while phase shifts are not. Thus, it is difficult to collect empirical data showing bifurcations under the same conditions. As a result, differentiating phase shifts from multiple stable states is not straightforward.
Fortunately, despite the requirement for extensive time series containing many shifts \cite{carpenter2001alternate}, there are several approaches that can be used to infer whether alternative attractors are involved in a shift \cite{scheffer2003catastrophic}. 1) Based on the principle that all attractor shifts imply a phase in which the system is speeding up as it diverges from a repeller, a statistical approach \cite{broock1996test} can be used. 2) Another approach compares the fit of contrasting models with and without attractor shifts \cite{carpenter1997dystrophy} or computes the probability distribution of a bifurcation parameter \cite{carpenter1997dystrophy}.
Significantly, some colonization events, such as those that occur in marine fouling communities \cite{sutherland1974multiple}, can be very persistent once established and difficult to replace until the cohort dies of old age. Unless the new state persists through additional generations by strengthening itself \cite{holmgren2001nino}, it seems inappropriate to describe such shifts as alternative stable states \cite{scheffer2003catastrophic}.

Although ``regime shift'' and ``alternative stable states'' differ from each other, as discussed above, they can both lead to abrupt phase transitions. Since, for some transitions, there is no apparent evidence to show whether they are related to alternative stable states or to regime shift
, various studies use them without making such distinctions \cite{scheffer2007regime}. In the following review of critical transitions, we also do not distinguish between them.

\noindent
\textit{\textbf{Critical slowing down.}}
As a system approaches but does not cross the tipping point, a phenomenon known in dynamical systems theory as ``critical slowing down' may occur \cite{wissel1984universal}. This phenomenon can be used as an early warning for the advent of the tipping point, which occurs over a range of bifurcations \cite{scheffer2009early}. At fold bifurcation points, the dominant eigenvalue characterizing the rate of change around equilibrium becomes zero, which implies that as the system approaches such a critical point, it becomes increasingly slow to recover from small perturbations \cite{van2007slow}.
Thus, the recovery rate after a small experimental perturbation can be used as an indicator of how close a system is to a bifurcation point, and such a perturbation is so small that it carries no risk of driving the system over the threshold.

\begin{figure}[!ht]
 \centering
\includegraphics[width=0.95\linewidth]{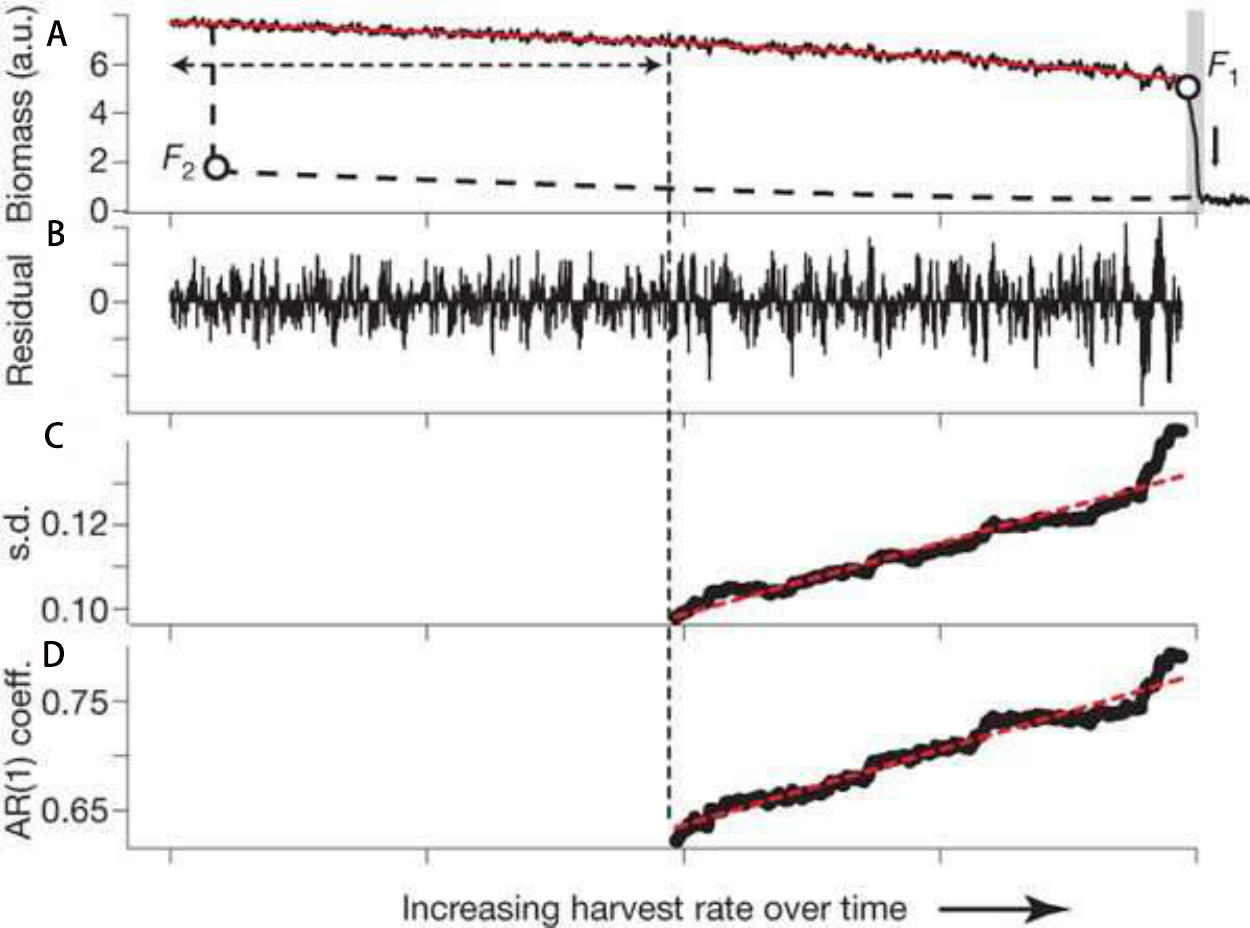} % figure 6
\caption{
Early warning signals for a critical transition in a time series generated by a model of a harvested population driven slowly across a bifurcation. (A) Biomass time series. (B) shows that the catastrophic transition is preceded by an increase in both the amplitude of fluctuation, expressed as standard deviation (s.d.) (C), and slowness, estimated as the lag-1 autoregression (AR(1)) coefficient (D), as predicted from theory.\\
\textit{Source:} Figure from \cite{scheffer2009early}.
}
\label{Auto_Correlation}
\end{figure}

A straightforward evaluation of critical slowing down by systematically testing recovery rates is suitable for theoretical models \cite{ovaskainen2002transient} but impractical or difficult to use for the monitoring of most natural systems. However, almost all real systems are persistently subjected to natural perturbations, and specific characteristic changes in the pattern of fluctuations in those systems are expected to occur as a bifurcation is approached. For example, since slowing down causes the intrinsic rates of change in the system to decrease, the state of the system at any given moment becomes increasingly similar to its past state. Thus, an increase in autocorrelation in the resulting pattern of fluctuations appears as the system approaches the tipping point \cite{ives1995measuring}.
Although different measurements can be used to quantify such an increase, the simplest way to quantify them is to look at lag-1 autocorrelation \cite{dakos2008slowing}, which can be directly interpreted as slowness of recovery in such natural perturbation regimes \cite{van2007slow}. It has been found that a marked increase in autocorrelation builds up long before the critical transition occurs (Fig. \ref{Auto_Correlation}) in both simple models \cite{dakos2008slowing} and complex and realistic systems \cite{lenton2008using}.

Another possible consequence of critical slowing down in the vicinity of a critical transition is increased variation in the pattern of fluctuations \cite{scheffer2009early}. Critical slowing down can reduce the ability to track the systems fluctuations \cite{berglund2002metastability} because it increases the standard deviation of the stationary distribution versus input fluctuations.

In short, the phenomenon of critical slowing down leads to three possible early-warning signals in the dynamics of a system approaching a bifurcation: (i) slower recovery from perturbations, (ii) increased autocorrelation, and (iii) increased variance \cite{scheffer2009early}.

Note that autocorrelation and variance are indirect measures of critical slowing down, and both are expected to increase before a critical transition. To determine whether either phenomenon is a result of measurement noise, researchers need to construct a null model for comparison. For example, Veraart et al. \cite{veraart2012recovery} found that the autocorrelation and the variance in population density increased as the population of cyanobacteria decreased toward a tipping point and that the trend was significant compared to the trends that were experimentally observed using 1000 datasets generated with a null model constructed based on the assumption that all variation in the measurements was uncorrelated and normally distributed measurement noise.

In summary, a complex network usually responds to external perturbations in a nonlinear and nonmonotonic way. Due to the existence of alternative stable states, a network may show very different states, even under the same conditions, making it extremely challenging to predict abrupt phase shifts or tipping points. Fortunately, as a system approaches the tipping point, the occurrence of the critical slowing down phenomenon brings about possible indicators, such as autocorrelation and variance, that effectively warn of the advent of tipping points.

\subsection{Analytical tools for network resilience} \label{1atfenr}
During the past 50 years, many analytical tools have been developed and used to analyze the resilience phenomenon or predict the tipping points in low-dimensional systems. Most of these tools focus on the equilibrium analysis of low-dimensional dynamical equations. Since the dynamic equations used represent dynamic models of specific systems, such as the vegetation-algae model in ecology \cite{scheffer1993alternative} and the simple positive feedback loop in biology \cite{xiong2003positive}, we review the progress on analytical tools of low-dimensional systems in the following four chapters according to the fields to which they belong. For large-scale networks, in addition to the unclear internal network dynamics, the lack of analytical tools has been another serious obstacle that restricts accurate network resilience prediction. However, thanks to recent advances in the development of powerful tools that can be used to deal with large-scale complex networks, various analytical tools for network resilience analysis have been developed.

\begin{SCfigure*}
 \centering
 \includegraphics[width=0.67\textwidth]{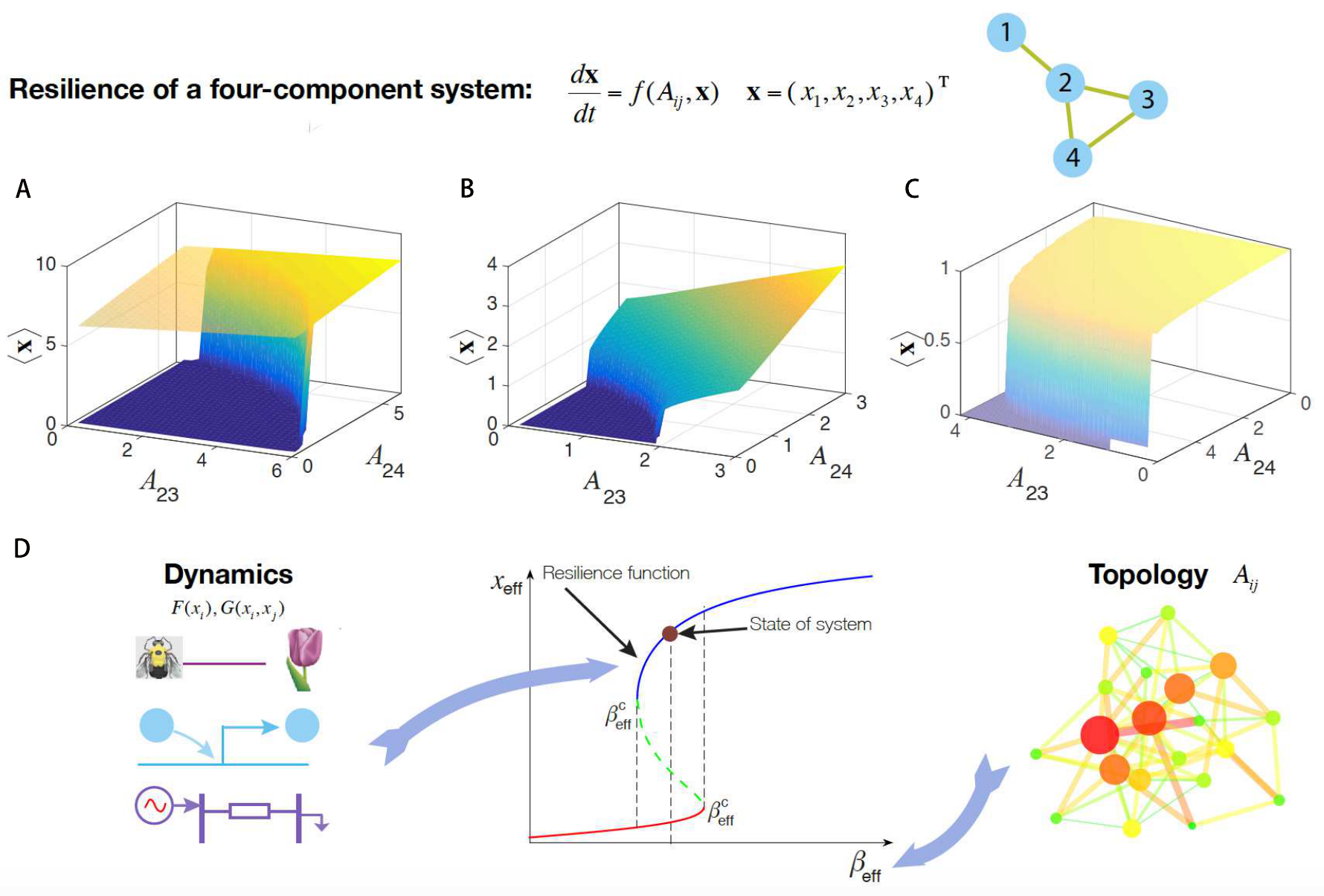}  % figure 7
\caption{
Network resilience in multidimensional systems.
(A-C) The 3D plots show the resilience plane for a four-node system.
(D) After applying the dimension-reduction method \cite{gao2016universal}, the multidimensional manifold shown in (A-C) collapses into a one-dimensional resilience function in $\beta$-space (D).
\textit{Source:} The figure is from \cite{gao2016universal}.}
\label{Barabasi_fig1}
\end{SCfigure*}

Gao et al. \cite{gao2016universal} recently developed a general analytical framework, named GBB reduction theory, for mapping the dynamics of high-dimensional systems into effective one-dimensional system dynamics. This model can not only accurately predict the system's response to diverse perturbations but also correctly locate the critical points at which the system loses its resilience. On the one hand, using the proposed dimension-reduction method, the patterns of resilience depend only on the system's intrinsic dynamics and are independent of the network topology. On the other hand, although topology changes do not alter the critical points, three key structural factors---density, heterogeneity and symmetry---could affect a system's resilience by pushing systems far from the critical points, enabling sustainability under substantial perturbations.
The study of universal resilience patterns in complex networks suggests potential intervention strategies that might be used to avoid the loss of resilience and principles for the design of optimally resilient systems that can cope successfully with perturbations.

In a multidimensional system, the dynamics of each component not only depend on the self-dynamics of the system but are also related to the interactions between the components and their interacting partners \cite{barzel2013universality, barzel2015constructing}.
The dynamic equation of a multidimensional system consisting of $N$ components (nodes) can be formally written as
\begin{equation}\label{MDDynamics}
\frac{dx_i}{dt}=F(x_i)+\sum_{j=1}^NA_{ij}G(x_i,x_j),
 \end{equation}
where $x_i(t)$ represents the activity of node $i$ at time $t$, $F(x_i)$ and $G(x_i,x_j)$ show the dynamical rules governing the system's components, and weighted adjacency matrix $A_{ij}$ captures the rate of interactions between all pairs of components.
Similar to the one-dimensional systems shown in Fig. \ref{OneDimensionalResilience}, the resilience of multidimensional systems can be captured by calculating the stable fixed point of equation (\ref{MDDynamics}). However, this point may depend on the changes in any of the $N^2$ parameters of the adjacency matrix $A_{ij}$. Moreover, there may be different forms of perturbations bringing changes to the adjacency matrix, for example, node/link removal, weight reduction or any combination thereof.
This means that the resilience of multidimensional systems depends on the network topology and on the forms of the perturbations that occur. For large-scale multidimensional models, it is impossible to predict resilience by direct calculation using equation (\ref{MDDynamics}). A framework based on dimension reduction addresses this challenge.

The activity of each node in a network is governed by its nearest neighbors through the interaction term $\sum_{j=1}^NA_{ij}G(x_i,x_j)$ of equation (\ref{MDDynamics}).
By using the average nearest-neighbor activity, we can obtain an effective state $x_{\rm eff}$ of the system as follows:
\begin{equation}\label{x_eff}
x_{\rm eff}=\frac{\mathbf{1}^{\rm T}A\mathbf{x}}{\mathbf{1}^{\rm T}A\mathbf{1}}=\frac{\langle s^{\rm out}x \rangle}{\langle s\rangle},
\end{equation}
where $\mathbf{1}$ is the unit vector $\mathbf{1}^{\rm T}=(1,...,1)^{\rm T}$,
$\mathbf{s}^{\rm out}=(s_1^{\rm out},...,s_N^{\rm out})^{\rm T}$ is the vector of outgoing degrees with $s_j^{\rm out} =\sum_{i=1}^NA_{ij}$,
$\mathbf{s}^{\rm in}=(s_1^{\rm in},...,s_N^{\rm in})^{\rm T}$ is the vector of incoming degrees,
the term of the right hand of the equation is $\langle s^{\rm out}x \rangle=\frac{1}{N}\sum_{i=1}^Ns_i^{\rm out}x_i$, and
$\langle s \rangle=\langle s^{\rm out} \rangle=\langle s^{\rm in} \rangle$ is the average weighted degree.

If adjacency matrix $A_{ij}$ has little correlation, then the multidimensional problem can be reduced to an effective one-dimensional problem by using the effective state $x_{\rm eff}$, which is
\begin{equation}\label{MultiD2OneD}
f(\beta_{\rm eff},x_{\rm eff})=\frac{x_{\rm eff}}{dt}=F(x_{\rm eff})+\beta_{\rm eff}G(x_{\rm eff},x_{\rm eff}),
\end{equation}
where $\beta_{\rm eff}$ is the nearest neighbor weighted degree that can be written as
\begin{equation}\label{beta_eff}
\beta_{\rm eff}=\frac{\mathbf{1}^{\rm T}A\mathbf{s}^{\rm in}}{\mathbf{1}^{\rm T}A\mathbf{1}}=\frac{\langle s^{\rm out}s^{\rm in} \rangle}{\langle s\rangle}.
\end{equation}
Therefore, the $N^2$ parameters of microscopic description $A_{ij}$ collapse into a single macroscopic resilience parameter $\beta_{\rm eff}$.
Any impact on the state of the system caused by the changes in $A_{ij}$ is fully accounted for by the corresponding changes in $\beta_{\rm eff}$, which are determined by the systems dynamics $F(x_i)$ and $G(x_i,x_j)$.
Figure \ref{Barabasi_fig1} shows that by mapping the multidimensional system into $\beta$-space, its response to diverse perturbations and tipping points can be accurately predicted.

\begin{SCfigure*}
 \centering
 \includegraphics[width=0.667\textwidth]{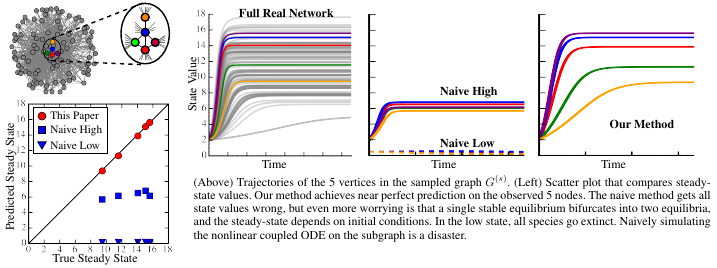}  % figure 8
\caption{ Predicting the steady-state abundances of 5 species
      interacting in a larger 97-species ecological network. The predictions
      are based only on the interactions of those five species (incomplete information). \\
\textit{Source:} The figure is from \cite{jiang2020true}.
}
\label{IncompleteDynamics}
\end{SCfigure*}

As stated above, the analytical tool is based on the assumption that the network is complete. However, many complex networks, such as gene regulatory networks and protein-protein interaction networks in biology, are incomplete. How to infer the resilience of an incomplete network is an essential question. Taking advantage of the mean-field approach in dimensional reduction, Jiang et al. \cite{jiang2020true} developed a tool that can be used to learn the true steady state of a small part of a network without knowing the entire network. Unlike the naive method in which the concerned nodes and isolates are subtracted from the other part of the network, the authors use a mean-field approximation to account for the impact of the other part of the network and summarize its impact using a resilience parameter $\beta_{\rm eff}$, as discussed above. The proposed tool can yield very close approximations to
the whole network's actual steady-state dynamics (Fig. \ref{IncompleteDynamics}). In contrast, the state-of-the-art method is the naive approach, which produces completely erroneous conclusions. Moreover, most real networks, primarily biological and ecological networks, are incomplete. This method can help us infer the true dynamics of these incomplete networks. Similarly, Jiang et al. \cite{jiang2020inferring} combined mean-field theory with combinatorial optimization to infer the topological characteristics, such as degree, from the observed incomplete networks.

In addition to the network-level resilience phenomenon, how does each node contribute to network resilience? Based on the equivalent one-dimensional model, Zhang et al. \cite{zhang2020resilience} derived a new centrality index, resilience centrality, and used it to quantify the ability of nodes to affect the resilience of the system. The node's resilience centrality is mainly determined by the degree and weighted nearest-neighbor degree of the node. This centrality performs better in prioritizing the node's importance in maintaining the system's resilience than do other centralities such as degree, betweenness and closeness. The proposed centrality metric enables us to design effective strategies for protecting real-world networks such as mutualistic networks and infrastructure systems.

In summary, based on these developed universal frameworks and using the methods for inferring network dynamics reviewed above, an increasing number of studies designed to predict resilience in complex networks can be conducted in the future. However, despite the advances that have been made in developing general analytical tools for evaluating network resilience, developing specific analytical tools that can be used to uncover vital resilience phenomena such as prediction of abrupt shifts or tipping points in real networks remains a challenge.

\subsection{Related concepts}
Two concepts, ``robustness'' and ``stability,'' are closely related to ``resilience''. They are commonly used to analyze the response of a system under changing conditions and are sometimes difficult to distinguish due to the lack of clear boundaries \cite{urruty2016stability}; in addition, their definitions may vary across contexts. For example, in the study of how a system responds to changes in a specific driver, Dai et al. \cite{dai2015relation} define resilience as the size of the basin of attraction and stability as the recovery rate. In the bacterial response to antibiotic treatment \cite{shade2012fundamentals}, resistance describes insensitivity to the treatment, and resilience defines the recovery rate. In particular, resilience and robustness have been used interchangeably in some of the literature on network analysis.
Here, we trace the original definitions of resilience and stability as they were first proposed in ecology \cite{holling1973resilience} and the robustness of complex networks \cite{callaway2000network}. We point out that these three notions describe the distinct properties of systems.

Resilience is probably the broadest of the three concepts discussed in this section \cite{urruty2016stability}, as it is the system's property of persistence, with probability of extinction being the result. Stability denotes a system's ability to return to a specific stable fixed point after a temporary disturbance. The more rapidly it returns and with the least fluctuations, the more stable it is. In this definition, stability is the system's property, and the degree of fluctuation around a specific state is its measure \cite{holling1973resilience}.
Resilience and stability are both defined in the context of network dynamics \cite{holling1973resilience, gao2016universal}, and the concept ``robustness'' is related to the static structure of a network; it measures the ability of the network to maintain its connectivity when a fraction of its nodes (links) are damaged \cite{albert2000error}.
Next, we further clarify the differences among the three concepts.

\subsubsection{Stability in complex systems}
The word ``stability'' is derived from the Latin term \textit{stabilis}, which means firm or steadfast. In a dynamical system, stability defines the system's ability to remain in an equilibrium state.
Stability has a rich history in ecology \cite{ives2007stability}. {\it Stability} was first defined as the constancy of a given attribute, regardless of the presence of disturbing factors \cite{may1972limit, may1972will}. For instance, stable ecological communities were those with relatively constant population sizes and compositions \cite{may1971stability}. The definition of stability was later expanded to include other properties of ecosystems, such as the maintenance of ecological functions despite disturbances \cite{turner1993revised} and the ability to return to the initial equilibrium state \cite{holling1973resilience}. These definitions lead to multiple interpretations of stability and overlap with the concept of resilience.
For example, for systems with alternative stable states, one concept of stability depends on the number of alternative stable states: systems that are more stable are those with fewer stable states \cite{ives2007stability}. Another concept of stability describes the ease with which systems can switch between alternative stable states, with more stable systems having higher barriers to switching \cite{ives2007stability}. The latter concept is quite the same as the meaning of resilience. Some of the literature even treats stability as a multifaceted notion and resilience as one component of stability \cite{donohue2016navigating}.
Since resilience itself is a multidimensional concept, integrating persistence, resistance, and the ability to recover/adapt, we trace it back to the original definitions of these two concepts \cite{holling1973resilience} and point out that resilience and stability describe distinct properties of systems.

The concept ``resilience'' concentrates on the boundaries of the domain of attraction, while ``stability'' focuses on equilibrium states. On the one hand, a system can be very resilient and still fluctuate considerably, i.e., have low stability. For instance, pest systems are highly variable in space and time; as open systems, they are greatly affected by dispersal and therefore have high resilience \cite{watt1968computer}. On the other hand, a stable system may have low resilience. For example, the commercial fishery systems of the Great Lakes are notably sensitive to disruption by humans because they represent climatically buffered, reasonably homogeneous and self-contained systems with comparatively low variability and hence high stability and low resilience \cite{holling1973resilience}.

\begin{figure}[ht!]
\centering
\includegraphics[width=0.48\textwidth]{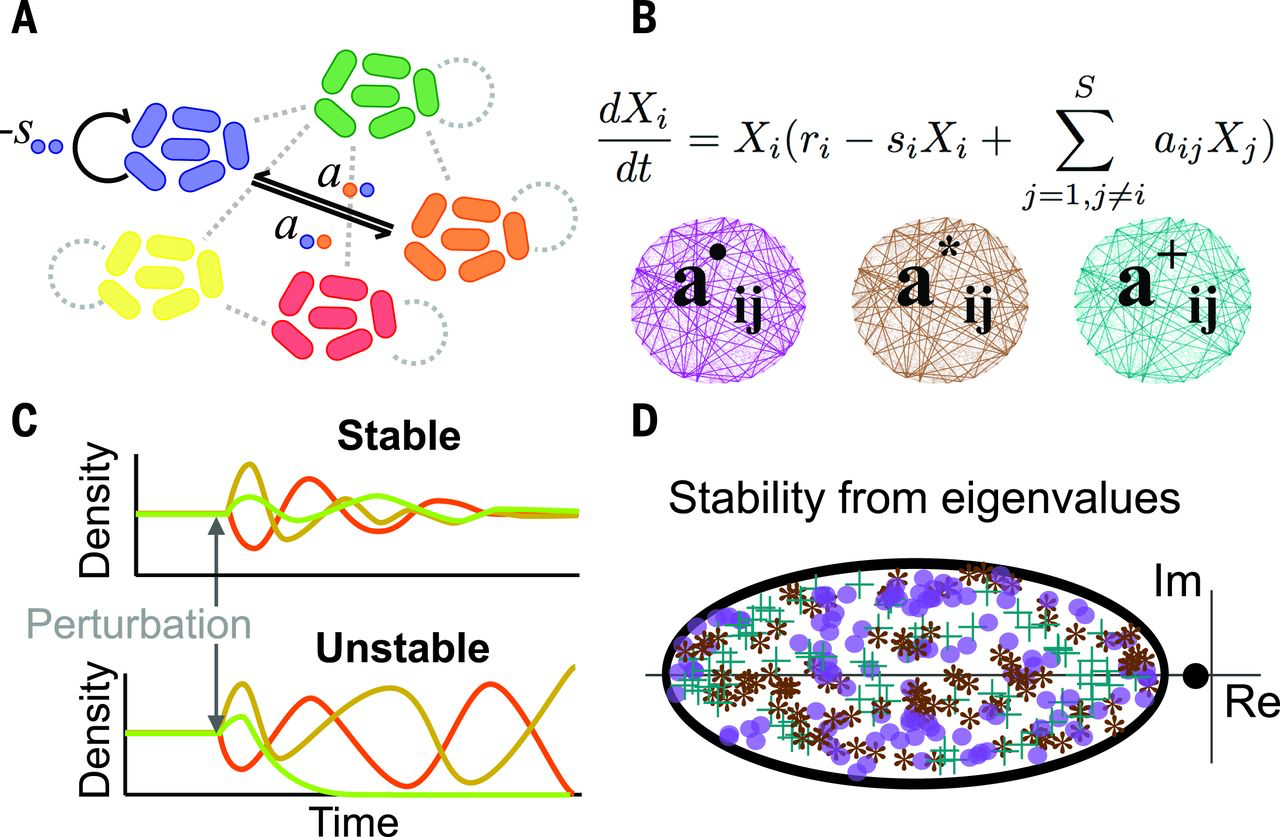} % figure 9
\caption{
Stability in a network of microbiota. (A) Ecological network theory captures networks of microbial species that interact with themselves ($-s$) and other genotypes ($a_{ij}$). (B) Coupled ordinary differential equations capture all possible combinations of connectivity and interaction types. Three sample networks are shown. (C) Communities that return to their previous densities after perturbation are classified as stable, those that return to their equilibrium more rapidly are categorized as more stable, and those that continue to diverge from the equilibrium are considered
unstable \cite{allesina2012stability}. (D) Linear stability analysis uses eigenvalues' real (Re) and imaginary (Im) parts.\\
\textit{Source:} The figure is from \cite{coyte2015ecology}.
}
\label{Microbia_Stability}
\end{figure}

As shown in Fig. \ref{Microbia_Stability}, various methods are used in stability analysis. On the one hand, nonlinear dynamical systems are usually modeled using coupled ordinary differential equations, and nonlinear stability analysis can be realized by observing how the states of variables evolve with time after perturbations. If a system can return to the original unperturbed state, then it is stable. Moreover, the faster the recovery is, the more stable the system.
On the other hand, a system's stability can be analyzed linearly, that is, by studying the eigenvalues of the adjacency matrix of the networked system \cite{butler2018stability}. The largest real part of the eigenvalues determines whether and how rapidly the system returns to its original state after perturbation \cite{allesina2012stability}. If it is negative, then the system is stable. The larger the absolute value is, the more quickly the system returns to stability. The imaginary parts of the eigenvalues predict the extent of oscillations in species densities during a return to equilibrium: larger imaginary components predict more frequent oscillations \cite{coyte2015ecology}.
Allesina et al. \cite{allesina2012stability} proposed analytical stability criteria for complex ecosystems, including different kinds of interactions: predator-prey, mutualistic and competitive.
Even stability criteria have been proposed, especially for ecosystems, and are widely applicable because they hold for any system of differential equations \cite{allesina2012stability}.

\subsubsection{Robustness in complex networks}
The term ``robustness' is derived from the Latin term \textit{Quercus Robur}, which means ``oak, a symbol of strength and longevity in the ancient world.
Robustness in a system characterizes the ability of the system to withstand failures and perturbations without loss of function. For instance, biologists define robustness as the ability of living systems to maintain specific functionalities despite unpredictable perturbations \cite{kitano2004biological}.
Notably, in network science, robustness is a concept related to the network's static topology, measuring its ability to maintain its connectivity when a fraction of its nodes (links) are damaged \cite{liu2015vulnerability}.

Mathematically, the robustness of a network is modeled as a reverse percolation process \cite{buldyrev2010catastrophic}.
Percolation theory models the process of randomly placing pebbles in a square lattice with probability $p$ and predicts the critical value $p_c$ at which a large cluster (called a ``percolating cluster') emerges.
At critical point $p_c$, a phase transition appears: many small clusters coalesce, forming a percolating cluster that percolates the whole lattice.

\begin{figure}[ht!]
\centering
\includegraphics[width=0.48\textwidth]{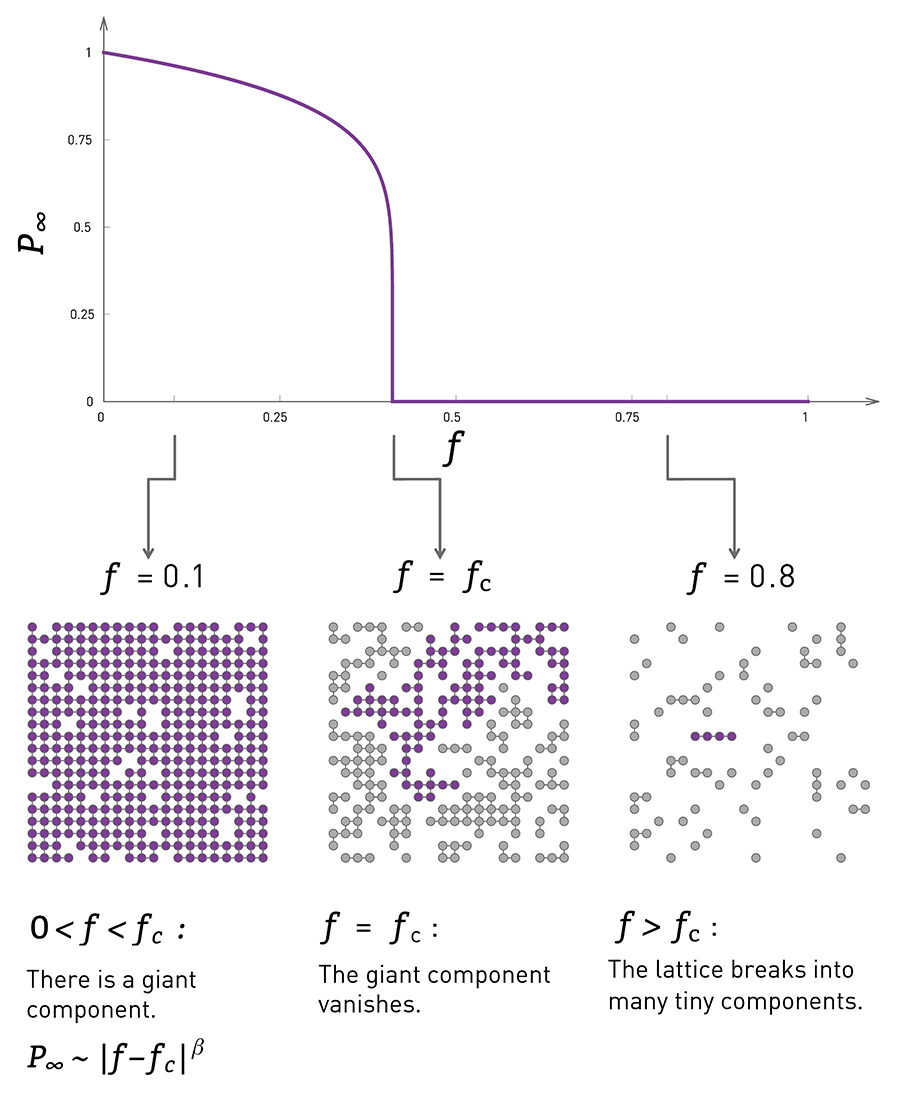}  % figure 10
\caption{
Network robustness is characterized by the size of the giant connected component after the random removal of a fraction, $f$, of its nodes.\\
\textit{Source:} The figure is from \cite{barabasi2016network}.
}
\label{Robustness_Network}
\end{figure}

We analyze a network's robustness by randomly removing a fraction, $f$, of nodes from the network and observing how the largest connected component size changes. When $f$ is small, node removal causes minor damage to the network, and the enormous connected component continuously exists in the network. When $f$ reaches a critical point, $f_c$, the giant connected component vanishes, as shown in Fig. \ref{Robustness_Network}, exhibiting that the largest cluster's size is finite compared to the whole network. The robustness of a network is usually either characterized by the value of critical threshold $f_c$ analyzed using percolation theory or defined as the integrated size of the largest connected cluster during the entire attack process \cite{gao2015recent}, which is $\int_0^1 P_{\infty}\;\mathrm{d}f$.

For a random network with arbitrary degree distributions $P(k)$, the largest connected component \cite{liu2017controllability} exists if
\begin{equation}\label{MolloyReed}
\kappa =\frac{\langle k^2 \rangle}{\langle k \rangle} >2,
\end{equation};
this is called the Molloy-Reed criterion \cite{cohen2000resilience}.
The random removal of a fraction, $f$, of nodes leads to changes in the degree distribution and parameter $\kappa$.
When parameter $\kappa <2$, the largest connected component disappears, and the network is fragmented into many disconnected components.
Based on the Molloy-Reed criterion, the critical percolation threshold is as follows:
\begin{equation}\label{criticalfc}
f_c=1-\frac{1}{\frac{\langle k^2 \rangle}{\langle k \rangle}-1},
\end{equation}
where $\langle k^2 \rangle$ and $\langle k \rangle$ are uniquely determined by the degree distribution $P(k)$.
For an Erd\H{o}s-R\'enyi (ER) network with an average degree of $\langle k \rangle$, the critical threshold is $f_c^{\rm ER}=1-1/\langle k \rangle$.
For a scale-free network with degree distribution $P(k) \sim k^{-\gamma}$, the threshold $f_c^{\rm SF}\rightarrow 1$ is $N\rightarrow \infty$ if $2<\gamma<3$;
this means that to fragment such a network, we must remove all the nodes \cite{cohen2000resilience}.
For example, the physical structure of the Internet ($\gamma \approx 2.5$) is impressively robust, with $p_c > 0.99$.
The study of the robustness of complex systems can help us understand the real world; for example, in biology and medicine, it can help us understand why some mutations lead to diseases while others do not, and it can enable us to make the infrastructures we use in everyday life more efficient and more robust.

\subsubsection{Robustness in a network of networks}
In the real world, systems are not isolated but are interdependent or interact with one another.
Taking the interactions between systems into account, many studies have attempted to understand the robustness of interdependent networks \cite{buldyrev2010catastrophic},
interconnected networks \cite{hu2011percolation},
multilayer networks \cite{radicchi2013abrupt},
multiplex networks \cite{baxter2012avalanche},
and a network of networks \cite{gao2012networks}.
In these systems, networks interact with each other and exhibit structural and dynamical features that differ from those observed in isolated networks.
For example, Buldyrev et al. \cite{buldyrev2010catastrophic} developed an analytical framework based on the generating function formalism, describing the cascading failures in two interdependent networks and finding a first-order discontinuous phase transition that is dramatically different from the second-order continuous phase transition found in isolated networks.
Parshani et al. \cite{parshani2010interdependent} studied a model closer to real systems; it consisted of two partially interdependent networks. They found that the percolation transition changes from first- order to second-order at a certain critical coupling as the coupling strength decreases.
Gao et al. developed an analytical framework to study the percolation of a tree-like network formed by $n$ interdependent networks (tree-like NON) \cite{gao2011robustness, gao2012networks} and discovered that while for $n=1$, the percolation transition is second-order, for any $n>1$ where cascading failures occur, a first-order (abrupt) transition occurs.
Liu et al. \cite{liu2016breakdown} found hybrid phase transitions in interdependent directed networks.

The robustness of networks is also related to their failure mechanisms. The studies reviewed above mainly focus on random failure. In real scenarios, most initial failures are not random but, rather, are due to targeted attacks on important hubs (nodes with high degree) or occur to low-degree nodes since important hubs are purposely protected \cite{huang2011robustness}.
Single real networks, like the Internet, are vulnerable to targeted attacks \cite{xu2021breakdown}. The simultaneous removal of several hubs breaks any network.
For coupled networks, Huang et al. \cite{huang2011robustness} proposed a mathematical framework for understanding the robustness of fully interdependent networks under targeted attacks; their model was later extended by Dong et al. \cite{dong2012percolation} to targeted attacks on partially interdependent networks. The latter authors developed a general technique that uses the random attack solution to map the solution onto the targeted attack problem in interdependent networks. Dong et al. \cite{dong2013robustness} further extended the study of targeted attacks on high-degree nodes in a pair of interdependent networks to the study of a network of networks. They found that the robustness of networks of networks can be improved by protecting essential hubs.

In most studies of network robustness, networks are treated as static and unweighted \cite{buldyrev2010catastrophic}. However, such an assumption may not apply to all real-world networks, which are usually dynamical and weighted.
For example, a traffic network is always topologically connected but may dynamically fail since the flux through the network could be zero. In addition, the link weight is also significant to the network's function. Taking the interactions between clownfish and sea anemones as an example, even if they are in the same pool and interactions between them are possible, clownfish may not find sea anemones when the water is seriously polluted, resulting in a low weight of their interactions.
If they lose protection from sea anemones, clownfish may be predated quickly, and sea anemones could also die without the food provided by clownfish. Although some studies of the ``dynamical robustness'' of networks \cite{tanaka2015dynamical} have been conducted, the studied dynamics are mainly limited to oscillations that are more related to synchronization \cite{bi2014explosive}. Thus, resilience is a broader concept that can be used to analyze the dynamic changes in networks' responses to perturbations.

In summary, a network's robustness is its ability to maintain its statistical topological integrity under node/link failures; it is usually characterized by the network's connectivities, such as the diameter \cite{jeong2000large} or size of the largest connected component.

\section{Tipping points in ecological networks}\label{Ecology}

The structure and function of the ecological systems formed by interacting species are subject to internal and external perturbations such as human-induced pressures, environmental changes \cite{holling1973resilience, scheffer2001catastrophic}, and species invasion. These ecological systems must respond to such perturbations by adjusting their activities in such a way as to retain their basic functionality when errors and failures occur. As we discussed in Sec.\ref{1rpids}, many systems have alternative stable states, where one is the desired state and another is undesired. Due to climate change and overexploitation by human activity, even a small perturbation may cause an ecosystem to cross the tipping point and reach an undesired state.
This transition between states uncovers alternative stable states \cite{knowlton1992thresholds, scheffer2001catastrophic} or phase shifts \cite{done1992phase}. 

The ability to predict the phase shift and tipping point of the alternative stable state is crucial for ecosystem management. Although both terms refer to changes in ecosystems that occur in response to disturbances, they represent different characteristic changes in dynamical systems. As shown in Fig. \ref{FourTypesShifts}, the following holds: 1) the existence of alternative stable states implies that at least two different stable equilibria can occur at the same parameter values (\textit{i.e.}, environment) over at least some of their range, where hysteresis phase transition \cite{petraitis2004detection} often appears; 2) a phase shift has a single state (although the state changes at the threshold) under all parameter values. Transitions between states are represented by dramatic qualitative changes in the former case and by simple quantitative changes in the latter case. It is an important question about the landscape of the system remains: whether it ``has a single valley or multiple ones separated by hills and watersheds'' \cite{may1977thresholds}. If the former is true, then the dynamical system has a unique attractor to which the system will continuously evolve from all initial conditions following any disturbance, showing no correlations with its past states. If the latter is true, then the state to which the system settles depends on the initial conditions; the system may return to the original stable state after small perturbations but may evolve into another attractor following large perturbations.

Therefore, it is crucial to acknowledge the existence of the multiple stable states phenomena in some mathematical models and empirical examples, illustrating the regime shifts in ecosystems (Sec. \ref{2msspie}). For the given system with multiple stable states, we review the regime shifts between stable states in both individual systems and networked systems in Sec. \ref{2eeoewmss}. Next, we summarize the recent research on predicting and controlling the resilience of ecosystems (Sec. \ref{2croe}). At the end of this section, we compare ecological resilience and engineering resilience in ecology.

\subsection{Multiple stable states phenomena in ecosystems}\label{2msspie}
Whether alternative stable states exist in nature has historically been the subject of debate \cite{sutherland1974multiple, scheffer1993alternative}.
The debate regarding what constitutes evidence for alternative stable states arose because there are two different contexts in which the term ``alternative stable states'' is used in the ecological literature \cite{beisner2003alternative}. One context excludes the effect of environmental change and regards the environment as fixed in some sense \cite{connell1983evidence}. The other context focuses on the effect of environmental change on the state of communities or ecosystems \cite{may1977thresholds}. For example, Connell et al. \cite{connell1983evidence} suggested that alternative stable states do not exist in systems that are untouched by humans, Sinclair et al. \cite{sinclair1995serengeti} pointed out that human beings are a part of system dynamics. Here, we do not treat the question of whether or not humans are a part of ecosystems but rather observe that humans do change the states of ecosystems \cite{gunderson2000ecological}.

Much of the literature over the last 50 years has shown increasing empirical support, gathered by ecologists, for the existence of alternative stable states \cite{sutherland1974multiple, may1977thresholds}.Transitions among stable states are used to describe many ecosystems, including semi-arid rangelands \cite{laycock1991stable}, lakes \cite{scheffer1993alternative}, coral reefs \cite{scheffer2001catastrophic}, and forests \cite{wilson1992positive}.Moreover, many theoretical models have been proposed to account for the alternative stable states that occur in ecosystems.

Next, we review the mathematical models that deal with multiple stable states.

\subsubsection{Mathematical models for multiple stable states}
Many theoretical models \cite{may1977thresholds} have demonstrated complex ecological system dynamics with multiple stable states. Simple theoretical analyses predict multiple stable states for 1) single-species dynamics via the Allee effect, 
2) two-species competitive interactions characterized by unstable coexistence, 
3) some predator-prey interactions, and
4) some systems combining predation and competition \cite{knowlton1992thresholds}. The theoretical models of multiple stable states are usually represented by a system of dynamical equations characterizing the evolution of species' states. If the equation describing the state's transformation in the ecosystem is nonlinear, then multiple stable states with all species present may exist. If the equations governing the species are linear, only one stable state exists in which all species are present. Nevertheless, other stable states in which some of the species are absent may exist \cite{lewontin1969meaning}.

For perturbations to state variables, state shifts have most often been achieved experimentally by the removal or addition of species \cite{sutherland1974multiple}. For example, overfishing is a classic case in which a new interior community state may arise simply through changes in the size of the fish population \cite{beisner2003alternative}. Two-species Lotka-Volterra competition is a case in which the interior coexistence equilibrium may be unstable and in which alternative states arise through the extinction of one population \cite{law1993alternative}.
Parameter changes \cite{scheffer2001catastrophic} may alter the location of a single equilibrium point or result in the transient destabilization of the current state, allowing the system to reach an alternative, locally stable equilibrium point that may not have existed before the parameter perturbations occurred.

These results support the notion that complex ecosystems maintain multiple stable states. In the following section, we review four representation models with multiple stable states in ecosystems.

\noindent
\textit{\textbf{Grazing ecosystems.}}
Assume that there is only one herbivore population with a constant density, $H$, in a grazing ecosystem \cite{may1977thresholds} that is sustained by vegetation, the biomass of which is $V$. The vegetation growth rate in the absence of grazing is $G(V)$, a function of $V$. The herbivores consume the vegetation at a net rate $C(V)=Hc(V)$, where $c(V)$ denotes the consumption rate per capita.
As the biomass $V$ increases, $G(V)$ first increases to a peak value and then decreases due to competition for resources. When $G(V)$ decreases to zero, the biomass $V$ reaches its maximum value $K$.
The per capita consumption function $c(V)$ increases with $V$ when $V$ is low and saturates to some constant $V_0$ due to the limited intake capacity and digestion rate. The maximum biomass $K$ of the vegetation and the saturated value $V_0$ of the herbivores are crucial factors in determining the final equilibrium point. We then define $\alpha=V_0/K$, thereby determining the number of stable states of a grazing ecosystem.

\begin{figure}[!ht]
\centering
\includegraphics[width=0.8\linewidth]{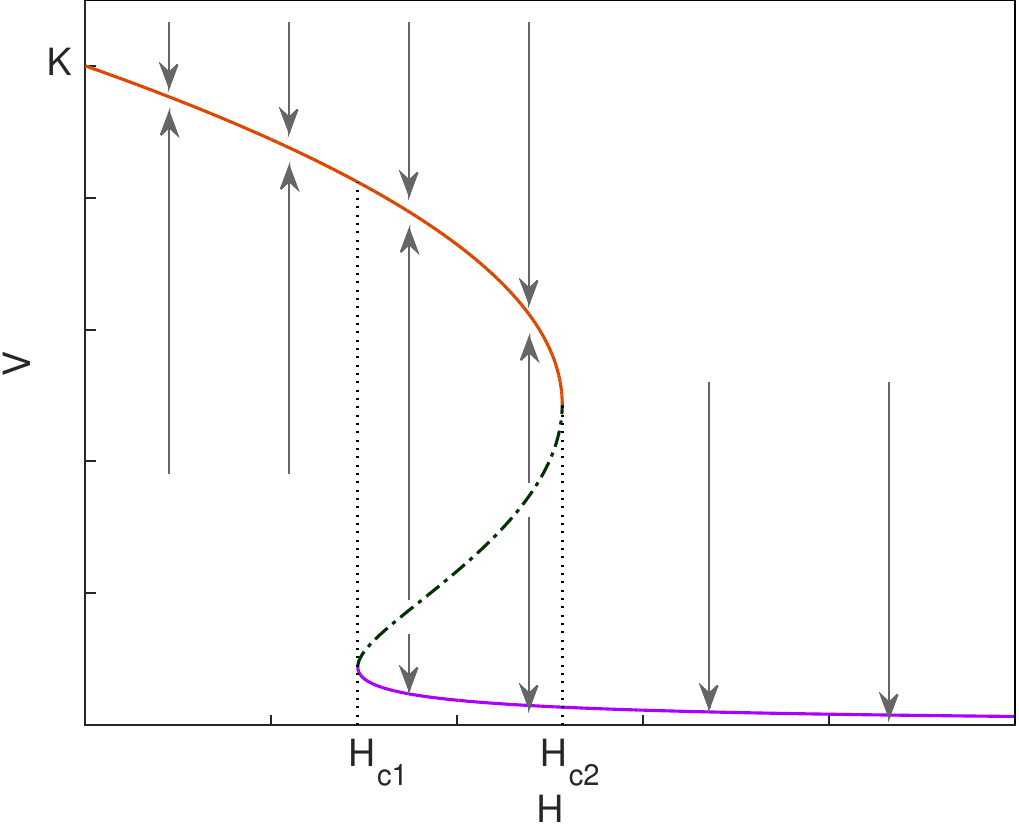}  % figure 11
\caption{
Equilibrium biomass $V$ is a function of the density, $H$, of herbivores with $\alpha=0.08$.
There are two resilience thresholds, $H_{c1}$ and $H_{c2}$, in the system.
For $H$ between the two thresholds, an unstable state (green dashed-dotted line) and two stable states appear, and the system may move to either the low or the high equilibrium depending on whether the initial $V$ value lies below or above the dashed-dotted ``breakpoint'' curve.\\
\textit{Source:} The figure has been modified from \cite{may1977thresholds}.
}
\label{Grazing_System_Fig}
\end{figure}

Functions $G(V)$ and $C(V)$ could have different explicit forms. Among them, the best known form of $G(V)$ is the logistic function $G(V)=rV(1-V/K)$, where $r$ is a specific growth rate describing how quickly $V$ approaches equilibrium, and $C(V)$ is the ``Type III'' consumption function $C(V)=\beta HV^2/(V_0^2+V^2)$ \cite{jorgensen2014encyclopedia}. The overall grazing model is then
\begin{equation}\label{GrazingSystems}
\frac{dV}{dt}=rV(1-\frac{V}{K})-\frac{\beta HV^2}{V_0^2+V^2}.
\end{equation}
Introducing the rescaled variables $X=V/K$, $\tau=rt$ and $\gamma = \beta H/rH$, the equation above takes the following form:
\begin{equation}\label{GrazingSystems_Rescale}
\frac{dX}{d\tau}=X(1-X)-\frac{\gamma X^2}{\alpha^2+X^2}.
\end{equation}
If the parameter value $\alpha < \frac{1}{3\sqrt{3}}$, then the system shows three equilibria: the low and high biomass equilibria are stable, and the intermediate biomass equilibrium is unstable.
Near the resilience threshold $H_{c2}$, as shown in Fig. \ref{Grazing_System_Fig}, a small increase in the stocking rate may move the system from the high biomass equilibrium to the low biomass equilibrium.

\noindent
\textit{\textbf{A minimal model.}}

\begin{figure}[!ht]
\centering
\includegraphics[width=0.7\linewidth]{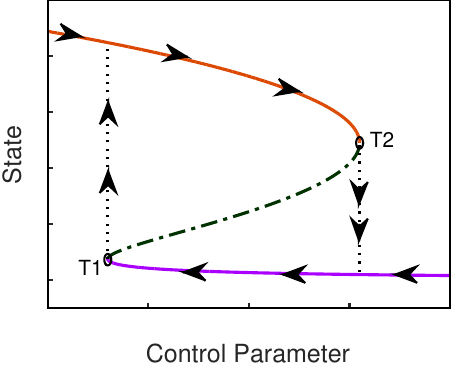}  % figure 12
\caption{
Hysteresis phase transition in an ecosystem.
If the stable state is at the upper branch and the control parameter slightly exceeds the threshold $T2$, then the system collapses rapidly to the lower branch, which we usually want to avoid in the real world. If the system is in a lower equilibrium branch and we wish to restore it to a higher equilibrium state, then we need to decrease the control parameter to a value smaller than threshold $T1$.\\
\textit{Source:} The figure has been modified from \cite{scheffer2001catastrophic}.
}
\label{hysteresisFig}
\end{figure}

The dynamic processes of ecosystems usually show a hysteresis transition; two examples of such transitions are desertification and lake eutrophication.
As shown in Fig. \ref{hysteresisFig}, if the system is on the upper branch, then once the control parameter exceeds the threshold $T2$, the system collapses into a stable state in the lower branch. To restore the state in the lower branch to a state in the upper branch, the control parameter must be lower than another threshold, $T1$, which is much smaller than $T2$. The following minimal model describes the change over time of such ``unwanted'' ecosystem properties $x$ \cite{scheffer2001catastrophic}:
\begin{equation}\label{MinimalModel}
\frac{dx}{dt}=a-bx+\frac{rx^p}{x^p+h^p},
\end{equation}
where $x$ is the state variable, and $r$ is the control parameter.
Parameter $a$ represents an environmental factor that promotes $x$, and parameter $b$ is the decay rate of $x$.
For $r=0$, the model has a single equilibrium at $x=a/b$. Otherwise, the last term representing the Hill function can cause alternative stable states.
Exponent $p$ determines the steepness of the switch occurring around the threshold $h$; the higher $p$ is, the stronger the hysteresis, as measured by the distance between thresholds.

\noindent
\textit{\textbf{Coral reef model.}}
After experiencing the mass disease-induced mortality of the herbivorous urchin \textit{Diadema antillanrum} in 1983 and of two framework-building species of coral, the health of reefs in the Caribbean has been greatly negatively affected, and the system is showing a phase change from a coral-dominated to an algae-dominated state.
Numby et al. \cite{mumby2007thresholds} discovered multiple stable states and hysteresis using a three-state analytical model that included corals, macroalgae and short algal turfs.

\begin{figure}
\centering
 \includegraphics[width=0.95\linewidth]{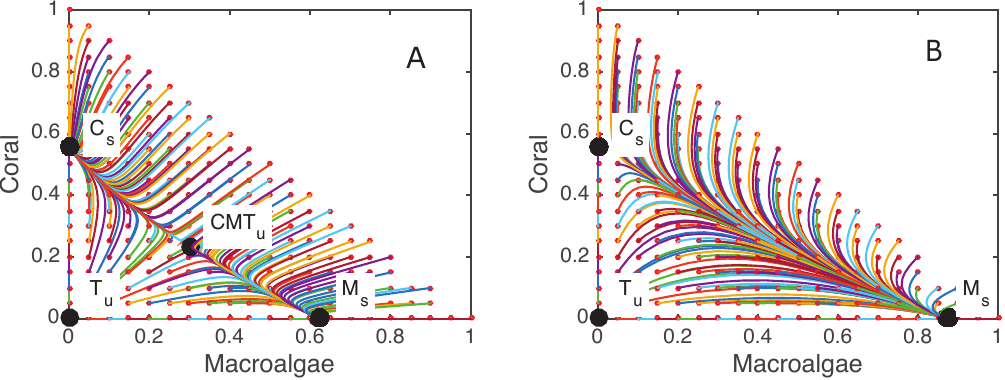}% figure 13
\caption{Equilibrium covers and trajectories over time of macroalgae and corals with grazing intensities $g$ of 0.3 (A) and 0.1 (B).
Equilibrium covers are represented by black circles.
The trajectories of system states are represented by lines beginning at different initial covers (red circles) and tending towards stable (denoted by subscript `s') rather than unstable (denoted by subscript `u') equilibria.
The parameters are set to $a=0.1$, $r=1$, $\gamma=0.8$ and $d=0.44$ \cite{blackwood2012effect}.\\
\textit{Source:} The figure has been modified from \cite{mumby2007thresholds}.
}
\label{Coral_fig}
\end{figure}

Let the coverage of corals, algal turf and macroalgae be denoted $C$, $T$ and $M$, respectively.
Assuming that the sum $T+M+C$ is constant and equal to one, the dynamics of the reef can be described by the following two equations:
\begin{equation}\label{CoralReefModel}
 \begin{aligned}
\frac{dM}{dt}=aMC-\frac{gM}{M+T}+\gamma MT,\\
\frac{dC}{dt}=rTC-dC-aMC,\\
 \end{aligned}
 \end{equation}
where $T$ can be expressed as $1-M-C$.
The term $-aMC$ indicates that macroalgae can overgrow corals, and $\gamma MT$ captures the phenomenon that macroalgae colonize dead coral by spreading vegetatively over algal turfs.
Coral dies naturally at the rate $dC$, and algal turfs arise when macroalgae are grazed ($gM/(M+T)$).
In addition, coral recruits and overgrows algal turfs at the combined rate $r$.
Figure \ref{Coral_fig} shows the phase plane of trajectories of the system from a given initial state to its equilibrium states, thereby revealing all the possible stable and unstable equilibria of the system.

\noindent
\textit{\textbf{A vegetation-algae model.}}
In shallow lakes, algal growth increases turbidity, while vegetation decreases turbidity. Such feedback among algae, vegetation, and turbidity may lead to the existence of multiple stable states.
Scientists employ multiple stable state models to understand the transitions between states in shallow temperate lakes; such models can show the regime shift between clear-water states dominated by vegetation and turbid states dominated by algae \cite{scheffer1990multiplicity, scheffer1993alternative}. Among the proposed models \cite{scheffer1990multiplicity}, a vegetation-algae model (also called the vegetation-turbidity model) \cite{scheffer1993alternative} captures the interactions between the growth of planktonic algae ($A$) and the abundance of vegetation ($V$), illustrating the potential for alternative equilibria in shallow lakes, as follows:
\begin{equation}\label{vegetation-algae}
 \begin{aligned}
 \frac{dA}{dt} & =rA\bigg(\frac{N}{N+h_N}\bigg)\bigg(\frac{h_v}{V+h_v}\bigg)-cA^2,\\
  V & =\frac{h_A^p}{A^p+h_A^p},\\
 \end{aligned}
 \end{equation}
where $r$ is the maximum intrinsic growth rate of algal turfs ($A$), and parameter $c$ is a competition coefficient.
Algal growth increases with the nutrient level ($N$) and decreases with vegetation abundance ($V$) in simple Monod relations; $h_N$ and $h_V$ denote the half-saturation constants. Note that the Monod function is a particular case of the Hill function \cite{heck1971statistical} with a power exponent of 1.
Vegetation abundance in a shallow lake decreases with algal biomass in a sigmoidal manner, with $h_A$ denoting the half-saturation constant.
The Hill coefficient of $p$ shapes the relation between vegetation abundance and algal biomass. A high $p$ value causes the change shape to approach a step function, representing the disappearance of vegetation from a shallow lake of homogeneous depth around critical algal biomass in cases in which turbidity makes the lake's average depth unsuitable for plant growth \cite{scheffer1993alternative}.

\begin{figure}[!ht]
\centering
 \includegraphics[width=0.95\linewidth]{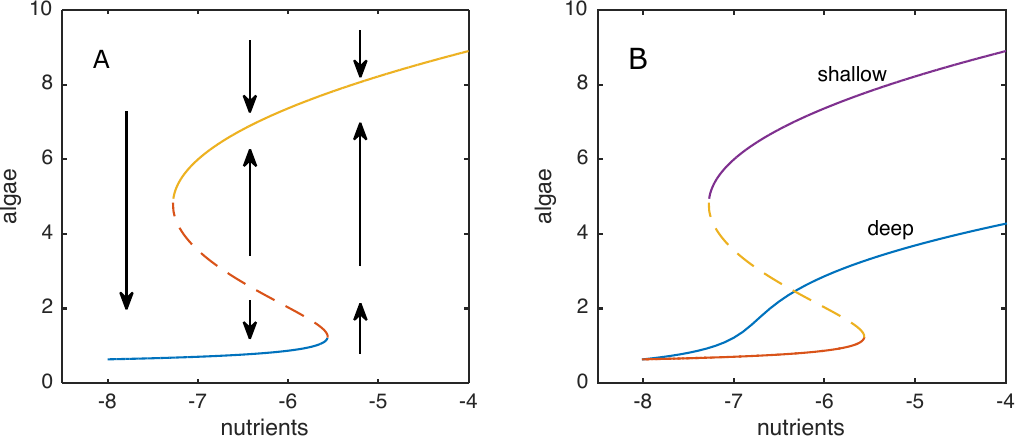} % figure 14
\caption{(A) An equilibrium state of algae abundance in shallow lakes as a function of nutrient level; the equilibrium shows multiple stable states. The solid lines represent stable equilibria, and the dashed line represents unstable equilibria.
(B) The equilibrium density of algae in deep lakes gradually decreases as the nutrient level decreases and does not show multiple stable states.\\
\textit{Source:} The figure has been modified from \cite{scheffer1993alternative}.
}
\label{vegetation-algae_model}
\end{figure}

In shallow lakes, the equilibrium density of algae changes following changes in the nutrient level, showing a catastrophic increase, as shown in Fig. \ref{vegetation-algae_model}A. The algal biomass has three equilibria over a certain range of nutrient levels; one of these is unstable, and two are stable.
In contrast, in deeper lakes, the vegetation abundance gradually decreases as turbidity increases (Fig. \ref{vegetation-algae_model}B) \cite{scheffer1993alternative}.
This finding indicates that the multiple stable states arising from the interactions captured by this model are limited to shallow lakes.

\subsubsection{Empirical examples of ecosystems with multiple stable states}\label{2eeoewmss}
\begin{figure}[!ht]
\centering
\includegraphics[width=0.95\linewidth]{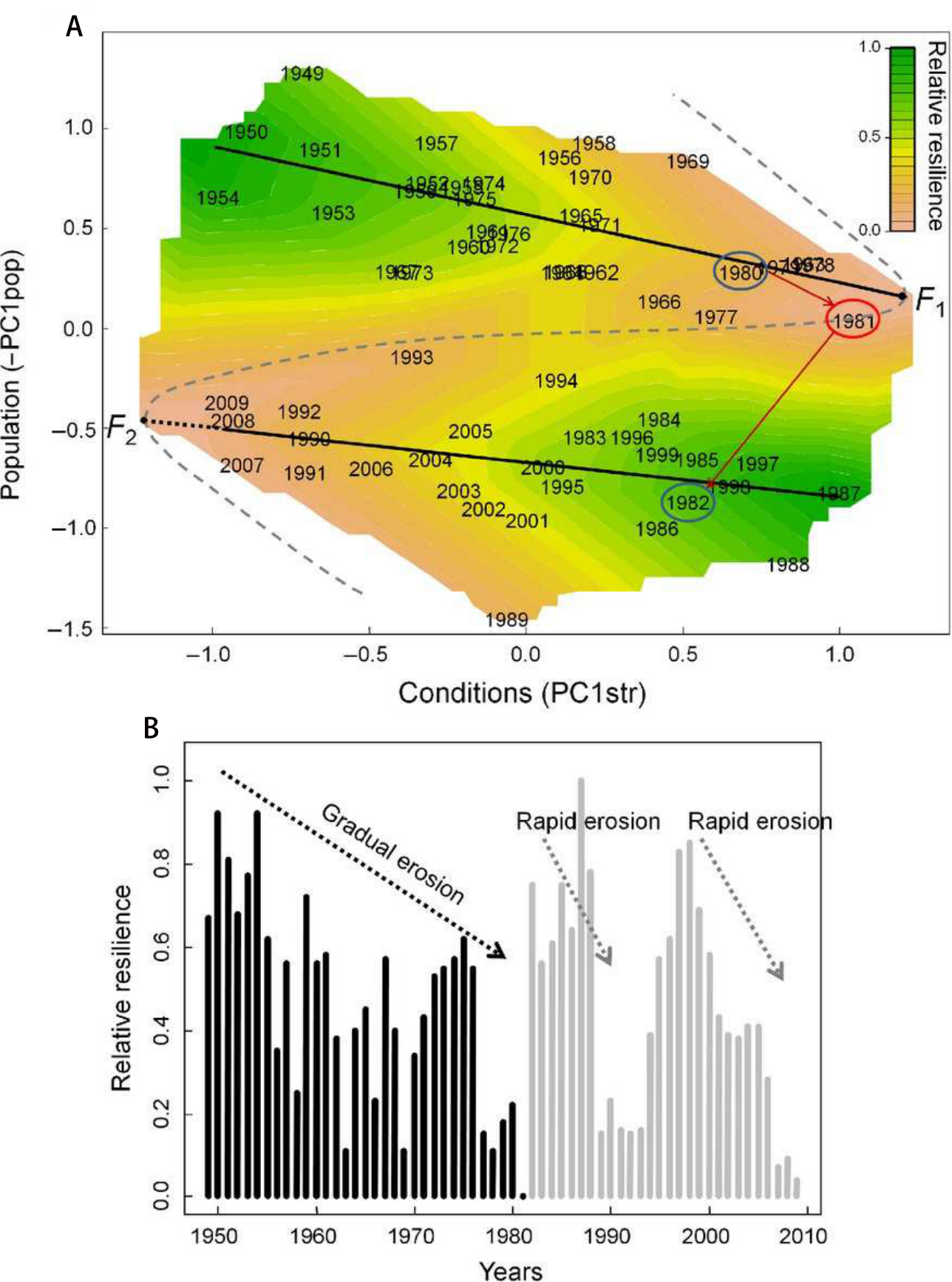} % figure 15
\caption{
Folded stability landscape and resilience assessment for Barents Sea cod (1949-2009). Within the empirical folded stability landscape for Barents Sea cod (A), continuous black lines mark the linear attractors, dotted black lines show the possible extension of the lower branch, dashed gray lines indicate the approximate position of the basin's borders, and F1 and F2 indicate the tipping points. Colors represent the relative resilience contour interpolated from the relative resilience in each year. Circles and arrows indicate the 1981 population shift. (B) indicates the relative resilience in each year; black and gray lines refer to old and new states, respectively.\\
\textit{Source:} The figure is from \cite{vasilakopoulos2015resilience}.
}
\label{FoldedBack_Cod}
\end{figure}

There are more theoretical studies than empirical studies of multiple stable states. However, moving from theory to practice is not straightforward. While the meaning of stability and equilibrium points is very clear in theory, it is not so in nature \cite{petraitis2004detection}.
Natural ecosystems with alternative stable states are expected to exhibit four {\bf key attributes} \cite{scheffer2003catastrophic}: 1) abrupt state shifts in time-series data; 2) sharp spatial boundaries between contrasting states or habitat units; 3) multimodal frequency distribution(s) of key variable(s), with each mode corresponding to an alternative stable ecosystem state; and 4) a hysteretic response to a changing environment \cite{moffett2015multiple}.
Despite these difficulties, the accumulated body of empirical evidence regarding multiple stable states shows that such states exist in natural ecosystems \cite{sutherland1974multiple}. For example, Vasilakopoulos et al. \cite{vasilakopoulos2015resilience} presented empirical evidence for the occurrence of a fold bifurcation in an exploited fish population, as shown in Fig. \ref{FoldedBack_Cod}.
In the following section, we review several other empirical examples of ecosystems \cite{gunderson2000ecological} in which multiple stable states have been postulated to exist.

\noindent
\textit{\textbf{Shallow Lakes.}}
The existence of qualitative differences in the states of lakes has long been recognized \cite{gunderson2000ecological}. Shallow lakes can have two alternative equilibria: a clear state dominated by aquatic vegetation and a turbid state characterized by high algal biomass \cite{scheffer1993alternative}. Many ecological mechanisms are probably involved, and each of these states is relatively stable due to interactions among nutrients, the types of vegetation present, and light penetration. The observed trends are  as follows:
1) increasing nutrient levels increase turbidity, 2) vegetation decreases and depends on turbidity, and 3) light penetration limits the growth of vegetation below certain depths.
In the clearwater state, which is dominated by macrophytes, vegetation can stabilize this state in shallow lakes up to relatively high nutrient loadings \cite{scheffer1993alternative}.
Once a system has switched to a turbid state dominated by phytoplankton, the system remains in such a state unless it experiences substantial nutrient reduction, which enables recolonization by plants.

\begin{figure}[!ht]
\centering
\includegraphics[width=0.45\textwidth]{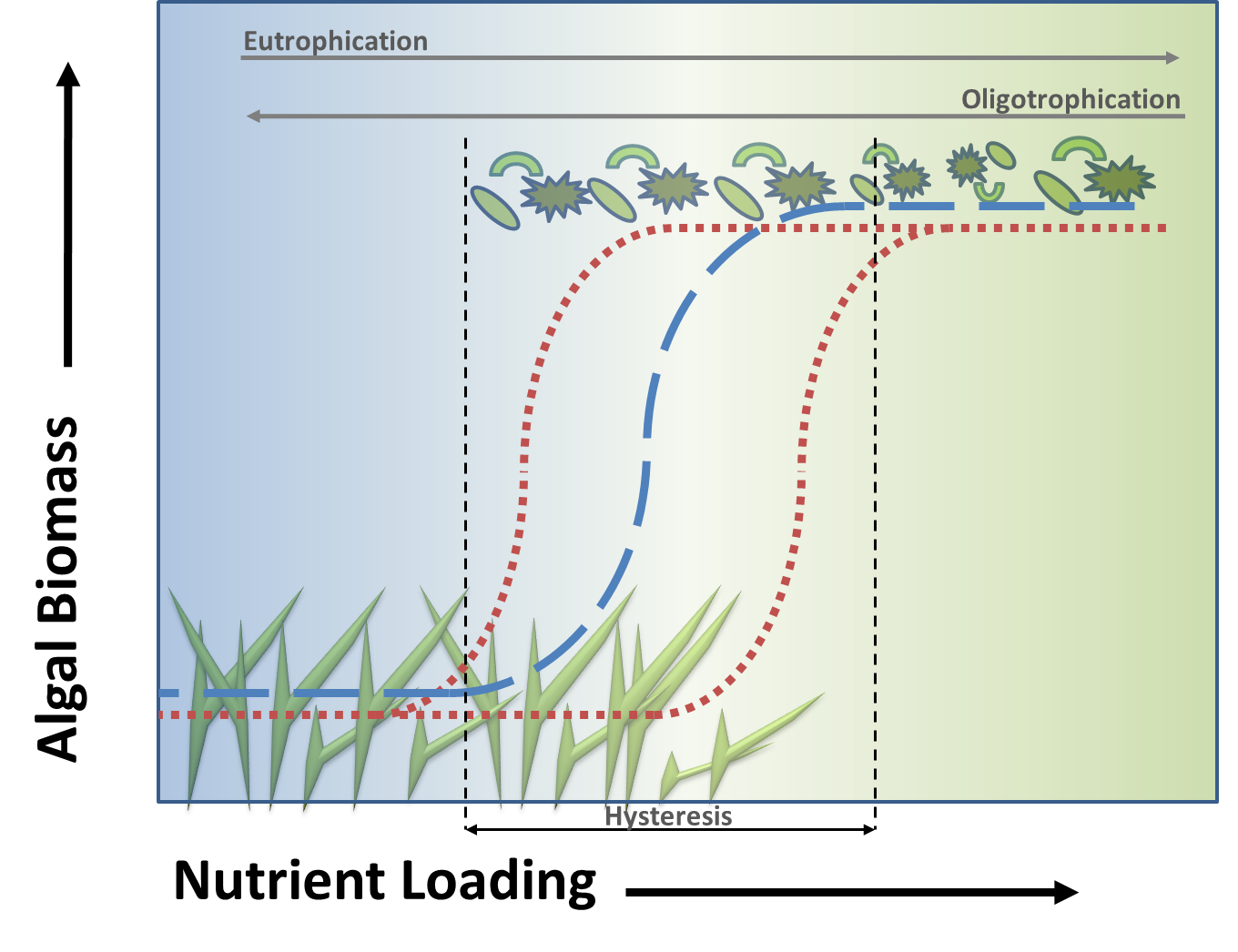} % figure 16
\caption{
Shallow lakes are often shown to respond nonlinearly to eutrophication (blue dashed line), and in many cases, hysteresis has been assumed (orange dotted line). Hysteresis is a nonlinear behavior in which the state of a lake depends not only on its present input but also on its prior state. As a result, two stable states can occur given the same conditions.\\
\textit{Source:} The figure is from \cite{janssen2014alternative, janssen2015research}.
}
\label{ShallowLake_Hysteresis}
\end{figure}

Scheffer et al. \cite{scheffer1993alternative} reviewed evidence for the shift in shallow temperate lakes between clearwater, macrophyte-dominated states, and turbid, phytoplankton-dominated states.
Transitions between states can be mediated by trophic relationships, where fish and nutrients are the primary drivers \cite{gunderson2000ecological}. On the one hand, the transition from a turbid to a clear state can be accomplished by decreasing stocks of planktivorous fish, which decreases predation on herbivorous zooplankton. Consequently, populations of herbivorous zooplankton increase, leading to an increase in herbivory and a reduction in phytoplankton biomass. In addition, increased light penetration and increased amounts of available nutrients result in the establishment of vegetation \cite{scheffer1997ecology}. On the other hand, shifts from the clear to turbid state can result from overgrazing of benthic vegetation by fish or waterfowl \cite{scheffer1997ecology}. Additionally, the points at which these two shifts occur are not the same, forming a hysteresis phase transition, as shown in Fig. \ref{ShallowLake_Hysteresis}.

\noindent
\textit{\textbf{Coral Reefs in marine ecosystems.}}
Multiple stable states exist not only in shallow lakes but also in other aquatic systems such as coral reefs, soft sediments, subtidal hard substrate communities, and rocky shores in marine ecosystems. Among these, coral reefs are the most well-known and best-documented cases \cite{petraitis2004detection}. Coral- and macroalgae-dominated states have long been known to exist on reefs \cite{hughes1994catastrophes}. Nevertheless, whether they can be called alternative stable states has been a highly controversial topic.
For example, Dudgeon et al. \cite{dudgeon2010phase} propose that the data from fossil and modern reefs support the phase-shift hypothesis with single stable states. Moreover, most studies of the transition from coral- to algae-dominated states do not distinguish between simple quantitative changes and dramatic qualitative changes associated with multiple stable states and hysteresis \cite{petraitis2004detection}.
Mumby et al. \cite{mumby2007thresholds} used a mechanistic model of the ecosystem to discover multiple stable states and verified it with Hughes' empirical data; they pointed out that both the theoretical model and the empirical data were far more consistent with multiple attractors than with the competing hypothesis of only a single coral attractor \cite{mumby2013evidence}.

\begin{figure}[!ht]
\centering
\includegraphics[width=0.46\textwidth]{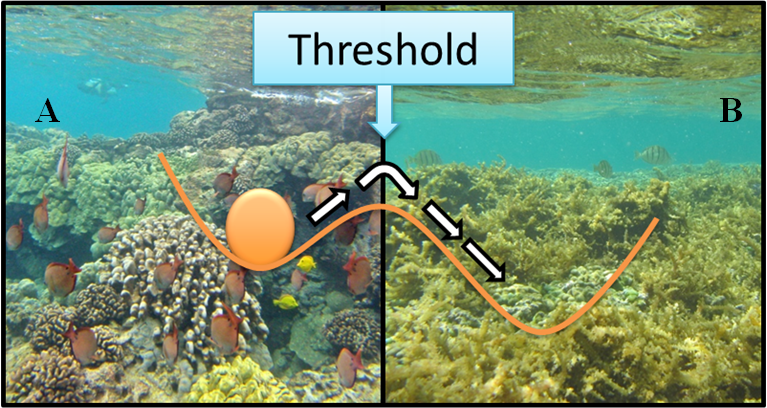} % figure 17
\caption{
If a sufficiently large disturbance occurs, then a coral-dominated state (A) crosses a threshold and transitions to an algae-dominated state (B). \\
\textit{Source:} The figure is from \cite{CoralReefGlossary}.
}
\label{CoralReef_RegimeShift}
\end{figure}

Multiple stable states in ecological systems occur when self-reinforcing feedback generates multiple stable equilibria under a given set of conditions \cite{heffernan2008wetlands}.
In the coral-dominated state, a decrease in coral cover liberates new space for algal colonization. Once maximum levels of grazing have been reached, further increases in the grazable area reduce the mean intensity of grazing and increase the chance that a patch of macroalgae will establish itself, ungrazed, from the algal turf. The resulting increase in macroalgal cover reduces the availability of coral settlement space and increases the frequency and intensity of coral-algae interactions. The resulting diminishing coral recruitment reduces the growth rate of corals and causes limited mortality, which, in turn, further reduces the intensity of grazing, thereby reinforcing the increase in macroalgae \cite{mumby2007thresholds}, leading to a stable macroalgae-dominated state, as shown in Fig. \ref{CoralReef_RegimeShift}.

\begin{figure}
\centering
\includegraphics[width=0.95\linewidth]{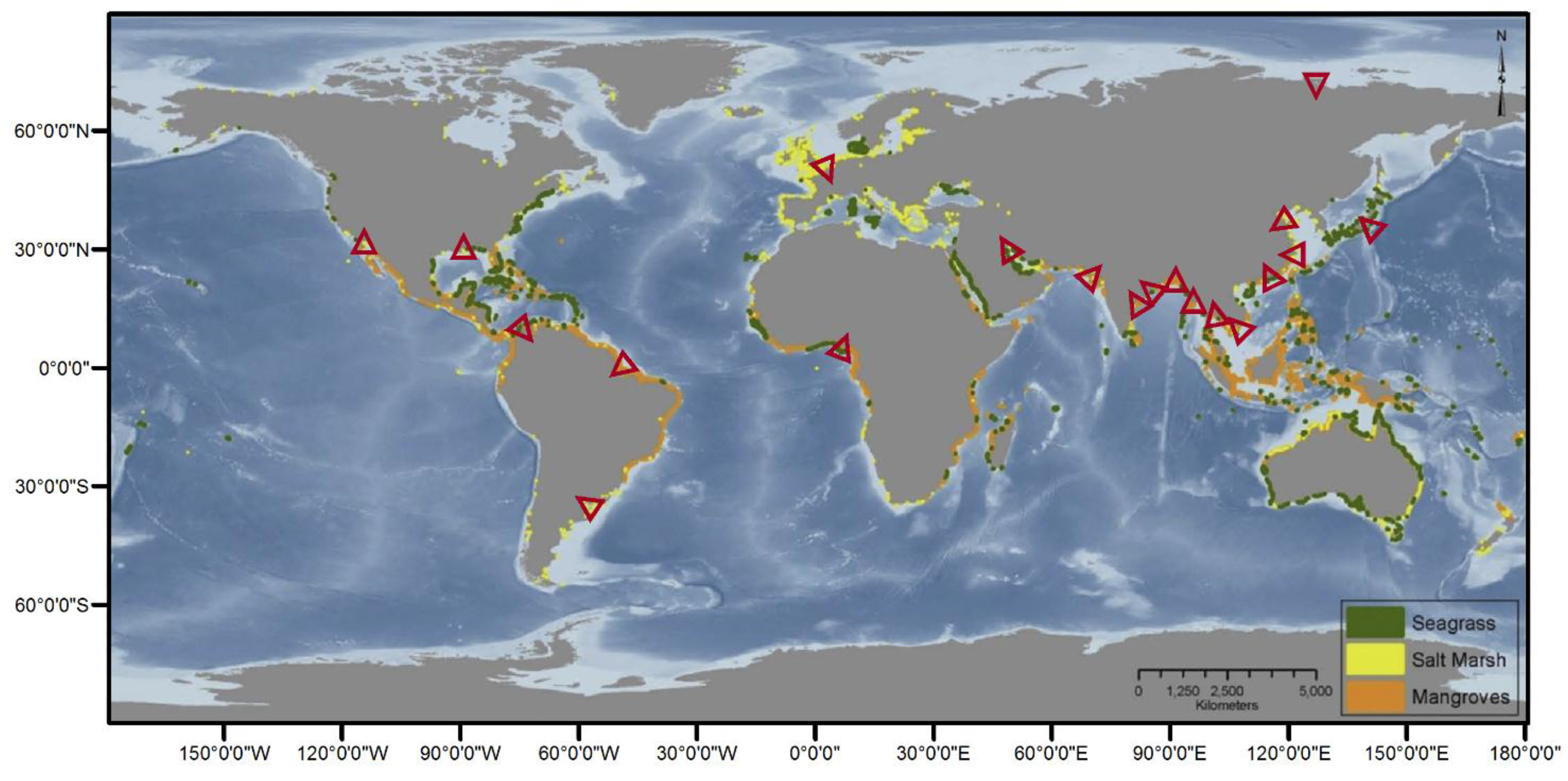} % figure 18
\caption{
Global distribution of seagrass meadows, salt marshes, mangroves, and major deltas. \\
\textit{Source:} The figure is from \cite{moffett2015multiple}.
}
\label{coastaltidalwetlands}
\end{figure}

\noindent
\textit{\textbf{Wetlands.}}
Each wetland is exposed to variations in soil, landscape, climate, water regime and chemistry, vegetation, and human disturbance.
Multiple stable states appear due to the interactions among plants, animals, and environmental changes.
For example, due to nonlinear and coupled ecological, hydrological, and geomorphological feedback, multiple stable states are established in different kinds of coastal tidal wetlands such as marshes, mangroves, deltas, and seagrasses \cite{moffett2015multiple}.
We show the global distribution of these four kinds of wetlands in Fig. \ref{coastaltidalwetlands}.
In salt marshes, field observations provide evidence for possible occurrences of alternative stable states: a bare sediment state, a vegetated state, and states with distinct marsh vegetation communities \cite{marani2013vegetation}.
For example, multimodal frequency distributions, abrupt state shifts, and hysteric responses make it likely that the system will persist in the new state rather than return to the prior state. This situation was also empirically observed in a case of pronounced changes in the elevation distribution of species in areas that shifted from bare flats to vegetated marshes in 1931$-$1992 in the Western Scheldt estuary \cite{wang2013does}.
In freshwater marshes and deltas, feedback between sedimentation and vegetation might result in the establishment of alternative stable ecosystem states. For example, after flooding disturbances, only plants with roots longer than a threshold length can survive; thus, the persisting root length is related to sedimentation \cite{wang2016biogeomorphic}.
In mangrove systems, one mechanism through which multiple stable states form, runaway sedimentation, has been identified in the field. Sediment may be preferentially deposited in large amounts. For example, the presence of mangrove biomass could promote the sudden deposition of sediment during cyclones, hurricanes, or tsunamis \cite{cahoon2003mass} and local sedimentation.
In seagrass meadows, the vegetated state could persist due to positive feedback: vegetation tends to decrease water turbidity, enabling submerged seagrasses to absorb enough light to perform photosynthesis, which stabilizes the bottom and reduces sediment resuspension \cite{carniello2014sediment}. In contrast, when the water depth increases, the vegetated stable state of a seagrass meadow can quickly and abruptly shift to a bare sediment state \cite{carr2010stability}.

\noindent
\textit{\textbf{Rangelands and woodlands.}}
On North American rangelands, lower successional stable states occur in sagebrush and other shrub-dominated vegetation types in the Great Basin. Laycock \cite{laycock1991stable} pointed out that in contrast to the assumption of a single stable state (climax), a vegetation type that is in a stable lower successional state usually does not respond to changes in grazing or even to the cessation of grazing, which is vital for ecosystem management.
On savanna rangelands, Walker \cite{walker1997resilience} and Ludwig et al. \cite{ludwig1996landscape} identified alternative stable states as woody/grass coverage and woody thicket, respectively
. A transition between these states is often triggered by grazing pressures that remove either drought-tolerant or perennial grasses \cite{ludwig1996landscape, walker1997resilience}. If grazing pressures are high, then the abundance of perennial grasses is decreased, leading to an increased abundance of woody plants. Once the woody community is established, fires burn less frequently, and the woody community persists for decades.

As we mentioned previously, human intervention is a significant cause of the presence of multiple stable states in ecosystems. For example, the woodlands of the Serengeti-Mara ecosystem in East Africa present multiple stable states as a result of elephant- and human-induced fires: without external perturbations, such as fires, elephants are unable to cause the vegetation to transition from woodland to grassland. Once a fire changes the ecosystem, the vegetation transitions from woodland to grassland, and elephants can cause it to remain in the grassland state \cite{dublin1990elephants}.

Next, we review works that present empirical and theoretical analyses and predictions regarding phase shifts in ecosystems.

\subsection{Regime shifts between multiple stable states}

\begin{figure}[!ht]
\centering
\includegraphics[width=0.6\linewidth]{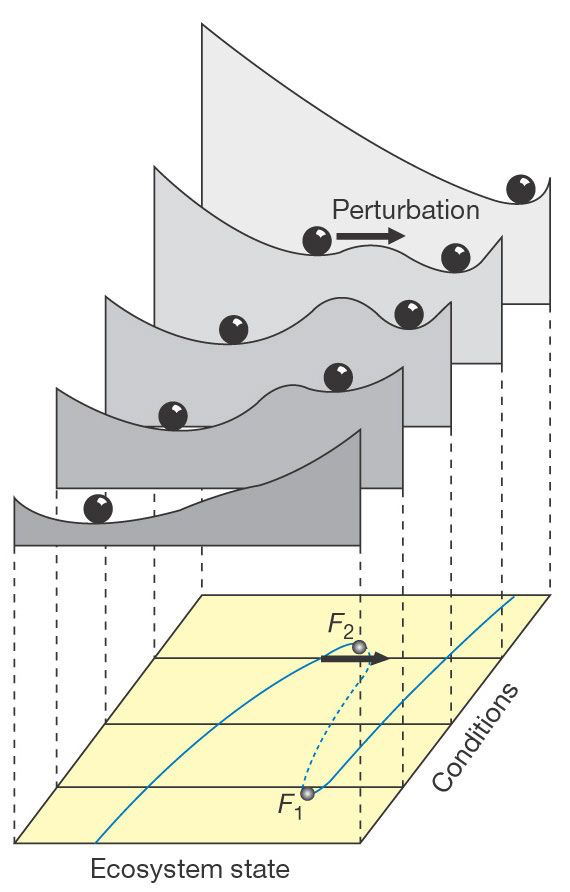} % figure 19
\caption{
External conditions affect the resilience of multistable ecosystems to perturbation.
The stability landscapes depict the equilibria and their basins of attraction at five different
conditions. Stable equilibria correspond to valleys; the unstable middle section of the
folded equilibrium curve corresponds to a hill. If the size of the attraction basin is
small, then resilience is low, and even a moderate perturbation may bring the system into
the alternative basin of attraction. \\
\textit{Source:} Figure from \cite{scheffer2001catastrophic}.
}
\label{AbruptShiftbeforeThreshold}
\end{figure}

Many ecosystems are exposed to internal and external perturbations. Perturbations can be gradual changes in climate, nutrient or toxic chemical loading, habitat fragmentation or biotic exploitation and even precipitating events such as hurricanes and earthquakes \cite{scheffer2001catastrophic}.
The responses of ecosystems to perturbations are nontrivial. 1) For the same perturbation, different ecosystems show different types of responses, which is consistent with our expectation;
2) the same ecosystem responds differently to different perturbations and may even differ in its response to a repetition of the same perturbation due to the existence of multiple stable states; and3) abrupt shifts can occur not only at the tipping point but also before the tipping point, depending on the strength of the disturbance and the size of the basin, which together define the resilience of a system \cite{holling1973resilience}.
If the disturbance is large enough, then the state of an ecosystem may cross the border of the attraction basins and enter another basin even when it has not yet reached the critical point.
When the basin is tiny, that is, when the system's resilience is low, a small perturbation could be sufficient to displace the ball enough to push it over the hill, resulting in a shift to the alternative stable state.
As shown in Fig. \ref{AbruptShiftbeforeThreshold}, since the size of the basin in which the system's state lies is low near the critical point, an abrupt shift in state may occur even before the critical point $F_2$ under a sufficiently large perturbation.

\subsubsection{Regime shift of individual ecosystems}

%Catastrophic shifts may occur in complex systems of various fields with a wide range, which has been attracting plenty of attention. For example, in medicine, there are spontaneous systemic failures like asthma attack \cite{venegas2005self} or epileptic seizures \cite{mcsharry2003prediction, litt2009epileptic}; in global finance, there is concern about systemic market crashes \cite{may2008complex}; in the earth system, abrupt shifts in ocean circulation or climate may occur \cite{lenton2008tipping}; and catastrophic shifts in rangelands, fish populations or wildlife populations may threaten ecosystem services \cite{scheffer2001catastrophic, scheffer2009early}.
The shift from one state to another may result from either a ``threshold or a ``sledgehammer effect \cite{barnosky2012approaching}. A state shift caused by a sledgehammer effect, such as the clearing of a forest using a bulldozer, usually conforms to our expectation. For threshold effects, in contrast, the critical threshold is reached as incremental changes accumulate, and the threshold value is generally not known in advance, so that state shifts resulting from threshold effects are usually not anticipated. In other words, undesired shifts between ecosystem states are caused by the combination of the magnitudes of external forces and the internal resilience of the system.
A sudden dramatic change need not be caused by a sudden sizeable external disturbance. When the system is close to the tipping point, even a tiny incremental change in conditions can trigger a tremendous shift; examples are the legendary straw that breaks the camels back'' and the capsizing of an overloaded boat when too many people move to one side \cite{scheffer2003catastrophic}. Abrupt phase transitions in ecosystems are increasingly common as a consequence of human activities that erode internal resilience; such activities include resource exploitation, pollution, land-use change, possible climatic impact and altered disturbance regimes \cite{folke2004regime, hughes1994catastrophes, steneck2013ecosystem}. 
Next, we show three typical examples of catastrophic shifts in ecosystems.

\noindent
\textit{\textbf{Fire in Australia.}}
In Australia, overhunting and the use of fire by humans some 30,000 to 40,000 years ago resulted in the elimination of large marsupial herbivores and created an ecosystem of fire and fire-dominated plants that continued to expand. This irreversibly switched the ecosystem from a more productive state that was dependent on rapid nutrient cycling to a less productive state in which there was slower nutrient cycling \cite{folke2004regime}. Another factor leading to the erosion of resilience in ecosystems is global warming \cite{graham2015predicting}, a result of the production of excessive carbon dioxide. Increasing temperatures modify key physiological, demographic and community-scale processes, decreasing the internal resilience of ecosystems.
The erosion of internal resilience renders ecosystems vulnerable to external disturbances \cite{steneck2013ecosystem} and can trigger a shift from one state to another undesired state.

\begin{SCfigure*}
 \centering
\includegraphics[width=0.75\textwidth]{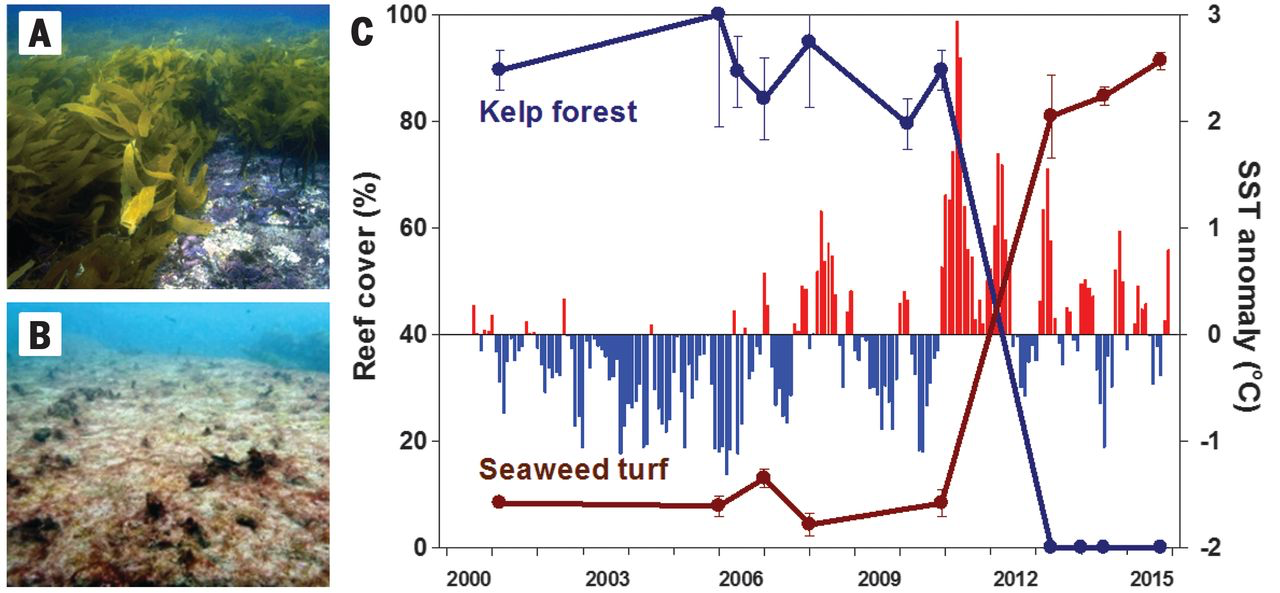} % figure 20
\caption{
Regime shift from kelp forests to seaweed turfs after the 2011 marine heat wave.
Kelp forests were dense in Kalbarri until 2011 (A); at that time, they disappeared from ~100 km of coastline and were replaced by seaweed turfs (B). (C) The habitat transition (lines) coincided with exceptionally warm summers that occurred in 2011, 2012, and 2013 (red bars), punctuating gradually increasing mean ocean temperatures over the past decades.\\
\textit{Source:} Figure from \cite{wernberg2016climate}.
}
\label{Regimeshift_kelpforests_turfs}
\end{SCfigure*}

\noindent
\textit{\textbf{Loss of kelp forests.}}
A marine heat wave caused the loss of kelp forests across ~2300 ${\rm km}^2$ of Australias
Great Southern Reef, forcing a phase shift to seaweed turfs \cite{wernberg2016climate}, as shown in Fig. \ref{Regimeshift_kelpforests_turfs}. Before December 2010, kelp forests covered more than ~70\% of the shallow rocky reefs on the midwestern coast of Australia \cite{wernberg2010decreasing}. However, only two years later, in early 2013, extensive surveys found a 43\% (963 ${\rm km}^2$) loss of kelp forests on the west coast. The previously dense kelp forests had disappeared, and a dramatic increase in the cover of turf-forming seaweeds was observed (Fig. \ref{Regimeshift_kelpforests_turfs}).
Wernberg et al. \cite{wernberg2016climate} deduced that this phase shift was caused by an extreme heat wave in which the temperatures exceeded a physiological tipping point for kelp forests beginning in 2011 and that the new kelp-free state was supported by reinforcing feedback mechanisms. In contrast, similar ecosystem changes have not been observed on the southwestern coast of Australia since the temperatures there remained within the thermal tolerance of kelp during the heat wave \cite{wernberg2013extreme}.
Many areas that experienced previous short-term climate change events, such as the large-scale destruction of kelp forests during the EI Ni${\rm \tilde{n}}$o Southern Oscillation event of 1982/83 \cite{dayton1984catastrophic}, have mostly recovered as environmental conditions returned to normal \cite{martinez2003recovery}. However, there are no signs of kelp forest recovery on the heavily affected reefs in western Australia \cite{bennett2015tropical}. Moreover, the current velocity of ocean warming is pushing kelp forests toward the southern edge of the Australian continent \cite{burrows2014geographical}. The local extinction of kelp forests would devastate lucrative fishing and tourism industries worth more \$10 billion Australian dollars per year \cite{bennett2016great} and endanger thousands of endemic species supported by the kelp forests of Australia's Great Southern Reef.

\begin{figure}[!ht]
 \centering
\includegraphics[width=0.95\linewidth]{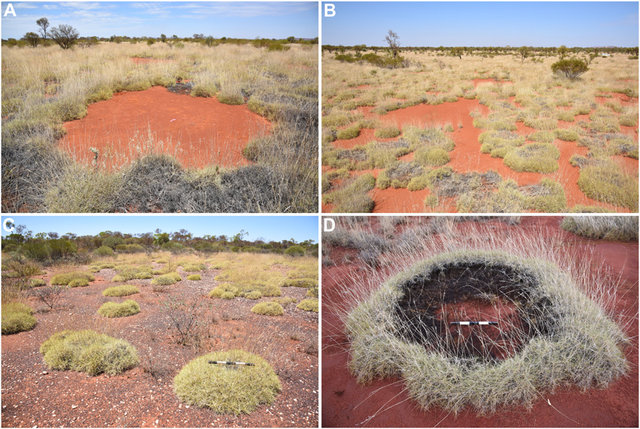} % figure 21
\caption{
Distinct pattern morphologies in the Triodia spinifex grassland. Gaps (A), labyrinths (B), spots (C), and rings (D).\\
\textit{Source:} The figure is from \cite{getzin2016discovery}.
}
\label{AridEcosystem}
\end{figure}

\noindent
\textit{\textbf{Self-organized patchiness in the arid ecosystem.}}
In an ecosystem, different attracting states can not only exist across different time scales \cite{rietkerk1997alternate, scheffer2001catastrophic, scheffer2003catastrophic} but can also coexist at the same time across different spatial scales; the latter is usually related to self-organized patchiness and the resource concentration mechanisms involved \cite{rietkerk2004self}. The most prominent example is the arid ecosystem \cite{klausmeier1999regular}, in which the self-organized patches differ in scale and shape and appear as gaps, labyrinths, rings and spots, as shown in Fig. \ref{AridEcosystem}.
This patchiness is a result of positive feedback between plant growth and the availability of water. A higher vegetation density allows for greater water infiltration into the soil and results in lower soil evaporation, and these effects may stabilize the vegetation state. Once the vegetation disappears, the bare soil is hostile to recolonization \cite{rietkerk1997alternate}.
Thus, the present state of vegetation in the system is determined not only by the soil-water distribution but also by biomass water feedback'' that is determined by the system's historical states \cite{rietkerk2004self}. More importantly, the vegetation state at some spatial locations may shift abruptly to a bare state if rainfall decreases beyond a threshold, which contributes to a more homogeneous bare state in arid ecosystems. However, increased rainfall may not result in recovery of the spotted vegetation state since the soil water under vegetated patches has already disappeared.
Similar patchiness across different spatial locations and catastrophic shifts have been observed in nutrient-poor savanna ecosystems \cite{lejeune2002localized} and in peatland ecosystems \cite{rietkerk2004putative}.
In addition to decreasing rainfall, overgrazing by cattle can also cause catastrophic shifts of a system from a spotted vegetation state to a decertified ecosystem state \cite{ludwig1996landscape}. Adequate grazing management and patchy crop production that conserves resources in marginally arable lands may help optimize productivity, thereby preventing such catastrophic shifts \cite{rietkerk2004self}.

Abrupt shifts can cause substantial loss of ecological and economic resources and may require drastic and expensive intervention to restore a desired state \cite{scheffer2001catastrophic}. Studies of the abrupt shifts and tipping points in ecosystems could help increase the understanding of the failure mechanisms involved and are crucial for ecosystem management. Thus, we should not only focus on the prevention of perturbations but also pay attention to sustaining the system in a large stability domain or pushing the system far away from the tipping points, thereby reducing the risk of unwanted state shifts.

\subsubsection{Regime shifts in networked systems}
Research on regime shifts is often confined to distinct branches of science, reflecting empirical,
theoretical \cite{carpenter2003regime}, or predictive approaches \cite{scheffer2009early}. Such approaches require in-depth knowledge of the causal structure of the system or high-quality spatiotemporal data. Hence, research on regime shifts has generally focused on the analysis of individual types of regime shifts, as previously discussed.
As humans increase their pressure on the planet, regime shifts are likely to occur more often, more severely, and more broadly \cite{rocha2018cascading}. 
Important questions to ask are whether regime shifts interact with one another, whether the occurrence of one increases the likelihood of another, and whether regime shifts at distant places correlate with each other.
All systems on the planet are closely intertwined across three dimensions: organizational level, space, and time. For example, the excessive use of fertilizers, an inadvertent result of growing more vegetables on land, could eutrophicate the downstream coastal waters that compromise food production from the ocean \cite{liu2015systems}; eutrophication is known to be one of three essential causes (eutrophication, warming, and consumption) of rapid decreases in marine biodiversity \cite{vonlanthen2012eutrophication}.
A variety of causal pathways connecting regime shifts have been identified, showing that the occurrence of a regime shift may affect the occurrence of another regime shift. For example, eutrophication is often reported as a regime shift preceding the development of hypoxia or dead zones in coastal areas \cite{diaz2008spreading}, and hypoxic events have been reported to affect the resilience of coral reefs to warming and other stressors in the tropics \cite{altieri2017tropical}.
Thus, there are potential interactions among regime shifts across systems, and these regime shifts should not be studied in isolation under the assumption that they are independent systems.

\begin{SCfigure*}
 \centering
\includegraphics[width=0.65\textwidth]{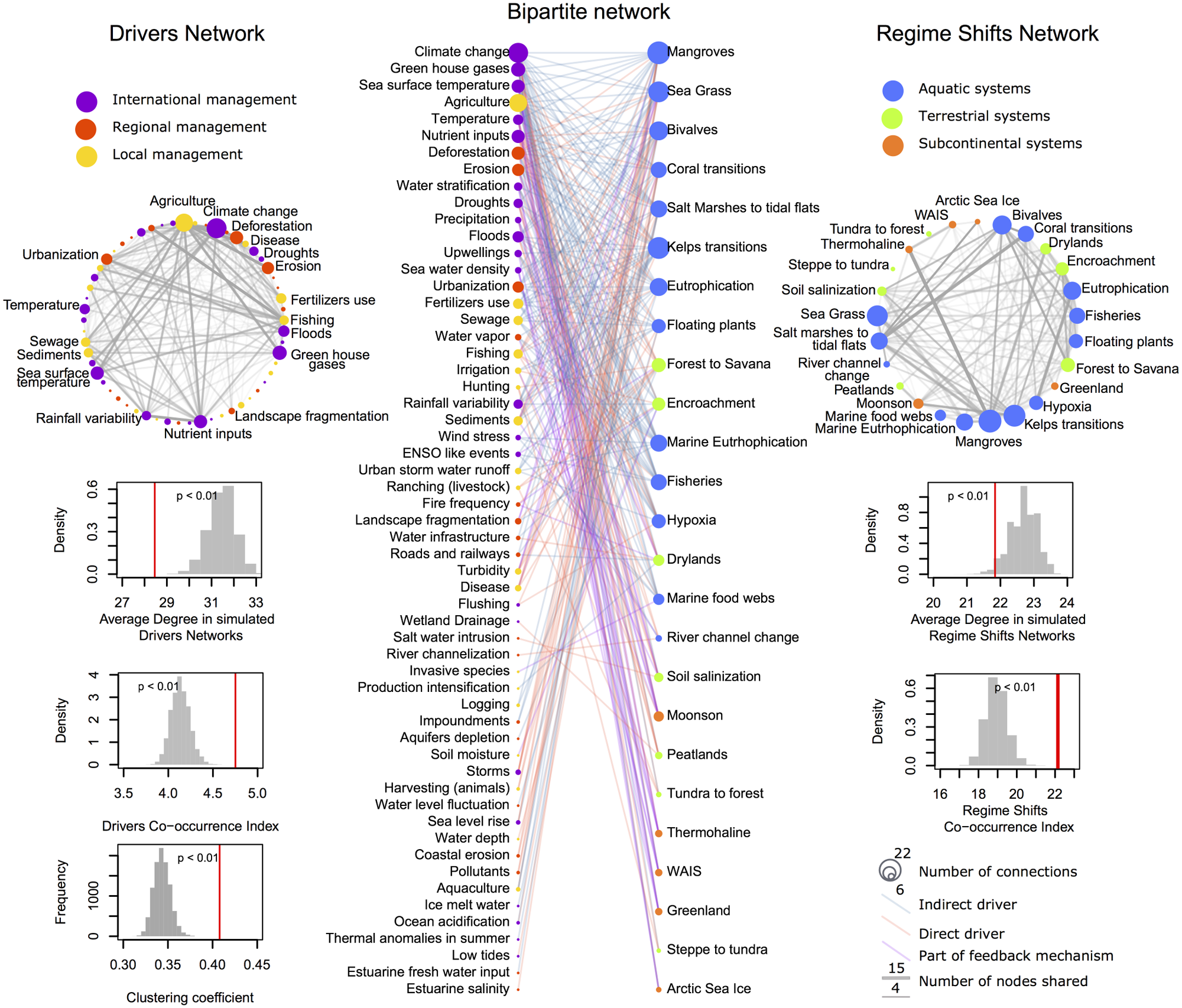} % figure 22
\caption{
Regime shifts-drivers network. In the center is the bipartite network of drivers (left) and regime shifts (right). On the right is the one-mode projection of regime shifts. The width of the links is scaled to the number of drivers shared, and the node size corresponds to the number of drivers per regime shift. On the left is the one-mode projection of drivers, with link width scaled to the number of regime shifts for which causality is shared and node size proportional to the number of regime shifts per driver. The structural statistical analysis is shown below each projection.\\
\textit{Source:} The figure is from \cite{rocha2015regime}.}
\label{BipartiteShiftDriver}
\end{SCfigure*}

Following the accumulation of empirical data on regime shifts, an open online repository of regime shift syntheses and case studies, called the regime shift database, has been built \cite{biggs2018regime}.
This database currently covers 35 types of regime shifts, more than 300 specific case studies based on a literature review of over 1000 scientific papers, and a set of 75 categorical variables that represent i) the main drivers of change; ii) the impact of regime shifts on ecosystem services, ecosystem processes and human well-being; iii) the land use, ecosystem type and spatial-temporal scale at which each regime shift typically occurs; iv) possible managerial options; and v) the assessment of the reversibility of the regime shift, the level of uncertainty related to the existence of the regime shift, and its underlying mechanism \cite{rocha2015regime}. These regime shift attributes can be used to fit statistical models that explore the role of cross-scale interactions \cite{rocha2018cascading}. For example, drivers include natural and human-induced changes that have been identified as directly or indirectly producing a regime shift \cite{nelson2006anthropogenic}. The same driver can induce different regime shifts.
Note that drivers and dynamics operate outside the feedback mechanisms of the system; thus, they are variables that are independent of the dynamics of the system. Direct drivers are those that influence the internal processes of the system or create feedback underlying a regime shift. Indirect drivers alter one or more direct drivers.
By mining the interactions between regime shifts and drivers, a bipartite network is constructed in Ref \cite{rocha2015regime}, from which two networks can be projected. The first is a network of drivers connected by the regime they caused. The second is a network of regime shifts connected by the drivers they share. These two projected networks are shown in Fig. \ref{BipartiteShiftDriver}. The analysis of regime-driver networks demonstrates that reducing the risk of regime shifts requires integrated action on multiple dimensions of global change across scales. Rocha et al. \cite{rocha2018cascading} give a more comprehensive analysis of cascading regime shifts within and across scales.

\begin{figure} [ht!]
 \centering
\includegraphics[width=0.95\linewidth]{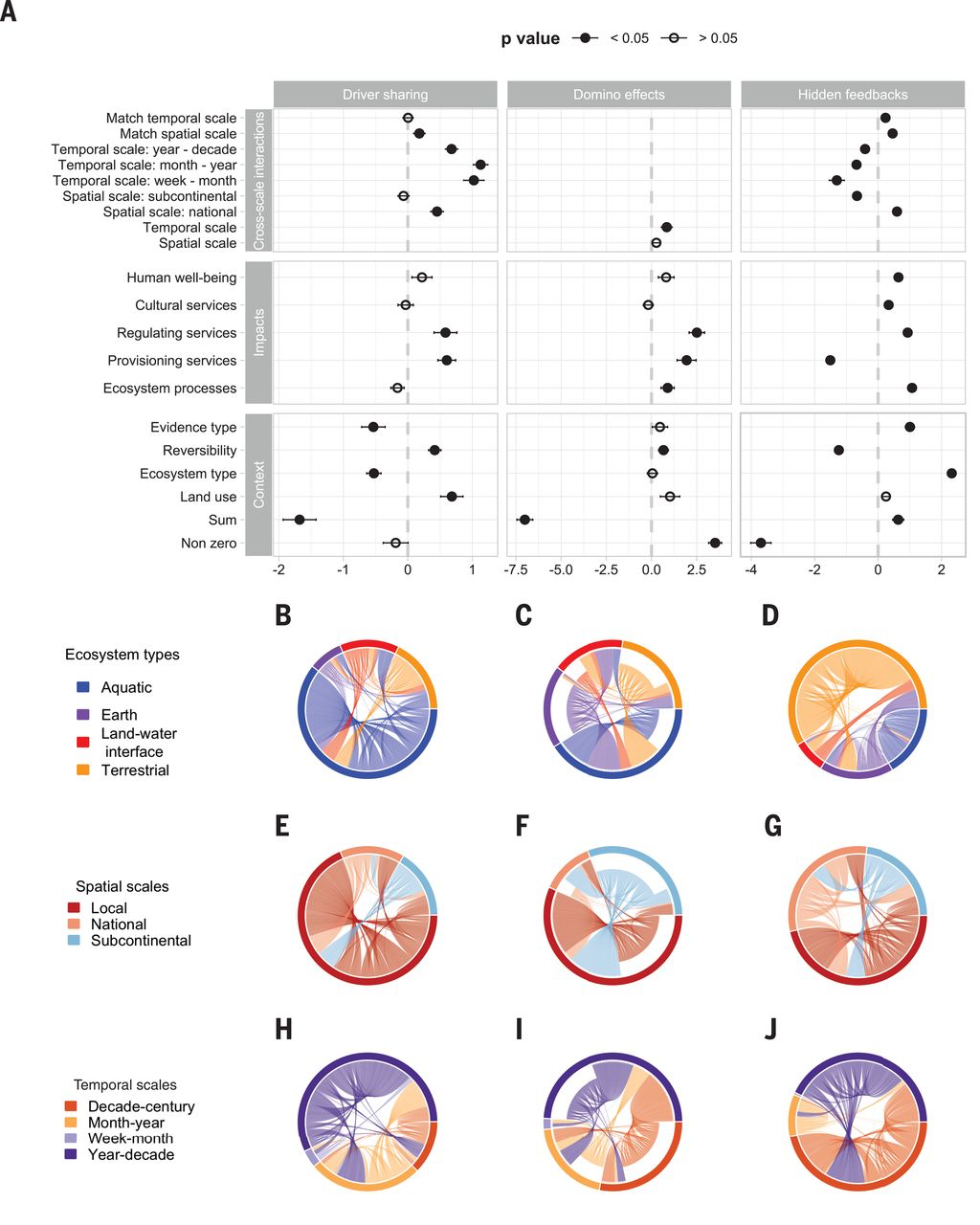} % figure 23
\caption{
Cascading effects across scales.
(A) Summary of the statistical results. (B to J) Circular plots showing the mixing matrices of cascading effects [driver sharing, (B), (E), and (H); domino effects, (C), (F), and (I); and hidden feedback, (D), (G), and (J)] according to ecosystem type and spatial or temporal scale.\\
\textit{Source:} The figure is from \cite{rocha2018cascading}.}
\label{RegimeShiftNetwork}
\end{figure}

Each entry in the regime shift database provides a literature-based synthesis of the key drivers and feedback underlying the regime shift and describes their impact on ecosystem services and human well-being and possible management options. For each regime shift, the database encodes the drivers and the underlying feedback into a causal loop diagram \cite{lane2008emergence}, which is a signed directed graph consisting of variables connected by arrows denoting causal influence \cite{rocha2015regime}.
Rocha et al. \cite{rocha2018cascading} use 30 regime shifts in a system in which complete synthesis exists. The causal loop diagrams have been curated to construct coupled regime shift networks, as shown in Fig. \ref{RegimeShiftNetwork}.
The authors merge pairs of regime shift networks using the following three types of connections between them:
(i) Driver sharing is the most common type of connection; the regime shifts connected through driver sharing are correlated in time and space but do not have to be independent \cite{rocha2015regime};
(ii) Domino effects occur when the feedback processes of one regime shift affect the drivers of another regime shift, creating a one-way dependency \cite{scheffer2012anticipating};
(iii) Hidden feedback is observed only when two regime shift networks are combined to generate new feedback that cannot be identified in the separated regime shifts \cite{liu2015systems}.

In analyzing the regime shift networks coupled through these three types of interaction, Rocha et al. \cite{rocha2018cascading} found that half of the regime shifts may have been causally linked at different scales \cite{scheffer2018seeing}.
The regime shift of networks coupled via driver sharing describes the co-occurrence patterns of 77 drivers across the 30 regime shifts analyzed. The drivers that most frequently co-occurred in this case were related to food production, climate change, and urbanization.
Regime shifts are more likely to share drivers when they use land in a similar way. Driver sharing is more likely to occur in dynamics that evolve faster in time when there is a progression of spatial scales.
Domino effects are not common; evidence of cross-scale interactions involving domino effects was only found in time but not in space.
Hidden feedback is more likely to occur in the range of decades to centuries and at the national scale.
The regime shift networks constructed in Ref. \cite{rocha2018cascading} enable researchers to systematically identify potential cascading effects.

The cascading effects among regime shifts have also been modeled as a network of tipping elements in Ref. \cite{kronke2019dynamics}; in the network presented there, each node represents one tipping element that is represented by a time-dependent quantity $x(t)$, which evolves according to the following autonomous ordinary equation:
\begin{equation}\label{TippingElementODE}
\frac{dx}{dt}=-a(x-x_0)^3+b(x-x_0)+r,
\end{equation}
where $r$ is the control parameter, and the coefficients $a, b>0$ and $x_0$ control the position of the system on the x-axis.
The interactions between tipping elements are modeled as a directed network of a linearly coupled system of ordinary differential equations.
\begin{equation}\label{TippingElementNetwork}
\frac{dx_i}{dt}=-a(x_i-x_0)^3+b(x_i-x_0)+r_i+d\sum_{j=1,j\neq i}^N a_{ij}x_j,
\end{equation}
where $d$ is the coupling strength, and $a_{ij}$ are elements in the adjacency matrix.
Such a network captures the cascading effects that occur due to interactions between tipping elements and has been applied to the Amazon rainforest. Although the presence of cascading effects usually means that the occurrence of one regime shift could trigger the occurrence of another regime shift, not all cascading effects reported in the literature are expected to amplify each other. For example, it has been reported that climate-tipping points can regulate each other and reduce the probability of regime shifts in forests \cite{gaucherel2017potential}.
Study of the interactions between regime shifts could help researchers develop methods for early-warning signal detection to predict regime shifts.

\begin{figure}[!ht]
 \centering
\includegraphics[width=0.95\linewidth]{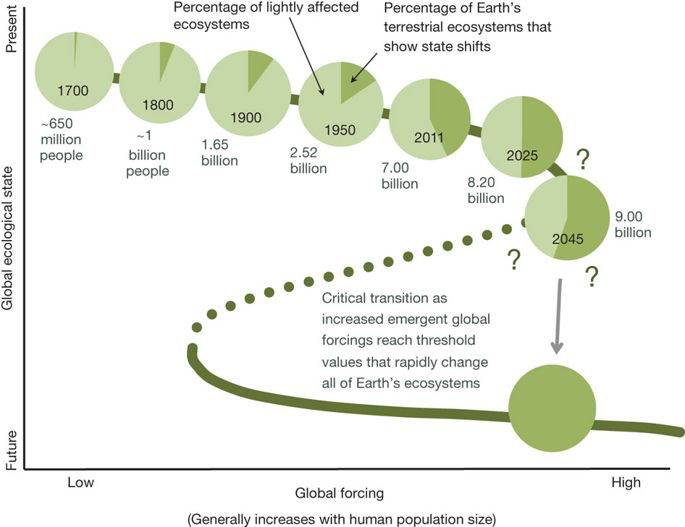} % figure 24
\caption{
Quantifying land use as one method of anticipating a planetary state shift.
The trajectory indicated by the green line represents a fold bifurcation with hysteresis. At each time point, light green represents the fraction of Earth's land that probably has dynamics within the limits characteristic of the past 11,000 years. Dark green indicates the fraction of terrestrial ecosystems that have unarguably undergone drastic state changes. A planetary state shift may occur assuming that there is conservative population growth and that resource use does not become any more efficient. \\
\textit{Source:} The figure is from \cite{barnosky2012approaching}.
}
\label{PlanetaryShift}
\end{figure}

Most studies of regime shifts focus on local ecological systems that evolve over short time spans \cite{scheffer2009early, drake2010early}. However, there are also planetary-scale critical transitions that operate over centuries or millennia \cite{barnosky2012approaching}, such as the `Big Five' mass extinctions \cite{barnosky2011has}, the Cambrian explosion \cite{marshall2006explaining}, and the recent transition from the last glacial to the present interglacial condition \cite{hoek2008last}. This transition caused a rapid warm-cold-warm fluctuation in climate between 14,300 and 11,000 years ago \cite{hoek2008last}. The significant biotic changes included the extinction of approximately half of the then-existing species of large-bodied mammals, several large bird and reptile species, and a few small animal species \cite{koch2006late}. A vital decrease in local and regional biodiversity occurred as geographic ranges shifted individualistically, resulting in novel species assemblages \cite{graham1996spatial}. Another significant change was a global increase in human biomass and human mobility to all continents \cite{barnosky2008megafauna}.
The collected evidence of these planetary-scale critical transitions suggests that global-scale state shifts are not the cumulative result of many smaller-scale events that originate in local systems. Instead, they require global-level forcing that emerges on the planetary scale and then percolates downwards to cause changes in local systems \cite{barnosky2012approaching}.
Currently, global-scale forcing mechanisms are induced by human population growth with attendant resource consumption \cite{steffen2011anthropocene}, energy production, consumption \cite{mcdaniel2002increased}, and climate change \cite{lenton2011early}; these factors cumulatively impose much stronger forcing than those that were active at the last global-scale state shift. Thus, the plausibility of a future planetary state shift appears high, as shown in Fig. \ref{PlanetaryShift}, which highlights the need to predict critical transitions by detecting early-warning signs.

\subsection{Predict and control the resilience of ecosystems}

Tipping points are critical thresholds in system parameters or state variables for which a tiny perturbation can lead to a qualitative change in the system \cite{kronke2019dynamics}.
Once a tipping point has been crossed, leading to a critical transition, it is extremely difficult or even impossible to restore the system \cite{barnosky2012approaching}. Thus, the prediction and prevention of unwanted critical transitions in ecosystems is a crucial challenge in ecology.
Critical transitions, or regime shifts, can result from ``fold bifurcations'' that show hysteresis \cite{scheffer2009early} or from more complex bifurcations \cite{hastings2010regime}.
In the latter case, there is no typical early-warning signals of the regime shift.
In contrast, regime shifts that occur due to fold bifurcations display preceding general phenomena that can be characterized mathematically. These include slowing recovery from perturbations and increasing variance of the patterns within state fluctuations \cite{scheffer2009early}, autocorrelation between fluctuations \cite{drake2010early}, asymmetry of fluctuations and rapid back-and-forth shifts (``flickering'') between states.

%\subsubsection{Early-warning signals}
\subsubsection{Predictive approaches for low-dimensional systems}
Predicting the tipping points at which critical transitions occur is extremely difficult because the state
of the system may show little change before it reaches the tipping point \cite{scheffer2009early}.
Fortunately, specific generic symptoms such as the ``critical slowing down'' discussed in Sec. \ref{Robustness} may occur in a broad class of systems as they approach a critical point, and these symptoms can be used for early-warning before the catastrophic shift occurs. These symptoms arise regardless of differences in the details of the system \cite{schroeder2009fractals}. Similar signals could indicate disparate phenomena such as the collapse of an overharvested population at ancient climatic transitions.

\noindent
\textit{\textbf{Skewness and flickering.}}
In the vicinity of a catastrophic bifurcation point, the asymmetry of fluctuations may increase \cite{guttal2008changing}. Taking the fold bifurcation as an example, an unstable equilibrium marks the border of the basin of attraction as the system approaches the attractor from one side, as shown by the dashed line in Fig. \ref{hysteresisFig}.
As the system approaches the bifurcation, the slope in the basin of attraction (Fig. \ref{Basin_Ball_scheme}) becomes less steep. Consequently, the system tends to remain at an unstable point but does not move to the opposite side of the stable equilibrium \cite{scheffer2009early}, leading to an increase in the skewness of the distribution of states. In addition, the system's state may moved back and forth between the basins of attraction of two alternative attractors under stochastic forcing, because the border becomes relatively lower when a critical transition is approaching.
This phenomenon is called flickering \cite{berglund2002metastability}; it is another early-warning signal, and the system may shift permanently to the alternative state if the underlying slow change in conditions persists.

\noindent
\textit{\textbf{Indicators in cyclic and chaotic systems.}}
The early-warning signs reviewed above exist in systems with underlying attractors that correspond to stable points but are not applicable to cyclic or chaotic systems; in those systems, critical transitions are associated with a different class of bifurcation \cite{kuznetsov2013elements}, and the early-warning signs are different. For example, the Hopf bifurcation, which marks the transition from a stable system to an oscillatory system \cite{strogatz2018nonlinear}, is signaled by critical slowing down \cite{chisholm2009critical}; close to the bifurcation, perturbations lead to long transient oscillations before the system settles into the stable state.

Another class of bifurcations is caused by intrinsic oscillations that bring the system to the border of the basin of attraction of an alternative attractor; such bifurcations are called basin-boundary collisions \cite{vandermeer1999basin}. They are usually not associated with particular properties related to stable or unstable points that can be analytically defined. However, the dynamics of the system may be expected to change characteristically before collisions occur; there may be increased autocorrelation between states, and the oscillations may become ``stretched'' \cite{rinaldi2000geometric}.
Additionally, there is the phenomenon of phase locking between coupled oscillators, and alternative attractors are often involved when the corresponding bifurcations are associated with critical slowing down \cite{leung2000bifurcation}. For example, increasing variance and flickering occur before an epileptic seizure, showing a phenomenon associated with the phase-locking of firing in neural cells.

\noindent
\textit{\textbf{Spatial patterns as early-warning signals.}}
Early-warning signals arise not only in time series but also in particular spatial patterns in the vicinity of a critical transition. The spatial term here is not limited to the physical distances between two elements such as the habitat patches in a fragment landscape \cite{hanski1998metapopulation} but also represents the functional associations or interactions between two entities, such as the connections between two people in a social network or between two functionally related financial markets \cite{scheffer2003slow}. In many systems, each entity tends to take a state similar to that of the entities to which it is connected. Moreover, phase transitions in such systems may occur in a way similar to the way in which phase transitions occur in ferromagnetic materials, where individual particles affect each other's spin \cite{schroeder2009fractals}. When such systems approach the tipping point, the distribution of the entities' states may change in particular ways, for example by showing a general tendency towards increased spatial coherence \cite{schroeder2009fractals} as measured by cross-correlation among entities.

Although many systems have similar early-warning signals, no general spatial patterns fit all systems. It is essential to know which class of system is involved as we interpret spatial patterns. For systems governed by local disturbance, the scale-invariant power-law structures found for large-scale parameters range vanish as a critical transition is approached \cite{kefi2007spatial}. In systems that display self-organized regular patterns \cite{turing1990chemical}, particular spatial configurations may arise in the vicinity of a critical transition. In the desert vegetation system, the nature of the pattern changes from maze-like to spots because of a symmetry-breaking instability \cite{rietkerk2004self} as the system nears a critical transition to a barren state.

Alternative stable states separated by critical thresholds also occur in ecosystems ranging from rangelands to marine systems \cite{laycock1991stable, gardner2003long}. Alternative stable states are usually related to the hysteresis transition, so it is difficult for a system to recover once the system enters a state by crossing a tipping point. For instance, as we mentioned previously, the recovery of an arid ecosystem \cite{klausmeier1999regular} from a barren state may require more rain than is needed to preserve the last patches since, in such a case, the concentration of soil water under the vegetated patches has already greatly decreased.

In summary, despite the lack of universal early-warning signals, effectively detecting them and taking actions that push the system far from its tipping points is very important in ecological management.

\subsubsection{Predictive approaches for networked systems}
There are many empirical examples of regime shifts with tipping points in ecosystems, and many theoretical models have been proposed to describe the mechanisms underlying these shifts.
However, prediction of the tipping points and regime shifts remains a challenge due to the extreme complexity of ecological systems, a complexity that is usually reflected in their high dimensionality (the presence of a large number of species and interactions), in their stochastic and nonlinear natural dynamics, and in the uncertainty of initial conditions or drivers \cite{moore2018predicting}.
Despite this complexity, it is possible to predict regime shifts through resilience indicators, e.g., the statistical measures of some key ecosystem variables \cite{jiang2019harnessing}.
For instance, ecologists have used a combination of models and observations from long-term datasets and short-term experiments to identify early-warning signals that appear before the critical transition occurs \cite{hastings2010regime}. In these models, initial conditions are informed by first principles and empirical data, the drivers are incrementally or dramatically altered, and the subsequent changes in the system are recorded. For ecosystems with certain types of dynamics, this approach could successfully detect early-warning signs. However, some other ecosystems, especially those that feature multiple attractors or the potential for chaos, exhibit abrupt changes with no advanced warning in the time series \cite{moore2018predicting}.
As we mentioned previously in the section on multiple stable states, simplified models of the system that include the essential components, interactions, and drivers and an element of stochasticity can be constructed \cite{moore2018predicting}. Examples include a minimal model of ecosystem catastrophic shits \cite{scheffer2001catastrophic}, the one-dimensional grazing ecosystems model \cite{may1977thresholds}, and the coral reef model \cite{mumby2007thresholds}. However, these models are built for low-dimensional systems, and they neglect the interactions between the studied components and other species. Thus, they do not apply to high-dimensional ecosystems.

As reviewed in Sec. \ref{Robustness}, a dimensional-reduction method \cite{gao2016universal} that maps the multidimensional system into an effective one-dimensional space has been developed, and the resilience phenomenon can thus be accurately predicted using this method.
Taking mutualistic networks as an example, the dynamics of such networks can be captured by the following equation \cite{gao2016universal}:
\vspace{-1cm}
\begin{strip}
\begin{equation}\label{MutualisitcDynamics}
\frac{dx_i}{dt}=B_i+x_i(1-\frac{x_i}{K_i})(\frac{x_i}{C_i}-1)+\sum_{j=1}^{N}A_{ij}\frac{x_ix_j}{D_i+E_ix_i+H_jx_j},
\end{equation}
\end{strip}
where $B_i$ is a constant influx due to migration, and the second term defines logistic growth \cite{hui2006carrying}, incorporating the Allee effect \cite{courchamp1999inverse};
the interaction term captures the symbiotic contribution of $x_j$ to $x_i$, which saturates when the populations are large.
For simplicity, the six parameters in the equation (\ref{MutualisitcDynamics}) are set as node-independent: $B_i=B$, $C_i=C$, $K_i=K$, $D_i=D$, $E_i=E$ and $H_i=H$.
Mapping the multidimensional equation (\ref{MutualisitcDynamics}) into one-dimensional dynamics based on GBB reduction\cite{gao2016universal}, we can obtain
\vspace{-1cm}
\begin{strip}
\begin{equation}\label{MutualisitcDynamics2OneD}
f(\beta_{\rm eff},x_{\rm eff})=B+x_{\rm eff}(1-\frac{x_{\rm eff}}{K})(\frac{x_{\rm eff}}{C}-1)+\beta_{\rm eff}\frac{x_{\rm eff}^2}{D+(E+H)x_{\rm eff}},
\end{equation}
\end{strip}
According to the rule of stability of motion, critical point $\beta_{\rm eff}^{c}$ can be calculated by the following equations:
\vspace{-1cm}
\begin{strip}
\begin{equation}\label{MDTo1DStabilityRule_Mutua}
 \left\{
 \begin{aligned}
 f(\beta_{\rm eff}, x_{\rm eff})=B+x_{\rm eff}(1-\frac{x_{\rm eff}}{K})(\frac{x_{\rm eff}}{C}-1)+\beta_{\rm eff}\frac{x_{\rm eff}^2}{D+(E+H)x_{\rm eff}}=0,\\
  \frac{\partial f(\beta_{\rm eff}, x_{\rm eff})}{\partial x_{\rm eff}}=-3\frac{x_{\rm eff}^2}{CK}+\frac{2(C+K)}{CK}x_{\rm eff}-1+\beta_{\rm eff}\frac{(E+H)x_{\rm eff}^2+x_{\rm eff}}{[D+(E+H)x_{\rm eff}]^2}<0.\\
 \end{aligned}
 \right.
 \end{equation}
\end{strip}
In addition, using equation $f(\beta_{\rm eff}, x_{\rm eff})=0$, we can describe $\beta_{\rm eff}$ as a function of $x_{\rm eff}$ as
\vspace{-1cm}
\begin{strip}
\begin{equation}\label{Mutua_Resilience}
\beta_{\rm eff}(x_{\rm eff})=-[B+x_{\rm eff}(1-\frac{x_{\rm eff}}{K})(\frac{x_{\rm eff}}{C}-1)]\frac{D+(E+H)x_{\rm eff}}{x_{\rm eff}^2},
 \end{equation}
\end{strip},
which is the inverse of the desired resilience function.
By inverting equation (\ref{Mutua_Resilience}), \textit{i.e.}, swapping the axes, the resilience function for this system can be graphically obtained; it predicts a bifurcating resilience function and a transition from a resilient state with a single stable fixed point, $\mathbf{x}^{\rm H}$, to a nonresilient state in which both $\mathbf{x}^{\rm H}$ and $\mathbf{x}^{\rm L}$ are stable.
The critical point of this bifurcation is predicted to be $\beta_{\rm eff}^{c}=6.79$.
Such a value is fully determined by the dynamics and is independent of the network topology $A_{ij}$.

Considering that mutualistic networks contain two different types of nodes, pollinators and plants, Jiang et al. \cite{jiang2018predicting} point out that a two-dimensional model is necessary to capture the bipartite and mutualistic nature of such networks. The authors use the letters $P$ and $A$ to denote plants and pollinators and $S_{P}$ and $S
_{A}$ to represent the number of plants and the number of pollinators, respectively, in the network.
The model is written as the following equation \cite{rohr2014structural}:
\begin{strip}
\begin{equation}\label{MutualisticWholeSystem}
 \left\{
 \begin{aligned}
 \frac{dP_i}{dt}=P_i \bigg( \alpha_i^{(P)}-\sum_{j=1}^{S_P}\beta_{ij}^{(P)}P_j+\frac{\sum_{k=1}^{S_A}\gamma_{ik}^{(P)}A_k}{1+h\sum_{k=1}^{S_A}\gamma_{ik}^{(P)}A_k}+\mu_P \bigg),\\
 \frac{dP_i}{dt}=A_i \bigg( \alpha_i^{(A)}-\kappa_i-\sum_{j=1}^{S_A}\beta_{ij}^{(A)}A_j+\frac{\sum_{k=1}^{S_P}\gamma_{ik}^{(A)}P_k}{1+h\sum_{k=1}^{S_P}\gamma_{ik}^{(A)}P_k}+\mu_A \bigg).\\
 \end{aligned}
 \right.
 \end{equation}
\end{strip}
The notation is as follows: $P_i$ and $A_i$ are the abundances of the $i$th plant and the $i$th pollinator, respectively; $\alpha$ is the intrinsic growth rate; and $\beta_{ii}$ and $\beta_{ij}$ are the parameters affecting intraspecific and interspecific competition, respectively.
Typically, intraspecific competition is stronger than interspecific competition, that is, $\beta_{ii}\gg \beta_{ij}$.
Parameters $\mu_P$ and $\mu_A$ describe the immigration of plants and pollinators, respectively,
and $\gamma$ quantifies the strength of the mutualistic interaction.

\begin{figure}
\centering
 \includegraphics[width=0.95\linewidth]{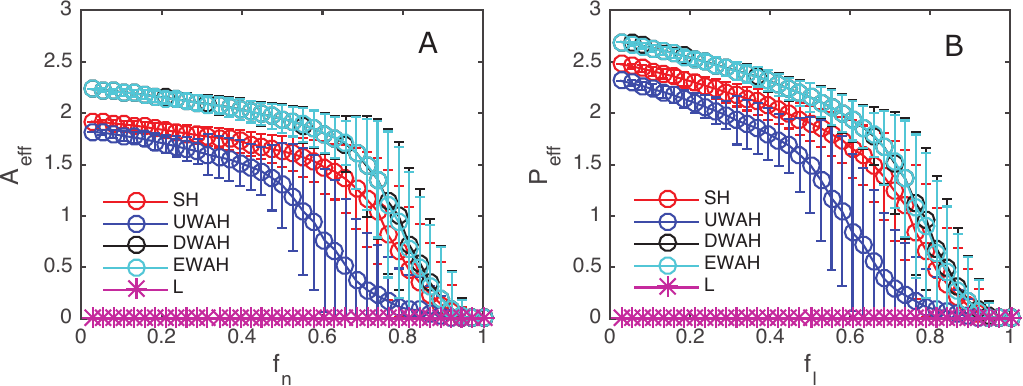} % figure 25
\caption{Resilience functions with tipping points in networks constructed from the data recorded at Tenerife, Canary Islands \cite{dupont2003structure}.
(A) Ensemble-averaged pollinator abundance with high initial values versus $f_n$, which is the fraction of removed pollinators. (B) Ensemble-averaged plant abundance with high initial values versus $f_l$, the fraction of removed mutualistic links.
Here, the parameters are $h=0.2$, $t=0.5$, $\beta_{ii}^{(A)}=\beta_{ii}^{(P)=1}$, $\alpha_i^{(A)}=\alpha_i^{(P)}=-0.3$, $\mu_A=\mu_P=0.0001$, and $\gamma_0=1$ and $\kappa=0$.
The figure has been reproduced from Ref. \cite{jiang2018predicting}.}
\label{Lai_PNAS_fig4}
\end{figure}

By assuming that the decay parameters for all the pollinators have an identical value $\kappa_i \equiv \kappa$ and that pollinators in the mutualistic network die one after another as a result of an increasingly deteriorating environment, the high-dimensional mutualistic network can be reduced to a dynamical system that contains two coupled ODEs (ordinary differential equations), one for the pollinators and another for the plants, which can be written as
\begin{strip}
\begin{equation}\label{Mutualistictwo-dimensionalModel}
 \left\{
 \begin{aligned}
\frac{dP_{\rm eff}}{dt}=\alpha P_{\rm eff}-\beta P_{\rm eff}^2+\frac{\langle \gamma_P \rangle A_{\rm eff}}{1+h\langle \gamma_P \rangle A_{\rm eff}}P_{\rm eff}+\mu,\\
\frac{dA_{\rm eff}}{dt}=\alpha A_{\rm eff}-\beta A_{\rm eff}^2-\kappa A_{\rm eff}+\frac{\langle \gamma_A \rangle P_{\rm eff}}{1+h\langle \gamma_A \rangle P_{\rm eff}}A_{\rm eff}+\mu,\\
 \end{aligned}
 \right.
 \end{equation}
\end{strip}
where dynamical variables $P_{\rm eff}$ and $A_{\rm eff}$ are the effective abundances of plants and pollinators, respectively.
Parameter $\alpha$ denotes the effective growth rate of the network, $\beta$ describes the combined effects of intraspecific and interspecific competition, $\kappa$ is the average species decay rate, and parameter $\mu$ accounts for the effects of migration of the species.
Two effective mutualistic interaction strengths, $\langle \gamma_P \rangle$ and $\langle \gamma_A \rangle$, can be obtained by averaging as ($i$), the unweighted average, ($ii$), the degree-weighted average, and ($iii$), the eigenvector-based average.
Figure \ref{Lai_PNAS_fig4} shows the ensemble-averaged pollinator and plant abundances in networks obtained by different averaging methods.
For these three averaging methods, method ($i$) works well for resilience functions without tipping points, and methods ($ii$) and ($iii$) perform well for resilience functions with tipping points, even in the presence of stochastic disturbances.

These model-driven approaches could trigger further studies that attempt to predict the tipping points in real high-dimensional networks, and the results, as reviewed above, could help us understand the general resilience patterns that are found in mutualistic networks. However, these methods cannot be applied directly to real complex networks since they were developed based on the assumption that the interactions between components are all positive. In real networks, there are both positive and negative interactions, such as cooperation (a positive interaction) and competition (a negative interaction), between species in ecological networks. How to predict tipping points in real ecological networks remains a major challenge.

\subsubsection{Control resilience of ecosystems}\label{2croe}
The ultimate goal of uncovering the mechanisms underlying tipping points is to develop biologically viable management/control principles and strategies that eliminate the tipping point, delay the occurrence of unwanted critical transitions\cite{jiang2019harnessing}, and restore a failed system to the desired state. These transitions include global ecological state transitions at the planetary scale \cite{barnosky2012approaching}, the shutdown of thermohaline circulation in the North Atlantic at the regional scale \cite{rahmstorf2002ocean}, and at the local scale, the switching of shallow lakes from clear to turbid waters and the global extinction of species \cite{scheffer1997ecology, drake2010early, dai2012generic}.
Due to the nonlinear nature of ecosystems, it is difficult to control them; controlling nonlinear dynamical networks remains an outstanding problem and is currently an active area of research \cite{dai2012generic}.
Especially with respect to the systems with critical transitions, Nishikawa et al. \cite{nishikawa2014controlling} investigated how small perturbations can be used to drive the system to the desired attractor. Vidiella et al. \cite{vidiella2018exploiting} demonstrated that in semiarid ecosystems, the phenomenon of an ecological ``ghost'' (a long transient phase during which the system maintains its stability) may be exploited to delay or prevent the occurrence of a tipping point \cite{vidiella2018exploiting}.

A tipping point transition is the consequence of gradual changes in the system that are caused by a slow drift in the intrinsic parameters and/or the external conditions. The intrinsic parameters include the species decay rate, the strength of mutualistic interactions, and the fraction of disappeared nodes and/or links in an ecological network; all of these can be altered by environmental conditions \cite{jiang2019harnessing}. Thus, these parameters are also called ``environmental parameters'' that determine the state of the ecological system. For instance, a sudden bloom of cyanobacteria in a lake or reservoir can be devastating because it kills fish on a large scale and poses significant toxicity risks to the environment. An effective way to prevent a bloom of cyanobacteria in a lake or a reservoir is to stop or significantly reduce nutrient inflow into the lake \cite{pace2017reversal}. Another example is the fisheries food web. If we could detect good early-warning signs in time, then regime shifts could be averted by a rapid reduction in angling and/or by gradual restoration of the shoreline \cite{biggs2009turning}.

\begin{figure}
\centering
 \includegraphics[width=0.95\linewidth]{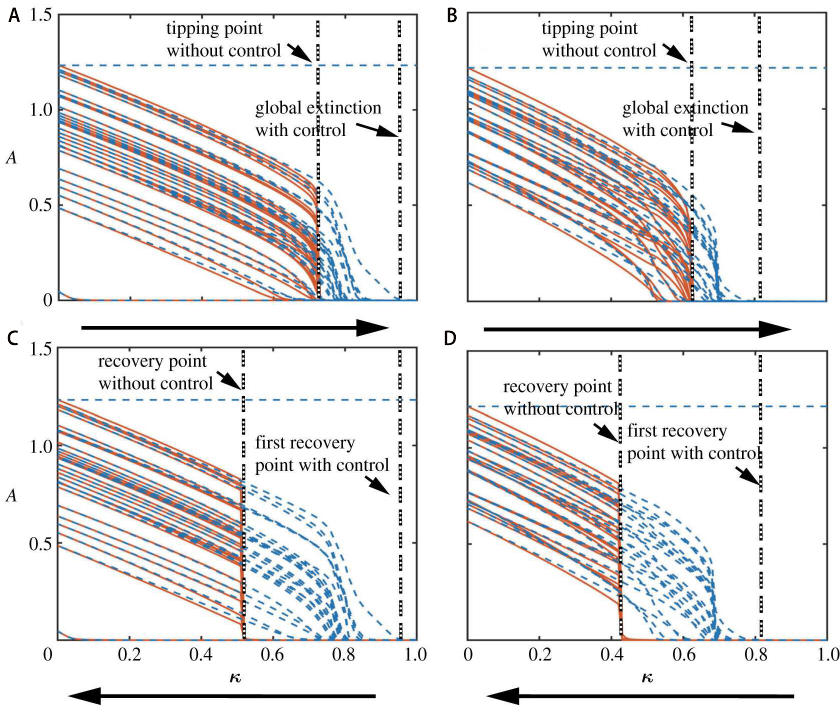} % figure 26
\caption{Managing a tipping point caused by an increase in the pollinator decay rate. The principle on which the management is based is maintenance of the abundance of the pollinator with the largest number of mutualistic connections (indicated by the light blue horizontal dashed line). Without abundance management, all pollinator populations collapse abruptly at a single value of $\kappa$ that represents a tipping point. However, with abundance management, the extinction process becomes gradual, effectively eliminating the tipping point.
The figure is from Ref. \cite{jiang2019harnessing}.}
\label{controlTippingpoint}
\end{figure}

\noindent
\textit{\textbf{Resilience loss prevention via tipping-point control.}}
Very recently, Jiang et al. \cite{jiang2019harnessing} investigated how tipping points in real-world complex and nonlinear dynamical ecological networks can be managed or controlled by altering the way in which species extinction occurs. An example could be replacing the massive extinction of all species with the gradual extinction of individual species as the environmental parameter continues to increase so that the occurrence of global extinction is substantially delayed. The authors focus on empirical bipartite pollinator plant networks whose dynamics are governed by mutualistic interactions \cite{rohr2014structural, jiang2018predicting}.
These networks are managed by choosing a ``targeted'' species and maintaining its abundance or by keeping the decay rate of this species unchanged as the environmental parameter increases.
This type of abundance management can eliminate the tipping point and delay the occurrence of total extinction, as shown in Fig. \ref{controlTippingpoint}. The duration of the delay depends on the particular species chosen as the target. All species can be ranked by the amount of delay they induce, and the amount of delay characterizes control efficacy. Their rankings are determined solely by network structure and are not associated with the intrinsic network dynamics \cite{jiang2019irrelevance}.

In the absence of abundance management, a hysteresis loop arises when attempts to restore a species population are made by improving the environment; \textit{i.e.}, this loop causes the environmental parameter to change in the direction opposite to the one that led to extinction. In such a case, to return the species abundances to the original level, the environmental parameter must be further away from the tipping point; \textit{i.e.}, the environment must be significantly more favorable than it was before the collapse. However, with abundance management, the hysteresis loop disappears, and species recovery begins when the species is at the point of global extinction. Notably, when the environmental parameter is the mutualistic interaction strength, species cannot recover without abundance management, yet full recovery can be achieved with it \cite{jiang2019harnessing}. Additionally, the species recovery point can be predicted reasonably well by a two-dimensional reduced model \cite{jiang2018predicting} derived under the condition that abundance control/management occurs. The management or risk mitigation strategy to maintain the abundance of certain pollinator species may be realized through the use of robotic pollinators \cite{chechetka2017materially} to expedite recovery \cite{rundlof2015seed}; this may help address the devastating problem of the relatively sudden disappearance of bee colonies that is currently occurring all over the world.

\begin{figure}[!ht]
\centering
\includegraphics[width=0.95\linewidth]{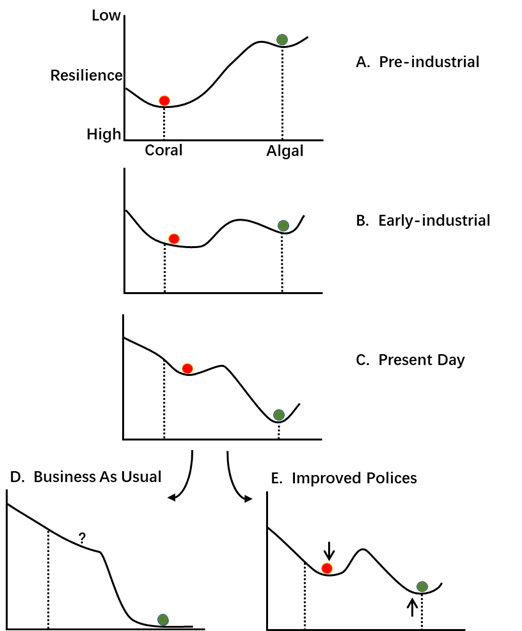} % figure 27
\caption{
Alternate states exist in coral reefs in different time periods. The condition and composition of the coral-dominated state has changed over time from a pristine state (the changes are indicated by the dashed vertical line above ``coral''). From (A) preindustrial times through (B) early industrial times to (C) the present, the coral-dominated reef state has become less common, and it could become uncommon in the future (D) under a ``business as usual'' scenario. However, if appropriate policies are implemented (E), then more reefs may be maintained in or shifted back to the coral-dominated state, with a concomitant reduction in the resilience of the algae-dominated state and increased resilience of the coral-dominated state.
This figure has been reproduced from Ref. \cite{graham2013managing}.
}
\label{resilience_management}
\end{figure}

\noindent
\textit{\textbf{Ecosystems restoration through interventions}}
In addition to studies that deal with preventing or delaying tipping points, another area of research related to resilience management focuses on recovering or restoring ecosystems that are already in an unwanted state or are prone to tipping points \cite{graham2013managing}. Here, we will review both (1) empirical and (1) theoretical studies of ecosystem restoration.

\noindent
(1) Empirical examples and ecosystem management.
Taking the system of coral reefs, which were in pristine condition in the preindustrial period, as an example, many coral reef ecosystems have been degraded to resilient assemblages that are no longer dominated by live coral, and reversing this degradation will require a reduction in human pressures on reefs and improved management of ecosystem processes that reverse the degraded condition and promote corals \cite{graham2013managing}.
The resilience of reef states can be expressed as deep or shallow valleys in a stable landscape, as shown in Fig. \ref{resilience_management}. Deeper valleys indicate higher levels of resilience, whereas shallow valleys are indicative of low resilience. As natural and anthropogenic drivers have changed reef systems, the coral-dominated state has become less resilient, while the algae-dominated state has become more resilient.
If we allow business to proceed as usual without management, then very few reefs will be maintained in a coral-dominated state. If appropriate policies are implemented, then more reefs can be sustained or returned to a coral-dominated (but non-pristine) state, providing society with critical goods and services. Nevertheless, recovering or rebuilding the community composition of coral reefs is very difficult. It requires scientists, policymakers, managers and resource users to cooperate in developing long-term commitments to improving reef management.
In addition to appropriate policies, it is also vital to act early since once a system has crossed a threshold, transition to a new stable state may take many years. Moreover, it has been found that exceptional weather events such as hurricanes may also create windows of opportunity for phase-shift reversals on coral reefs \cite{graham2013managing}. If we restrict fishing \cite{macneil2015recovery}, establish networks of herbivore management areas \cite{chung2019building}, and take advantage of shocks, then it may be easier to rebuild a coral-dominated state in coral reefs.

\noindent
(2) Theoretical approaches and ecosystem revival.
Once a networked system shifts to an undesired state, reversing the topological damage, namely retrieving the lost nodes/edges or strengthening the weakened interactions, may not guarantee the spontaneous recovery to the desired state. Indeed, many systems remain in the dysfunctional state, despite reconstructing their damaged topology since they exhibit a hysteresis phenomenon. Sanhedrai et al. develop a two-step recovery scheme to address this challenge: first - topological reconstruction to the point where the system can be revived, then dynamic interventions to reignite the system's lost functionality~\cite{sanhedrai2022reviving}. Applied to a range of nonlinear network dynamics, they identify a complex system's recoverable phase, a state in which a microscopic intervention can reignite the system, \textit{i.e}.\ controlling just a single node. A failed system may also recover to their desired states under stochastic environmental conditions\cite{ma2021universality}. Recently, Cheng et al. show that nucleation theory can be employed to advance resilience restoration in spatially-embedded ecological systems. They find that systems may exhibit single-cluster or multi-cluster phases depending on their sizes and noise strengths. They also discover a scaling law governing the restoration time for arbitrary system sizes and noise strengths in two-dimensional systems.

Furthermore, since all the systems on the planetary scale are becoming ever more networked and interdependent due to human interventions, there is a growing need to focus on managing resilience in ecosystems worldwide, considering both studies discussed above. It is also vital to embrace the novel conditions and propose realistic goals for resilience management suitable for rapidly changing environments. Thus, adaptive governance \cite{schultz2015adaptive} has been suggested as a practical approach, as it rests on the assumption that landscapes and seascapes must be understood and governed as complex social-ecological systems rather than as ecosystems alone.

\subsection{Ecological resilience and engineering resilience}
In the discussion of multiple stable states (or alternative stable states), a crucial feature termed ``resilience'' has attracted considerable attention. Resilience refers to the ability of a system to retain its basic functionality when internal changes or external perturbations occur. In 1973, Holling \cite{holling1973resilience} introduced the notion of resilience and used it to characterize the degree to which a system can endure perturbations without collapsing or shifting to a new stable state.
Since then, different authors have used ``resilience'' in different ways, leading to a great deal of confusion about this term.
As shown in Fig. \ref{Basin_Ball_scheme}, Beisner et al. \cite{beisner2003alternative} associated resilience with the features of a basin that act to retain the states of species (the location of the ball). These features are the steepness of the slope and the width of the basin. From one perspective, the steepness of the sides of the basin affects the time it takes for the ball to return to the lowest point in the basin and the rate at which the system returns to a single steady or cyclical state following a perturbation. This point of view is reflected in the term ``engineering resilience'', which was used by Holling et al. \cite{peterson1998ecological}. Engineering resilience focuses on the behavior of a system when it remains within the stable domain that includes this steady state \cite{peterson1998ecological}, meaning that it concentrates on stability near an equilibrium steady state.
From another perspective, the width of the basin also affects the movement of the ball, and the ball can only move out of the basin if its movement undergoes sufficiently large perturbations.
The degree of perturbation of the state variables affects the likelihood that the ball escapes from the basin. This point of view is related to ``ecological resilience'' \cite{holling1996engineering}, which emphasizes conditions that lead to a state that is far from any equilibrium steady state; under such conditions, instabilities can flip a system into another regime of behavior.

Engineering resilience is probably the most frequently invoked meaning or definition of resilience \cite{martin2011regional}. It assumes that the system is in equilibrium before disturbances are introduced, and the system's resilience is defined in terms of its stability when it is close to its steady state. Hence, engineering resilience denotes the system's ability to return to its preperturbed steady state \cite{song2015integrating}.
The concept of engineering resilience measures the time it takes for a system to recover to its preperturbed state 
or the relative change in its recovery back to equilibrium after a disturbance.
For example, the resilience triangle paradigm based on lost functionality and recovery time has been used to quantify a system's resilience \cite{zobel2011representing}.
Rose \cite{rose2007economic} used the time-dependent aspects of recovery in the definition of dynamic resilience. Ouyang et al. \cite{ouyang2015resilience} assessed the resilience of interdependent infrastructure systems during a period that included a damage propagation stage and a recovery stage.
In the psychological literature, engineering resilience has been used to characterize an individual's capacity to rebound or ``bounce back'' to his or her original state following stressful experiences \cite{smith2008brief}.
In natural systems, resilience has been used to explain equilibrium states related to water column disturbances in microbiology, the dynamic restoration of coastal dunes, and the restoration of critical ecosystem services \cite{shade2011resistance}.

Ecological resilience, or system resilience, quantifies a system's ability to undergo a disturbance without changing its structure, identity, or functions \cite{holling2002resilience}. It focuses on the magnitude of the disturbance that a system can withstand without shifting to another regime \cite{martin2011regional}. Hence, the corresponding system simultaneously monitors and reorganizes the processes that govern the system's behavior while it accommodates or resists the disturbance \cite{holling2002resilience}.
Note that this term implies the potential existence of multiple stable states. If the perturbation is small, then the system could return to the preperturbation state. However, if a shock perpetuates changes in conditions that exceed some intrinsic threshold, then the system changes regimes such that the structure or function of the system becomes fundamentally different. Moreover, when a threshold is crossed, returning to the previous state is difficult \cite{standish2014resilience}.
For example, many ecological systems may undergo abrupt transitions to alternative regimes \cite{scheffer2001catastrophic}. Such transitions are often irreversible and widespread in human health \cite{venegas2005self}, the economy \cite{perrings1998resilience} and the environment \cite{may1977thresholds}.

In the psychological literature, ecological resilience has been recognized as an individual's capacity to be robust, to demonstrate confidence in one's strengths and abilities, and to be stoic, resourceful, and determined as one navigates through key challenges across and within one's life \cite{golubovich2014safety}. In ecology and biology, ecological resilience is represented by bacterial responses to water column disturbances \cite{shade2011resistance}, the strong responses related to receptors and their mechanisms of action \cite{chaumot2012molecular}, and the permanence of farm systems \cite{paniagua2013farmers}.
The concept of ecological resilience implies a need to predict the critical points (or ``tipping points") at which the system loses its resilience and to uncover the mechanisms underlying such critical transitions. In short, engineering resilience focuses on efficiency, constancy, and predictability, while ecological resilience focuses on persistence, change, and unpredictability \cite{holling1996engineering}. They are alternative paradigms that characterize two different fundamental aspects of a system's ability to maintain its functionality.

Engineering resilience is most useful for a system that exists near a single or global equilibrium condition. In the tradition of engineering, systems are designed with a single operating objective, which is usually associated with a single steady state \cite{waide1976engineering}. This means that we can always repair a system that can return to its regular operation, even under large perturbations.
Resilience in fields such as engineering, physics, control system design, and material engineering refers to engineering resilience \cite{gunderson2000ecological}.
For a system that contains multiple stable states, ecological resilience is a more meaningful measure, compared to engineering resilience, for ecological systems. In ecology, resilience has been defined as ``a measure of the persistence of systems and of their ability to absorb change and disturbance and still maintain the same relationships between populations or state variables'' \cite{doring2015resilience}. Note that there are some other studies about engineering resilience in ecological systems.
Once the disturbances exceed a tipping point, the system shifts to another regime.

%\begin{comment}
\section{Phase transitions in biological networks}\label{Biology}

Living organisms owe their existence to biochemical processes that form various types of biological networks such as gene regulatory networks, protein interaction networks, metabolic networks at the molecular level \cite{liu2014detection}, cell interaction networks at the cellular level \cite{kirouac2009cell}, and disease networks \cite{huttlin2017architecture} at the phenotypic level. All of these biological networks evolve dynamically and may have inherent nonlinearity. Their responses to internal signals or external perturbations are usually not gradual but show switch-like behavior \cite{laurent1999multistability}. It is found that biological networks may shift abruptly from one state to another during processes such as cell fate induction~\cite{chang2011systematic} and the onset of diseases or cancer \cite{yu2016physical}. Such state transitions can be modeled as a stability landscape \cite{scheffer2001catastrophic} and analyzed using the ecological resilience framework \cite{holling1973resilience}. As discussed in Chapter \ref{Ecology}, abrupt shifts in the state are usually associated with the existence of multiple stable states \cite{beisner2003alternative}. This section reviews (1) the research about multiple-stable-states phenomena exhibited in biological networks (Sec. \ref{BioSecBistability}), including mathematical models and empirical evidence; (2) various levels/scales of biological systems (Sec. \ref{BioSecResilience}), such as microscopic genetic circuits, mesoscopic organisms, and macroscopic complex multicellular organisms; and (3) prediction of resilience loss in biological systems (Sec.\ref{3Indicatorresilience}).

\subsection{Bistability in biological systems}\label{BioSecBistability}
Similar to the multiple stable states in ecosystems discussed in Chapter \ref{Ecology}, a system-level property called ``bistability'' (or, more generally, multistability) universally exists in all biological systems \cite{gardner2000construction}. This property may be of particular relevance to biological systems that switch between discrete, alternative stable states; generate oscillatory responses; or ``remember'' transitory stimuli \cite{angeli2004detection}.
Uncovering the mechanisms underlying ``bistability'' is crucial for understanding basic cellular and biochemical processes, such as cell cycle progression, cellular differentiation, and cellular apoptosis \cite{ghaffarizadeh2014multistable}, and the onset of disease and cancer, as well as in the origin of new species \cite{wilhelm2009smallest}.
Due to the prevalence and fundamental importance of bistability in biochemical systems, many theoretical and experimental studies have attempted to identify the necessary conditions for a signal transduction pathway to exhibit bistability \cite{ferrell2002self}, which is an emergent phenomenon in networks of biochemical reactions rather than a property of single molecules or single reactions.

\subsubsection{Generators of biological bistability}\label{BioSecBistabilityGener}

In cell signaling \cite{liu2014identifying}, most (or perhaps all) biochemical reactions are reversible. For instance, DNA methylation and DNA demethylation occur, and proteins are phosphorylated and dephosphorylated, respectively. However, many biological transitions, such as cell differentiation \cite{wang2009bistable}, the cell cycle, and immune stress response \cite{bhattacharya2010bistable}, are essentially irreversible. A crucial question has been raised: how might reversible reactions lead to practically irreversible changes to cell fate \cite{ferrell2002self}?

Monod and Jacob \cite{monod1961general} proposed that the answer to the above question lies in the way in which gene regulatory systems are wired; in such systems, feedback loops are required mechanisms for producing biological bistability and irreversibility. Different types of signal transduction circuits, such as pairs of natural transcriptional repressors wired to inhibit one another \cite{lebar2014bistable} and positive feedback loops composed of activators sharing the same opposing repressors in gene regulation, can achieve such goals.
For example, let us consider two gene products, P1 and P2, each of which inhibits the transcription of the gene encoding the other product. Thus, the system has a stable state in which P1 is on and P2 is off or an alternative stable state in which P1 is off and P2 is on. Once either stable state has been established, the system could remain in such a state until an external stimulus pushes the system to transition to the alternative stable state. Similar behavior is presented in positive feedback loops (mutual activation toggle-switch motifs \cite{sabouri2008antagonism}); a system of that type can toggle back and forth between a state in which both A and B are off and a state in which both A and B are on \cite{ferrell2002self}, thereby showing the ability to ``remember'' a transient differentiation stimulus long after the triggering stimulus has been removed.

\begin{figure}[!ht]
\centering
\includegraphics[width=0.95\linewidth]{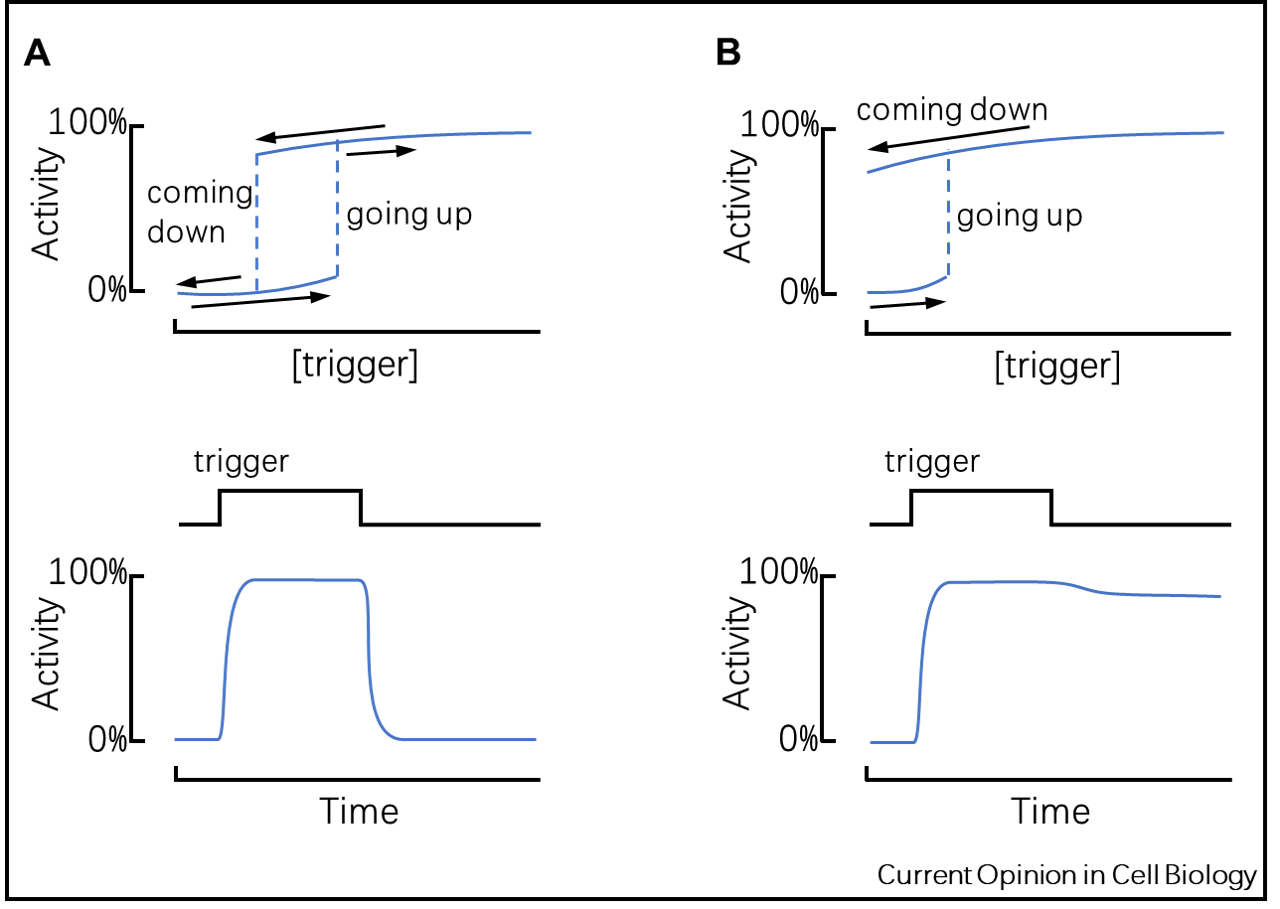} % figure 28
\caption{
 Hysteresis and irreversibility in bistable signaling circuits. (A) Any bistable
circuit should exhibit some degree of hysteresis; this means that different stimulus/response curves are obtained depending on the system's previous state. (B) Irreversibility is achieved when a bistable system has very strong feedback.\\
\textit{Source:} The figure has been modified from \cite{ferrell2002self}.
}
\label{Hysteresisirreversibility}
\end{figure}

Feedback alone does not guarantee that a system is bistable. A bistable system must also have some type of nonlinearity within the feedback circuit \cite{ferrell2002self}. Feedback circuit enzymes must respond to their upstream regulators in an ``ultrasensitive'' manner in which continuously varying signals are converted into discrete outputs (ON or OFF responses) \cite{ferrell1996tripping}. A bistable circuit always exhibits some degree of hysteresis, indicating that the system has alternative stable states in the presence of a specific stimulus. The response to this stimulus is related to the system's previous state; it is more difficult to flip the system between states than to maintain the system in its flipped state \cite{ferrell2002self}, as shown in Fig. \ref{Hysteresisirreversibility}A. In such a case, the bistable switch is a two-way switch (the system can be switched back and forth between ON and OFF states); this is characteristic of metabolic pathways such as the lac operon \cite{sabouri2008antagonism}. If the feedback in a circuit is sufficiently strong, then the system may exhibit irreversibility (Fig. \ref{Hysteresisirreversibility}B). Such a system may remain in the flipped state indefinitely after the triggering stimulus is removed. In this case, the bistable switch is a one-way switch (once flipped, it cannot be turned back); switches of this type play major roles in developmental transitions such as apoptosis.

Although feedback regulation has been considered a prerequisite for bistable behavior \cite{wang2009bistable}, it is not a necessary component of switching phenomena. For example, 
in protein kinase cascades, bistability and hysteresis can arise solely through a distributive kinetic mechanism of two-site mitogen-activated protein kinase (MAPK) phosphorylation and dephosphorylation with no apparent feedback \cite{markevich2004signaling}.
In summary, epigenetic bistability appears to be at the heart of decisive, irreversible biological state transitions such as cell differentiation and the cell cycle \cite{pomerening2003building}. Transitions of this type arise from feedback loops and ultrasensitivity \cite{wilhelm2009smallest} or from particular network topology with a specific range of parameters.

\subsubsection{Mathematical models of biological bistability}

Computational modeling and the theory of nonlinear dynamical systems allow one to describe bistability and understand why it occurs \cite{ferrell2011modeling}. Below, we review two classical mathematical models that were developed to show how bistability arises.

\begin{SCfigure*}
\centering
\includegraphics[width=0.6\textwidth]{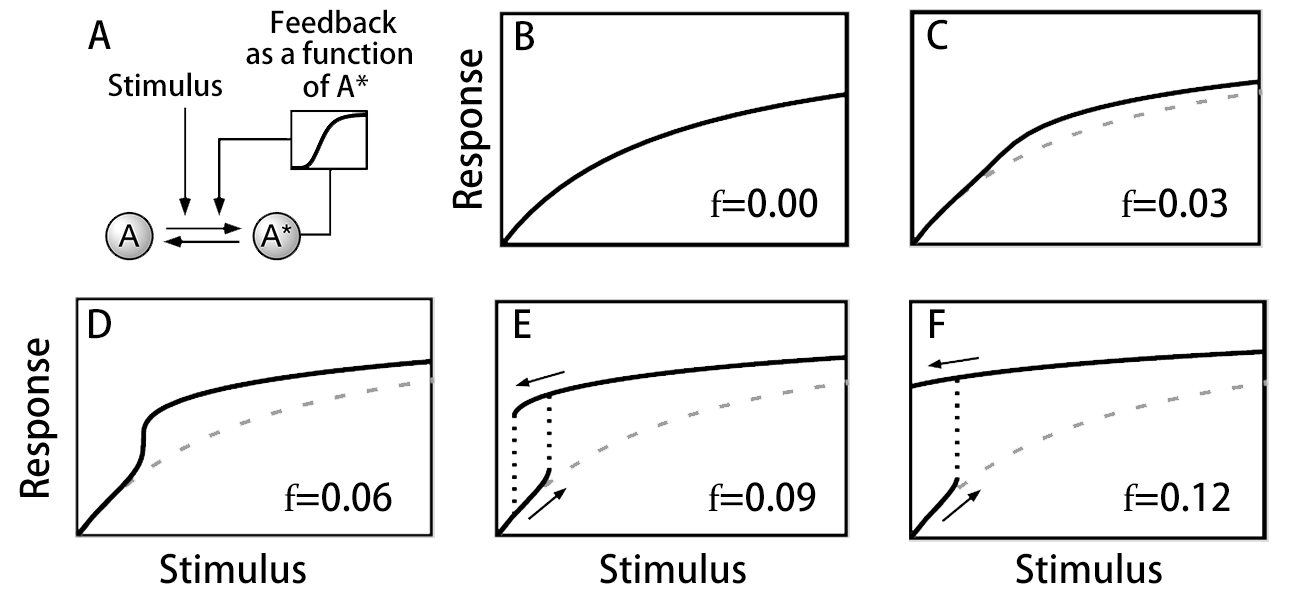} % figure 29
\caption{
A simple positive feedback loop and its stimulus-response curve for different feedback strengths. (A) Genetic circuit of a positive feedback loop. (B-F) Stimulus-response curves calculated numerically from Eq. \ref{SimplePosFeebback} under different feedback strengths $f$; the other parameters were set as $n=5$, $K=1$ and $k_{\rm inact}=0.01$.\\
\textit{Source:} The figure is from \cite{xiong2003positive}.
}
\label{positivefeedbackLoop}
\end{SCfigure*}

\noindent
\textbf{A simple positive feedback loop (a one-ODE model).} As discussed above, the most common mechanism for generating bistability is the existence of positive feedback loops that direct two mutually exclusive cell states. A positive feedback loop may be formed from one or two signaling proteins; the simplest such loop consists of a single signaling protein that can be reversibly switched between an inactive form (A) and an active form (A*). As shown in Fig. \ref{positivefeedbackLoop}A, the transition between A and A* is assumed to be regulated by an external stimulus and by positive feedback between the transition and the stimulus. This process can be modeled by the following ordinary differential equation (ODE) \cite{xiong2003positive}:
\begin{equation} \label{SimplePosFeebback}
\begin{split}
\frac{d[{\rm A*}]}{dt}= & \{{\rm stimulus}\times([{\rm A_{tot}}]-[{\rm A*}])\} \\ 
                 & +f\frac{[{\rm A*}]^n}{K^n+[{\rm A*}]^n}-k_{\rm inact}[{\rm A*}]
\end{split}
\end{equation}
in which the first term represents the basal transcription rate due to an external stimulus, the second term denotes the effect of the positive feedback, which is modeled by a nonlinear Hill equation in which $K$ is the effector concentration needed for half-maximum response, $n$ denotes the Hill coefficient, and parameter $f$ represents the strength of the feedback. The last term represents the inactivation of A* at a degradation rate $k_{\rm inact}$.

The shape of the nonlinear stimulus-response curve described by Eq. \ref{SimplePosFeebback}
 depends on the value of $f$ (the strength of the feedback). When $f=0$, there is no feedback in the system, and it responds in a way that follows a monostable smooth Michaelian curve, as shown in Fig. \ref{positivefeedbackLoop}B. As feedback strength $f$ increases, more nonlinearity, or ultrasensitivity, is introduced, and the stimulus-response curve acquires a sigmoidal shape; however, the system is still monostable, as shown in Fig. \ref{positivefeedbackLoop}C and \ref{positivefeedbackLoop}D. If $f$ continues to increase, then the system shows hysteresis and becomes bistable for some stimulus values, as shown in Fig. \ref{positivefeedbackLoop}E. Eventually, the feedback becomes so strong that the system's response is no longer reversible, as shown in Fig. \ref{positivefeedbackLoop}F.
This simple model can successfully explain the bistability or irreversibility of many biological processes, such as the maturation of \textit{Xenopus oocytes} \cite{xiong2003positive}, induced osteogenic differentiation of a myogenic subclone \cite{wang2009bistable}, and the white-opaque switch in Candida albicans \cite{huang2006bistable}.

\begin{figure}[!ht]
\centering
\includegraphics[width=0.95\linewidth]{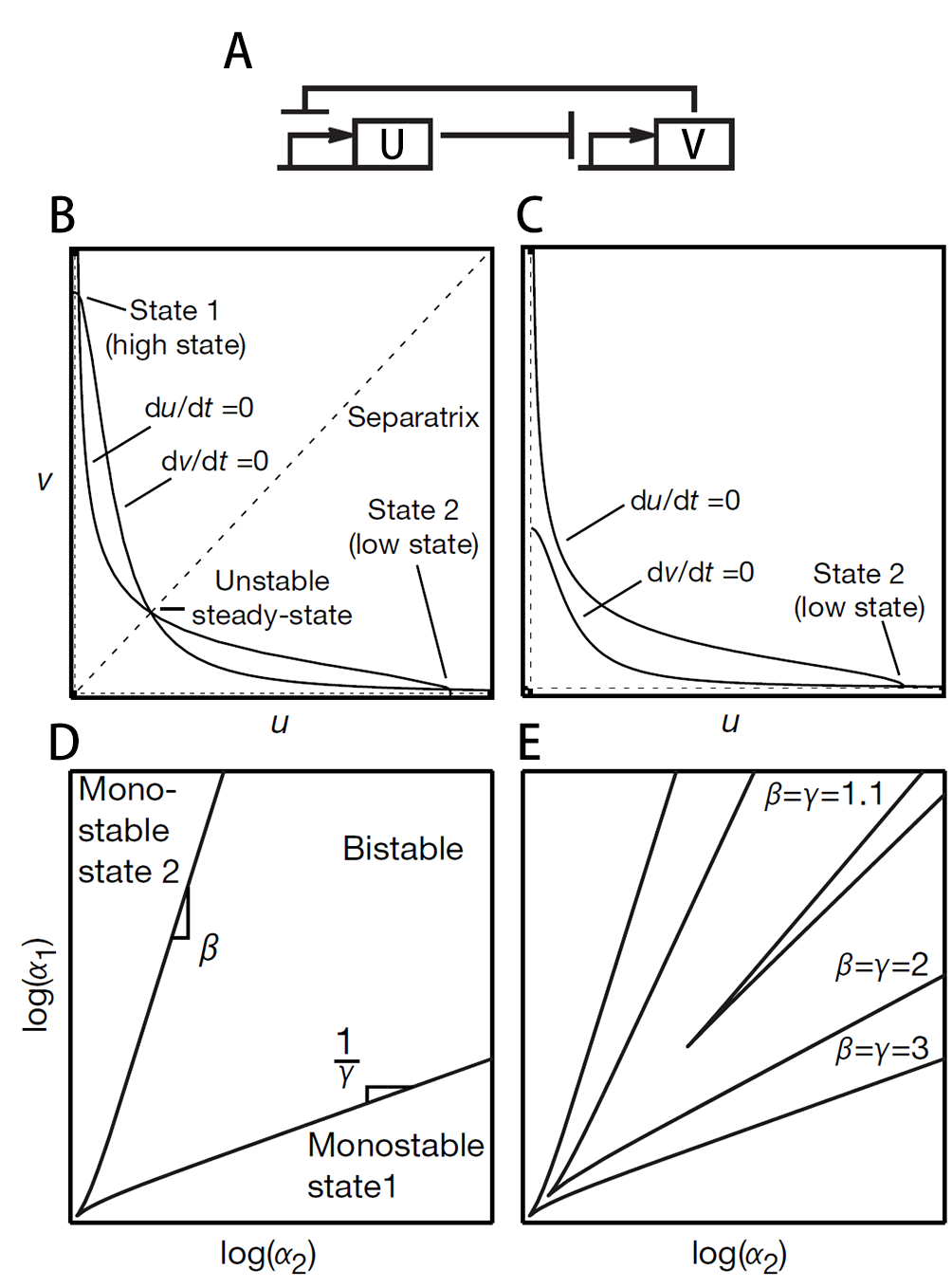} % figure 30
\caption{
Geometric structure of the two-ODE model. (A) Schematic illustration of the mutual repression network. (B) One unstable and two stable states appear when there are balanced rates of synthesis of the two repressors. (C) Monostable states arise when the rates of synthesis of the two repressors are unbalanced. (D) The bistable region. The lines mark the transition (bifurcation) between bistability and monostability. The slopes of the bifurcation lines are determined by exponents $\beta$ and $\gamma$ for large $\alpha_1$ and $\alpha_2$. (E) Reducing the cooperativity of repression ($\beta$ and $\gamma$) reduces the size of the bistable region. Other parameters are set to $k_1=k_2=1$ and $d_1=d_2=1$\\
\textit{Source:} The figure has been modified from \cite{gardner2000construction}.
}
\label{toggleswitch}
\end{figure}

\noindent
\textbf{A mutually inhibitory network (a two-ODE model).}	
As we previously mentioned, in addition to positive feedback loops, a double-negative feedback loop, also called a mutually inhibitory network, can generate bistability \cite{ferrell2002self}. Such a loop consists of two signaling proteins, repressors U and V, that inhibit each other's expression or activation, as shown in Fig. \ref{toggleswitch}A. Given a certain stimulus, the system can switch between two distinct states: one in which there is a high expression level of repressor U and a low expression level of repressor V and another in which there is a low expression level of repressor V and a high expression level of repressor U. The nonlinear dynamics of this system can be captured by the following two-ODE model, which is also called the ``toggle switch'' model \cite{gardner2000construction}:
\begin{equation} \label{twoODEmodel}
\begin{split}
\dfrac{dU}{dt}=\dfrac{\alpha_1}{1+k_1V^\beta}-d_1U\\
\dfrac{dV}{dt}=\dfrac{\alpha_2}{1+k_2U^\gamma}-d_2V
\end{split}
\end{equation}
The first terms in the equation represent the effect of negative feedback loops, and the second terms denote degradation/dilution of the repressors. $U$ and $V$ represent the concentrations of the two repressors; $\alpha_1$ and $\alpha_2$ represent the maximal production rate of repressors $U$ and $V$, respectively; $\beta$ and $\gamma$ are the cooperativity coefficients of repressors $V$ and $U$, respectively; parameters $k_1$ and $k_2$ describe the repression strengths, which are determined by the repressors' rates of binding to and dissociation from the relevant promoters; and parameters $d_1$ and $d_2$ represent the degradation rates of the repressors.

The equilibrium solutions of this model can be found by drawing the nullclines ($dU/dt=0$ and $dV/dt=0$). Suppose that the system has balanced synthesis rates of the two repressors and that the cooperativity coefficients $\beta, \gamma>1$. In this case, the nullclines are sigmoidal in shape and intersect at three points, indicating that the system exhibits bistability with one unstable and two symmetrical stable states, as shown in Fig. \ref{toggleswitch}B. However, if the synthesis rates are imbalanced, then the nullclines intersect only once, producing a monostable state (Fig. \ref{toggleswitch}C). The slopes of the bifurcation lines are determined by the cooperativity coefficients $\beta$ and $\gamma$. The size of the bistable region increases as the rate of repressor synthesis ($\alpha_1$ and $\alpha_2$) increases (Fig. \ref{toggleswitch}D) and decreases when $\beta$ and $\gamma$ decrease (Fig. \ref{toggleswitch}E). Note that $\beta$ and $\gamma$ represent the degrees of cooperativity in the binding. The case $\beta, \gamma=1$ corresponds to linear repression and uncooperative binding of monomers. In contrast, $\beta, \gamma>1$ requires the cooperative binding of two or more repressor proteins \cite{warren2005chemical} that must form a polymer or multiple repressors to cooperatively bind to promoters that possess more than one operator site \cite{schultz2008extinction}. Thus, cooperative binding is necessary to create double negative feedback loops that result in alternative stable states. 
	
\begin{SCfigure*}
\centering
\includegraphics[width=0.667\textwidth]{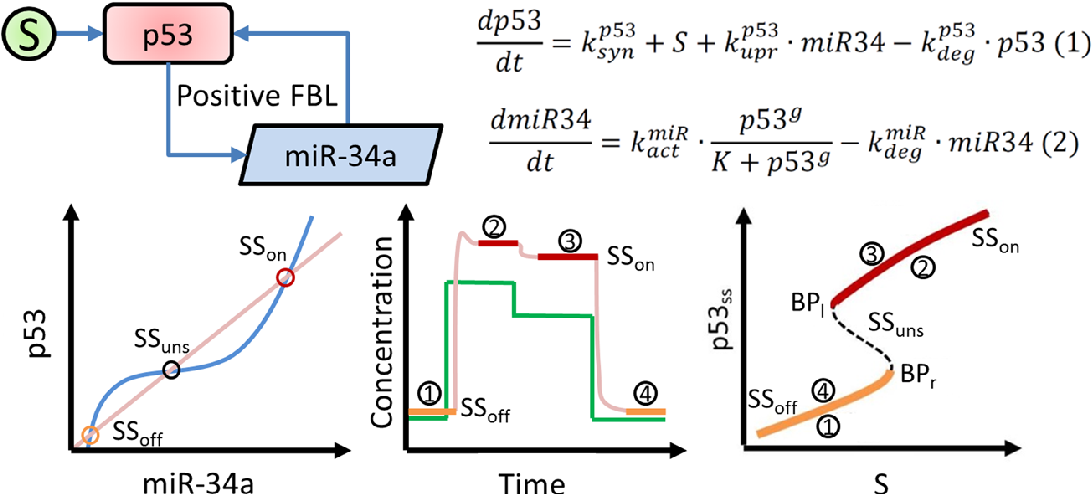} % figure 31
\caption{
Bistability in the p53/miR-34a feedback loop demonstrated using a mathematical model. After drawing the nullclines of two equations, one unstable and two stable states appear to show bistability. The middle bottom plot shows the evolution of p53 (the red line) and S (the green line) with time, and the bifurcation plot shows different steady states of p53 (p53ss) as a function of different intensities of S.\\
\textit{Source:} The figure is from \cite{lai2016understanding}.
}
\label{p53_miRNA_loop}
\end{SCfigure*}

Mutual inhibitory networks have been widely used to describe bistable phenomena that occur during the growth and development of organisms. For instance, the decision between epiblast and primitive endoderm fate in mammalian embryonic stem cells can be described by a simple mutual repression circuit modulated by FGF/MAPK signaling \cite{schroter2015fgf}. The lysogenic-lytic bistable switch in bacteriophage $\lambda$ can be described by similar models \cite{bednarz2014revisiting}. Furthermore, such models can be used to engineer artificial gene networks in mammalian cells. The developed epigenetic circuitry can switch between two stable transgene expression states after transient administration of two alternate drugs \cite{kramer2004engineered}, enabling precise and timely molecular interventions in gene therapy.

In addition to the two basic mathematical models reviewed above, various derived models have been proposed to describe bistability or irreversibility in various biological processes. For example, the bistability in p54 stable states can be explained by a positive feedback loop composed of the transcription factor p53 and a microRNA, miR-34a, in which p53 upregulates the transcription of miR-34a. Moreover, in turn, miRNAs indirectly upregulate p53 expression by repressing SIRT1, a negative regulator of p53 \cite{lai2012modeling}. Such a process can be modeled by a two-dimensional ODE model \cite{lai2016understanding}, as shown in Fig. \ref{p53_miRNA_loop}. Martinez et al. \cite{martinez2018bistable} constructed a minimal delay-differential equation (DDE) model with a region of bistability that causes the \textit{subtilis} biofilms to jump from a stable state to an oscillatory growth attractor when perturbations occur.
Wang et al. \cite{wang2009bistable} used a bistable switch model to analyze the observed differentiation behavior of human marrow stromal cells. Bala et al. \cite{bala2018bistability} proposed a modification to an existing mathematical model of mitosis-promoting factor control in \textit{Xenopus oocyte extract} and used MPF as a bifurcation parameter, giving rise to bistability in the MPF activation module. These theoretical studies of biological bistability have advanced our understanding of the possible mechanisms that underlie critical cellular processes such as the cell cycle and cell differentiation.
Next, we review the empirical studies on bistability in the growth and development of cells.

\subsubsection{Empirical studies of biological bistability} \label{EmpiricalBistability}
Bistable switches are sufficient for encoding more than two cell states without rewiring of the circuitry \cite{fang2018cell}; this has been found experimentally to be at the core of determination of state transitions in the cell cycle and cell differentiation.

\begin{figure*}
\centering
\includegraphics[width=0.95\linewidth]{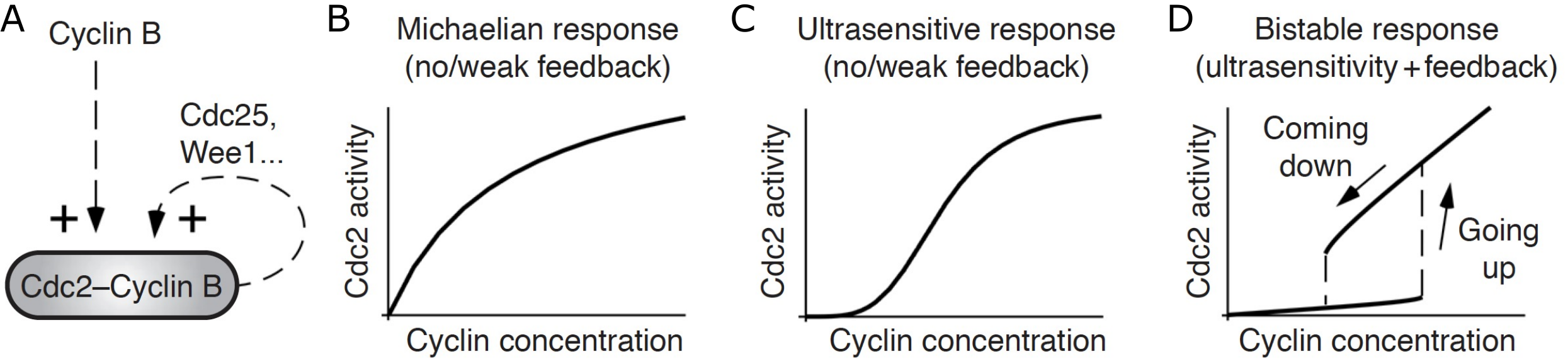} % figure 32
\caption{
Cdc2 responds to the cellular concentration of nondegradable cyclin in three ways. (A) A Michaelian response is expected if cyclin has directly activated Cdc2. (B) An ultrasensitive response can arise from multistep activation mechanisms due to the presence of stoichiometric inhibitors or the occurrence of saturation effects. (C) A bistable response can arise from a combination of ultrasensitivity and positive feedback.\\
\textit{Source:} Figure from \cite{pomerening2003building}.
}
\label{Feedback_ultrasensitivity}
\end{figure*}

\noindent
\textbf{Bistability in the cell cycle}
The cell cycle of early-stage embryos provides an example of a robust biological oscillator \cite{pomerening2005systems}. Sustained oscillations are generated by interlinked feedback loops \cite{coudreuse2010driving}. In the essential negative feedback loop, the cell-division cycle protein kinase Cdc2 activates the anaphase-promoting complex (APC), leading to the destruction of cyclin and the inactivation of Cdc2 \cite{pomerening2005systems}. Under some circumstances, a long negative-feedback loop is by itself sufficient to produce oscillations \cite{goldbeter2002computational}. However, operating alone, the negative feedback loop in the Cdc2/APC system alone may produce damped oscillations \cite{pomerening2005systems}. Pomerening et al.\cite{pomerening2005systems} experimentally demonstrated the presence of positive feedback loops in the Cdc2/APC system. In one of these positive feedback loops, Cdc2 mediates the activation of Cdc25 and inactivation of Wee1 and Myt1, enabling the system to function as a bistable system \cite{novak1993numerical} by toggling between two discrete alternative stable states and showing hysteresis, as shown in Fig. \ref{Feedback_ultrasensitivity}. The introduction of bistability ensures that the Cdc2/APC system produces sustained oscillations without approaching a stable steady state. 
Next, we show the mechanism through which bistable switches govern mitotic control in the cell cycle.

\begin{SCfigure*}
\centering
\includegraphics[width=0.8\textwidth]{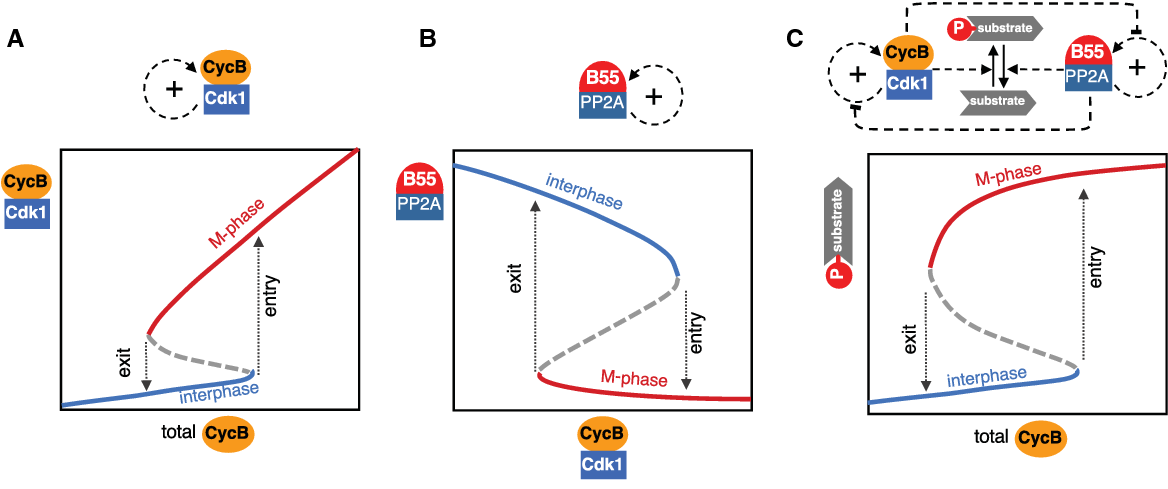} % figure 33
\caption{
Bistable switches in mitotic control. Schematic signal-response diagram for (A) Cdk1 autoactivation, (B) PP2A:B55 feedback regulation and (C) mitotic substrate phosphorylation by interlinked kinase-phosphatase switches
.\\
\textit{Source:} The figure is from \cite{rata2018two}.
}
\label{Mitotic_Control}
\end{SCfigure*}

Cdc2 (also referred to as Cdk1) and cyclins form a stoichiometric complex and play a key role in the control of the G2/M transition of the cell cycle \cite{doree2002cdc2}. Positive feedback in the Cdk1 activation loop, the major Cdk1-counteracting phosphatase PP2A:B55, and feedback regulation work as bistable switches. They generate different thresholds for the transition between interphase and M phase cell cycle states. As shown in Fig. \ref{Mitotic_Control}, there are distinct thresholds for mitotic entry and mitotic exit; this provides robustness of the M phase state and prevents cells from reverting to the interphase state in the noisy cellular environment \cite{rata2018two}.
In addition, the two distinct states, interphase and M phase states, are created by the action of bistable switches; they can be stabilized by positive feedback loops that prevent the cell from coming to rest in an intermediate transitional state \cite{oikonomou2010frequency}.

\noindent
\textbf{Bistability in cell fate switching}
Cell fate decision-making is the process through which a cell commits to a differentiated state in growth and development. In this process, the bistable switch mechanism is prevalent in directing two mutually exclusive cell fates \cite{angeli2004detection}. One typical example is the maturation of \textit{Xenopus} oocytes; the immature oocyte represents a default fate, and the mature oocyte represents an induced fate \cite{abrieu2001interplay}. At the biochemical level, oocyte maturation is controlled by p42 MAPK and cyclin B/Cdc2, which are known to be organized into positive feedback loops \cite{ferrell2002self, xiong2003positive}. For example,
Mos activates p42 MAPK through the intermediacy of MEK. Once active, p42 MAPK promotes the accumulation of Mos \cite{gotoh1995initiation}. Through such positive feedback, p42 MAPK makes an all-or-none, bistable response to progesterone or to microinjected Mos \cite{ferrell1998biochemical}. Xiong et al. \cite{xiong2003positive} provided experimental evidence that the p42 MAPK and Cdc2 systems, with their strong positive feedback loops, produce bistability and generate irreversible biochemical responses after the cell receives transient stimuli \cite{xiong2003positive}.

\begin{SCfigure*}
\centering
\includegraphics[width=0.6\textwidth]{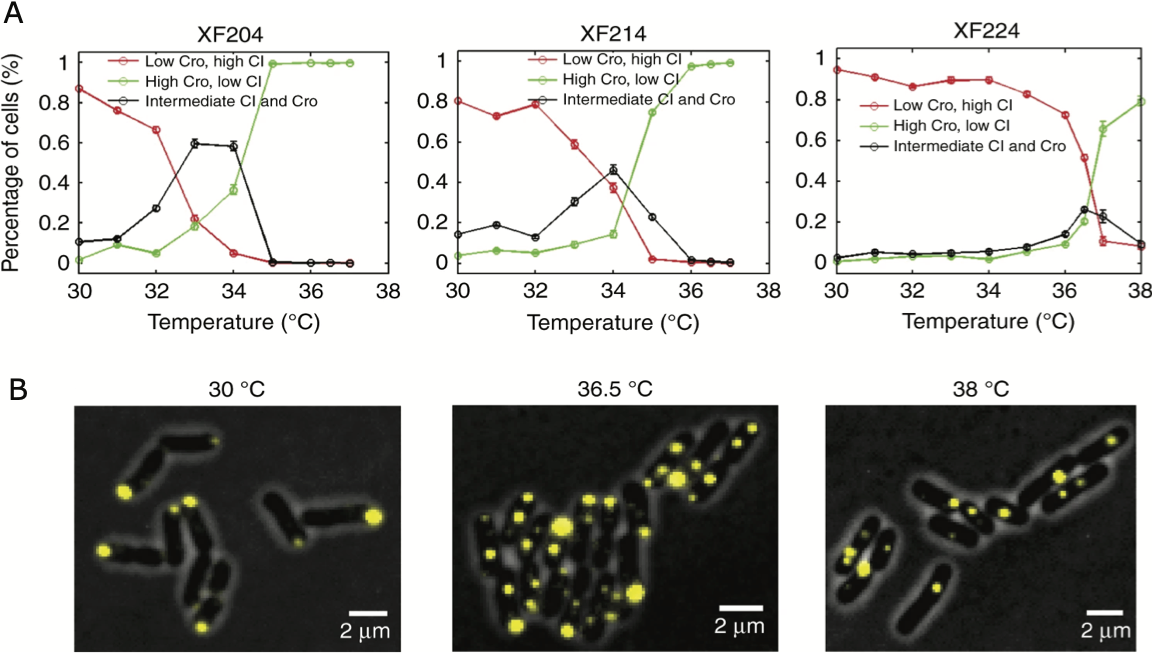} % figure 34
\caption{
Expression levels of CI and Cro in strains XF204, XF214, and XF224 at different temperatures result in more than two expected cell populations.
(A) Percentages of cells expressing only CI (red), only Cro (green), or both CI and Cro (black) in strains XF204, XF214, and XF224 at different temperatures. (B) Representative fluorescence images of XF224 cells showing CI expression (yellow pole-localized Tsr-Venus-Ub spots) and Cro expression (yellow quarter/midcell-localized LacI-Venus-Ub spots) at low, intermediate, and high temperatures. The fluorescence images are overlaid with phase-contrast cell images (gray).\\
\textit{Source:} The figure is from \cite{fang2018cell}.
}
\label{Lambda_CI_Cro}
\end{SCfigure*}

Another example of experimentally verified bistability in cell differentiation is the classic bistable bacteriophage $\lambda$ switch \cite{st2008determination}. This switch is composed of two mutually repressive TFs, CI and Cro; the levels of expression of these TFs determine cell fate developmental decision-making as the bacteriophage infects its host \cite{trinh2017cell}. Expression of CI but not that of Cro confers lysogenic growth, and expression of Cro but not that of CI confers lytic growth \cite{fang2018cell}. Fang et al. \cite{fang2018cell} experimentally demonstrated the emergence of two new expression states in a model of the bistable switch of bacteriophage $\lambda$. They constructed strains XF204, XF214 and XF224, in which the expression level of Cro follows the order ${\rm [Cro]_{XF224}<[Cro]_{XF214}<[Cro]_{XF204}}$. The investigators quantified the expression levels of CI and Cro in individual cells of the three strains at different temperatures for more than 20 generations. As shown in Fig. \ref{Lambda_CI_Cro}, typical bistable behavior was observed. At low temperatures, cells predominately exhibited low Cro and high CI levels (red lines). They flipped to high Cro and low CI levels (green lines) at high temperatures. Between the two extremes, cells with high Cro and high CI levels (black lines) also appeared due to the different rates of switching of the expression levels of Cro and CI \cite{fang2018cell}.

In addition to experimental studies of cell fate switches in unicellular organisms and viruses, many studies show that bistability commonly exists during cellular differentiation of multicellular organisms. For example, Wang et al. \cite{wang2009bistable} used a human bone marrow stromal cell subclone to study myogenic and osteogenic differentiation. They found that BMP2-induced osteogenic differentiation of these cells exhibits a threshold effect and an all-or-none response that could be successfully analyzed using a bistable switch model \cite{ozbudak2004multistability}. Bhattacharya et al. \cite{bhattacharya2010bistable} showed that two mutually repressive feedback loops could generate a bistable switch capable of directing B cells to differentiate into plasma cells. In the process, the differentiated cells must execute certain biological functions that require them to have a low dedifferentiation rate \cite{cai2007dedifferentiation}. The mechanism that prevents differentiated cells from dedifferentiating is positive feedback in cell signaling pathways. Ahrends et al. \cite{ahrends2014controlling} showed that ultrahigh feedback connectivity exists in mammalian preadipocytes; these cells support more than six positive feedback loops that together lead to a slow rate of differentiation.

In summary, bistability is generally considered a control mechanism for biological processes; it is valuable in many situations, including cell differentiation \cite{ chang2006multistable, goldbeter2007sharp, wang2009bistable}, lytic-lysogenic switching in bacteriophage $\lambda$ \cite{tian2004bistability}, immune stress response \cite{bhattacharya2010bistable}, and the lactose utilization network in Escherichia coli \cite{ozbudak2004multistability}. Bistability allows cells to respond to stimuli by undergoing a discrete transition to a single state chosen from among two or more distinct stable states. Knowing the bistability range and the conditions of responses within an organism helps us understand the inherent mechanisms through which these processes occur and advances our knowledge of biomolecular controllers.

\subsection{Resilience at multiple levels of biological networks}\label{BioSecResilience}
Resilience in a biological network is characterized by the ability of the network to maintain its functions and adjust its dynamics to ensure steady output when exposed to external and internal perturbations. Accurate prediction of the output of a dynamic biological network after perturbations is crucial for understanding signal fidelity in biological networks and designing noise-tolerant gene networks \cite{pedraza2005noise}. 
Next, we review studies on resilience in biological networks at multiple levels ranging from simple genetic circuits to complex multicellular organisms.

\subsubsection{Resilience of genetic circuits in the presence of molecular noise} \label{CircuitNoiseSec}
The theoretical and empirical studies of biological bistability reviewed in Section \ref{BioSecBistability} are nearly all deterministic. Here, we do not consider the common presence of stochastic fluctuations that are induced by extrinsic and intrinsic noise \cite{raser2005noise}. Extrinsic noise is communicated by exogenous sources, such as oscillatory cascades that regulate the progression of the cell cycle \cite{vujovic2019notch} and environmental stressors \cite{mitosch2017noisy}. Intrinsic noise can be interpreted as random fluctuations within an individual cell. Typically, these fluctuations alter the intensity of a signal, leading to an altered stoichiometric relationship between the input and the output signals \cite{raser2005noise}. Such fluctuations are inherent properties of transcriptional, posttranscriptional and translational dynamics \cite{balazsi2011cellular}. For example, replication-transcription conflict \cite{garcia2016transcription} and RNA polymerase backtracking mediated by R-loop formation \cite{skourti2014double} are stochastic events that are caused by intrinsic noise in an individual cell \cite{vujovic2019notch}. Intrinsic noise, or molecular noise, commonly exists in cells and enables the phenotypic diversification of completely identical cells that are exposed to the same environment \cite{balazsi2011cellular}.

\begin{SCfigure*}
\centering
\includegraphics[width=0.667\textwidth]{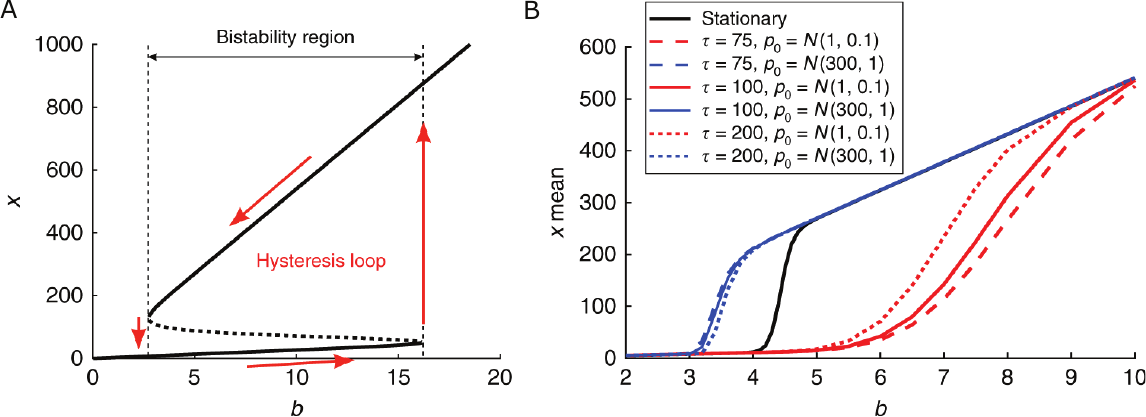} % figure 35
\caption{
Hysteresis in deterministic versus stochastic systems. (A) Hysteresis loop of a deterministic self-regulatory system without considering molecular noise. (B) Transient hysteresis in a stochastic self-regulatory system. $\tau$ is associated with the time scale of protein degradation. Slow transients lead to multiple mean states, resulting in transitory hysteretic behavior.\\
\textit{Source:} The figure is from \cite{pajaro2019transient}.
}
\label{Hysteresis_Stochastic}
\end{SCfigure*}

The inherent stochasticity of biochemical processes is inevitable and arises due to the random nature of chemical reactions within a cell \cite{van1992stochastic}. When only a few molecules of a specific type are present in a cell, the stochastic effect can become prominent \cite{raser2005noise}. For example, the feedback loops (toggle switch) that generate bistability, reviewed in the previous section, can randomly switch between two states in the presence of noise \cite{gardner2000construction}. Furthermore, fluctuations in small systems, especially in systems that involve components with low molecular concentrations, can lead to additional states that are essentially unstable \cite{schultz2008extinction}. An example of such instability is provided by the empirical study of the classic bistable bacteriophage $\lambda$ switch \cite{fang2018cell} reviewed above. The switch based on the expression levels of two mutually repressive transcription factors (CI and Cro) not only controls two stable states (low, high and high, low) but also shows two unstable states (high, high and low, low) that have different probabilities. Such stochastic dynamic behavior of a gene regulatory network is governed by a chemical master equation that describes the evolution in time of the probability distribution of the system state \cite{pajaro2019transient}. As shown in Fig. \ref{Hysteresis_Stochastic}, a simple gene regulatory motif demonstrates deterministic bistability and hysteresis if molecular noise is ignored. However, stochastic hysteresis loops with multiple mean states appear with different probabilities, and the probabilities of these states under different initial conditions can be simulated because they follow Gaussian distributions $N(\mu, \sigma)$ with mean $\mu$ and standard deviation $\sigma$.

The role of molecular noise in biological networks could be described as the ``two sides of a coin'' \cite{skourti2014double}. On the one hand, the induction or amplification of genetic noise is a vital evolutionary prosurvival strategy in unicellular and multicellular organisms, as it can foster phenotypic heterogeneity in a population \cite{vujovic2019notch}. On the other hand, stochastic fluctuations may prevent the biological network from functioning adequately and may limit the ability to control cellular dynamics biochemically \cite{pedraza2005noise, lestas2010fundamental}. The good side of noise can only be demonstrated on a considerable evolutionary time scale and in large biological populations, while its detrimental side constantly influences the molecular activities that occur in cells. Life in the cell is a complex battle between randomizing and correcting statistical forces, and many control circuits have evolved to eliminate, tolerate or exploit noise \cite{lestas2010fundamental}. For example, negative feedback can suppress noise \cite{paulsson2004summing}, while positive feedback can stabilize differentiated states in cells \cite{balazsi2011cellular}.

Despite the presence of molecular noise, the switch-like behavior that occurs in biological networks is not destroyed. For example, bistable switches in protein interaction networks operate reliably despite the stochastic effects of molecular noise \cite{sabouri2008antagonism}. The reason for this is that circuits consisting of transcription factors and microRNAs can not only show biological bistability, as discussed in Section \ref{BioSecBistability}, but can also confer resilience to biological processes in the presence of intrinsic and extrinsic noise \cite{lai2016understanding}. These characteristics are accomplished through noise buffering \cite{siciliano2013mirnas}, which employs a microRNA-based mechanism that stabilizes gene expression and hence decreases variation in gene expression \cite{lai2016understanding}. Additionally, circadian networks can oscillate reliably in the presence of stochastic biochemical noise. When cellular conditions are altered, the ability to resist such disturbances imposes strict constraints on the oscillation mechanisms underlying circadian periodicity in vivo \cite{barkai2000circadian}.

The resilience of genetic circuits to stochastic noise relies mainly on the underlying connections between molecules. These connections create a complex genetic network that contains multiple nested feedback loops. For example, upon induction of cell differentiation, nested feedback loops prevent the established phenotypes from being reversed even in the presence of significant fluctuations \cite{acar2005enhancement}. Two positive feedback loops are involved. One is mediated by the cytoplasmic signal transducer Gal3p, which generates bistability in cell differentiation. Moreover, the parallel loop mediated by the galactose transporter Gal2p increases the difference in expression between the two states, enhancing the induced cell differentiation state. Another example is the circadian oscillation generated by the negative regulation exerted by a protein on the expression of its own gene \cite{young2001time}. Intercellular coupling can increase the resilience of this regulation to molecular noise \cite{gonze2006circadian}. In addition, the cooperativity in repression discussed in Section \ref{BioSecBistabilityGener}, coupled with transcriptional delay, can enhance biological systems' resilience \cite{gonze2006circadian, gupta2013transcriptional}. Transcriptional delay is another intrinsic property of genetic circuits; it results from the sequential nature of protein synthesis and the time required for transcription factors to move to their target promoters \cite{josic2011stochastic, li2009effects}. Gupta et al. \cite{gupta2013transcriptional} found that additional delay significantly increases the mean residence times of systems in states that are close to stable and that it stabilizes bistable gene networks.

\subsubsection{Resilience of unicellular organisms under environmental stress} \label{BiologyUnicellularSec}

Resilience in unicellular organisms such as yeast, cyanobacteria and other microbes refers to their ability to survive a disturbance \cite{meredith2018applying}. It can be achieved either by absorbing the effects of a disturbance without undergoing a notable change or through cooperative growth that enables the community of microorganisms to recover its abundance. The ability to tolerate a disturbance depends primarily on the traits associated with individual cells; conversely, the ability to recover depends mainly on traits associated with the population. Notably, in microbiology, the former (insensitivity to a disturbance) is defined as resistance, and the latter (the ability to recover) is called resilience \cite{carvalho2019antibiotic}. Since the term ``resilience'', as first proposed, included both the ability to withstand perturbations and the that to recover from them \cite{holling1973resilience} (see Chapter \ref{Ecology}), we use ``resilience'' to refer to both these abilities.
Next, we review two typical examples of resilience in unicellular organisms exposed to environmental stresses.

\noindent
\textbf{Cooperative growth of budding yeast in sucrose.}
The budding yeast \textit{Saccharomyces cerevisiae} is a single-celled eukaryotic fungus that uses oxygen to release energy from sugar. The concentration of sugar in the yeast's environment affects the rate of yeast growth; up to a point, higher sugar concentrations result in faster growth (yeast cannot grow at extremely high sugar concentrations, but this is beyond the scope of our discussion here) \cite{d2006effects}. In an experiment \cite{dai2013slower}, laboratory populations of budding yeast were grown in sucrose, and a daily dilution was performed in which a fraction (for example, 1 in 500 or a dilution factor of 500) of the cells was transferred to fresh medium. As the dilution factor increased (and the sucrose concentration decreased), the yeast populations decreased and collapsed at a tipping point, showing an abrupt phase shift. The tipping point is the point at which the cooperative growth of yeast cells occurs; such cooperative growth creates positive feedback loops.

\begin{SCfigure*}
\centering
\includegraphics[width=0.7\textwidth]{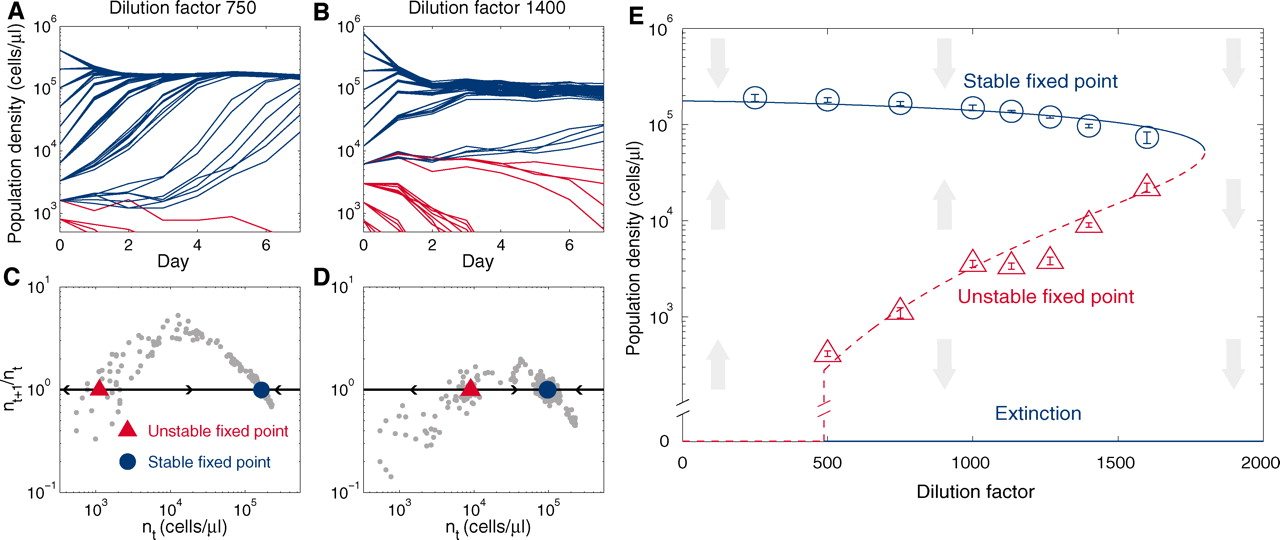} % figure 36
\caption{
Cooperative growth of yeast in sucrose leads to bistability and to a fold bifurcation. (A to D) Individual populations started at different initial densities were grown in 2\% sucrose with daily dilutions into fresh medium. Small populations below a critical density became extinct (red traces), whereas larger populations converged (blue traces) and maintained a stable density. (E) The stable and unstable fixed points measured in the experiments are shown as symbols.\\
\textit{Source:} The figure is from \cite{dai2012generic}.
}
\label{Bistability_Yeast}
\end{SCfigure*}

Budding yeast consume sucrose but must break it down into glucose and fructose before they can transport it across their cell walls. Yeasts break down sucrose by secreting an enzyme known as invertase \cite{koschwanez2011sucrose}. Sucrose is hydrolyzed into glucose and fructose in an extracellular process that can be shared between yeast cells, enabling them to work cooperatively. Such cooperation improves cell survival in yeast populations by allowing the cells to efficiently process sucrose  \cite{celiker2013cellular}. In the experiment in which yeast populations were grown in sucrose and daily dilutions at certain dilution factors were performed, bistability appeared in cultures that had been started at a wide range of initial cell densities \cite{dai2012generic}. As shown in Fig. \ref{Bistability_Yeast}, cultures that began below a critical cell density went extinct, whereas those started at higher initial densities survived and reached a finite stable fixed point. A fixed point is recognized at which the ratio of population densities on subsequent days $n_{t+1}/n_t=1$, where $n_t$ is the population density on day $t$ ($t=1, 2, ..., 6$). Of the cultures that were started at densities near the critical density (an unstable fixed point), some populations survived due to cooperative growth, and others became extinct. The cooperation between yeast cells enables the cells to break down sucrose more efficiently, and more cells grow, leading to positive feedback so that the population can survive at a nonzero stable fixed point. In contrast, if there is insufficient cooperation initially, then weak positive feedback can drive the population to collapse.

For more efficient cooperation, budding yeasts aggregate and build communities or grow by incomplete cell separation, forming undifferentiated multicellular clumps. The cells within the clumps cooperate in collecting food and, at low sucrose concentrations, have a growth advantage over equal numbers of single cells. Clumped cells are able to grow when sucrose is scarce, whereas single cells cannot. In addition, clumps containing more cells grow faster than smaller clumps \cite{koschwanez2011sucrose}; this finding suggests a possible origin of multicellularity and demonstrates one of the advantages of evolving from a unicellular organism to a complex multicellular organism. Such multicellularity is an essential mechanism used by yeast populations to retain resilience against fluctuations in sucrose concentrations \cite{hope2017experimental}.

\noindent
\textbf{Bistable response of cyanobacteria exposed to increasing light levels.}
Cyanobacteria are blue-green bacteria that are abundant in the environment. They are among the world's most crucial oxygen producers and carbon dioxide consumers \cite{schuergers2016cyanobacteria}. For cyanobacteria, light is the primary source of energy. Phycobilisomes and photosystems absorb light and convert it into chemical energy through photosynthesis \cite{montgomery2014regulation}. The color (wavelength) and intensity (irradiance) of light affect the growth of cyanobacteria. For example, the growth rates of cyanobacteria are similar under orange and red light but are much lower under blue light \cite{luimstra2018blue}. It is important to note that increasing the light intensity over a certain range increases the growth rate of cyanobacteria; above this intensity, light is harmful to cyanobacteria and may even lead to the collapse of cyanobacteria populations \cite{veraart2012recovery}. Understanding how cyanobacteria respond to light can improve photosynthetic efficiency and overall resilience.

\begin{figure}[!ht]
\centering
\includegraphics[width=0.95\linewidth]{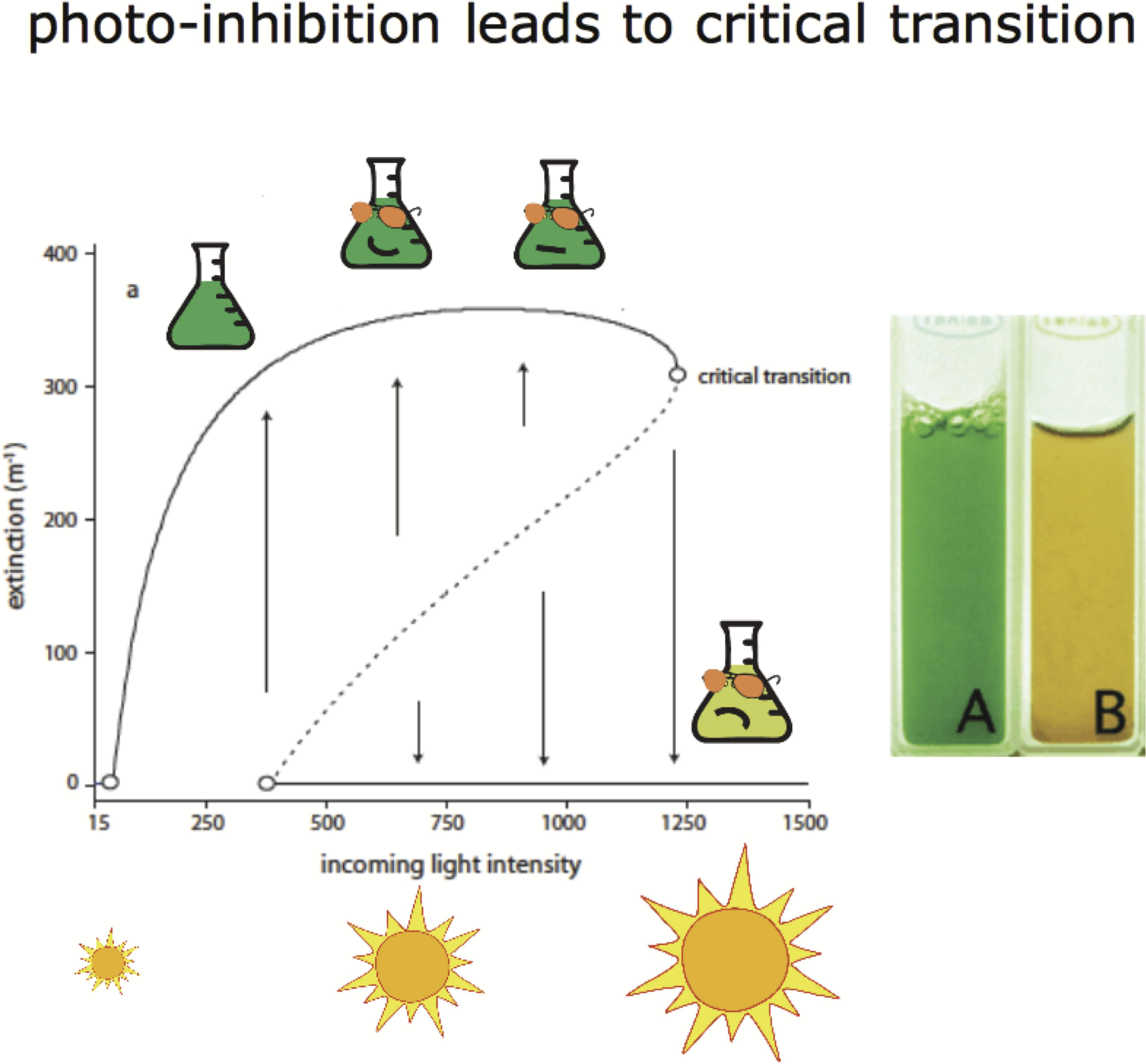} % figure 37
\caption{
Critical transition in a cyanobacteria population as light intensity increases.\\
\textit{Source:} The figure is from \cite{photoinhibition}.
}
\label{BlueLight}
\end{figure}

The phenomenon in which the rate of photosynthesis decreases with increasing light is caused by photoinhibition. Photoinhibition manifests itself as a series of reactions that inhibit the activity of the photosystem; it is apparent in cyanobacteria populations and phytoplankton species that are sensitive to high light levels \cite{gerla2011photoinhibition}. The existence of photoinhibition forces cyanobacteria to carefully balance the harvesting of sufficient photons to maximally drive photosynthesis with avoidance of the damaging effects of excess energy capture \cite{wiltbank2019diverse}. This balance is achieved by phototaxis. Positive phototaxis is cell movement toward light, and negative phototaxis is movement away from light. If the light is strong and extensive and there is no location to which cyanobacteria can move to escape the light, then the cyanobacteria shade one another. Mutual shading can ameliorate light stress and promote the growth of cyanobacteria; this, in turn, encourages more mutual shading, creating a positive feedback loop \cite{veraart2012recovery}. Such positive feedback leads to the presence of alternative stable states with respect to the cell density of cyanobacteria populations (Fig. \ref{BlueLight}). A population at low density cannot provide sufficient shading to protect itself against photoinhibition. Hence, that population becomes extinct. However, population densities above a threshold allow the population to create conditions in which turbidity is sufficient to suppress photoinhibition, allowing the population to establish itself \cite{gerla2011photoinhibition}.

\begin{figure}[!ht]
\centering
\includegraphics[width=0.95\linewidth]{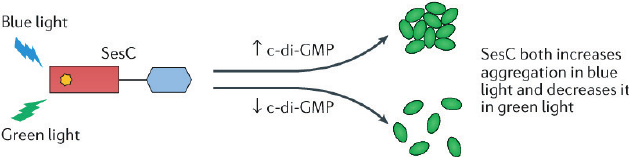} % figure 38
\caption{
Cell aggregation: shielding from light of harmful wavelengths. Whether increasing or decreasing aggregation occurs depends on the color of the light.\\
\textit{Source:} The figure is from \cite{wiltbank2019diverse}.
}
\label{Cran_Aggregation}
\end{figure}

Similar to the formation of cell clumps in the yeast population discussed above, cyanobacteria also aggregate to provide the shading required to protect their photosynthetic machinery from damage by excessive light \cite{zilliges2008extracellular}. Unlike yeast cells in a clump, which usually stick together, the tightness of cyanobacterila cell aggregation increases or decreases as the light conditions change. As shown in Fig. \ref{Cran_Aggregation}, cells aggregate in blue light, which is harmful to them; however, when exposed to green light, cells relax their clumps to optimize light capture \cite{wiltbank2019diverse}. Thus, the resilience of cyanobacteria populations can be enhanced by the regulation of cell-cell contact.

\noindent
\textbf{Bacterial antibiotic responses.}
The extensive use of antibiotics has resulted in a situation in which multidrug-resistant pathogens have become a severe menace to human health worldwide. Antibiotics affect bacterial cell physiology at many levels, and bacteria respond to antibiotics by changing their metabolism, gene expression, and possibly even their mutation rate \cite{mitosch2014bacterial}; all of these changes affect their rate of growth and ability to survive. Bacterial cells may survive antibiotic treatment because individual cells become tolerant or adapt to the treatment or because the population recovers by recolonization, reproduction or rapid regrowth after some cells are killed \cite{meredith2015collective}. Such mechanisms can be characterized according to the resistance-resilience framework in microbiology \cite{shade2012fundamentals}; in this framework, resistance is defined as insensitivity to treatment, while resilience is defined as the time required for a community to recover its former composition and functions after treatment \cite{carvalho2019antibiotic}. The involvement of biological mechanisms related to antibiotic resistance and antibiotic resilience has been attracting increasing attention from the scientific community \cite{shade2012fundamentals}, which has contributed to the development of new treatment strategies to cope with and prevent the increased prevalence of resistant pathogenic bacteria \cite{mitosch2014bacterial}.

\begin{SCfigure*}
\centering
\includegraphics[width=0.667\textwidth]{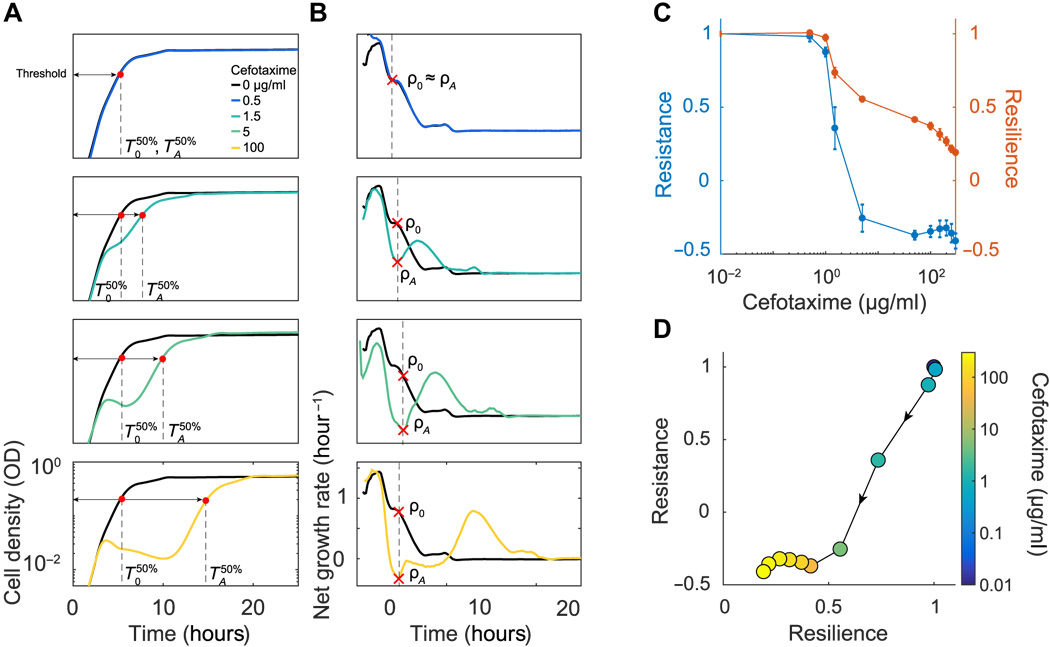} % figure 39
\caption{
Quantifying antibiotic resilience and antibiotic resistance. (A) When no
antibiotic is added, the population transitions nearly exponentially to its carrying capacity (black curve). When the antibiotic concentration is increased, the time required for the population to recover increases, and the population's resilience decreases. (B) The population's net growth rate quantifies its resistance. (C) Resistance and resilience as functions of cefotaxime concentration. (D) The resistance-resilience map defines a phenotypic signature.\\
\textit{Source:} The figure is from \cite{meredith2018applying}.
}
\label{Resistance_Resilience_Antibiotic}
\end{SCfigure*}

Meredith et al. \cite{meredith2018applying} applied the resistance-resilience framework to the analysis of bacterial pathogens that produce extended-spectrum $\beta$-lactamases (ESBLs). ESBLs are becoming increasingly prevalent and can degrade many $\beta$-lactam antibiotics, the most widely used class of antibiotics. In the absence of antibiotic treatment, a population of ESBL-producing bacteria grows approximately exponentially until the growth rate decreases due to limited nutrient availability; the population size then stabilizes at the carrying capacity, as shown by the blue and black curves in Fig. \ref{Resistance_Resilience_Antibiotic}A. The time needed for a population to reach 50\% of its carrying capacity is denoted as $T^{50\%}$. If antibiotics are introduced, then the time required for the bacterial population to reach its carrying capacity increases; the higher the antibiotic concentration is, the longer is the time required for the population to reach its carrying capacity, as shown by the green and yellow lines in Fig. \ref{Resistance_Resilience_Antibiotic}A. Antibiotic resilience is defined as the inverse of the treated population's $T^{50\%}$ ($T^{50\%}_A$) normalized to the untreated population's $T^{50\%}$ ($T^{50\%}_0$); it is written as
\begin{equation}\label{AntibioticResilience}
{\rm Resilience}=\frac{T^{50\%}_0}{T^{50\%}_A}.
\end{equation}
Antibiotic resistance is defined as the ratio of the minimum net growth rate of a treated population ($\rho_A$) to the net growth rate of an untreated population growing at the same time under the same conditions ($\rho_0$); it takes the form
\begin{equation}\label{AntibioticResistance}
{\rm Resistance}=\frac{\rho_A}{\rho_0}.
\end{equation}
As shown in Fig. \ref{Resistance_Resilience_Antibiotic}B, the higher the antibiotic concentration is, the less resistant the population. The resistance-resilience framework can be used to visualize the shift in a population's antibiotic response; once the population undergoes a crash, resistance is minimized, and resilience dominates the population's survival (Fig. \ref{Resistance_Resilience_Antibiotic}C). This resistance-resilience framework effectively reveals the phenotypic signatures of individual bacterial strains (Fig. \ref{Resistance_Resilience_Antibiotic}D) when treated with $\beta$-lactams. The analysis reveals that effective treatment minimizes both resistance and resilience.

In summary, resilience in unicellular organisms depends on the organism's genotype, the underlying molecular connections within individual cells, and the cooperation among cells. The latter can improve cell growth, which in turn encourages more cooperation. This creates positive feedback and leads to bistability in the density of unicellular populations. Multicellular organisms can be viewed as having evolved to take advantage of the benefits of clumping or clustering of cells. Thus, the insight gained from the study of unicellular organisms can be extended to multicellular organisms.

\subsubsection{Potential landscapes of cellular processes in complex multicellular organisms}

\begin{figure}[!ht]
\centering
\includegraphics[width=0.95\linewidth]{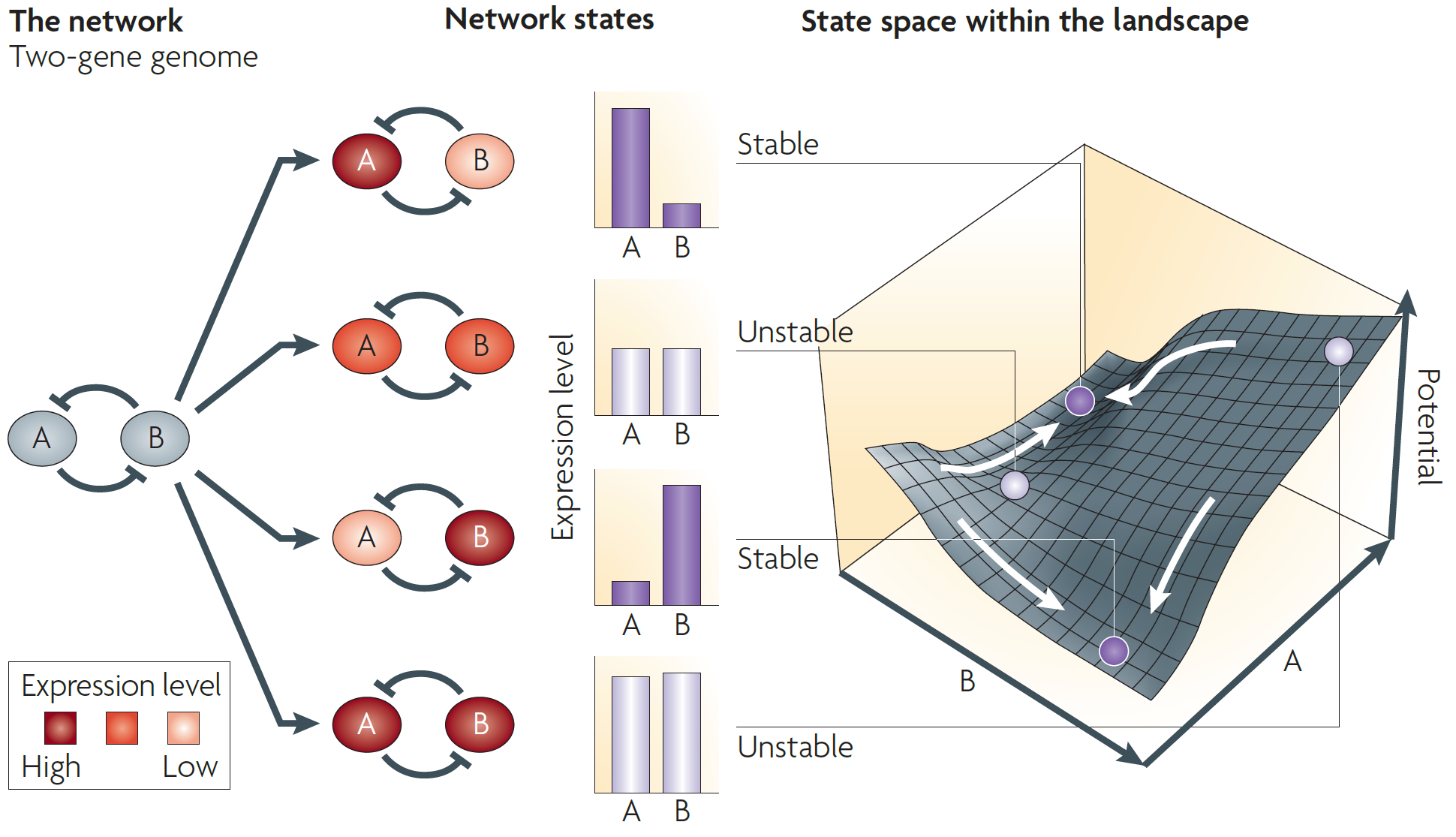} % figure 40
\caption{
Quasi-potential landscape of a double negative feedback loop \cite{fang2018cell}, which has been empirically found to have four stable states occupied with different probabilities.\\
\textit{Source:} The figure is from \cite{brock2009non}.
}
\label{network_landscape}
\end{figure}

The quasi-potential landscape has provided a useful tool for understanding phase transitions within biological systems, especially those related to the resilience of complex multicellular organisms \cite{li2014landscape}. To draw the quasi-potential landscape of a biological network, we may need to find quasi-potential functions of that network. Generally, the dynamics of a biological network can be described as a series of continuous differential equations \cite{zhou2012quasi}:
\begin{equation}\label{BioNetworkDynamic}
\frac{d\boldsymbol{x}}{dt}=\boldsymbol{F}(\boldsymbol{x}),
\end{equation}
where $\boldsymbol{x}(t)=(x_1(t),x_2(t),\dots,x_N(t))^T$ represents $N$ system variables (e.g., gene expression, biological molecular concentrations, and other variables) and $\boldsymbol{F}(\boldsymbol{x})=(F_1(\boldsymbol{x}),F_2(\boldsymbol{x}),\dots,F_N(\boldsymbol{x}))^T$ describes the ``forces'' that act to change the states of the corresponding variables. In closed equilibrium systems in which there is no significant exchange of energy, materials, or information with the outside environment, such as protein folding, the local dynamics can be determined by the gradient of the interaction potential energy \cite{zhou2012quasi}:
\begin{equation}\label{PetentialEnergy}
\boldsymbol{F}_i(\boldsymbol{x})=-\frac{\partial{U}(\boldsymbol{x})}{\partial x_i},
\end{equation}
and the potential function $U(\textbf{x})$ can be directly inferred as
\begin{equation}\label{Petentialfunction}
U(\boldsymbol{x})=-\sum_{i=1}^{n}\int{\boldsymbol{F}_i(\boldsymbol{x})dx_i}.
\end{equation}
However, most biological systems are nonequilibrium open systems, and their dynamics cannot be fully captured by pure gradients (Eqs. \ref{PetentialEnergy} and \ref{Petentialfunction}). Fortunately, we can use the master equation \cite{gardiner1985handbook} discussed in Section \ref{CircuitNoiseSec} to describe the evolution in time of the probability that the system will remain in each state. 
Next, we review studies of the reconstruction and analysis of the landscapes of cellular processes such as the cell cycle, cell differentiation, and disease progress in multicellular organisms.

\begin{SCfigure*}
\centering
\includegraphics[width=0.667\textwidth]{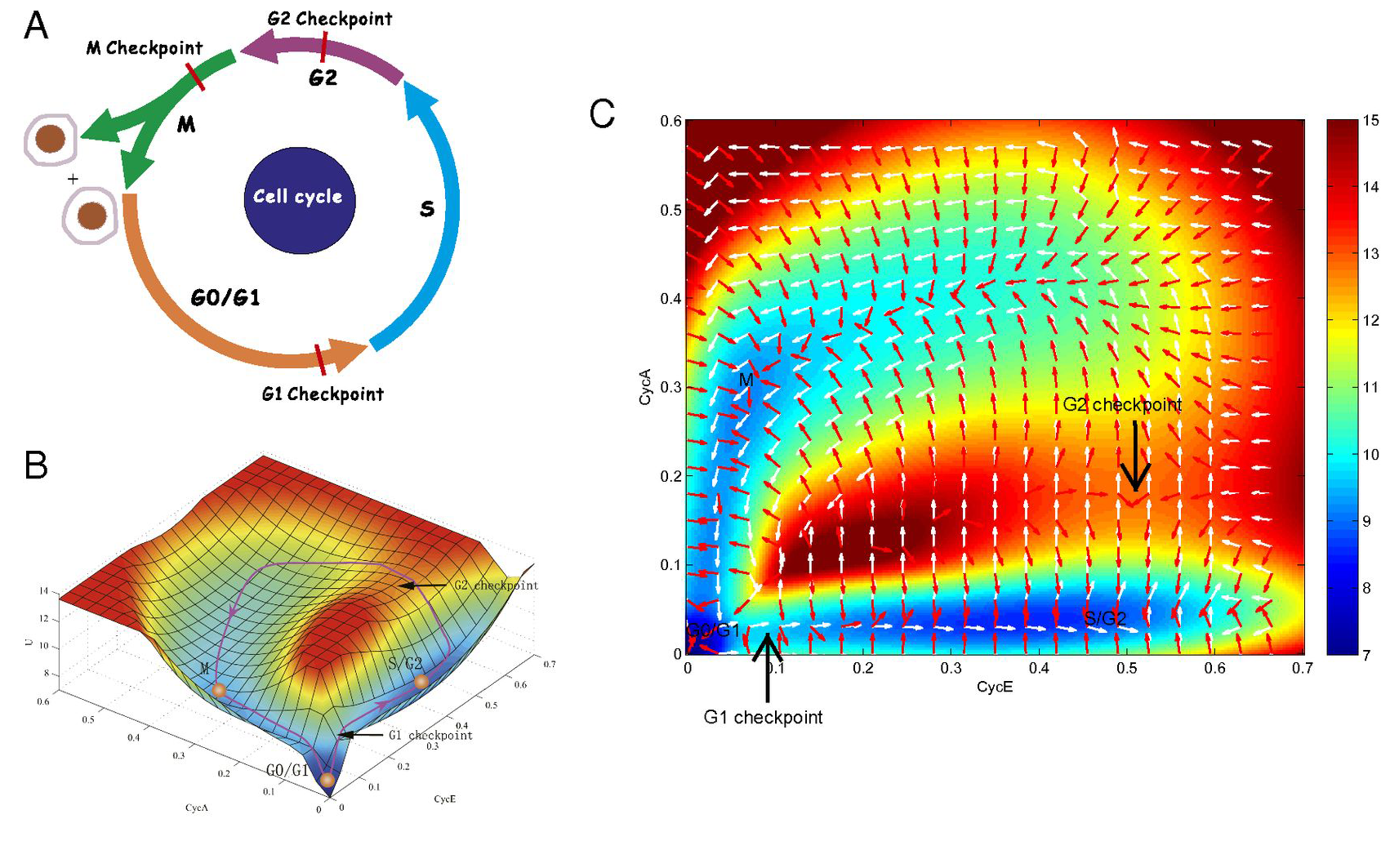} % figure 41
\caption{
Potential landscape of a complete cell cycle network. (A) The four phases of the cell cycle with three checkpoints: G1, S, G2, and one M phase. (B) Global landscape of the cell cycle that contains three phases and two checkpoints. (C) The 2D landscape, where white arrows represent probabilistic fluxes, and red arrows represent the negative gradients of potential.\\
\textit{Source:} The figure is from \cite{li2014landscape}.
}
\label{cellcycle2}
\end{SCfigure*}

\noindent
\textbf{Mexican hat landscapes of cell cycles.}
In Section \ref{EmpiricalBistability}, we reviewed the simple genetic circuit in the cell cycle and showed that it could generate bistability. However, the complete cell cycle process is far more complex than the outline that was presented there. The cell cycle consists of a series of events that occur within a cell during its replication and division. As shown in Fig. \ref{cellcycle2} A, it comprises several distinct phases: G1 phase (resting), S phase (DNA synthesis), G2 phase (interphase), and M phase (mitosis) \cite{li2014landscape}. The activation of each phase requires the proper progression and completion of the previous phase, which is monitored by cell cycle checkpoints. Many studies have been conducted to uncover the mechanisms underlying the cell cycle process, both in unicellular organisms such as budding yeast \cite{wang2008potential} and in multicellular organisms such as \textit{Xenopus laevis} \cite{zhang2018exploring} and mammals \cite{li2014landscape}. It has been found that the landscape and flux framework can be effectively used to explain the entire process of the cell cycle and its checkpoints \cite{wang2008potential}. For example, Zhang et al. \cite{zhang2018exploring} quantified the underlying landscape and flux of the embryonic cell cycle of \textit{Xenopus laevis}. The authors also described the corresponding Mexican hat landscape, which displayed several local basins, and barriers along the oscillation path were revealed. The local basins characterize the different phases of the \textit{Xenopus laevis} embryonic cell cycle. The local barriers represent the checkpoints at which global quantification of the cell cycle occurs. In addition, through global sensitivity analysis of landscape and flux, the key elements for controlling the cell cycle speed are identified, which helps in designing effective strategies for drug discovery for cancer.

\begin{figure}[!ht]
\centering
\includegraphics[width=0.95\linewidth]{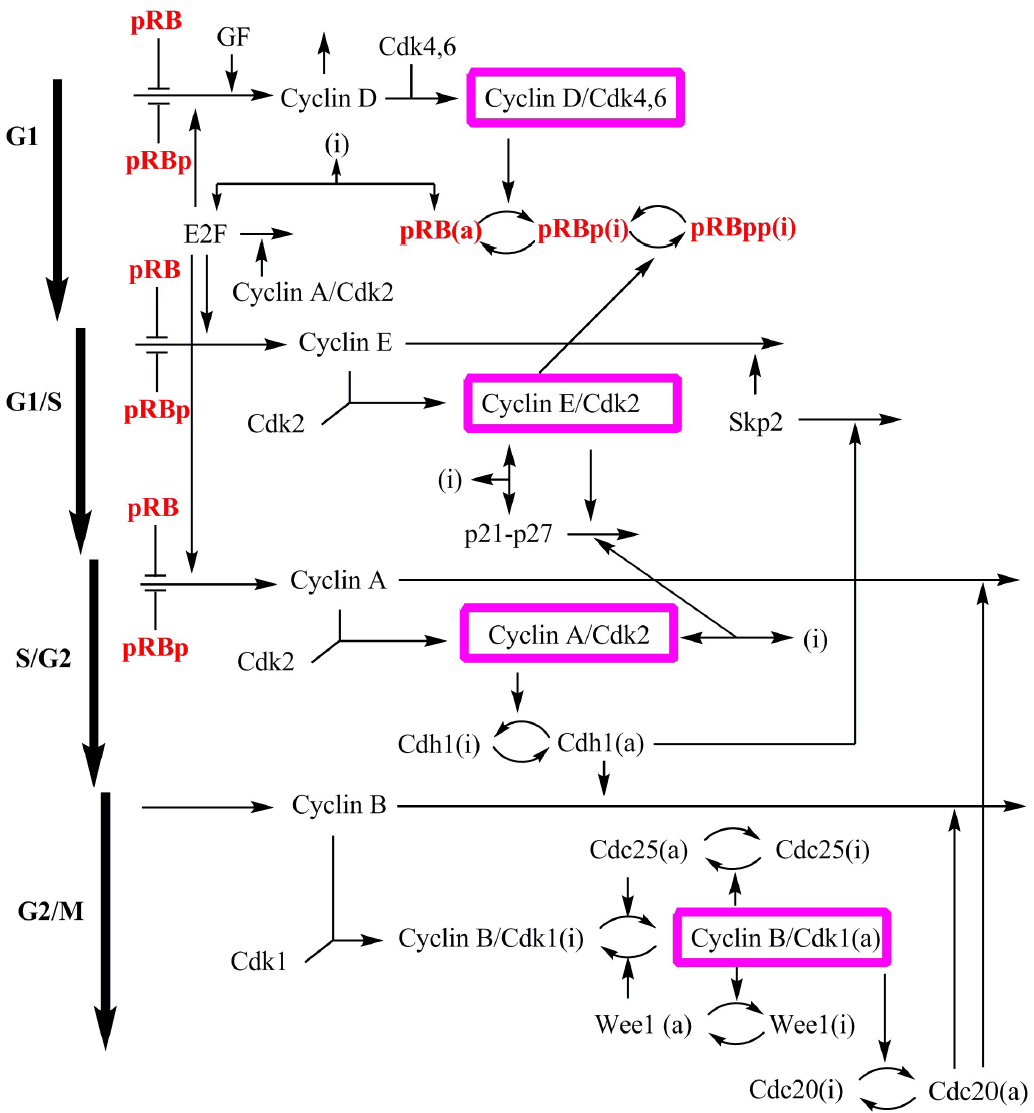} % figure 42
\caption{
Diagram of the mammalian cell cycle network. The network includes four major complexes formed by cyclins and cyclin-dependent kinases (cyclin/CDks): cyclin D/Cdk4-6, cyclin E/Cdk2, cyclin A/Cdk2, and cyclin B/Cdk1. Together, these complexes determine cell cycle dynamics. Cell cycle progression is controlled by mutual repression regulation between the tumor suppressor retinoblastoma protein (pRB) and the transcription factor E2F.\\
\textit{Source:} The figure is from \cite{li2014landscape}.
}
\label{cellcyclenetwork}
\end{figure}

Another typical example is the well-studied mammalian cell cycle network and the detailed diagram is shown in Fig. \ref{cellcyclenetwork}. The network involves four major complexes formed by cyclins and cyclin-dependent kinases (cyclin/CDks), centered on cyclin D/Cdk4-6, cyclin E/Cdk2, cyclin A/Cdk2, and cyclin B/Cdk1. Together, they determine the cell cycle dynamics \cite{li2014landscape}, which can be described by a set of 44 nonlinear ordinary differential equations. Li et al. \cite{li2014landscape} used the probability landscape and flux to determine the main driving force for the dynamics in the mammalian cell cycle network. The landscape directly reflects the steady-state probability distribution $P_{ss}$, which is obtained by using a self-consistent mean field approximation \cite{wang2010potential}, and the potential landscape $U$ ($U=-lnP_{ss}$), giving the weight of each state. Projection of such a 44-dimensional landscape onto a two-dimensional state space yields a landscape of the mammalian cell cycle system in terms of two key proteins, CycE and CycA. This landscape, shown in Fig. \ref{cellcycle2} B, has a Mexican hat shape. In the landscape, the red-colored region represents high potential (there is a low probability that the system will reach this region), and the blue-colored region along the ring of the Mexican hat has low potential (there is a high probability that the system will reach this region). The low-potential blue region forms a circle of oscillation trajectory that guarantees the resilience of cell cycle oscillation dynamics. The progression of a cell along the cycle path is determined by two driving forces: potential barriers for deceleration and curl flux for acceleration. The curl probability fluxes are shown as white arrows in the 2D landscape (Fig. \ref{cellcycle2}C), and the negative gradients of the potential landscape are represented by red arrows. The force from the negative gradient of potential attracts the cell cycle into the oscillation ring. The flux drives the cell cycle oscillations along the ring path. Furthermore, along the ring with heterogeneous potential, there are three basins of attractions corresponding to three cell cycle phases (G0/G1, S/G2 and M) and two barriers representing two checkpoints (the G1 and G2 checkpoints).

The potential landscape can help us understand the role of attractors in whole cell cycle dynamics quantitatively. This landscape also provides a simple physical explanation for the mechanism of cell cycle checkpoints. In addition, the influence of external or internal perturbations on the resilience of the cell cycle could be learned by performing global sensitivity analysis \cite{li2014landscape}. Such analysis quantifies the changes in the barrier heights, period, and flux when parameters (regulation strengths or synthesis rates) are changed. By selecting the parameters that are highly influential in determining the genes in and regulation of the network, key elements or wirings that control the stability and progression of the cell cycle are identified. The results can also be verified through experiments, leading to predictions and potential anticancer strategies.

\begin{figure}[!ht]
\centering
\includegraphics[width=0.95\linewidth]{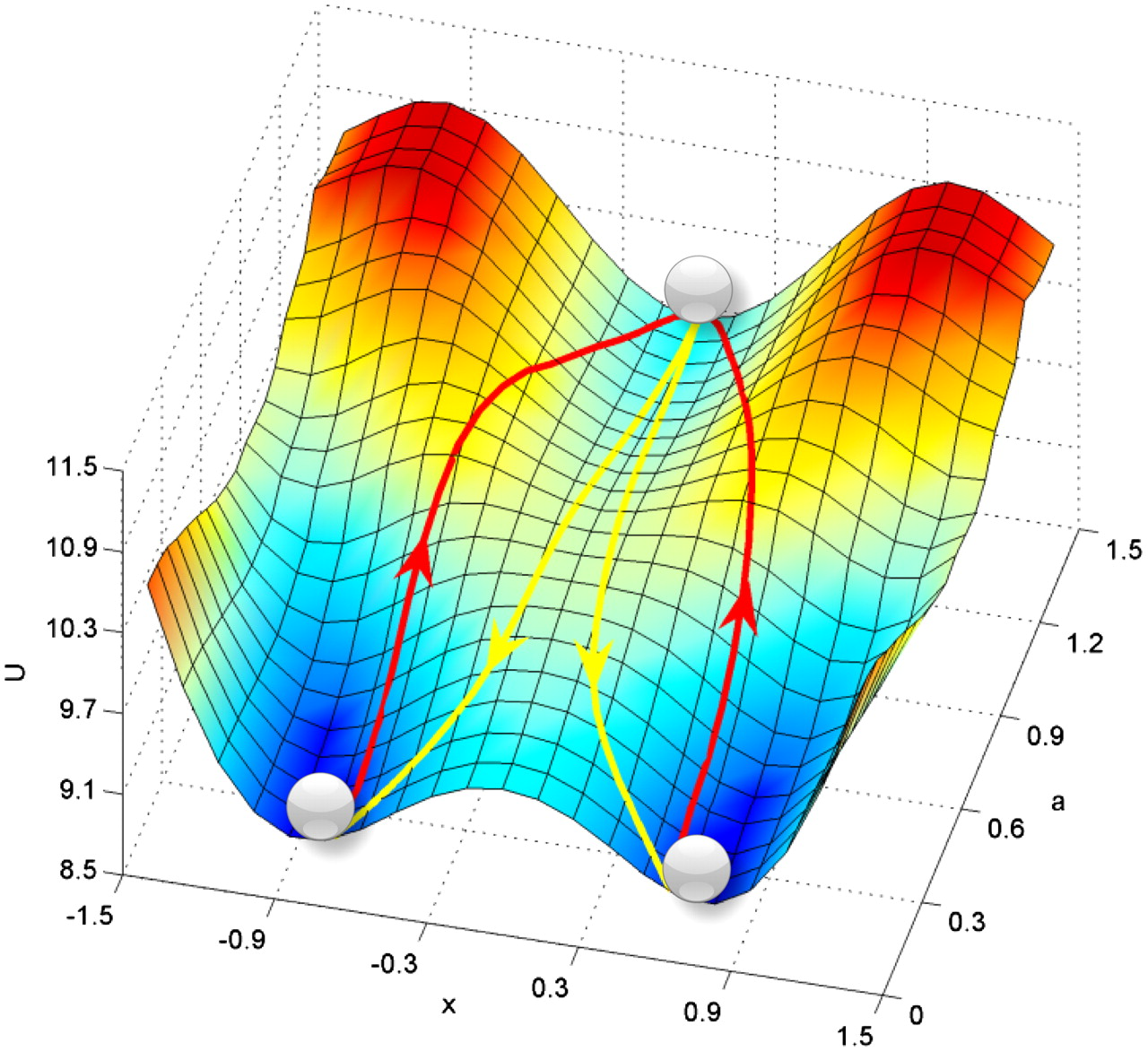} % figure 43
\caption{
Waddington's quantified developmental landscape and pathways\\
\textit{Source:} The figure is from \cite{wang2011quantifying}.
}
\label{Landscape_MinimalSystem}
\end{figure}

\noindent	
\textbf{Epigenetic landscapes for cell fate induction.} In cell fate induction, a cell progresses from an undifferentiated state to one of a number of discrete, distinct, differentiated cell fates. Waddington's epigenetic landscape \cite{waddington2014strategy}, in which cells are represented by balls rolling downhill through a landscape of bifurcating valleys, is probably the most famous and powerful metaphor that has been used to describe this process \cite{ferrell2012bistability}. Waddington's landscape begins with a single valley, representing the single undifferentiated steady state. As time goes on, alternative valleys representing individual differentiated states appear. Each valley in the landscape represents a possible cell fate, and the ridges between the valleys maintain the cell fate once it has been chosen. In Waddington's landscape, the undifferentiated state is unstable, which triggers the differentiation process. This is true for stem cells during embryonic development. However, in adults, multipotent undifferentiated cells are stable. Wang et al. \cite{wang2011quantifying} developed a framework to quantify Waddington's landscape for a simple gene regulatory circuit that governs a binary cell fate decision module through the construction of an underlying probability landscape for cell development. The circuit consists of two mutually inhibitory transcription factors, the interactions of which can be described by minimal system equations \cite{huang2007bifurcation}. In such a quantified landscape, the undifferentiated state can remain stable, but the cell has a small, finite chance (induced by fluctuations) to climb up from the basin of attraction and escape to one of the differentiated states (Fig. \ref{Landscape_MinimalSystem}).

\begin{SCfigure*}
\centering
\includegraphics[width=0.67\textwidth]{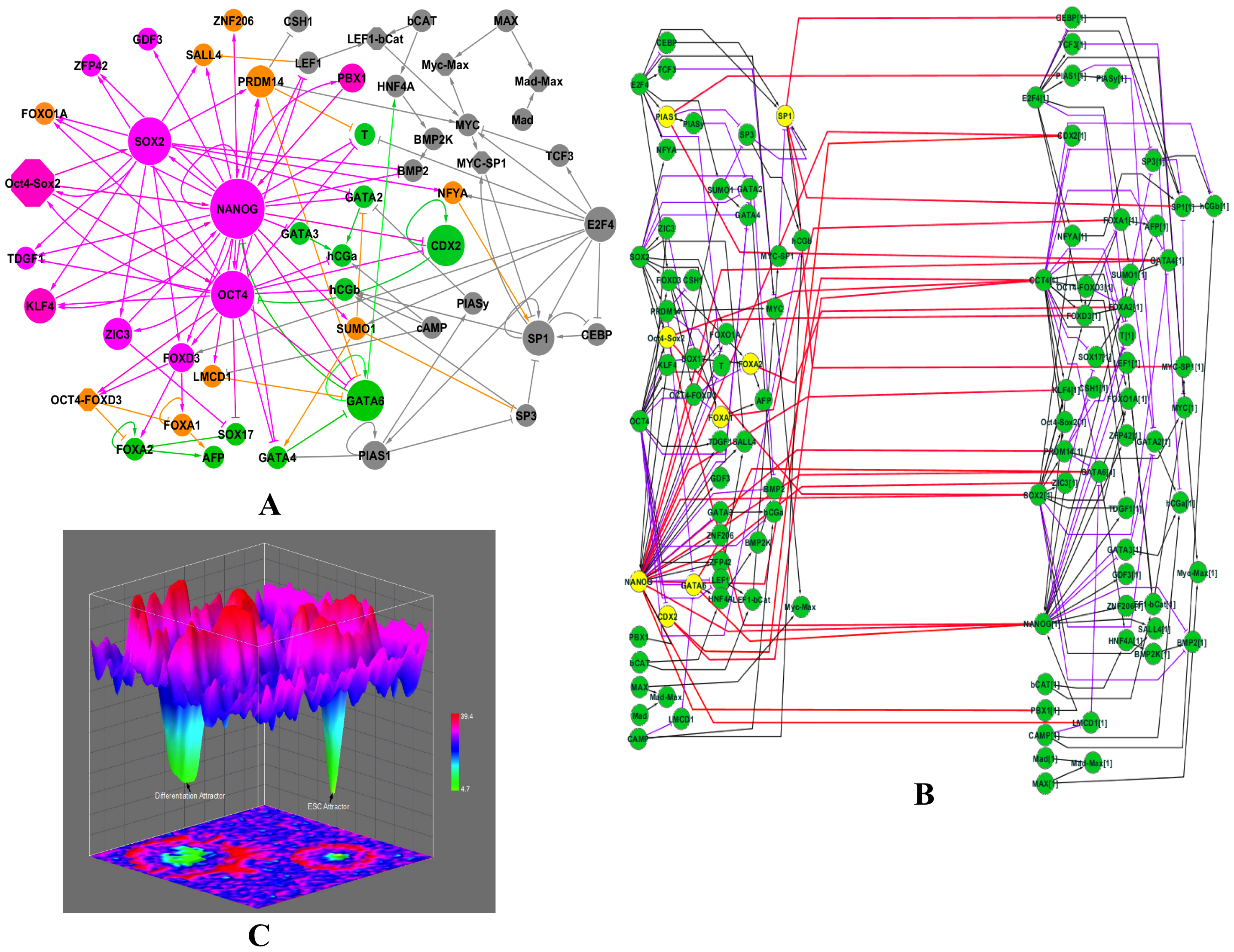} % figure 44
\caption{
Epigenetic landscape for the human embryonic stem cell network. (A) Genetic network regulating the self-renewal and differentiation of hESCs. (B) Bayesian network (2TBN) model of the genetic network in hESCs. (C) Illustration of the cell-state potential landscape. The color represents the potential of the cell state. The higher the potential is, the smaller the probability that the cell will reach that particular cell state.\\
\textit{Source:} The figure is from \cite{chang2011systematic}.
}
\label{HumanStemCell_Landscape}
\end{SCfigure*}

In Fig. \ref{Landscape_MinimalSystem}, the depths of two valleys that represent potential distributions along the trajectories starting from the same bifurcation point appear quite even. However, due to the presence of complexity and noise (such as the molecular noise discussed in Section \ref{CircuitNoiseSec}), the landscape that illustrates the cell developmental processes of multicellular organisms may be quite bumpy, and stochastic state transitions may occur during the cell differentiation process. For example, Wang et al. \cite{wang2010potential} considered the effect of stochastic fluctuations in a canonical gene circuit. Using the Fokker-Planck equation to reconstruct the potential landscape, they found that the system could hop from one stable branch (e.g., the pluripotent state) to another (e.g., either one of two differentiated states) with some probability, even when it did not reach bifurcation. The higher the noise level is, the higher the probability of random state transitions. Fortunately, organisms have developed mechanisms to prevent this stochastic effect \cite{acar2005enhancement}. 
Next, we present a specific example of cell differentiation.

Chang et al. \cite{chang2011systematic} estimated the epigenetic landscape of a genetic network involved in regulating pluripotency and human embryonic stem cell differentiation. This genetic network (Fig. \ref{HumanStemCell_Landscape}A) was constructed by starting with a set of marker genes of pluripotency and differentiation lineages. The authors then determined the regulatory paths between specific pairs of genes.
 The resulting network features direct regulatory interactions among 52 nodes and includes three key regulators of embryonic stem cells (Oct4, NANOG and Sox2), six protein complexes, and genes with expression marks the differentiation lineages. To calculate the probability of each cell state, each protein in the network is considered a binary variable that is either active or inactive. The genetic network is then transformed into a dynamic Bayesian network parameterized by the Monte Carlo Markov chain method (Fig. \ref{HumanStemCell_Landscape}B). This network is used to simulate the evolving stochastic characteristics of the system. Since the exact path through which extracellular signals activate or repress the production of transcription factors in human embryonic stem cells is unknown, an estimation of the landscape was needed. Chang et al. \cite{chang2011systematic} chose to manipulate the expression levels of three key regulators (Oct4, Sox2 and NANOG), mimicking extracellular conditions. The joint probabilities of all the nodes in the network with Oct4/Sox2/NANOG set to various activity levels were calculated. Moreover, the integration of these probabilities was used to estimate the network landscape. As shown in Fig. \ref{HumanStemCell_Landscape}C, two states have significantly higher probabilities than the others. The states represent human embryonic stem cells and their differentiated states. In the human embryonic stem cell state, all embryonic stem cell markers are active (1), and all differentiation markers are inactive (0). In the differentiated state, the activity compositions are reversed. These two states are separated by barrier states with smaller probabilities, preventing noise from causing transitions between cell types.

In addition, through global sensitivity analysis of parameters and of connections between genes in the human stem cell developmental network, Li et al. \cite{li2013quantifying} quantitatively predicted which connection links or nodes (genes) are critical for cellular differentiation and reprogramming. The results can be directly tested from the experiments. The identified key links can be used to guide differentiation designs and reprogramming tactics.

\noindent			
\textbf{Quantifying the landscape underlying cancer networks.}
Cancer is believed to be a genetic disease that arises from the accumulation of multiple genetic and epigenetic alterations \cite{yu2019landscape}. It has long been recognized as an evolutionary process \cite{armitage1954age}, the physical mechanisms of which underlie cellular transitions from the normal to the cancer state can be effectively studied using the idea of ecological resilience \cite{korolev2014turning}. Landscape analysis is a crucial tool for quantifying and visualizing  transitions between normal and cancer states \cite{brock2009non}. 
Next, we review studies that reveal the landscape of cancer systems at both the molecular and cellular levels.

\begin{figure}
\centering
\includegraphics[width=0.95\linewidth]{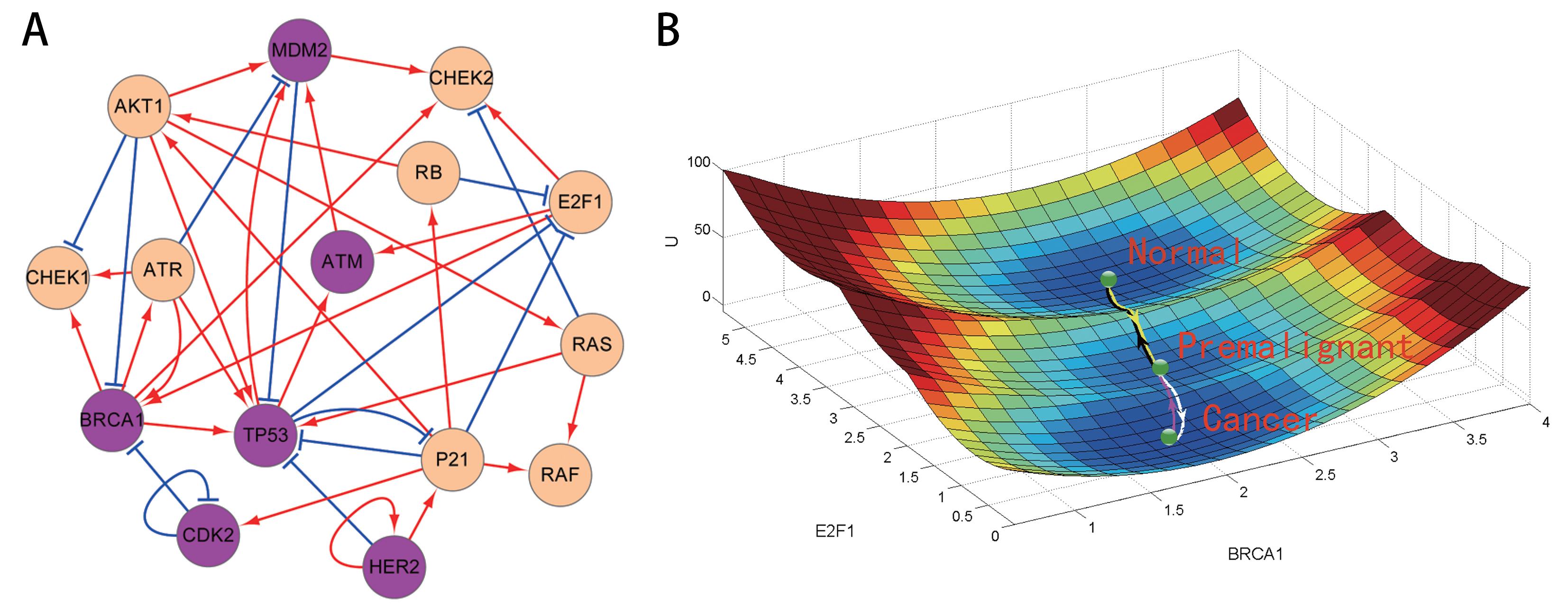} % figure 45
\caption{
(A) Diagram of the gene regulatory network of breast cancer, which contains 15 nodes (genes) and 39 edges (26 activation interactions and 13 repression interactions). (B) Tristable landscape of the breast cancer gene regulatory network.\\
\textit{Source:} Figure from \cite{yu2016physical}.
}
\label{BreastCancerLandscape}
\end{figure}

Yu et al. \cite{yu2016physical} constructed a reliable gene regulatory network for breast cancer. This network consisted of 15 genes crucial for breast cancer and 39 regulatory relationships among them. As shown in Fig. \ref{BreastCancerLandscape}A, this gene regulatory network contains four oncogenes (BRCA1, MDM2, RAS, and HER2), three tumor suppressor genes (TP53, P21, and RB), five kinases (CHEK1, CHEK2, AKT1, CDK2, and RAF) essential for the maintenance of cell cycle regulation, two genes (ATM and ATR) important for signal transduction, and a transcription factor (E2F1). The interactions between genes include both activation and repression. The temporal evolution of the dynamics of this gene network is determined by the driving force of gene regulation defined as follows:
\begin{equation}\label{BreastCancerDynamic}
\frac{dX_i}{dt}=F_i=-K_i*X_i+\sum_{j=1}^{m_1}\frac{a_j*X_j^n}{S^n+X_j^n}+\sum_{j=1}^{m_2}\frac{b_j*S^n}{S^n+X_j^n},
\end{equation}
where $X_i$ represents the level of expression of gene $i$ $(i=1, 2, ..., 15)$, and the three terms on the right side of the equation represent self-degradation, activation and repression. Parameters $K$, $a$ and $b$ are constants; $S$ represents the threshold of the sigmoid function; $m_1$ ($m_2$) is the number of nodes that activate (repress) node $i$; and $n$ is the Hill coefficient. Based on 15 dynamic equations (one for each gene or protein in the network) and the self-consistent mean field approximation, the steady-state probability distribution $P_{ss}$ is obtained together with the potential landscape $U=-lnP_{ss}$. Since it is difficult to visualize the landscape in a 15-dimensional space, Yu et al. \cite{yu2016physical} projected the landscape onto a 2-dimensional subspace that spans the expression levels of BRCA1 (an oncogene of breast cancer) and E2F1 (a biomarker of breast cancer).

As shown in Fig. \ref{BreastCancerLandscape}B, there are three attractor basins on the phenotypic landscape; these three basins represent the normal, premalignant and cancer states. In the normal state, cell growth, cell cycle arrest and apoptosis obey the rules they normally follow. In the premalignant state, the cells grow and exhibit some abnormal features that resemble certain cancer characteristics. In the cancer state, cell growth becomes uncontrollable, and the cells eventually spread to other organs of the body. The progression of breast cancer can be seen as involving transitions between different state basins. Transitions between the normal and the premalignant state are almost reversible, while those between the premalignant state and the cancer state are irreversible; this clearly illustrates the mechanisms of cancerization \cite{yu2019landscape}. In addition, global sensitivity analysis shows that changing the strengths of the key regulatory components of the breast cancer gene regulatory network can allow for the landscape topography to move in preferred directions that are beneficial for returning cancer cells to the normal state.

In addition to the epigenetic landscape of cancer derived from the gene regulatory network (molecular layer) \cite{yu2016physical}, the cancer-immune system interaction network (cellular layer) is also important for understanding tumorigenesis, the development of cancer and the impact of immunotherapy. Li et al. \cite{wenbo2017uncovering} constructed a cancer immune network consisting of 13 cells (a cancer cell and 12 types of immune cells) and 13 cytokines and thereby revealed the underlying mechanism of cancer immunity based on landscape topography. In the landscape, three steady states (normal, low cancer, and high cancer states) appear. When certain cell-cell interactions occur, limited cycle oscillations emerge; such oscillations are common in the immune system \cite{stark2007oscillations} and provide a physical view of tumorigenesis and cancer recovery processes.
	
In summary, landscape and flux theory has provided a powerful tool for quantifying and visualizing the dynamic transitions between different stable states that occur during various important biological processes in multicellular organisms, such as the cell cycle \cite{zhang2018exploring}, cell differentiation \cite{yu2017epigenetic}, and the initiation and development of disease \cite{yu2019landscape}. Based on the epigenetic landscape, key elements and links are identified through global sensitivity analysis, which quantifies how much perturbation steady stable states can withstand.

\subsection{Indicators of resilience in biological systems}\label{3Indicatorresilience}
Due to the presence of bistability or multistability, a biological system can shift abruptly from one state to an alternative stable state at a tipping point \cite{veraart2012recovery}. Once the shift has occurred, reversing it may be difficult \cite{hutchings2004marine}. Thus, finding early warning signals before the catastrophic shift occurs has important implications for preventing biological population collapses and the onset of diseases by effective human intervention. 
Next, we review studies on detecting effective indicators as early warning signals before biological population collapse and disease onset.

\subsubsection{Early warning signals before biological populations collapse}
As discussed in Section \ref{BiologyUnicellularSec}, due to the cooperative growth of budding yeast in sucrose, bistability and a fold bifurcation appear when the dilution factor is changed \cite{dai2012generic}. As shown in Fig. \ref{Bistability_Yeast}, fold bifurcation occurs when the stable and unstable fixed points ``collide''. Populations exposed to higher dilution factors always collapse, as extinction is the only stable state. At lower dilution factors near the bifurcation, the population becomes less resilient because the basin of attraction around the stable state (characterized by the distance between the stable and unstable fixed points) decreases in size, increasing the chance of extinction due to stochastic perturbations. Dai et al. \cite{dai2012generic} tested the resilience of yeast populations at different dilution factors to a 1-day salt shock with sodium chloride and found that populations at low dilution factors were able to recover from the perturbation. Conversely, those at high dilution factors went extinct.

When a biological population approaches a tipping point, its rate of recovery from small perturbations tends to zero \cite{veraart2012recovery}. Thus, in yeast populations in the vicinity of a bifurcation \cite{dai2012generic}, the critical slowing down phenomenon is directly observed. Since in this vicinity populations become more vulnerable to disturbance and more time is needed for them to recover from small perturbations, the system becomes more correlated with its past, leading to an increase in autocorrelation between density fluctuations at different time points. Among more than 46 replicate yeast populations, the magnitude of the fluctuation in population density over a period of five days was shown to increase as the dilution factor increased, and the standard deviation and the coefficient of variation of the fluctuation also both increased \cite{dai2012generic}. These three indicators based on critical slowing down are found to be good early warning signals. However, skewness, a suggested early warning signal that is not based on critical slowing down, measures the asymmetry of fluctuations in population density \cite{guttal2009spatial}and is not a good warning signal for yeast population collapse \cite{dai2012generic}.

\begin{figure}[!ht]
\centering
\includegraphics[width=0.95\linewidth]{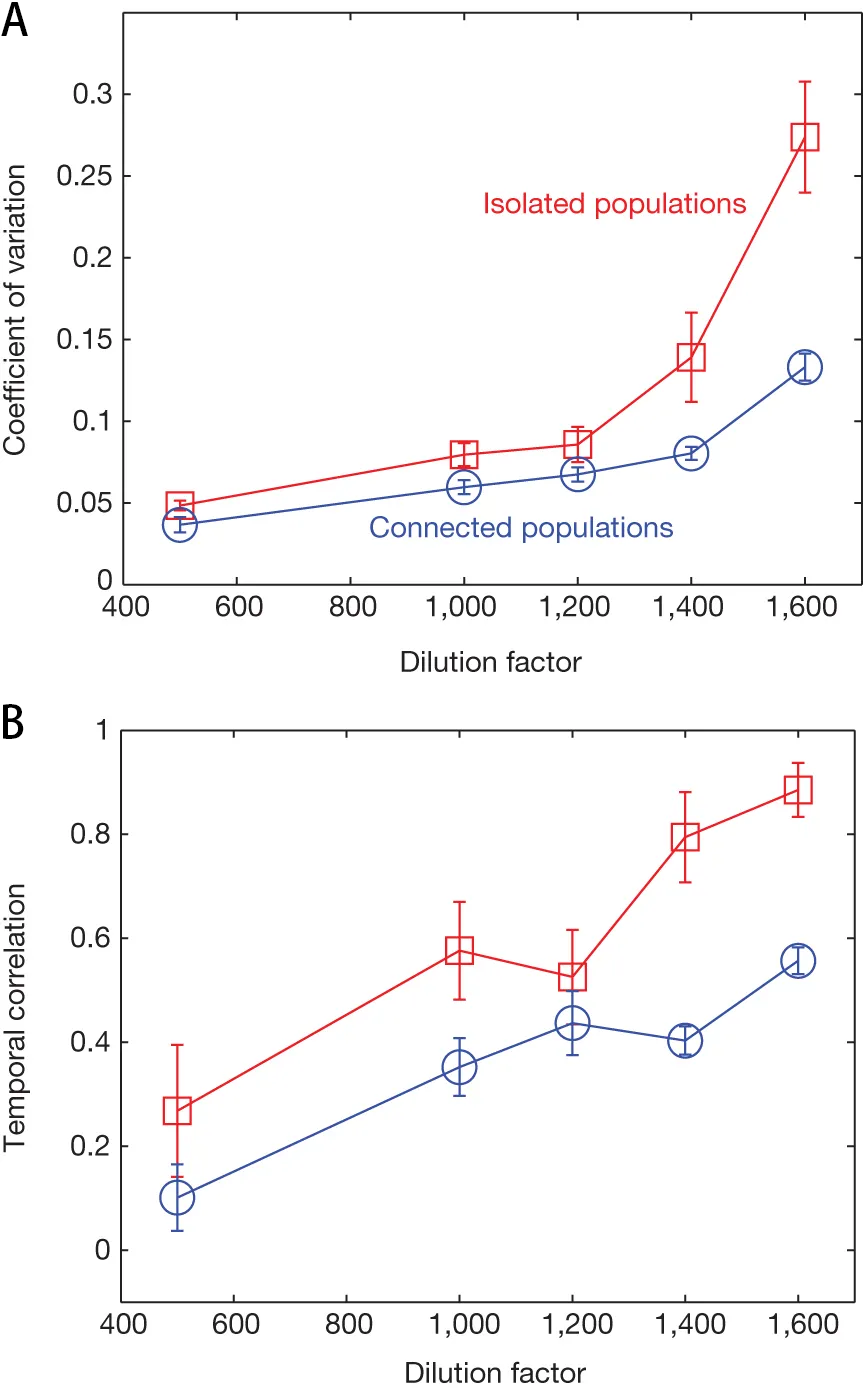} % figure 46
\caption{
Early warning signals based on fluctuations are suppressed in connected populations. The coefficient of variation (A) and the temporal correlation (B) of both isolated populations (red squares) and connected populations (blue circles) increased before the tipping point occurred. The signals were suppressed in the connected populations, possibly due to the averaging effect of dispersal.\\
\textit{Source:} The figure is from \cite{yu2016physical}.
}
\label{EarlySignalYeast}
\end{figure}

The three indicators discussed above (autocorrelation, the standard deviation, and the coefficient of variation) based on critical slowing down are measured from the time series of population density, which requires observation over a long time span. Dai et al. \cite{dai2013slower} identified indicators based on spatial structure as early warning signals for yeast population collapse. They spatially extended the yeast populations \cite{dai2012generic}, and spatial coupling between local populations was introduced by transferring 25\% of each local population to each of its nearest neighbors. Consistent with critical slowing down, clear increases in the coefficient of variation and the autocorrelation of connected populations were observed as the system approached the tipping point. However, the magnitudes of the increases in the coefficient of variation and the autocorrelation were smaller than those observed in the isolated populations (Fig. \ref{EarlySignalYeast}). Such suppression of the two leading indicators in connected populations is caused by the averaging effect of dispersal. In an extreme case in which ten populations are mixed completely every day, the populations show almost no increase in variation before the tipping point is reached. However, spatial coupling introduces another warning indicator, recovery length, which is the distance necessary for connected populations to recover from spatial perturbations. The recovery length increases substantially before population collapse occurs, suggesting that the spatial scale of recovery can provide a superior warning signal before tipping points arise in spatially extended systems \cite{dai2013slower}. Field-based evidence for spatial signatures of critical slowing down in natural conditions was described by Rindi et al. \cite{rindi2017direct} in a marine benthic system approaching a tipping point at which the system shifts from a canopy-dominated to a turf-dominated state.

In addition to signals based on critical slowing down, trait variation, especially a change in fitness-related phenotypic traits, has been identified as an early warning signal of population collapse \cite{clements2018indicators}. For instance, significant decreases in the mean body size of individuals in stressed populations occur prior to collapse or extinction of the populations 
\cite{clements2016including}. When the cyanobacterial population approaches a tipping point, the photosystem II quantum yield (an indicator of chlorophyll concentration) decreases significantly \cite{veraart2012recovery}. Berghof et al. \cite{berghof2019body} showed that there is genetic variation in resilience indicators based on body weight deviations in laying chickens. The maturation schedules of cod populations under fishing stress change significantly before the populations collapse \cite{dakos2019ecosystem}. In addition, recent studies show that combining trait-based and traditional density-based indicators can not only provide early warning signals that are significantly more reliable but also generate reliable signals earlier than the use of abundance data alone \cite{clements2016including}. For instance, mean body size or variation in body size combined with fluctuations in population density could provide a more sensitive predictor of critical transitions in protist populations than either of the two indicators alone \cite{clements2016including}.

Thus, before biological populations collapse, statistical indicators based on critical slowing down, e.g., temporal and spatial fluctuations in population density, increased skewness and phenotypic trait variation, are efficient early warning signals. On the one hand, as discussed above, not all indicators always perform well in certain population collapses. For example, skewness is not a good indicator of yeast population collapse \cite{dai2012generic}, and variation in population density fluctuation does not increase before cyanobacteria populations collapse due to increases in light intensity \cite{veraart2012recovery}. On the other hand, some indicators may appear at the same time in the same system. For example, Drake et al. \cite{drake2010early} conducted an experiment using replicate laboratory populations of Daphnia magna exposed to a controlled decline in environmental conditions. Four statistical indicators, i.e., coefficient of variation, autocorrelation, skewness, and spatial correlation in population size, all showed evidence of the approaching bifurcation as early as 110 days (8 generations) before the transition occurred.

\subsubsection{Critical slowing down as an early warning signal of the onset of disease}
\begin{SCfigure*}
\centering
\includegraphics[width=0.667\textwidth]{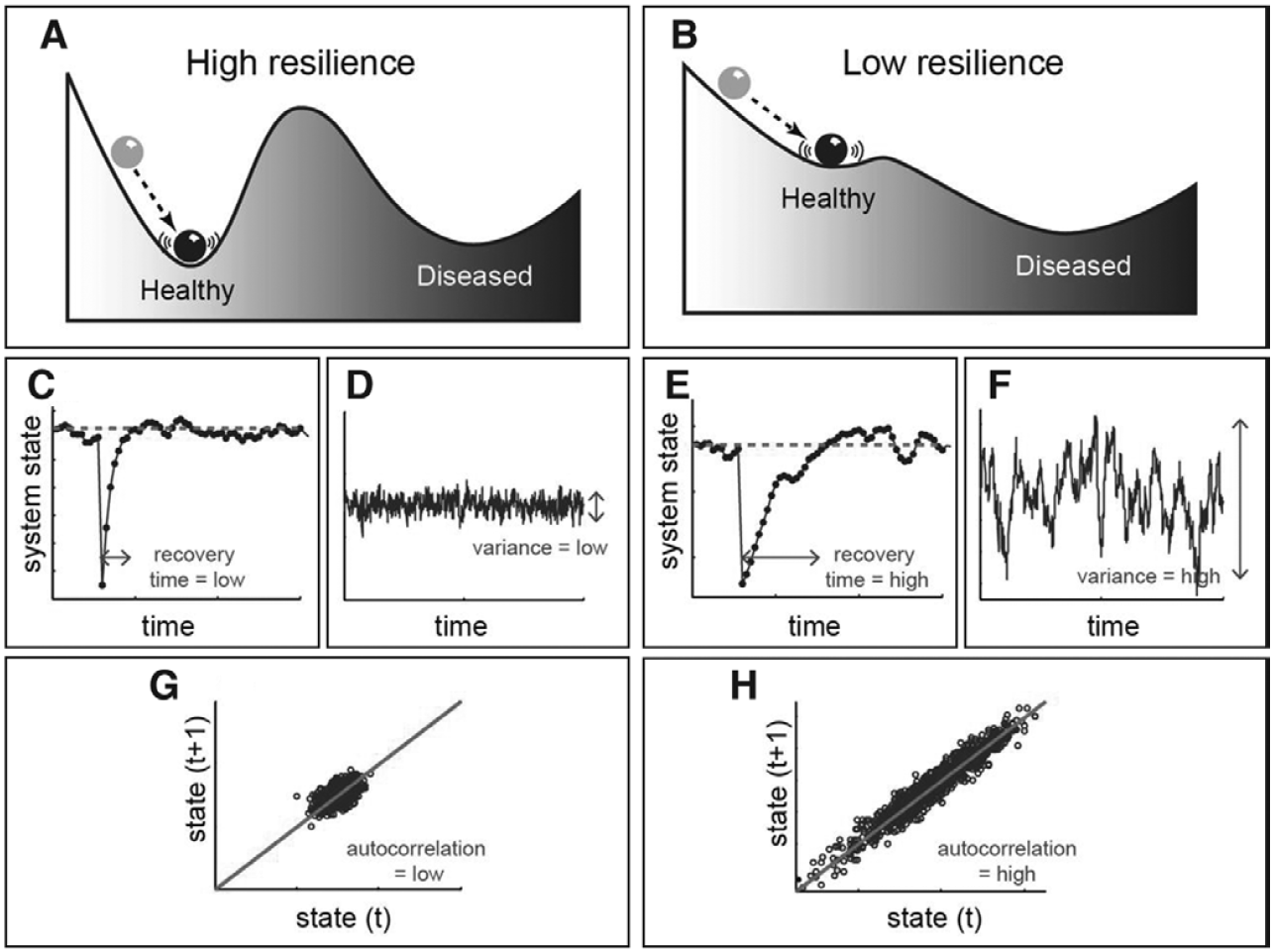} % figure 47
\caption{Critical slowing down is a generic indicator that a patient has lost resilience in the sense that the patient may shift more easily from his or her current ``healthy'' state into an alternative ``diseased'' state. Compared to cases with high resilience (A, C, D and G) that are far from the tipping point, if the patient is close to the tipping point (B), three statistical indicators, recovery time (E), variance (F), and autocorrelation (H), of critical slowing down all increase.\\
\textit{Source:} The figure is from \cite{rikkert2016slowing}.
}
\label{csd}
\end{SCfigure*}

In humans and animals, resilience is the capacity to be minimally affected by disturbances or to rapidly return to the state that was maintained before exposure to a disturbance. Less resilient people or animals are expected to be more susceptible to environmental perturbations, which may lead to disease or death \cite{berghof2019body}. Due to the inherent complexity and nonlinear dynamics of biological systems, transitions from health to disease are usually not continuous but instead involve sudden shifts in system states \cite{trefois2015critical}. Treating a person or an animal as a dynamical system, ``healthy'' and ``diseased'' are two alternative stable states that can be modeled as two basins in the stability landscape (Fig. \ref{csd}A and B). When the dynamics of an individual have high resilience and are far from the tipping point, it is difficult to move the ball (representing the state of the individual) from ``healthy'' to ``diseased'' since the basin representing the ``healthy'' state is deep and wide. As the dynamics of an individual approach a tipping point, the ``healthy'' basin becomes shallow and narrow, increasing the chance that a transition from ``healthy'' to ``diseased'' will occur due to stochastic fluctuations. Once the catastrophic shift occurs, it is difficult to reverse. Therefore, early diagnosis is important in medicine, and a trend from diagnosing disease to predicting disease has emerged \cite{balling2019diagnosing}, especially for chronic diseases. Increasing evidence shows that critical slowing down could be a source of efficient early warning signals for the onset of diseases. 
Next, we review examples of the detection of early warning signals prior to disease onset.

Epilepsy is a central nervous system disorder in which abnormal brain activity occurs, causing seizures or periods of unusual behavior and sensations and sometimes loss of awareness \cite{namazi2016signal}. A national survey of the UK population found that epilepsy and seizures can develop in any person at any age \cite{holmes2019attitudes}. Epilepsy makes living a normal life difficult because seizures begin and end suddenly and are not easy to predict \cite{trefois2015critical}. Since the 1970s, researchers have been seeking ways to predict the occurrence of seizures from the electroencephalograms (EEGs) of epilepsy patients. For example, Meisel et al. \cite{meisel2012scaling} found that a Hopf bifurcation may occur near seizure onset, and increased variance in spiking patterns of individual neurons has been proposed as an early warning signal for detecting the onset of a sudden epileptic seizure. However, most of the previous studies on prediction that yielded rather promising results were recently found to be irreproducible \cite{mormann2016seizure}. For instance, Wilkat et al. \cite{wilkat2019no} analyzed long-term, multichannel recordings of brain dynamics in 28 subjects with epilepsy and found no clearcut evidence for critical slowing down prior to 105 epileptic seizures. However, prior to the critical transition between the ictal and postictal states of an epileptic seizure (the self-termination of the seizure), human brain electrical activity at various spatial scales exhibits the common dynamical signature of an impending critical transition, indicating critical slowing down \cite{kramer2012human}.

Clinical depression is a severe mood disorder characterized by a wide array of symptoms such as inability to sleep, low mood, loss of interest in activities that were previously enjoyed, and suicidal tendencies \cite{trefois2015critical}. These symptoms are correlated and form a network. For example, a person may become depressed through the following causal chain of feelings and experiences: stress $\rightarrow$ negative emotions $\rightarrow$ sleep problems $\rightarrow$ anhedonia \cite{borsboom2013network}. In the network of symptoms, positive feedback loops exist, such as worrying $\rightarrow$ feeling down $\rightarrow$ even more worrying or feeling down $\rightarrow$ engaging less in social life $\rightarrow$ feeling even more down \cite{bringmann2013network}. Such positive feedback loops can cause a system to have alternative stable states, and gradually changing external conditions may cause the system to approach a tipping point. Thus, the onset and remission of clinical depression may occur suddenly. Leemput et al. \cite{van2014critical} showed for a large group of healthy individuals and patients that the probability of an upcoming shift between a depressed and a normal state is related to statistical indicators of critical slowing down. The authors monitored the time series of emotion scores for four observed variables representing four emotional states: cheerful, content, sad and anxious. Among the general population sample, 13.5\% of subjects showed transitions from normal states to clinically depressed states. In individuals who are close to a tipping point, both temporal autocorrelation and variance of fluctuations in emotional scores are higher than in individuals who are far away from emotional transitions (Fig. \ref{csd}). This difference suggests that critical slowing down could be an early warning signal of the onset and termination of depression. In addition, Wichers et al. \cite{wichers2016critical} directly observed increasing early warning signal patterns prior to individuals' critical transitions to depressive symptoms based on long-term (10 times a day over 239 days) emotion monitoring data. The ability to anticipate transitions between healthy and diseased states could prove beneficial in terms of the timing and magnitude of treatment interventions, two abilities that are essential in health care optimizing.

In addition to applications in epilepsy and clinical depression, early warning signals, especially those based on critical slowing down, have been found to occur prior to the onset of other diseases. For example, Quail et al.\cite{quail2015predicting} detected early warning signals that predicted the onset of abnormal alternating cardiac rhythms. They treated embryonic chick cardiac cells with a potassium channel blocker; such treatment leads to the initiation of alternating rhythms, and the transition is associated with a period-doubling bifurcation. When the system approaches the bifurcation, its dynamics slow down, and noise amplification and oscillations in the autocorrelation function appear in the aggregate interbeat intervals. On return maps, the slope of the line that relates the current interbeat interval to the following interbeat interval can indicate how far the system is from the transition. Hsieh et al. \cite{hsieh2014changing} developed a statistical indicator based on a probabilistic risk assessment framework and used it to predict and assess ozone-associated decrements in lung function. They proposed a composite indicator as a predictor; the indicator includes standard deviation, coefficient of variation, skewness, autocorrelation, and coefficient of spatial correlation. Tambuyzer et al. \cite{tambuyzer2014interleukin} found that increases in the amplitude of fluctuations in interleukin-6 levels in individual pigs can be used as an indicator of the infection state.

Critical slowing down does not always indicate an upcoming critical transition for noisy biological systems. Due to the decreased stability of the attractor, systems may exhibit flickering between two states until the alternative attractor eventually gains stability and becomes the new stable state \cite{balling2019diagnosing}. This has been observed in the onset of paroxysmal atrial fibrillation \cite{rikkert2016slowing} and epilepsy \cite{kramer2012human}. 
In conclusion, the development and experimental validation of early warning signals for the onset of diseases is a promising direction that may have future therapeutic applications related to the prediction of therapeutic responses and clinical outcomes and the design of personalized treatments.
	
\subsubsection{Dynamical network biomarkers in the progression of complex diseases based on gene expression data}
The studies on detecting early warning signals of critical transitions in biological systems discussed above are based on specific phenotypic data. According to the central dogma, gene expression is the cornerstone of all cellular activities inside human and animal cells. In addition, due to the development of high-throughput technologies, massive amounts of gene expression data have been accumulated \cite{balling2019diagnosing}. It is crucial and necessary to evaluate effective early warning signals for critical transitions in biological processes, especially the development of diseases, based on gene expression profiles. In 2012, Chen et al. \cite{chen2012detecting} derived a theoretical index based on a dynamical network biomarker (DNB) that serves as a general early-warning signal indicating an imminent bifurcation or sudden deterioration before a critical transition. The authors validate the relevance of DNB to diseases using related experimental data and functional analysis.
Next, we review studies in which DNBs were identified based on time-series gene expression data and used to predict critical transitions in biological processes.

\begin{figure*}
\centering
\includegraphics[width=0.8\textwidth]{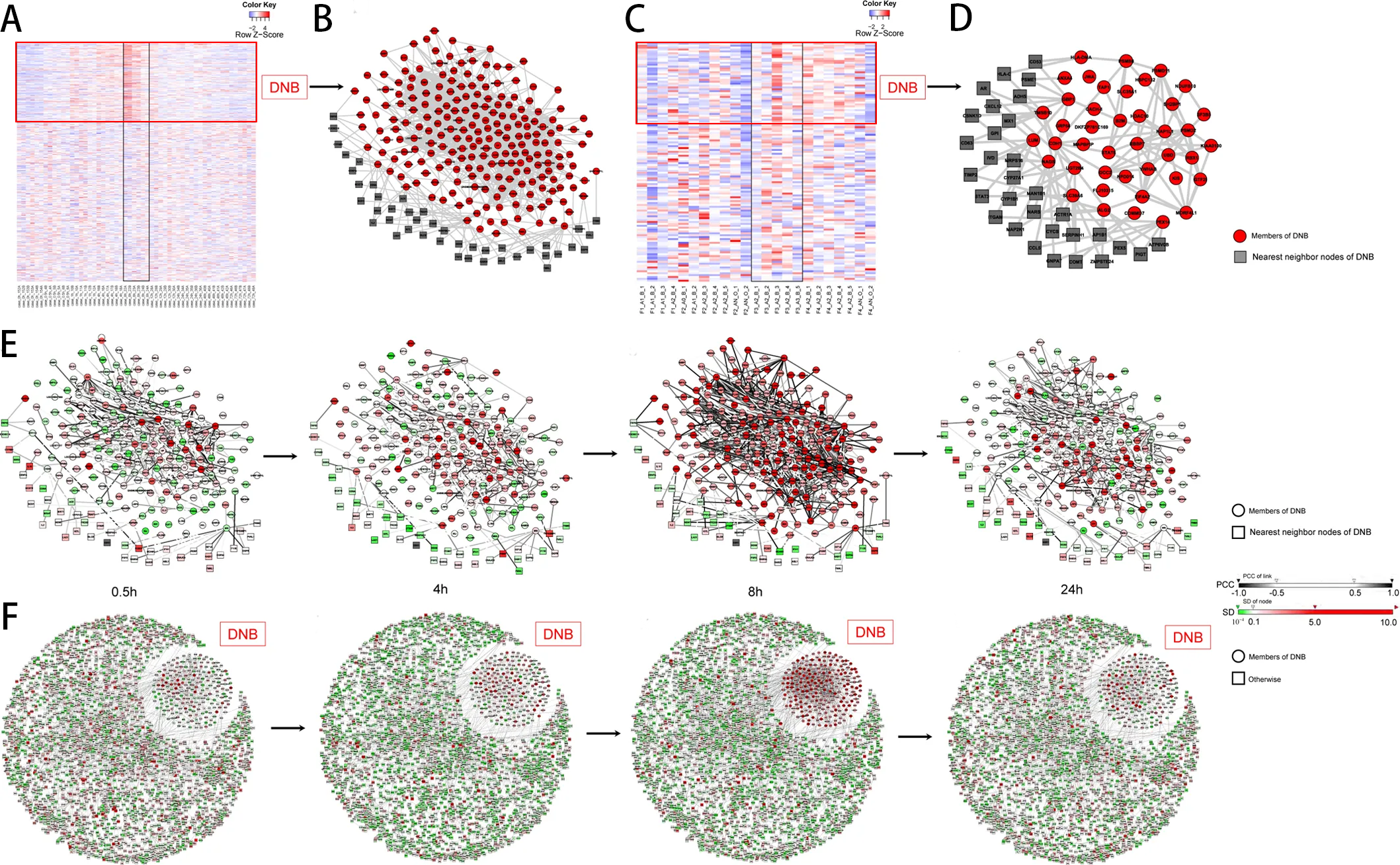} % figure 48
\caption{
DNBs as early warning signals for two complex diseases.
(A) and (C) show the expression profiles of detected DNBs (red rectangle) and other randomly selected non-DNBs from the acute lung injury and HBV-induced liver cancer datasets, respectively. (B) and (D) are the corresponding gene co-expression networks; red nodes represent the DNBs, and their nearest neighbors are shown as gray nodes. For acute lung injury, the dynamic evolution of the network structure for the identified DNB subnetwork is shown (E), and the whole mouse network including DNBs (F) is visualized; the 8-h time point is the critical point.\\
\textit{Source:} The figure is from \cite{chen2012detecting}.
}
\label{DNB}
\end{figure*}

Given the gene expression profiles of numerous genes from several samples or across different experimental conditions, the correlations between gene expression levels can be calculated and used to map a gene co-expression network \cite{stuart2003gene}. The expression levels of specific genes may change significantly during the development of a complex disease; some examples of such genes are oncogenes \cite{jeong2014overexpression} and tumor suppressor genes that act as unigenes \cite{zhu2019identification} in many cancers. When the levels of expression of these genes change, the correlations between genes also change, leading to a change in network structure. Through an analysis of the nonlinear dynamics of gene expression near the bifurcation point, Chen et al. \cite{chen2012detecting} demonstrated that there exists a group of genes (or more generally molecules) that act as DNBs and showed that when a system reaches the predisease state, the Pearson correlation coefficients between DNBs ($PCC_d$) increase while the correlations between DNBs and non-DNBs ($PCC_o$) decrease. Additionally, the average standard deviations of DNBs ($SD_d$) drastically increase in a manner that is coincident with the critical slowing down phenomenon \cite{van2014critical}. As shown in Fig. \ref{DNB}, the connections between DNBs become intense, and the standard deviations of the DNBs' states increase when the gene co-expression network is in the predisease state. Chen et al. \cite{chen2012detecting} further proposed the composite index $I=\frac{SD_d\cdot{PCC_d}}{PCC_o}$, which increases significantly in the predisease state and has been shown to be related to the onset of complex diseases such as acute lung injury and liver cancer \cite{liu2013dynamical}.

Detecting DNBs through the implementation of such a framework \cite{chen2012detecting} requires the calculation of reliable Pearson correlation coefficients based on multiple case samples at each time point. This limits the application of this framework to individual predisease state prediction. Due to the strongly fluctuating and correlated nature of DNBs in the predisease stage, their normalized expression levels have a double-peak distribution; in contrast, non-DNBs have a single-peak distribution in the predisease stage. Thus, Liu et al. \cite{liu2014identifying} used Kullback-Leibler divergence, which measures the difference between two data distributions, to formulate a DNB single-sample score that can be used to identify the predisease state of a disease based on a single case sample. This facilitates early diagnosis before the onset of the disease state or the occurrence of severe deterioration. In addition, near the bifurcation point, DNB biomolecules exhibit significantly collective behavior with fluctuations, resulting in an increase in local entropy. Therefore, an index called the single-sample landscape entropy score (SLE score) can predict the critical transition of biological processes \cite{liu2019single}.
Next, we show the applications of such DNB-based methods to the detection of early warning signals for critical transitions in the development of diseases and other biological processes.

\textbf{Signaling the onset and deterioration of diseases.} Diabetes, one of the most common chronic diseases, has two major subtypes: type 1 diabetes is caused by failure of the pancreas to produce enough insulin due to the loss of beta cells \cite{atkinson1999nod}, and type 2 diabetes results from insulin resistance, which makes it impossible for blood sugar to enter cells \cite{hayden2002islet}. The progression of both subtypes of diabetes encompasses multiple stages (i.e., a healthy stage, a predisease stage, and a disease stage), and clinical diagnosis of the disease is usually made in the disease stage, a point at which the disease is challenging to reverse \cite{trefois2015critical}. To detect the predisease stage of type 2 diabetes, Li et al. \cite{li2013detecting} applied a DNB-based method \cite{chen2012detecting} to temporal-spatial gene expression data obtained from rats in different stages of type 2 diabetes. They found two different critical states during type 2 diabetes development, and these were characterized as responses to insulin resistance and severe inflammation. They also showed that most DNB genes, particularly the core genes, tend to be located upstream of biological pathways, indicating that DNB genes act as causal factors rather than by producing downstream molecules that change the transcriptional activity of other genes. In the case of type 1 diabetes, two DNBs have been identified as early warning signals for two critical transitions leading to peri-insulitis and hyperglycemia in nonobese diabetic mice \cite{liu2013detecting}. Moreover, Zeng et al. \cite{zeng2014deciphering} identified the modules present at the predisease stage based on dynamical network biomarkers that can serve as warning signals for the predisease state.

Influenza, an infectious disease caused by the influenza virus, spreads worldwide, leading to global respiratory illness and even death. The development of a single-sample-based DNB method makes it possible to predict influenza in advance at the individual level. Based on temporal gene microarray data of human influenza infection caused by the H3N2 virus, Liu et al. \cite{liu2014identifying, liu2019single} successfully identified predisease samples from individuals before severe disease symptoms had emerged in those individuals. In addition, influenza displays seasonal collective outbreaks. To identify the early warning signals that precede an influenza outbreak, Chen et al. \cite{chen2019detecting} collected historical longitudinal records of flu-caused hospitalization from 278 clinics distributed in 23 wards in Tokyo, Japan, and from 225 clinics distributed in 30 districts in Hokkaido, Japan, from 2009 to 2016. They constructed a network based on the actual locations of wards and their adjacency relationships. By applying a local network DNB-based index to predict influenza outbreaks in each ward, Chen et al. \cite{chen2019detecting} detected an early warning signal that provided an average 4-week window lead before each seasonal outbreak of influenza.

Acquired drug resistance of cancer cells is considered the primary reason patients fail to respond to cancer therapies. Generally, acquired drug resistance can be regarded as a biological network evolutionary process that enables the system to adapt to the drug environment \cite{liu2018hunt}. According to the biological features of the time-dependent progression of MCF-7 breast cancer cells exposed to tamoxifen, the process of acquiring drug resistance can be divided into three stages: nonresistance, preresistance (or the tipping point) and a resistance state. DNB network alterations occur before tamoxifen resistance is observed, and these alterations follow the appearance of mutated genes \cite{liu2018hunt}. Furthermore, distant metastasis of cancer cells is the leading cause of cancer death. Detection of the tipping point before metastasis occurs is critical in preventing further irreversible deterioration.

DNB-based methods have been used to detect early warning signals of lung cancer \cite{liu2019single} and of metastasis of other cancers \cite{liu2017quantifying}. Yang et al. \cite{yang2018dynamic} recently analyzed time-series gene expression data obtained in a spontaneous pulmonary metastasis mouse HCCLM3-RFP model using a DNB-based method and identified CALML3 as a core DNB member. This protein has been further verified as a suppressor of metastasis; thus, it represents a prognostic biomarker and a therapeutic target in hepatocellular carcinoma.

\textbf{Identifying the tipping points in other biological processes.} The DNB-based method can be used to predict the critical transitions in the development of diseases as well as the critical transitions in other complex biological processes based on time-series gene expression data. In the cell population-based view, the cell differentiation process has been seen as a stereotyped program leading from a single progenitor to a functional cell. Cell fate induction requires broad changes in gene expression profiles at the single-cell level, and cell-to-cell gene expression stochasticity could play a key role in differentiation. Mojtahedi et al. \cite{mojtahedi2016cell} showed that commitment of blood progenitor cells to the erythroid or myeloid lineage is preceded by destabilization of their high-dimensional attractor state. This destabilization causes differentiating cells to undergo a critical state transition. At the point at which commitment to a specific fate occurs, a peak in gene expression variability appears \cite{buganim2012single}. Thus, the structure of the gene co-expression network changes significantly. Using the DNB method, a group of DNB-encoding genes whose expression shows increased fluctuations and correlations that can be used as early warning signals for differentiation in primary chicken erythrocytic progenitor cells have been identified \cite{richard2016single}. The critical differentiation state of MCF-7 breast cancer cells can also be identified using the DNB method. This approach provides an opportunity to interrupt and prevent the continuing costly cycle of managing breast cancer and its complications.

Immunotherapy using antibodies that block immune checkpoints is an emerging success story for some patients with cancer. However, the majority of patients gain no benefit but experience considerable toxicity from such therapy \cite{lesterhuis2017dynamic}. Many efforts have been made to identify biomarkers that respond to immune checkpoint blockade in cancer \cite{lesterhuis2017dynamic}. However, none of the biomarkers identified to date respond to such blockades reliably enough to be approved for routine use. Fortunately, the therapeutic response to immune checkpoint blockade represents a critical state transition of a complex system. DNB-based methods have shown their ability to predict transitions from a predisease state to a disease state \cite{liu2019single}. Thus, DNB-based methods may be a potential tool for predicting immune checkpoint blockade. Lesterhuis et al. \cite{lesterhuis2017dynamic} proposed that these dynamic biomarkers could help practitioners identify responses in patients who otherwise appear to be nonresponders, thus making it possible to preserve such responses during immunotherapy and facilitating the identification of new therapeutic targets for combination therapy.

In other words, DNBs are a group of molecules that display strongly correlated activities and fluctuations \cite{yang2018dynamic} that serve as early warning signals for the onset and deterioration of complex diseases as well as indicators of other biological processes such as cell fate induction. The identification and monitoring of DNBs provide a new way to delineate the mechanisms that underlie various biological processes and can help in the scheduling of treatments for complex diseases.

Based on a summary of the existing studies on resilience in biological networks, we conclude that bistability, or multistability, exists universally in biological networks at different levels ranging from genetic circuits at the molecular level through unicellular populations and disease networks at the phenotypic level. Bistability and multistability are generated through the underlying feedback loops and ultrasensitivity \cite{pomerening2003building} or by particular network topology \cite{craciun2006understanding}. Due to the existence of alternative stable states, biological networks can respond digitally to external or internal stimuli. Once the stimulus exceeds a tipping point, critical transitions appear, and the biological network shifts abruptly from one state to another in a way that is difficult to predict. Nevertheless, we can identify resilience indicators as effective early warning signals for critical transitions in biological systems. These indicators include the coefficient of variance, autocorrelation based on critical slowing down, and dynamical network biomarkers. Studies of resilience in biological networks can help us understand complex biological systems, design effective therapeutic methods, and find more health management applications. 
Next, we review the literature on the resilience of socioecological systems and social systems.

\section{Behavior transitions in animal and human networks}\label{Social}

As in ecological systems (chapter \ref{Ecology}) and biological networks (chapter \ref{Biology}), the resilience of a social network, which is defined as its ability to cope with perturbations, shocks, and stress, can take on different forms in various settings. Resilience requirements can range from the preservation of the entire system or the sustainability of its structure and operational ability. We consider resilience in the three general settings described below.

In the first setting, we consider the social systems of humans, which are a subject of traditional sociology. In this setting, we acknowledge two different notions of resilience. One focuses on something that we call the {\it cultural resilience} of a society, which requires that the people in a society evolve their opinions and beliefs in a smooth and orderly manner. The other type of resistance, termed here as {\it survival resilience}, is concerned with preserving a social system. Intuition suggests that this type of resilience requires that a society or community respond to external stress or challenges.

The second setting considers the resilience of social animals and their communities. In this setting, we examine only survival resilience, which focuses on species and preservation of their communities in the presence of environmental change and competition from other species and communities. 

The third setting focuses on a type of resilience related to two strongly linked subsystems: social networks and ecological systems. As a species, humans have been extremely successful in spreading and in dominating all ecosystems on Earth. However, the further unrestricted growth of the human population and the increased exploitation of the living and mineral resources of the Earth may threaten the stability of the global ecosystem and make it difficult to preserve human civilization. We refer to this complex system as the ``planetary socioecological system'' to underscore its global range. The goal is to increase the {\it survival resilience} of this complex system in the face of the increasingly limited resources of our planet and the rapid changes in the environment caused by the increasing burden of human activities.

\noindent
{\bf Cultural resilience in social networks.}
Sociology was initially a leader and an early adopter of social network analysis. The first example of such an analysis focused on a friendship network in a primary school class of 1880-81 and was conducted by the German teacher Johannes Delitsch using mixed methods ~\cite{Heidler2014relationship}. A more advanced approach, similar to modern social network analysis and called {\it sociometry}~\cite{Moreno1938statistics}, is discussed in Ref. \cite{Freeman2004development}. This approach was used to analyze the interactions among inmates in a prison \cite{Moreno1932application}. However, data collection issues posed a significant barrier to sociometric analyses. Data collection was laborious and imprecise, and the results were prone to misinterpretation. A breakthrough came in the 1990s when internet and wireless-based interactions became widespread. These interactions provided easily collected datasets that were scalable to millions and even billions (e.g., Facebook) of members.
Network analysis has become a gold mine for social scientists. It has also opened network analysis to statistical physics, including novel applications of the classic Ising model~\cite{Newman1999scaling} and newer network models~\cite{barabasi2016network}. Alternative approaches developed in a new branch of study called {\it computational sociology} use agent-based computer simulations \cite{peng2015collective}. Together, these developments have revolutionized sociology \cite{lazer2009life}.

Social analyses demonstrate that network structure can enhance or weaken a network's survival resilience. For instance, the communities in a social network play a vital role in its resilience \cite{magis2010community}. Their strengths and dynamics are closely related to the social capital accumulated by community members \cite{aldrich2015social}. In turn, these strengths are essential for the resilience of a network when it must respond to a crisis or disaster \cite{cutter2014geographies}. In general, the more significant the social capital accumulated within a community is, the stronger the overall resilience of such networks compared to a homogeneous system in which there are weak communities or no communities at all.

In the context of social systems, cultural resilience focuses on avoiding drastic and disruptive changes in the prevailing opinions and beliefs held by the members of a social system. This type of resilience is studied using models of the social interactions that enable opinions to evolve and innovation to spread through the influence and persuasion exerted by personal interactions. Thus, these models do not consider subjugation of societies by military or police forces. The models that represent these processes from the perspective of individuals include the voter model \cite{nardini2008s}, the naming game (NG) \cite{xie2011social}, the threshold model (TM) \cite{granovetter1978threshold}, and their variants. Other models that focus on interactions that lead to group behavior, such as flock/swarm behavior \cite{komareji2013resilience}, are also applicable to the simulation of social network dynamics.

As orderly and evolutionary changes are often necessary to enable social systems to adapt to an evolving environment, such changes might occur so rapidly and be so disruptive that they result in loss of cultural continuity and cultural resilience. Hence, studies that focus on discovering conditions under which drastic shifts in the prevailing opinions and beliefs held by individuals in a social system arise are of great importance. Equally important is the ability to derive from system parameters the critical points at which such shifts appear. Identifying measurable early warning signals or the distance to the tipping point is vital to the development of resilience policies and disaster avoidance strategies.

In naming game studies, $N$ agents hold a set of opinions on a specific topic and send these opinions to each other. One round of a game consists of $N$ steps in which one agent is uniformly randomly selected as a speaker and one of that agent's neighbors is uniformly randomly selected as a listener. The speaker uniformly randomly selects one of his/her opinions and sends it to the listener. If the listener has this opinion in his or her set, then both the speaker and the listener keep the sent opinion and purge all others. Otherwise, the listener adds the sent opinion to his/her set. In this way, local opinion majorities may become global ones. Achieving a global consensus is unlikely unless most agents hold an opinion in common in the initial state. However, critical tipping points arise in the presence of so-called committed agents (also referred to as zealots)~\cite{lu2009naming}. These agents never change the single opinion that they initially hold and promote it whenever they assume the role of speaker.
The critical point value is a function of the fraction of the population belonging to each of the committed minorities and the total number of such communities. A small number of committed communities among which one is sufficiently dominant in size guarantees that the social system will rapidly reach a consensus state~\cite{lu2009naming} consistent with the opinion of the dominant committed minority \cite{xie2011social, xie2012evolution}. Conversely, a large number of small committed communities also guarantees a consensus state. However, the consensus opinion is independent of the initial size of the corresponding committed minority \cite{pickering2016analysis}, and this may create messy or even disruptive transitions. Other outcomes are quite possible \cite{xie2012evolution}. Committed agents can also be defined in the voter model \cite{mobilia2007role}.

Inspired by the binary naming game model, a three-state social balance model with an external deradicalization field has been studied~\cite{singh2016competing}. The mean-field analysis conducted in that study demonstrates the existence of a critical value of the external field strength. This value separates a weak external field under which the system exhibits a metastable fixed point from a solid external field under which there is only one stable fixed point. At the critical value, the field is at a saddle point. The dynamic illustrated by the three-state social balance model is similar to the naming game dynamic, demonstrating that an external field that influences the entire network at the same time does not change the system's dynamics significantly.

\noindent
{\bf Survival resilience.}
In addition to studies of cultural resilience at the population level, social resilience is also studied at the lowest level of social systems, i.e., the individual level. Such studies focus on how the outer environment triggers fluctuations in the lives of individuals. This research has more substantial and deeper roots in psychology, especially in developmental psychology, than in sociology or ecology. Positive psychology, including family-school partnership and community support, has been observed to foster resilience in individuals \cite{bryan2005fostering}. More recently, the role of social capital in communities to which individuals belong has been studied, often in the context of an individual's reactions to traumatic events \cite{almedom2005social}. A natural extension of such studies is the impact of community resilience on health and human development. Such studies focus on the codevelopment of resilience theory in the context of socioecological systems (SESs) \cite{almedom2005social}. Another example of such codevelopment focuses on urban resilience during responses to disaster, terrorism, or other disturbances \cite{eakin2017opinion}; it concentrates on one aspect of survival resilience.

A study of the structural root causes of the resilience of consensus in dynamic collective behavior is presented in Ref. \cite{komareji2013resilience}. The authors construct a dynamic signaling network and use it to analyze the controllability of group dynamics. The system is a small-world network. Resilience (i.e., the alignment of opinions) in the presence of exogenous environmental noise is studied. The authors found that resilience depends strongly on the out degrees of the nodes. The group exhibits a higher level of resilience when larger out degrees are present. In addition, when a single, giant, strongly connected component survives the disturbance, it often self-organizes into a large-scale coherent alignment of individuals.

Social contagion theory describes peer effects, interpersonal influences, and other events or conditions that may result in the emergence of crowd behavior. Based on an analysis of empirical datasets \cite{christakis2013social}, the authors describe the regularities that they discovered. Those discoveries motivate the authors to propose that human social networks may exhibit a ``three-degrees-of-influence'' property. They also reviewed statistical approaches they have used to characterize interpersonal influences related to diverse phenomena such as obesity, smoking, cooperation, and happiness.

The community's resilience in the presence of disastrous events such as violent acts of terrorism is studied in Ref. \cite{dynes2005community}. The author observes that the community is the locus of response to disaster in a way that depends on the amount of ``social capital" that has been accumulated through interactions between members. The author analyzes the ways in which the capabilities that the community already possesses for dealing with disasters can be enhanced. One example is that having policies that address the threat of terrorism can increase the resilience of communities facing such a threat.

Studying the resilience of social systems poses a unique challenge compared to studying the resilience of systems in the natural sciences, in which systems and the interactions among their elements are unambiguously and often formally defined. This challenge arises because the social sciences rely on informal descriptions of social networks and their relationships. Such descriptions originated in traditional sociology, but sociologists have recently also embraced sociophysical models. Most of those, however, are of unproven validity.

\subsection{Cultural resilience of social systems}\label{sc}
Traditional approaches to social systems use the macroscopic scale at which all interhuman interactions can be reduced to a set of mutual standards and patterns that are characteristic of interactions within groups and institutions of an underlying social system. Relatively stable and frequently arising group types such as marriages, families, corporations, and religious groups are often elevated to formal or legal organizational status. An example of such an approach is presented in Ref. \cite{parsons2017social}, which introduces a scheme called AGIL that defines four core functions that collectively underlie the stability and survival of a social system. Function (A) defines adaptation that enables a system to succeed in its physical and social environment and gradually transform the environment to better satisfy its needs. Function (G) defines goal attainment that the system uses to achieve its primary goals. Function (I) defines integration methods that enable a system to coordinate and regulate its components' interrelationships as it strives to create a cohesive whole. Finally, function (L) defines the latency of the feedback a system receives and uses to furnish, maintain, and renew itself and to motivate its individuals to perform their roles according to social and cultural expectations. At such a high level of generality, social resilience is defined by Parsons' general theory in which intra- and intersystemic relationships are characterized by cohesion, consensus, and order imposed by the abovementioned four core functions. The goal is to represent the current status of social norms and rules and thereby to provide a framework for analyzing the system's dynamics. 

The opposite approach focuses on understanding resilience by observing the interactions of individuals endowed with roles and attributes; this approach makes it possible to analyze the dynamics of human interactions and motivates the development of social network approaches. As discussed in the previous section, cultural resilience in social systems focuses on the continuous and orderly evolution of beliefs that avoid disruptions and discontinuities in culture. The prevailing approach to studying the dynamics of consensus formation~\cite{lu2009naming} in social systems relies on modeling elementary interactions between individuals and observing what stable or semistable consensus emerges from the simple rules regarding elementary interactions and the underlying social network structure. Accordingly, we review studies of cultural resilience by discussing approaches that are based on the most popular rules for agent interaction.

\subsubsection{Naming game model}

\begin{figure}[ht!]
        \centering
        \includegraphics[width = 0.9\linewidth]{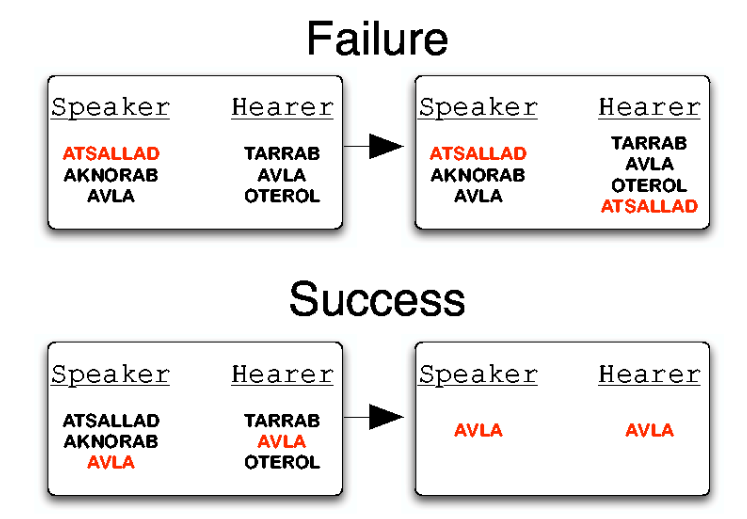} % figure 49
        \caption{Example of the naming game interaction rule. \\
        \textit{Source:} The figure is from \cite{baronchelli2008depth}.}
        \label{fig:ng_demo}
 \end{figure}

The naming game model was initially introduced to account for the emergence of a shared vocabulary through social/cultural learning \cite{baronchelli2008depth}. However, over time, the naming game has become an archetype for linguistic evolution and mathematical social and behavioral analysis \cite{pickering2016analysis}. Figure~\ref{fig:ng_demo} shows an example of the naming game interaction rules.
The most exciting property of the naming game is that it is a minimal model that employs local communications and that captures the generic and essential features of the agreement process that occurs in networked agent-based systems. Examples of such systems are a group of robots, the emergence of shared communication schemes, and the development of a shared key for encrypted communications.

As briefly mentioned earlier, in the naming game, agents perform pairwise interactions to reach an agreement regarding the name assigned to a single object. The time unit is the time needed to execute $N$ interaction steps, where $N$ is the agent population size. In each step, a ``speaker'' is first chosen in a uniformly random manner. Then, again uniformly randomly, a ``listener'' is chosen from among the speaker's neighbors as defined by the underlying communication topology. Thus, the interactions are limited to pairs of neighboring nodes. The speaker uniformly randomly selects a word from its vocabulary and sends it to the listener. If the listener's vocabulary contains this word, the interaction is termed ``successful'', and both nodes collapse their vocabularies to this one word. Otherwise, the communication is ``unsuccessful,'' and the listener adds the received word to its dictionary.
In the context of the spread of opinion, each node's vocabulary represents the opinion that the node supports. Therefore, in this context, the number of opinions is often limited to two, leading to a binary version of the naming game. In such a setting, an agent who holds opinion A and draws from his friend opinion B enters a mixed state in which he considers both opinions. Next-neighbor interaction then allows for this node (this agent) to choose the last received opinion as its unique opinion. This state of hesitation introduces specific resistance to changing the opinion that a person holds immediately upon hearing another opinion from a neighbor.

The basic properties of the naming game were established in Ref. \cite{baronchelli2008depth}. The authors analyzed the role of system size in scaling, the convergence of the system to a stable or a semistable state as a function of both the vocabularies of the agents and the total number of words generated by them, and the disorder transition from the network point of view. Later, a novel method that substantially reduces interagent communication cost while preserving the semiglobal robust consensus was proposed in Ref. \cite{meng2019event} for a class of nonlinear uncertain multiagent systems. Therein, the conditions that guarantee robust consensus were revealed, and Zeno-behavior was theoretically excluded. A method of this type was successfully used in the coordinated control of multiple unmanned surface vessels \cite{liu2019collective}.

\begin{figure}[ht!]
        \centering
        \includegraphics[width=0.95\linewidth]{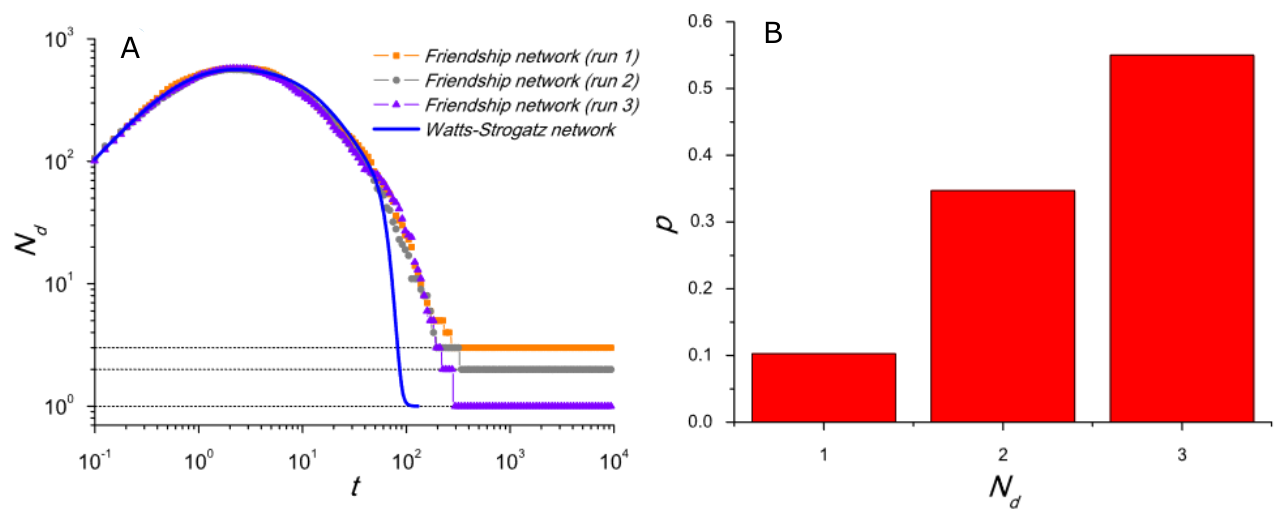} % figure 50
        \caption{(A). The number of different opinions $N_d$ versus time for different high-school friendship networks and the Watts-Strogatz network. (B). The relative frequency of final configurations with $N_d$ different opinions for the same high school friendship networks as in (A) based on 10,000 independent runs. \\
\textit{Source:} The figure is from \cite{lu2009naming}.}
        \label{fig:lu2009_res}
\end{figure}

A study of the dynamics of the original naming game in empirical social networks was presented in Ref. \cite{lu2009naming}. The initial number of opinions was equal to the number of interacting agents. The study focused on the impact that communities in the underlying social graphs have on the process of convergence towards consensus. The authors' main conclusion is that networks with solid community structures make it difficult for or prevent the system from reaching global agreement. The evolution of the naming game in these networks maintains clusters of coexisting opinions practically indefinitely. The authors also investigate how agent-based network strategies facilitate convergence to global consensus. Figure~\ref{fig:lu2009_res} plots the number of opinions surviving to the end of simulation along the time axis.

The naming game model assumes that only one interaction occurs at each time instance; therefore, only two different nodes are active in the relationship at any given time, one as a speaker and another as a listener.
In some extensions of the naming game, the model groups are modified. In one variant, there is no restriction on the number of active nodes, so anyone can be both a speaker and a listener simultaneously. This variant is suitable for modeling competition between groups within a peer-to-peer network \cite{gao2014naming}. As expected, more intensive interactions enable the global consensus to arise earlier than it does in the regular naming game, as does the presence of a large number of initial opinions within the competing groups. The original naming game assumes an initial condition in which each agent creates its own word for an observed phenomenon, and this was a natural assumption in the case of the initial applications. However, in the case of opinion spread, limited size vocabulary or even simply binary vocabulary could be sufficient to model the spread of opinions, and it would offer the benefit of simplifying the analysis. Such simplifications are helpful for analysis of transitions to the stable state that occur in the naming game.

\subsubsection{Committed minorities}

An essential change in the convergence time arises when committed agents are allowed. Without committed agents, the rules of the naming game allow the listener to remain committed to his opinion through only a single interaction. Indeed, upon receiving the same opinion in two interactions, the agent would change its current opinion to the new opinion. The idea of a committed agent is that its resistance to changing its opinion extends to an infinite number of interactions with friends who send this agent an opinion that it does not hold. In short, committed agents hold their initial opinions forever. Nevertheless, they eagerly send it out when selected as speakers.
With a certain percentage of this type of agent, we would expect a consensus change to occur with high probability and speed. Numerous studies have shown that this percentage is significantly lower than half of the population.
Scenarios of this type make it possible for a minority of committed members to enforce a consensus on the majority of society. In Ref. \cite{xie2011social}, the authors demonstrated that a small fraction, $p$, of randomly distributed committed agents could rapidly reverse the prevailing majority opinion in a population. Specifically, when the committed fraction increased beyond a critical value $p_c$ of approximately $10\%$ for fully-connected graphs, there was a dramatic decrease in the time needed for the entire population to adopt the committed opinion. To simplify the analysis, the binary (also called two-word) agreement model was used. In this model, only two opinions, $A$ and $B$, are defined. Therefore, the binary model uses only three states: $\{A\}$, $\{B\}$, and $\{A,B\}$. It is a minimal model of opinion competition. Tab.~\ref{tab:xie2011_table} lists the interactions that occur in the binary model. The following mean-field equations represent the dynamics of a system that features these interactions:

\begin{equation}
  \begin{array}{ll}
  \frac{dn_{A}}{dt} &= -n_{A}n_{B} + n^{2}_{AB} + n_{AB}n_{A} + \frac{3}{2}pn_{AB},\\
  \frac{dn_{B}}{dt} &= -n_{A}n_{B} + n_{AB}n_{B} - pn_{B},
  \end{array}
  \end{equation} \label{BinaryModelInteractions}
where $p$ is the fraction of the committed nodes in state $\{A\}$ in the population; the fractions of uncommitted nodes in states $\{A\}$ and $\{B\}$ in the population are denoted as $n_{A}$ and $n_{B}$, respectively; and the fraction of the nodes in the mixed state $\{A,B\}$ is $n_{AB} = 1 - p - n_{A} - n_{B}$.

\begin{table}[ht!]
\centering
\begin{tabular}{|c|c|} \hline
Before interaction & After interaction \\ \hline
A $\overset{A}{\rightarrow}$  A & A - A \\ \hline
A $\overset{A}{\rightarrow}$  B & A - AB\\ \hline
A $\overset{A}{\rightarrow}$  AB & A - A\\ \hline
B $\overset{B}{\rightarrow}$  A & B - AB \\ \hline
B $\overset{B}{\rightarrow}$  B & B - B\\ \hline
B $\overset{B}{\rightarrow}$  AB & B - B\\ \hline
AB $\overset{A}{\rightarrow}$  A & A - A \\ \hline
AB $\overset{A}{\rightarrow}$  B & AB - AB\\ \hline
AB $\overset{A}{\rightarrow}$  AB & A - A\\ \hline
AB $\overset{B}{\rightarrow}$  A & AB - AB \\ \hline
AB $\overset{B}{\rightarrow}$  B & B - B\\ \hline
AB $\overset{B}{\rightarrow}$  AB & B - B\\ \hline
\end{tabular}
\caption{All possible interactions in the binary agreement model. The left column shows the opinions of the speaker (first) and the listener (second) prior to the interaction; the opinion voiced by the speaker during the interaction is shown above the arrow. The column on the right shows the states of the speaker-listener pair after the interaction.\\
\textit{Source:} The table is from \cite{xie2011social}.}
\label{tab:xie2011_table}
\end{table}

Fixed-point and stability analyses of the above mean-field equations show that for any value of $p$, the consensus state in the committed opinion ($n_A = 1 - p, n_B = 0$) is a stable fixed point of the mean-field dynamics. However, when $p$ is below the critical fraction $p_c$, two additional fixed points appear; one of these is an unstable fixed point (a saddle point), whereas the second is stable and represents an active steady state in which $n_A$, $n_B$ and $n_{AB}$ are all nonzero (except in the trivial case in which $p = 0$).

\begin{figure}[ht!]
    \centering
    \includegraphics[width = \linewidth]{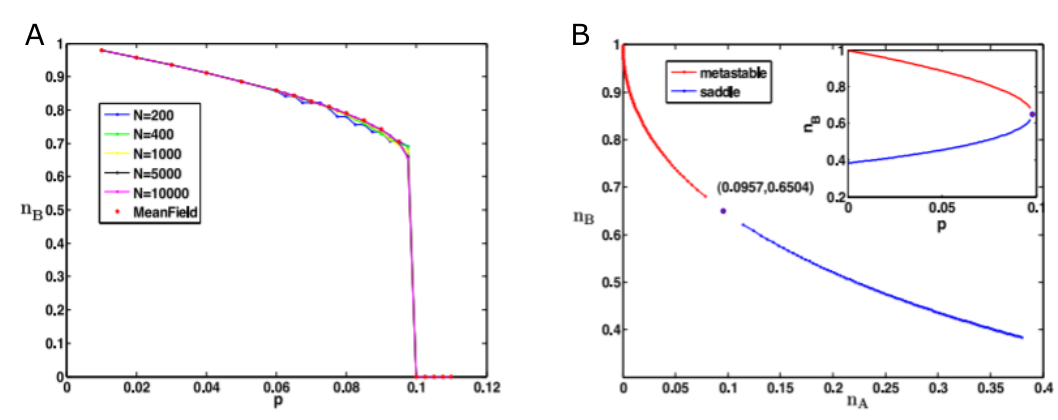} % figure 51
    \caption{ (A) The steady state fraction $n_B$ of nodes in state $\{B\}$ as a function of fraction $p$ of committed nodes in state $\{A\}$ for complete graphs of different sizes, conditioned on survival of the system. The simulation results are averaged over 100 realizations of the binary agreement dynamics. (B) Movement of the stable fixed point and the saddle point in a phase space as a function of fraction $p$ of committed nodes; the graph shows that in this case, $p_c=0.0957$. The inset shows the fraction of nodes in state $\{B\}$ at the stable (red) and unstable (blue) fixed points as $p$ is varied.\\
\textit{Source:} The figure is from \cite{xie2011social}.}
    \label{fig:xie2011_res}
\end{figure}

Figure~\ref{fig:xie2011_res}(A) illustrates the impact of the size of the finite network on the steady-state fraction $n_B$ of the nodes in state $\{B\}$ with the critical fraction $p_c$ of committed nodes in state $\{A\}$ compared to the mean-field approximation. Figure~\ref{fig:xie2011_res}(B) displays a plot of the movement of the stable and saddle points as a function of the fraction p of committed nodes.

Time $T_c$ needed to reach consensus can also be computed using a quasi-stationary approximation, indicating that the survival probability decays exponentially. The precise dependence of consensus time on $p$ can also be obtained for $p < p_c$ by considering the rate of exponential growth of $T_c$ with $N$. In other words, assuming $T_c \sim \exp(\alpha(p)N)$ yields 
  
  \begin{equation}
    T_c(p < p_c) \sim exp((p_c - p)^{\gamma}N).
  \end{equation}
  
\begin{figure}[ht!]
        \centering
        \includegraphics[width = \linewidth]{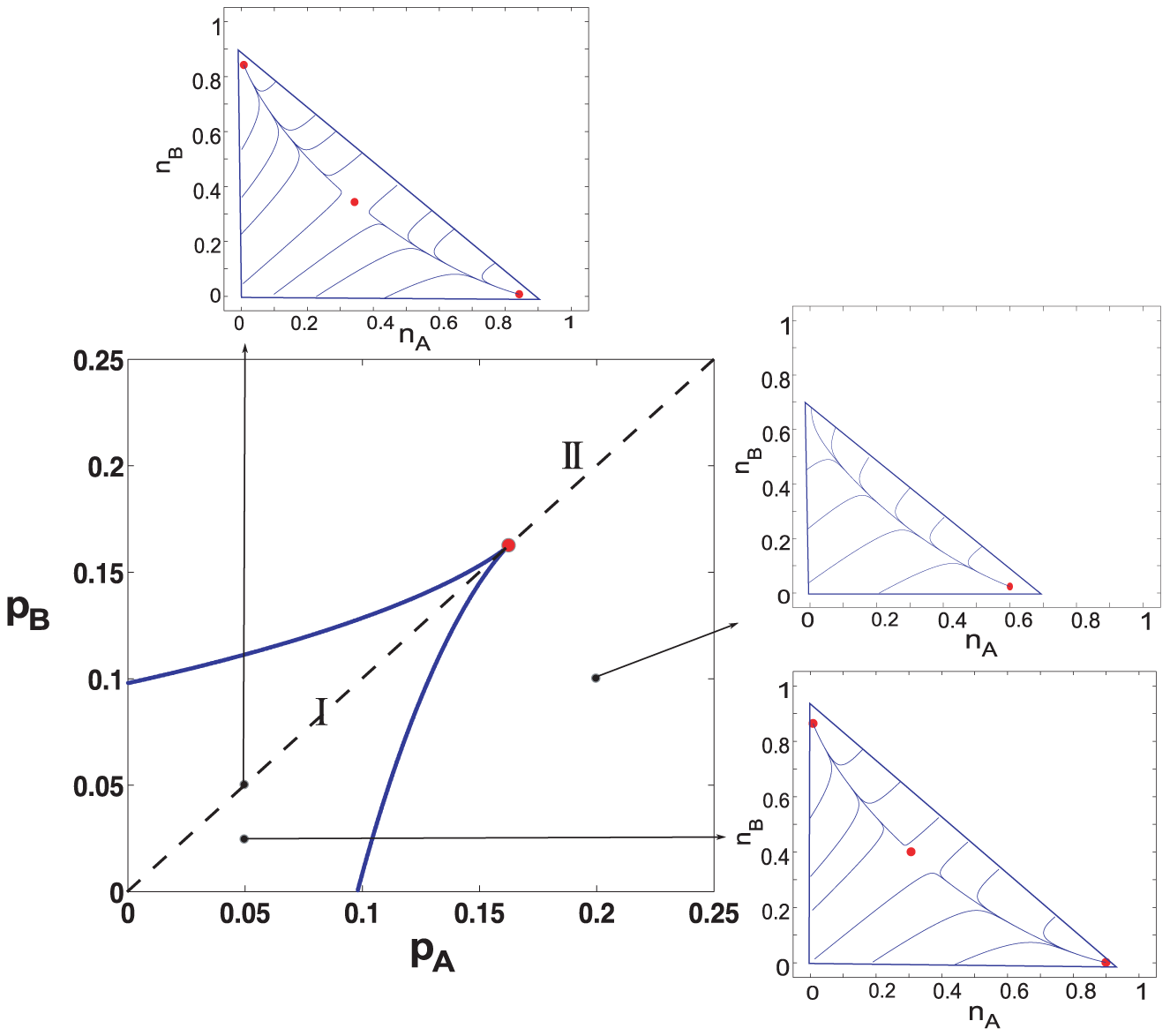} % figure 52
        \caption{Phase diagram obtained by integrating the mean-field equations. The two lines indicate saddle-node bifurcation lines; these lines form boundaries between two regions with markedly different behavior in phase space. For any values of parameters within the beak, denoted as region I, the system has two stable fixed points separated by a saddle point. Outside the beak, in region II, the system has a single stable fixed point. The saddle-node bifurcation lines meet tangentially and terminate at a cusp bifurcation point. \\
 {\it Source:} Figure from \cite{xie2012evolution}.} \label{fig:compcomm}
      \end{figure}

The simulation results for the case in which the underlying network topology is chosen from an ensemble of ER random graphs are also presented in Ref. \cite{xie2011social} for a system of the given size $N$ with the given average degree $\langle k \rangle$. In this case, the qualitative features of the evolution of the system are the same as those observed in the case of the complete graph. Having the above result, a natural question arises regarding what occurs in more general cases of opinion evolution in which more than one group is committed to distinct, competing opinions. The simplest case involves two groups and two opinions, $A$ and $B$; the two groups constitute fractions $p_A$ and $p_B$, respectively, of the total population. Such a case was studied in Ref. \cite{xie2012evolution}. The authors applied the mean-field version of the model. In the asymptotic limit of network size and neglecting fluctuations and correlations, the system can be described by the following mean-field equations:
\begin{equation}
\begin{split}
\frac{dn_A}{ft} = -n_An_B+n^2_{AB}+n_An_{AB}+1.5p_An_{AB}-p_Bn_A,\\
\frac{dn_B}{ft} = -n_An_B+n^2_{AB}+n_Bn_{AB}+1.5p_Bn_{AB}-p_An_B.
\end{split}
\end{equation}
The phase diagram of this system in parameter space $(p_A,p_B)$ consists of two regions (see Fig. \ref{fig:compcomm}), one in which two stable steady states coexist and another in which only a single stable steady state exists. These two regions are separated by two fold-bifurcation (spinodal) lines, which meet tangentially and terminate at a cusp (the critical point). 

 The simplified binary naming game model in which the initial state contains nodes that can have one of only two opinions was studied in Ref. \cite{zhang2011social}. The authors investigate consensus formation, establish the asymptotic consensus times and provide asymptotic solutions for the binary naming game. A six-dimensional ODE analytically captures the dynamics of the binary naming game model with committed agents~\cite{zhang2012analytic}. The authors showed that the tipping points for social consensus decreased when the sparsity of the network increased. The impact of committed agents on the number of opinions persisting in the network and on the scaling of consensus time was investigated in Ref. \cite{zhang2014opinion}.

Further investigations of the value of tipping points show that they depend not only on the percentage of minority agents and the density of the network connectivity but also on the distribution of speaker activities over time \cite{doyle2017effects}. The main conclusion of these studies was that a group with a higher density of short waiting times between its members' speaking activities enforced a consensus on its opinion more readily and quickly than did a group with the same overall speaking activity but a lower initial intensity of speaking.

\begin{figure}
    \centering
    \includegraphics[width=\linewidth]{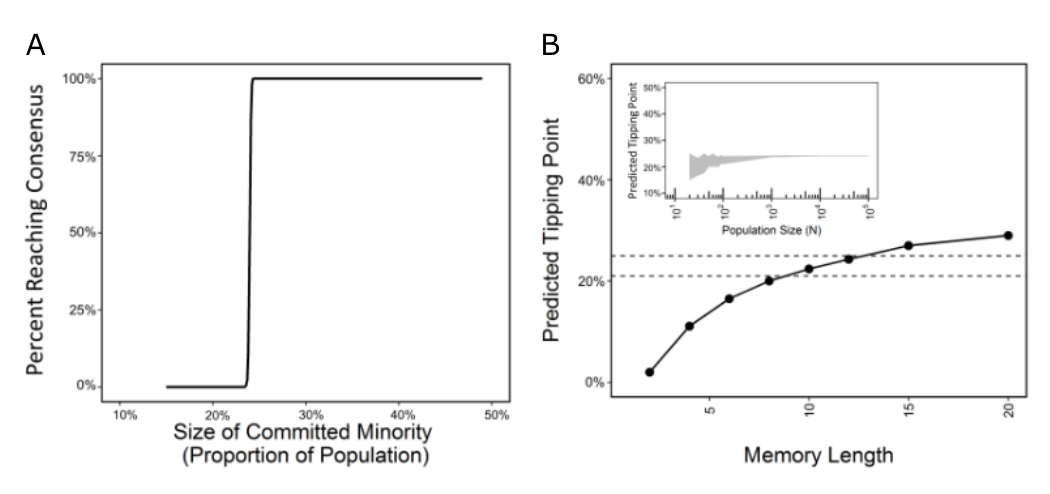} % figure 53
    \caption{(A). Theoretical modeling of the proportion of outcomes in which the alternative behavior is adopted by $100\%$ of the population. In this system, there are $N$=1000 agents, $T$=1000 interactions, and $M$=12 memories that can be used to store past interactions used in the agents' decisions. (B). The value of the predicted critical fraction is shown as a function of the individuals' average memory length, $M$. The dashed lines indicate the range covered by the experimental trials, showing the largest unsuccessful minority (21\%) and the smallest successful minority (25\%). Although the expected size of the critical fraction increases as $M$ increases, this relationship is concave, allowing the predicted tipping point to remain well below 50\% even for $M>100$. The inset shows the effect of increasing population size on the precision of prediction of the critical fraction of the committed minority $C$ when $M=12$ and $T=1000$. For $N<1000$, small variations in the predicted tipping point emerge due to stochastic variations in individual behavior. The shaded region indicates the range of $C$ over which the trials succeed frequently but without certainty. Above this region, for larger $C$, the probability of success approaches 1; for $C$ below this region, the likelihood of success goes to 0. \\
\textit{Source:} The figure is from \cite{centola2018experimental}.}
    \label{fig:centola}
  \end{figure}

A recent empirical study of tipping points in a system in which there is a minority of committed agents is presented in Ref. \cite{centola2018experimental}. The authors showed that the theoretically predicted dynamics of critical fractions did emerge within an empirical system of social coordination based on an artificially created system of evolving social conventions. The authors first synthesized the diverse theoretical and observational accounts of tipping point dynamics and used this synthesis to derive theoretical predictions for the size of an effective critical fraction of committed agents. A simple model of strategic choice was used in which agents decide which opinion to adopt by choosing the option that yielded the greatest expected individual reward given their history of social interactions. The model predicted a sharp transition in the collective dynamics of opinions as the size of the committed minority reaches a critical fraction of the population (see Fig.~\ref{fig:centola}). Two parameters determined the theoretical predictions for the size of the critical fraction: length $M$ of individual memory of the past interactions and population size, $N$. Inspection of these parameters shows that the predicted value of the tipping point changes significantly with the individuals' average memory length $M$ (see Fig.~\ref{fig:centola}).

\begin{figure}[ht!]
    \centering
    \includegraphics[width=\linewidth]{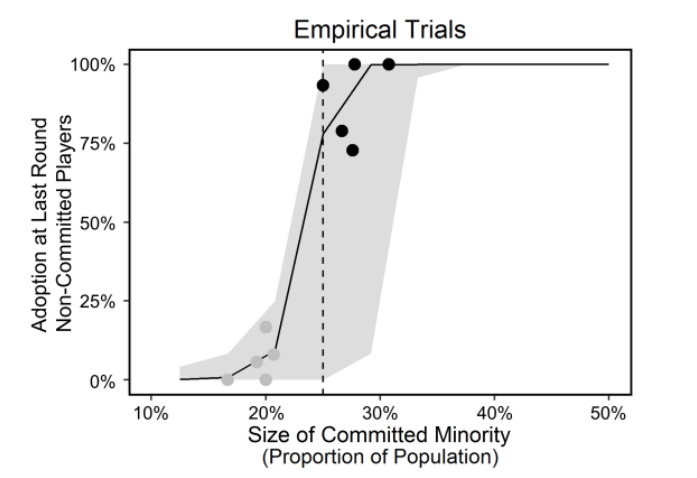} % figure 54
    \caption{Final success levels from all trials (gray points indicate trials with $C < 25\%$; black points indicate trials with $C \geq 25\%$) with 95\% confidence intervals $N=24$, $T=45$, $M=12$ (gray area indicates 95\% confidence for trials with $C \geq$ 25\%). Also shown is the theoretically predicted critical fraction (solid line shows results averaged over 1,000 replications). The dotted line indicates $C=25\%$. The theoretical model of critical fractions provides a good approximation of the empirical findings. For short time periods ($T < 100$), the prediction of the size of the critical fraction is not exact (ranging from $20\% < C < 30\%$ of the population); however, over longer time periods ($T > 1000$), the transition dynamics become more precise (solid line).\\
\textit{Source:} The figure is from \cite{centola2018experimental}.}
    \label{fig:centola_1}
  \end{figure}

  The authors recruited 194 subjects via the World Wide Web and clustered them into online communities for the experiments. Figure \ref{fig:centola_1} shows a summary of the final adoption levels across all trials, along with expectations based on the introduced empirically parameterized theoretical model, with 95\% confidence intervals. The figure compares these observations to numerical simulations of the theoretical model using population sizes and observation windows comparable to those used in the experimental study ($N=24$, $T=100$, $M=12$). The theoretically predicted critical fraction values from this model fit the experimental findings well.

Some of the variability in the critical fraction results shown in Fig.~\ref{fig:centola}, which can be explained by one crucial parameter, the strength of the agent's commitment. Considering different scenarios, the commitment strength can either be controlled \cite{niu2017impact} or not controlled \cite{centola2018experimental} in the experiments. These authors analyzed the impact of commitment strength, defined as the number of subsequent interactions with speakers with opposite opinions needed for the committed agent to change his opinion. For traditional committed agents, this strength is infinite. The authors allow the commitment strength to be set to any value and to change as a result of interactions with other agents, introducing a {\it waning commitment}. This analysis shows that commitment strength affects the size of the critical fraction of the population that is necessary to achieve a minority-driven consensus. Increasing the commitment strength decreases the size of the critical fraction needed for waning commitment, and waning commitment can occur as a result of interactions with agents who hold the opposing opinion. Conversely, increasing commitment can strengthen through interactions with agents who hold the same opinion. The size of the critical fraction decreases as the commitment strength increases. Suppose the strength of commitment is distributed randomly among committed nodes according to a distribution. In that case, the higher the standard deviation of the distribution is, the larger is the critical fraction with waning commitment needed to achieve a minority-driven consensus, and the smaller is the fraction with increasing commitment needed for such a consensus. Assuming that the participants in the experiments~\cite{centola2018experimental} have different strengths of commitment using the variable commitment strength~\cite{niu2017impact} or changing their commitment strengths over time might enable the model to find an excellent match to the experimental results~\cite{centola2018experimental}. 

Two generalizations aim to investigate the sensitivity of naming game dynamics to nodes in the multi-opinion state~\cite{thompson2014propensity}. The first allows the speaker to make asymmetric choices regarding the opinion to send, and the second enables the listener to retain its mixed state even if the received opinion is in agreement with the listener's state. The authors study the impact of each of these two generalizations on the system dynamics. Hence, this version of the Naming Game model gains two continuous parameters. Both parameters vary listener-speaker interactions at the individual level. The first parameter, called {\it propensity}, biases the choice of speakers with mixed opinions by defining the probability $p\neq 0.5$ that the speaker will choose a particular one of the available opinions to send to the listener. The second parameter, {\it stickiness}, allows the listener to hold a mixed opinion, receive a matching opinion from the sender and retain its current mixed opinion with a certain predefined probability $s$. The authors use the ``listener-only changing opinion at interaction'' version of the naming game~\cite{marvel2012encouraging}. The so-generalized naming game preserves the existence of critical thresholds defined by the fractions of the population belonging to committed minorities. Above this threshold, a committed minority causes a rapid convergence to consensus that occurs over a period of time that is proportional to the logarithm of the network size, even when other parameters such as propensity, stickiness or both influence the system dynamics. However, the two introduced parameters cause bifurcations of the system's fixed points, leading to changes in the system's consensus.

\begin{figure}[ht!]
    \centering
     \includegraphics[width=0.9\linewidth]{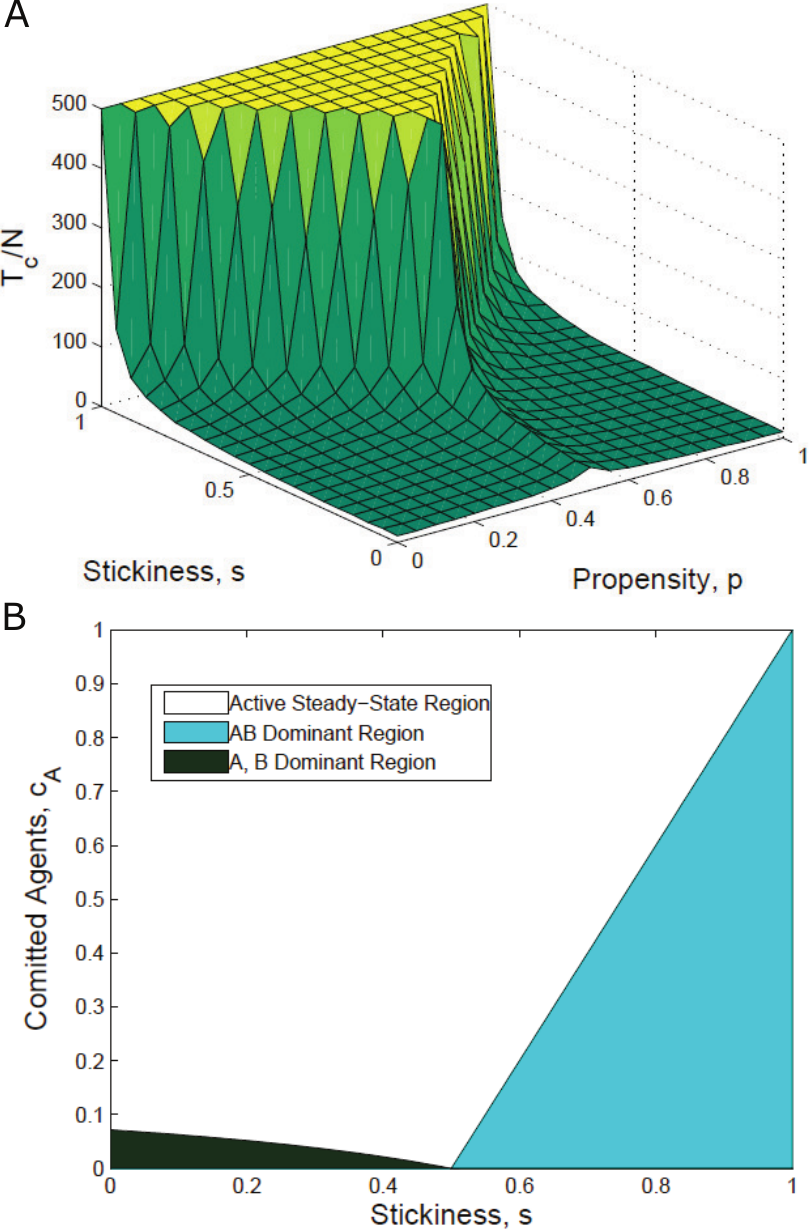} % figure 55
    \caption{The left surface shows the average ratio of consensus time $T_c$ to network size $N$ as a function of propensity $p$ and stickiness $s$. The lighter the color is, the longer the consensus time is. Both $p$ and $s$ influence the shape of the surface. The right panel shows regions of global stability of the system $c_A $-$s$ plane, where $c_A$ denotes the fraction of agents committed to an opinion, and $s$ denotes the stickiness to the mixed opinion.\\
\textit{Source:} The figure is from \cite{thompson2014propensity}.}    
    \label{fig:thompson}
  \end{figure}

Figure~\ref{fig:thompson} summarizes the most interesting findings from the abovementioned study. With stickiness, $s>0.5$, a new stable region arises in which a large number of neutrals (agents holding mixed opinions) can effectively prevent the committed agents from producing a consensus.

\begin{figure}[ht!]
    \centering
    \includegraphics[width=\linewidth]{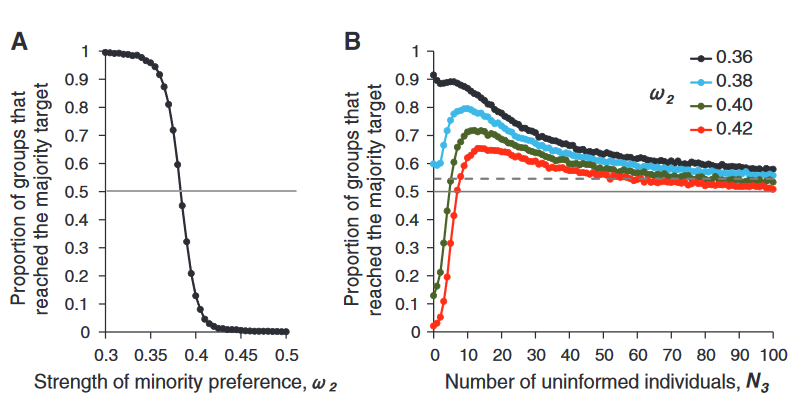} % figure 56
    \caption{(A) Without uninformed individuals, as the minority increases its preference strength, it increasingly controls group opinion. (B) In the presence of a sufficient fraction of uninformed individuals, the minority can no longer achieve the majority by increasing its preference strength.\\
\textit{Source:} Figure from \cite{couzin2011uninformed}.}
    \label{fig:couzin2011_fig1}
  \end{figure}

The experiments on social animals can validate these findings \cite{couzin2011uninformed}. The authors use a group of cows to demonstrate that a strongly opinionated minority can dictate group choice over a wide range of conditions. However, the presence of uninformed individuals (nodes holding a mixed opinion in terms of the naming game model vocabulary) spontaneously inhibits this process, returning control to the noncommitted majority. The results presented in the above reference highlight the role of uninformed individuals in achieving democratic consensus amid internal group conflict and informational constraints. Figure~\ref{fig:couzin2011_fig1} shows a comparison of the final majority reached in a system without uninformed individuals with the final majorities reached in systems with various numbers of uninformed individuals. In the presence of a sufficient fraction of uninformed individuals, the minority can no longer make its opinion the majority opinion even by increasing the strength of minority preference.

Furthermore, the properties of the opinions themselves and the preferences of the committed agents also affect the size of the critical fraction. One opinion has a fixed stickiness and investigate how the critical size of the fraction of agents with the competing opinion required to tip the entire population varies as a function of the competing opinion's stickiness~\cite{doyle2016social}. The authors analyze this scenario by developing a complete-graph topology through simulations and through a semi-analytical approach, thereby defining an upper bound for the critical minority fraction of the population.

Systems with high Shannon entropy have higher consensus times and lower critical fractions of committed agents than do low-entropy systems\cite{pickering2016analysis}. The authors also show that the critical number of committed agents decreases as the number of opinions increases and that it increases with each opinion's community size. The most surprising result is that when the number of committed opinions is based on the social system size, the critical size needed for minority-driven consensus is constant. In such a case, if the social system size tends to infinity, the critical fraction tends asymptotically to zero. The results lead to the conclusion that committed minorities can more easily enforce their opinions on highly diverse social systems, demonstrating that such systems are inherently unstable.

\subsubsection{Agent interaction models}
\begin{figure}
    \centering
    \includegraphics[width = \linewidth]{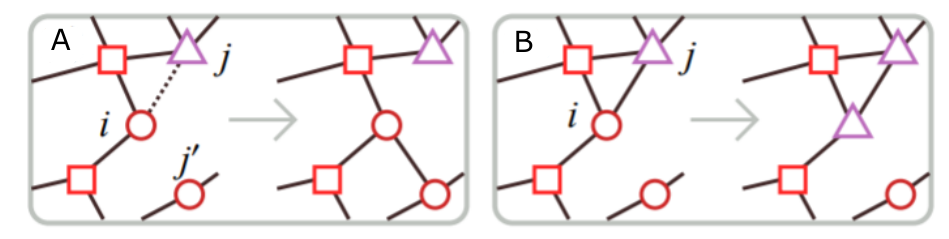} % figure 57
    \caption{A diagram of the model, with vertex shapes representing opinions. At each time step, the system with probability $\phi$ is updated according to panel (A) and
    with probability $1 - \phi$ according to panel (B). In
    panel (A), a vertex $i$ is selected uniformly randomly, and one of its edges (in the case shown, the edge $(i, j)$) is rewired to a new vertex $j'$ that holds the same opinion as $i$. In panel (B), vertex $i$ adopts the opinion of
    one of its neighbors (in the case shown, this opinion is denoted $j$.\\
\textit{Source:} The figure is from \cite{holme2006nonequilibrium}.}
    \label{fig:holme2016_illustration}
  \end{figure}

Other, less popular, models of agent interaction and opinion evolution have been introduced. For instance, in Ref. \cite{holme2006nonequilibrium}, the authors presented a simple model of opinion evolution in social systems. This model, which is shown in Fig.~\ref{fig:holme2016_illustration}, combines two rules: (i) individuals form their opinions in such a way as to maximize agreement with the opinions of their neighbors, and (ii) network connections are created between nodes holding the same opinion. 
  
We would expect the resilience of an opinion system such as that described above to increase over time because agents are adapting their opinions to agree with the opinions of their neighbors. However, this also increases the latency of interactions. Conversely, the formation of new connections reduces the latency by grouping the like-minded agents together and thereby speeding up consensus formation.

\noindent
{\bf Threshold model.}
The threshold model is used to represent the spread of novelty. In this model, nodes are in one of the two states: the old opinion and the new opinion~\cite{granovetter1978threshold}. The most crucial challenge is finding the smallest number of nodes, called initiators or seeds, that, when initiated with a new opinion, can cause this opinion to spread to the entire network. Any individual holding an old opinion switches to the new opinion when a predefined fraction $\phi$ of its neighbors holds the new opinion. This model is similar in some ways to the binary naming game model. Nevertheless, the semantics of opinion change differ from those in that model, and they are one-directional because once having switched to the new opinion, a node cannot turn back. In the simplest case of this model, all nodes have the same adoption threshold. In that version of the model, the cascade size triggered by a set of initiators is a function of the initiator fraction \cite{singh2013threshold}. The authors conclude that for a high threshold, there is a critical initiator fraction that, when crossed, assures that a global cascade will occur. The authors also observe that communities can extend opinion spreads and explore different initiator selection strategies.

\begin{SCfigure*}
        \centering
        \includegraphics[width =0.7\textwidth]{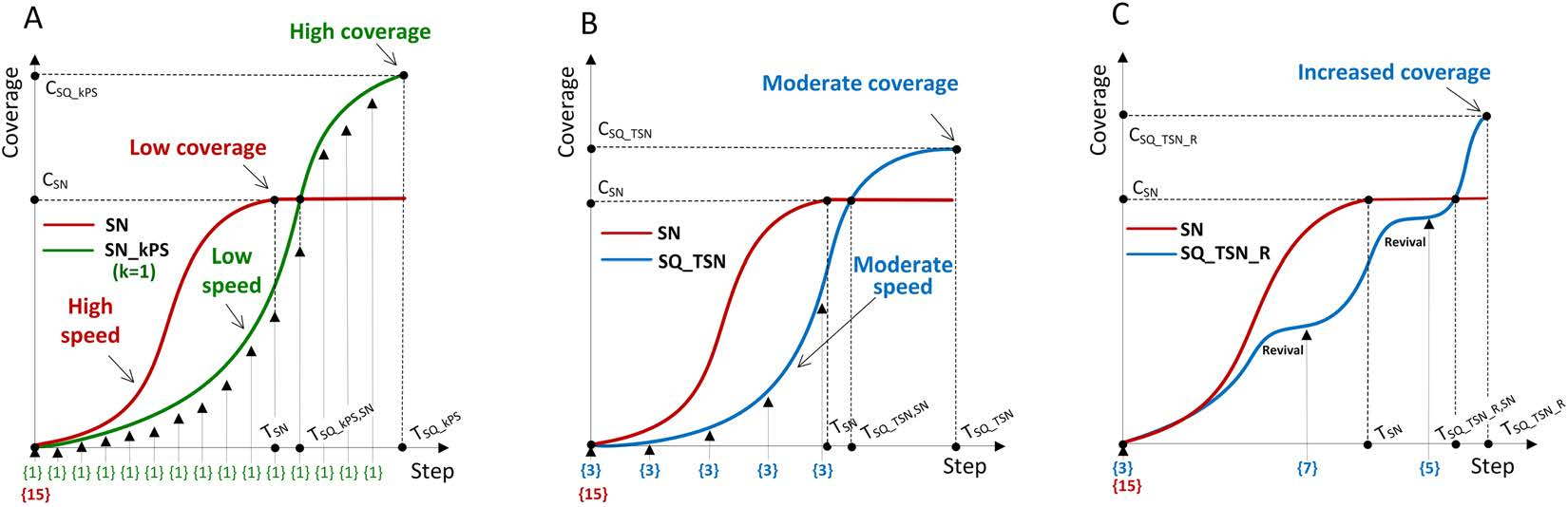} % figure 58
        \caption{Balancing speed and coverage for single-stage and sequential seeding. (A) $k$ per stage sequential seeding strategy (SN\_kPS) with $k = 1$ compared with the single-stage approach (SN). (B) Sequential strategy based on the reference time (SQ\_TSN) compared with the single-stage approach (SN). (C) Sequential strategy with revival mode (SQ\_TSN\_R) compared with the single-stage approach (SN). \\
 {\it Source:} Figure from \cite{ Jankowski2017Balancing}.} \label{figbalancing}
      \end{SCfigure*}

The commonly used activation in this model consists of selecting and initiating all seeds that hold the new idea at once; this results in the fastest spread but may not cause the highest fraction of the network nodes to transition to the new idea. Therefore, some seed initiation approaches use a sequence of initiation stages \cite{Jankowski2017Balancing, Jankowski2018Probing}. At later stages, sequential strategies avoid seeding highly ranked nodes that have already been activated by diffusion spreading between seeding stages. The gain arises when a saved seed is allocated to a node that is challenging to reach via diffusion. The experimental results \cite{Jankowski2017Balancing} indicate that regardless of the seed ranking method used, sequential seeding strategies deliver better coverage than single-stage seeding in approximately 90\% of cases. Longer seeding sequences tend to activate more nodes, but they also extend the duration of diffusion. Figure \ref{figbalancing} shows that various sequential seeding variants resolve the trade-off between coverage and diffusion speed separately.

The results show that sequential seeding is nearly always better than its single-stage equivalent when the same parameters are used. The global results for all networks, strategies, and parameters show better results than does sequential seeding in 95.3\% of simulation cases. The results obtained from simulations demonstrate that the improvement can also exceed 50\% with the use of the same number of seeds that are used in single-stage seeding.

The above conclusions were reached in experiments in which both simulations were making arbitrary decisions; therefore, the results did not always reflect a fair comparison. To avoid that effect, a coordinated execution of randomized choices enables a precise comparison of different algorithms in general\cite{Jankowski2018Probing}. The results of comparisons of sequential seeding with single activation seeding using this new approach reinforce the idea that each newly activated node at each stage of spreading that attempts to activate its neighbors uses the same random value to decide which neighbor to activate.

Using this new approach, the authors prove that sequential seeding delivers spread coverage that is at least as good as that obtained using single-stage seeding. Moreover, under modest assumptions, sequential seeding performs provably better than single-stage seeding when the same number of seeds and the same node rankings are used. Another interesting result is that, surprisingly, applying sequential seeding to a simple degree-based selection leads to higher coverage than that achieved by the computationally expensive greedy approach that is considered to be the best heuristically. Figure \ Ref {fig:coordinated-exc} compares the performances of sequential seeding and greedy heuristics and shows the theoretical limit.

      \begin{figure}[ht!]
        \centering
        \includegraphics[width = \linewidth]{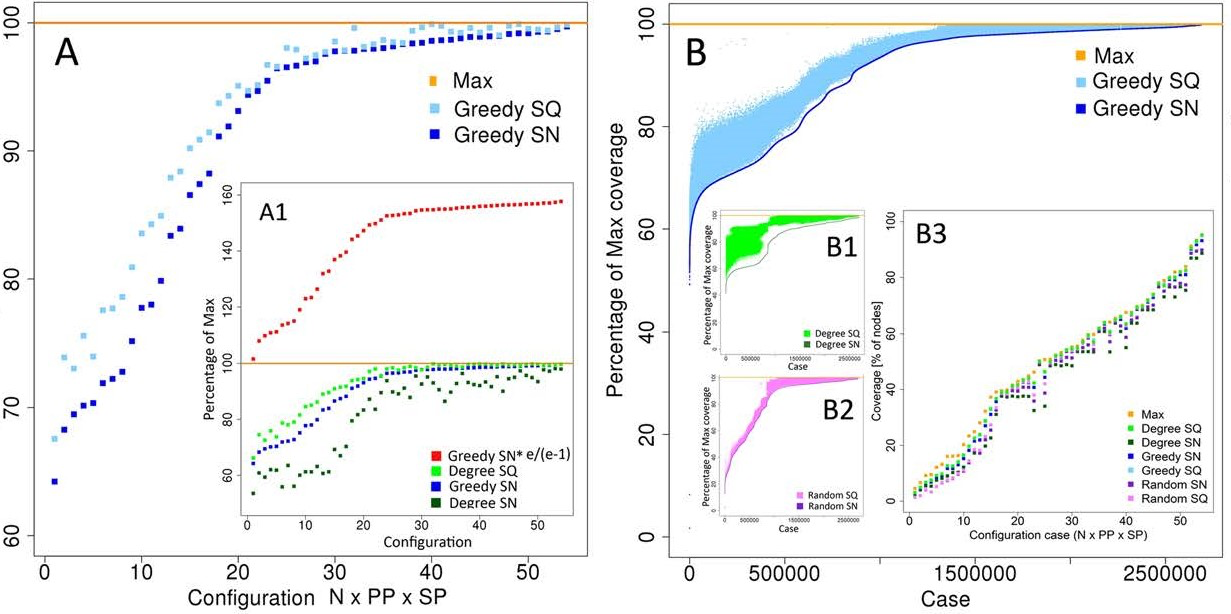} % figure 59
        \caption{Performance comparison on undirected networks. (A) Averaged performance of sequential SQ and single-stage SN seeding using a greedy heuristic for node selection. The results show fractions of the maximum coverage $C_{Max}$ as a function of the network size $N$, the probability $PP$ of propagation across each edge, and the fraction of seed selection percentage $SP$ of nodes selected as seeds. (A1) Performance of sequential SQ and single-stage SN seeding using degree-based ranking. The results are compared with the maximum coverage achieved and its upper bound as a function of the configuration parameters, $N$, $PP$, and $SP$. To achieve coordinated execution, random binary choices for each edge of whether to propagate or not propagate information across this edge are made at simulation initialization and applied to all compared executions. (B) Coverage of the sequential method SQ as a percentage of $C_{Max}$ for single-stage SN and max seeding using greedy node selection. (B1), (B2) Sequential seeding performance SQ achieved using single-stage SN and max seeding using random and degree-based node selections, respectively. (B3) Performance of sequential SQ and single-stage SN seeding represented as the percentage of activated nodes within the network (coverage) for random seed selection, degree-based ranking and greedy seed selection in comparison with the maximum coverage as a function of the individual configurations. \\
 {\it Source:} Figure from \cite{ Jankowski2018Probing}.} \label{fig:coordinated-exc}
      \end{figure}

A more challenging model in which nodes have different thresholds sampled from a given distribution of threshold values with known average and variance is studied in Ref. \cite{karampourniotis2015impact}. The authors investigate the behavior of cascade size in the presence of different numbers of initiators and various threshold distributions for both synthetic and real-world networks. The authors observe that the tipping point behavior of the cascade size in terms of the deviation of the threshold distribution changes into a smooth crossover when the initiator set is sufficiently large. They demonstrate that for a specific value of the threshold distribution variance, opinions spread optimally. In synthetic graphs, the spread asymptotically becomes independent of the system size, and global cascades can arise merely through the addition of a single node to the initiator set.

\noindent      
{\bf Voter model.}
The voter model is likely the simplest model of the spread of opinions with dynamics that may lead to consensus~\cite{Sood2005voter}. In this model, each node in the underlying network selects one of the two opposing opinions defined in the model. Starting from a random initial distribution of opinions, the model evolves dynamically in steps. At each step, a node without an opinion is uniformly randomly selected and assigned the opinion of a uniformly arbitrarily chosen neighbor with an opinion. The model dynamic is simple; the opinion that is initially held by the majority is likely to bring all nodes to consensus on this opinion.

An exciting extension of this model enables the node not only to update its opinion but also to break and establish connections with other nodes~\cite{nardini2008s}. The authors also introduce further model variations, such as the ``direct voter model'' and the ``reverse voter model''. They also use the memory-based naming game style opinion change in which no agent can move directly from one opinion to another but must transit through an intermediate state. The introduced mechanisms are studied using a mean-field analysis. The authors show that slight modifications of the interaction rules can have drastic consequences on the global behavior of opinion formation models in the case of dynamically evolving networks. The mean-field analysis accounts for differences due to the asymmetric coupling between interacting agents and the asymmetry of their degrees. The necessity for agents to transit via an intermediate state before changing their opinions strongly enhances the trend towards consensus. Allowing the interacting agents to bear more than one opinion simultaneously makes this a variant of the naming game model and makes rapid consensus to a single opinion possible.

\noindent
{\bf Axelrod model.}
An exciting variant of the Axelrod model~\cite{Axelrod1997dissemination} was introduced and studied in Ref.~\cite{singh2012accelerating}. The authors study a variant of the Axelrod model in a network with a homophily-driven rewiring rule imposed. In this model, network nodes represent individuals endowed with sets of $F$-independent attributes that define every node's state. Each attribute can take one of $q$ distinct traits represented by integers in the range $[0, q-1]$. Initially, each attribute of each node uniformly randomly chooses its value from among $q$ integers.

The network is interconnected randomly as an Erdos-Renyi (ER) random graph with the given average degree $\langle k \rangle$. At each time step, the system uniformly randomly selects a node $i$ and then selects its neighbor $j$. The similarity between nodes $i$ and $j$ is then computed by counting the number of attributes for which $i$ and $j$ possess the same trait. If the similarity is equal to or greater than the given threshold, the influence step is executed.
Node $j$ adopts the trait of node $i$ for an attribute that is uniformly randomly chosen from among the attributes on which nodes $i$ and $j$ currently disagree. Otherwise, the rewiring step dissolves the link between $i$ and $j$ and replaces it with a link from $i$ to node $k$, a node that is uniformly randomly selected from among the nodes that are not connected.

The authors show that with these dynamics and in the presence of a fraction of committed agents that exceeds the critical value, the consensus time becomes a logarithmic function of network size $N$.
Moreover, the authors demonstrate that slight changes in the interaction rules can produce strikingly different results in the scaling behavior of the consensus time $T_c$. However, all the interaction rules
tested in the study qualitatively preserve the benefits gained from the presence of committed agents.

\subsection{Survival resilience in animal social systems} \label{sa}  

Analogous to the social systems of humans are the social systems of social animals (ants, honeybees, and cows). These systems are also within the scope of this chapter. Studies in this area include studies of foraging behavior \cite{beekman2001phase}, nest building \cite{toffin2009shape}, and coping with stress \cite{doering2018social}. The resilience of social insects depends on three crucial elements of their system infrastructure: transportation networks, supply chains, and communication networks~\cite{middleton2016resilience}. To assure resilience, the system may use three different approaches: resistance, redirection, and reconstruction.

\subsubsection{Social foraging and nest building} 

A study of foraging behavior in ant colonies that focuses on the transition between disordered and ordered foraging in Pharaoh's ant is presented in Ref. \cite{beekman2001phase}. The authors found that small colonies forage in a disorganized manner, while large colonies transition to pheromone-based foraging.

The theoretical model derived from those empirical findings is defined as follows:

\begin{equation}
\begin{array}{cl}
\frac{dx}{dt}
&= (\text{ants beginning to forage at feeder}) \\
& \quad - (\text{ants losing pheromone trail}) \\
&= (\alpha + \beta x)(n - x) - \frac{sx}{s + x},
\end{array}
\end{equation}
where $x$ is the total increase in the number of ants walking to a single food source, $n$ represents the colony size, and constants $\alpha$, $\beta$, and $s$ denote the probability per minute per individual ant of finding food through independent searching, the probability per minute per individual ant of arriving at the food source via a pheromone-marked trail, and the maximum rate at which an ant can leave the trail, respectively.

\begin{figure}[ht!]
  \centering
  \includegraphics[width = \linewidth]{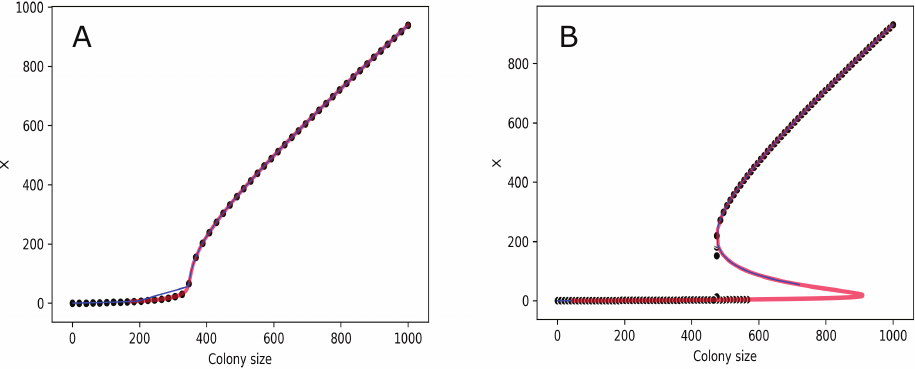} % figure 60
   \caption{Predicted total number of ants, $x$, walking to a single food source as a function of colony size, $n$, when the feeder is found (A) frequently ($\alpha = 0.021$) and (B) infrequently ($\alpha = 0.0045$) by independently searching ants. The other parameters, $\beta = 0.00015$ and $s = 10$, are fixed. The blue line shows the theoretically unstable points, and the red line shows the theoretically stable points. The black dots represent the results obtained in the simulation.\\
\textit{Source:} Figure modified from \cite{beekman2001phase}.}
  \label{fig:beekman_02}
\end{figure}

This leads to a cubic equation for $x$ and to equilibrium solutions at which $dx/dt = 0$. Setting $s = 10$ and using two values of alpha $0.021$ and $0.0045$ results in the plots shown in Fig.~\ref{fig:beekman_02}.

The authors observe that hysteresis arises for low independent searching ability $\alpha$ and intermediate values of colony size $n$. This directly demonstrates that foraging in Pharaoh's ants is influenced by colony size in a nonlinear and discontinuous way. This transition in foraging is strongly analogous to the first-order transition in physical and social systems.
 
\begin{figure}[ht!]
   \centering
   \includegraphics[width=\linewidth]{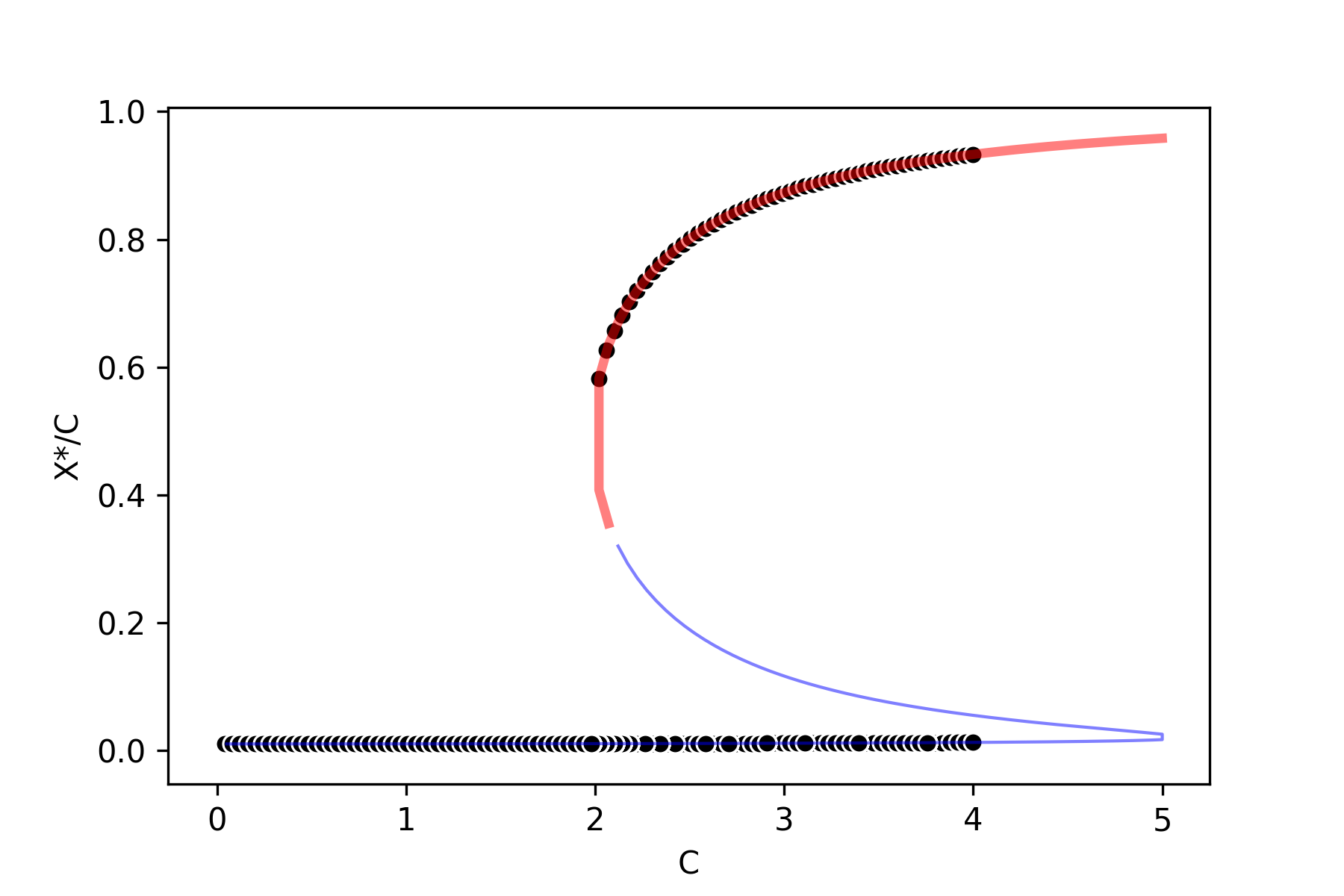} % figure 61
   \caption{Attractors and dynamics in a model of foraging in which a function $X^*/C$ is used to model the ratio of the stable number of forages $X^*$ to the actual number of foragers $C$ in one-source bee foraging systems. The plots are obtained using $f=1 and \alpha=0.01$. The blue line shows the theoretically unstable points, and the red line shows the theoretically stable points. The black dots show the results of the simulation.\\
\textit{Source:} The figure has been modified from \cite{loengarov2008phase}.}
   \label{fig:loengarov_01}
 \end{figure}

 Theoretical modeling of honeybee foraging dynamics in one-source and multisource settings is presented in Ref. \cite{loengarov2008phase}. The stable number of foragers depends on the bee concentration and the scouting rate. By fixing the scouting price, the author analyze the dynamics of the proportion of potential foragers foraging in one-source and two-source settings. The results show that both phase transitions and bistability arise in the proposed model. The simulation results for the one-source case are shown in Fig.~\ref{fig:loengarov_01}.
 
\begin{figure}[ht!]
   \centering
   \includegraphics[width=0.5\linewidth]{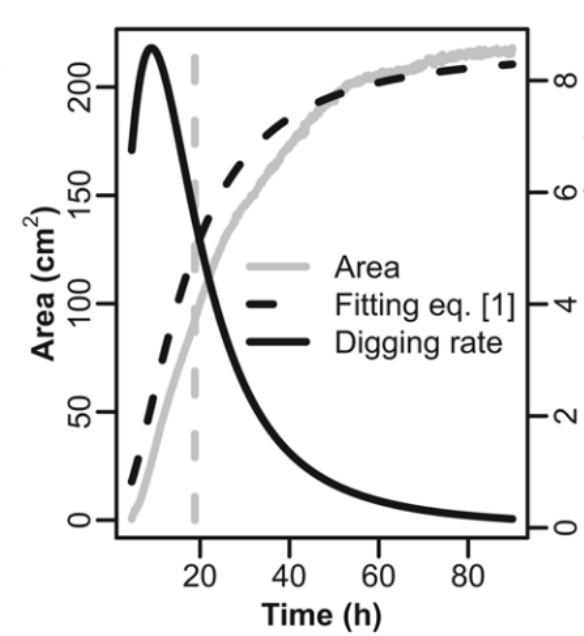} % figure 62
   \caption{Example of experimental results. The gray dashed vertical line represents the occurrence of the morphological transition that separates the first and second growth stages.    The evolution of the nest area and the digging rate are plotted with the fitting parameters set as follows: $A_M = 218.10 \text{cm}^2$, $\alpha = 1.95$, $\beta = 16.3 \text{h}$, and $r^2 = 0.99$. The gray line represents the area, the black dashed line corresponds to the fitting curve, and the black curve represents the digging rate. The digging rate is calculated as the derivative $\frac{dA}{dt}$ of the fitted area.\\
\textit{Source:} The figure is from \cite{toffin2009shape}.}
   \label{fig:toffin_fig_a}
 \end{figure}

 The process of collective ant nest building is studied in Ref. \cite{toffin2009shape}. To analyze the diversity in nest shape, the authors conducted two-dimensional nest-digging experiments under homogeneous laboratory conditions in which shape diversity emerges only from digging dynamics. A stochastic model highlights the central role of density in shape transition. Figure~\ref{fig:toffin_fig_a} shows the experimental results. The digging dynamic follows the equation

\begin{equation}
 A = \frac{A_M t^{\alpha}}{\beta^{\alpha} + t^{\alpha}},
 \end{equation}
where $A$ (in $\text{cm}^2$) is the excavated area (i.e., the nest area), $t$ (in hours) is the time elapsed since the beginning of nest digging, $A_M$ is the maximal area of the nest that is ultimately dug out, $\alpha$ represents the level of cooperation among the ants, and $\beta$ is the time value when $A = 0.5 A_M$.

\subsubsection{Social animals' response to stress}
\begin{figure}[ht!]
   \centering
   \includegraphics[width=\linewidth]{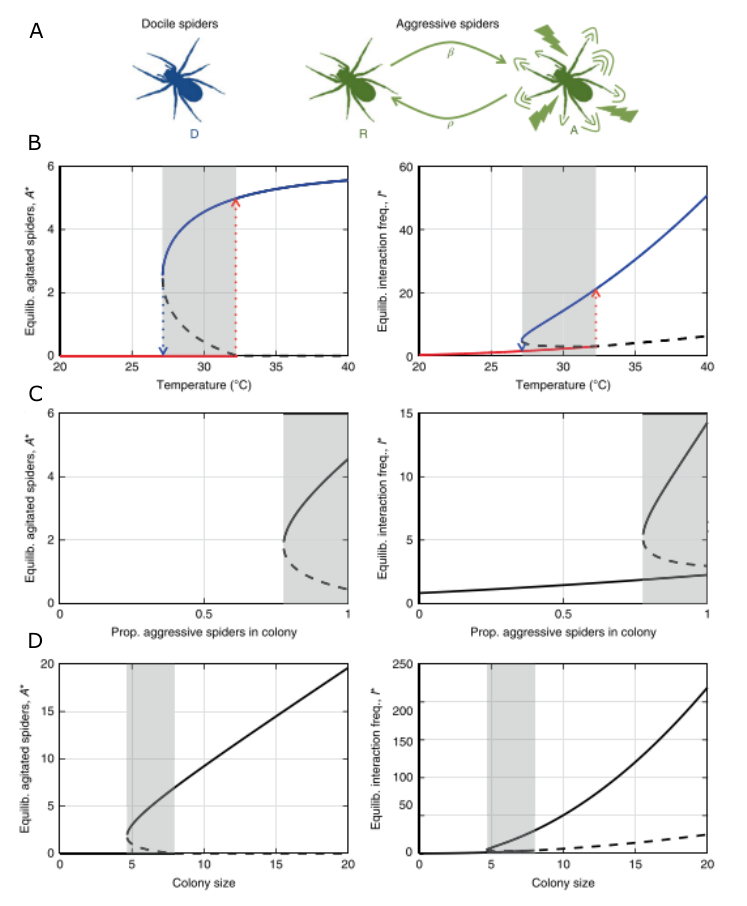} % figure 63
   \caption{(A). The authors developed a three-state model describing the interactions between docile and aggressive spiders that allowed for transitions of agitation state among aggressive spiders. (B). For a six-spider colony ($N = 6$), the model predicted one, two or three equilibria as a function of temperature. (C). An example in which the appearance of hysteresis depends upon colony composition. (D). An example in which colony size determines the shape of hysteresis. \\
\textit{Source:} The figure is from \cite{doering2018social}.}
   \label{fig:doernig_01}
 \end{figure}

The factors that contribute to the appearance of abrupt tipping points in animal societies in response to stress are studied in Ref. \cite{doering2018social}. The authors of that study constructed an analytical model and used it to study how the composition of a society in terms of the personalities of its members alters the propensity of the society to shift from a calm to a violent state in response to thermal stress. The model was evaluated based on the results of subjecting experimental societies of the spider \textit{Anelosimus studiosus} to heat stress. The authors demonstrate that both colony size and the personalities of the members of the society influence the timing of and recoverability from sudden transitions in the social state.
Figure~\ref{fig:doernig_01} illustrates the developed three-state model for social tipping points in social spiders.
 
Experiments and field studies were used in Ref. \cite{wood2019evolving} to investigate the social and ecological factors that affect cold tolerance in range-shifting populations of the female-polymorphic damselfly \textit{Ischnura elegans} in northeast Scotland. Range shifting, i.e., poleward shifts in geographical range, occurs when the climate warms and the geographical positions of some species' thermal range limits change as the species expand their ranges within these new thermal boundaries.

\begin{figure}[ht!]
   \centering
   \includegraphics[width=\linewidth]{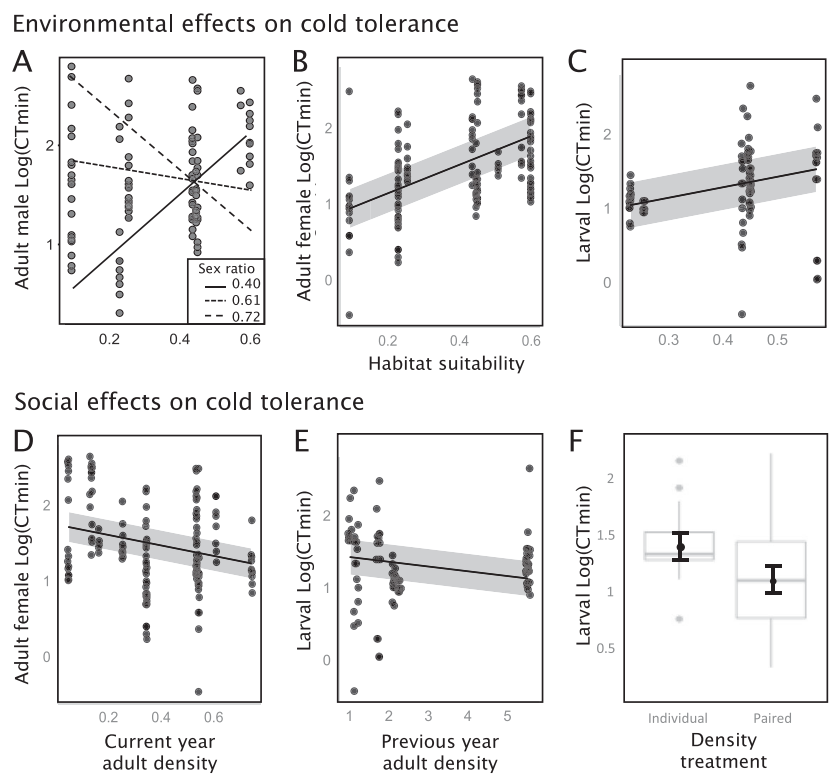} % figure 64
   \caption{Environmental (A-C) and social (D-F) determinants of cold tolerance in adult and larval \textit{Ischnura elegans} captured near the elevation range limit of the species in Scotland. The x-axis ``Habitat suitability'' value is computed using a maximum entropy model for species distribution \cite{phillips2004maximum}. With higher habitat suitability, the model predicts the area to be more suitable for living. ``Density treatment'' in (F) denotes different scenarios of social pressure - whether animals are placed in the environment alone or with other members of the species.\\
\textit{Source:} The figure is from \cite{wood2019evolving}.}
   \label{fig:wood_01}
 \end{figure}

The authors also consider the consequences of changing cold tolerance for evolutionary change. Both environmental and social effects on cold tolerance and female color morph frequency were recorded. Manipulation of animal density in the laboratory provided experimental evidence that social interactions directly influence cold tolerance. The authors suggest that there is a broader need to consider the role of evolving social dynamics in reciprocally shaping both the thermal physiology of individuals and the thermal niches occupied by their species. Figure~\ref{fig:wood_01} shows the results of the study.

\begin{figure}[ht!]
  \centering
  \includegraphics[width=\linewidth]{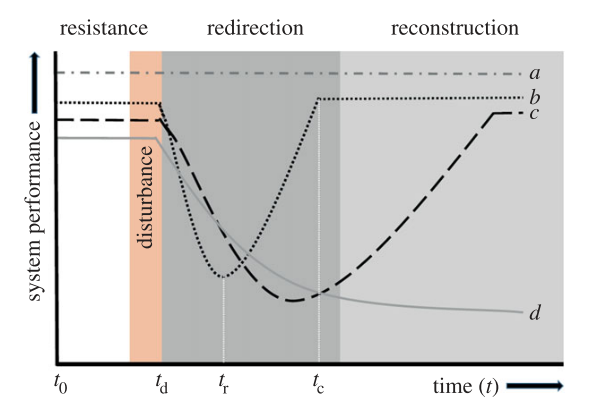} % figure 65
  \caption{Potential responses of a system to a disturbance. System performance is an experiment-specific measure of system functionality (harvesting rate, brood production rate, and traffic flow). Time ($t_0$) is the time at which the experiment begins; predisturbance ($t_d$) indicates the beginning of the disturbance, $t_r$ indicates the beginning of the recovery phase in the system, and $t_c$ indicates the point at which recovery is complete (i.e., system performance has returned to predisturbance levels). (A) A system that has invested in resistance and, as a result, does not experience a decrease in functionality after the disturbance. (B) A system that is using redirection. Although there is an initial decrease in performance, it is rapidly mitigated by rerouting flows using the existing infrastructure. (C) A system that primarily uses reconstruction-based resilience strategies. Since reconstruction requires the construction of new infrastructure, this system takes longer to recover predisturbance performance. (D) A nonresilient system that does not recover its predisturbance performance. \\
\textit{Source:} The figure is from \cite{middleton2016resilience}.}
  \label{fig:middleton_01}
\end{figure}

The resilience of three critical social insect infrastructure systems (transportation networks, supply chains, and communication networks) is reviewed in Ref. \cite{middleton2016resilience}. The authors describe how systems differentiate investment in three approaches to resilience: resistance, redirection, or reconstruction (cf. Fig.~\ref{fig:middleton_01}). The authors also observe that human infrastructure networks rapidly decentralize and enhance their interconnections, thus becoming more like social insect infrastructures. Therefore, human infrastructure management might learn from social insect research. In turn, the latter can use the mature analytical and simulation tools developed for the study of human infrastructure resilience. The essential difference between the two design strategies is that, unlike changes in human social systems, changes in animal social systems~\cite{beekman2001phase} can only occur through evolution of the social dynamics of particular species.

A review of the tipping points that arise in the dynamics of animal societies is provided in Ref. \cite{pruitt2018social}. The authors begin by listing concepts that are directly related to social tipping points, including behavioral states, environmental parameters, attractors and basins of attraction, and perturbations. Among the social properties most relevant to tipping points, they list relatedness and group size, key individuals, behavioral diversity, social organization, and prior experience. Finally, among the types of social tipping points, they list social scale tipping points, metabolic tipping points, and social and cognitive tipping points.

\subsection{Resilience of the planetary socioecological system}\label{se}
Socioecological systems (SESs), also known as human-environmental systems (HESs), or ``coupled human-and-natural systems'' have been widely studied recently. Such systems are ubiquitous in agriculture, water use, terrestrial and aquatic systems, the global climate system, and elsewhere \cite{bauch2016early}. A socioecological analysis of resilience enables us to study people-environment interactions across varying dimensions, times, and scales \cite{stokols2013enhancing}.

Most of the current serious attempts to integrate the social dimension into such studies are associated with works focusing on resilience. The large number of sciences involved in such exploratory attempts has led to the discovery of new links between social and ecological systems. Recent advances include an enhanced understanding of social processes like social learning and social memory, mental models and knowledge-system integration, visioning and scenario building, leadership, agent groups, social networks, institutional and organizational inertia and change, adaptive capacity, transformability, and systems of adaptive governance that allow the management of essential ecosystem services \cite{folke2006resilience}.

Integrating the natural science and the social science aspects within one framework has been a crucial challenge in such studies. Whereas the concept of a ``system'' is practically universal in the natural sciences, definitions of a system vary across social sciences and their branches, placing these areas of study outside the realm of the natural sciences.

One possible approach to this challenge is presented in Ref. \cite{olsson2015resilience}. The authors of that study suggest using an institutional lens to integrate the social and natural aspects of joint resilience. This is proposed because institutions may become methodological linchpins for integrating the social and natural sciences for the sake of sustainability. Moreover, progress in the social sciences and a core understanding of the evolution of social systems enables researchers to evaluate the influence of different institutions on socioecological system resilience more formally than they have been able to do in the past. For example, global and international institutions can promote and support the building of collaborations among countries and thereby enhance resilience by negotiating global agreements and treaties \cite{dutting2010building}. National and local governments can support the organizational resilience of industries and thereby strengthen the social response to social unrest \cite{gittell2006relationships}. Resource-dependent industries can support the resilience of social systems by sustaining their workforces in times of social stress~\cite{marshall2007resource}. Labor market resilience may increase the resilience of social systems against recession. Economic resilience may increase the resilience of social systems to disasters \cite{giannone2011market}.

\begin{figure}[ht!]
    \centering
    \includegraphics[width=\linewidth]{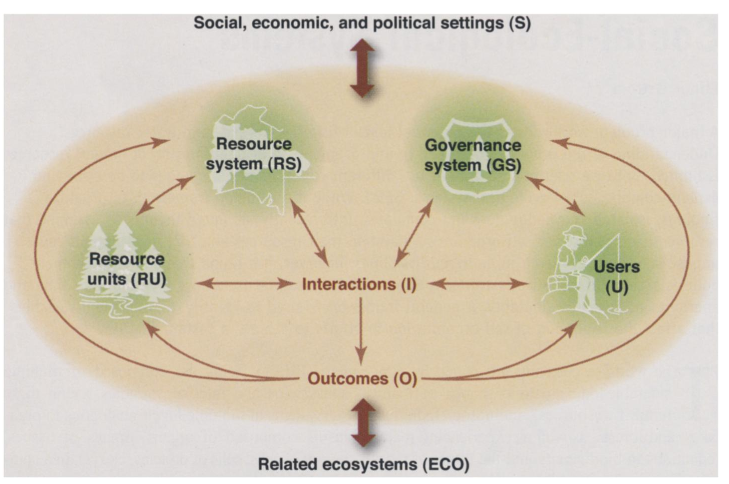} % figure 66
    \caption{Core subsystems in a framework for analyzing socioecological systems. \\
\textit{Source:} The figure is from \cite{ostrom2009general}.}
    \label{fig:ostrom2009_fig1}
  \end{figure}

Sustainability is a topic that is often studied jointly with resilience in the context of SESs. In Ref. \cite{ostrom2009general}, the authors present and use a general framework to identify ten subsystem variables that affect the likelihood of self-organization in efforts to achieve a sustainable SES (see Fig.~\ref{fig:ostrom2009_fig1}).

 Three related attributes of SESs determine their future trajectories: resilience, adaptability, and transformability. Resilience (the capacity of a system to absorb disturbance and reorganize while undergoing the changes needed to retain its necessary function, structure, identity, and feedback) has four components: latitude, resistance, precariousness, and panarchy; these are most readily portrayed using the metaphor of a stability landscape \cite{walker2004resilience}.
  
 The origin of the resilience perspective and an overview of its development to date are presented in Ref. \cite{folke2006resilience}. With roots in one branch of ecology and the discovery of multiple basins of attraction in ecosystems in the 1960s-1970s, this perspective has inspired social and environmental scientists to challenge the dominant ``stable equilibrium'' view.

\begin{figure}[ht!]
    \centering
    \includegraphics[width=\linewidth]{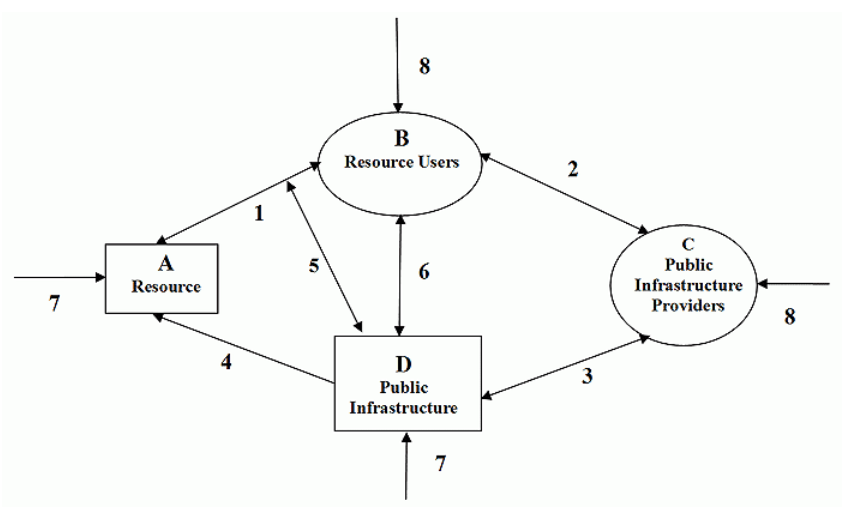} % figure 67
    \caption{A conceptual model of a socioecological system. The entities are defined in Table~\ref{anderies2004_table1}, and the links are detailed in Table~\ref{anderies2004_table2}.\\
\textit{Source:} Figure from \cite{anderies2004framework}.}
    \label{fig:anderies2004_fig1}
  \end{figure}

The institutional configurations that affect interactions among resources, resource users, public infrastructure providers, and public infrastructure are discussed in Ref. \cite{anderies2004framework}. We summarize the results in Fig.~\ref{fig:anderies2004_fig1}, which illustrates the framework, in Tab.~\ref{anderies2004_table1}, which presents the entities involved, and in Tab.~\ref{anderies2004_table2}, which depicts the actual links involved. The authors propose a framework that helps identify SES's potential vulnerabilities to disturbances. The authors posit that the link between resource users and public infrastructure providers is key to the robustness of SESs, although this link has frequently been ignored in the past. 

  \begin{table*}
    \scriptsize
    \begin{tabularx}{\textwidth}{lXX}
    \hline
     Entities & Examples & Potential Problems \\ 
    \hline
     A. Resource & Water source / Fishery & Uncertainty complexity / Uncertainty \\ \\
     B. Resource users & Farmers using irrigation & Stealing water, getting a ``free ride'' on maintenance \\
     				  & Fishers harvesting from inshore fishery & Overharvesting \\ \\
     C. Public infrastructure providers & Executive and council of local users' associations & Internal conflict or indecision about which policies to adopt \\
      & Government bureau & Information loss\\ \\
    
     D. Public infrastructure & Engineering works & Wear out over time \\ \\
     
     Institutional rules & Memory loss over time, deliberate cheating & \\ \\
     
     External environment & Weather, economy, political system & Sudden changes or slow unnoticeable changes \\
    \hline 
    \end{tabularx}
    \caption{From \cite{anderies2004framework}. Entities involved in socioecological systems}
    \label{anderies2004_table1}
\end{table*}
  
The intellectual roots and core principles of social ecology are traced in Ref. \cite{stokols2013enhancing}. The authors demonstrate how these principles enable a broader conceptualization of resilience. This conceptualization may be found in much of the following literature.

\subsubsection{Early warning signals in socioecological systems}

A coupled HES that is close to a tipping point is generally far less resilient to perturbations than is the same system when it is far from such points \cite{bauch2016early}. In \cite{bauch2016early}, the authors show that a coupled HES can exhibit a richer variety of dynamical regimes than can the corresponding uncoupled system. Thus, early warning signals can be ambiguous because they may herald either collapse or preservation. In addition, the authors observe that human feedback can partially mute the early warning signals of a regime shift or cause the system to evolve toward and perpetually remain close to a tipping point.

The coupling of the environmental dynamics models and the social dynamics model presented in Ref. \cite{bauch2016early} can be summarized as follows:

\begin{enumerate}
      \item The forest ecosystem is modeled as 
      \begin{equation}
      \frac{dF}{dt} = RF(1-F) - \frac{hF}{F+s},
      \end{equation}
      where $R$ is the net growth rate, $s$ is the supply and demand parameter, $h$ is the harvesting efficiency, and $F$ is a function of time representing the size of the forest.

\item The social dynamics model is defined as
\begin{equation}
      \frac{dx}{dt} = kx(1-x)\triangle U - (- kx(1-x)\triangle U),
      \end{equation}
where $\triangle U$ is the utility gain of changing opinion, $x$ is the function of time representing the proportion of the population adopting the opinion, and $k$ is the social learning rate.
      
      \item The coupling is represented by the following equations:
      
      \begin{equation} 
      \left\{
      \begin{aligned}
      & \dot F = RF(1-F) - \frac{h(1-x)F}{F+s}, \\
      & \dot x = kx(1-x)[d(2x-1) + \frac{1}{F+c} - \omega],
      \end{aligned}
       \right.
      \end{equation} 
       where $c$ is the rarity valuation parameter, $d$ is the social norm strength, and $\omega$ is the conservation cost.
    \end{enumerate}

\begin{table*}
\centering
\scriptsize
\begin{tabularx}{\textwidth}{XXX} 
\hline
Link & Examples & Potential Problems \\\hline
(1) Between resource and resource users                & Availability of water at time of need / availability of fish                      
  & Too much or too little water / too many uneconomic fish / too many valued fish                          \\ \\
(2) Between users and public infrastructure providers   & Voting for providers / Contributing resources / Recommending policies / Monitoring performance of providers 
   & Indeterminacy / Lack of participation / Free-riding / Rent seeking / Lack of information \\ \\
(3) Between public infrastructure providers and public infrastructure 
   & Building initial structure / Regular maintenance & Overcapitalization or undercapitalization  \\ 
   & Monitoring and enforcing rules                  & Shirking disrupting temporal and spatial patterns of resource use / Cost / corruption\\ \\
(4) Between public infrastructure and resource          & Impact of infrastructure on the resource level     & Ineffective \\ \\
(5) Between public infrastructure and resource dynamics        
   & Impact of infrastructure on the feedback structure of the resource/ Harvest dynamics       / Ineffective, unintended consequences \\ \\
(6) Between resource users and public infrastructure    & Coproduction of infrastructure itself,     
   maintenance of works, monitoring and sanctioning & No incentives / Free-riding           \\ \\
(7) External forces on resource and infrastructure      & Severe weather, earthquake, landslide, new roads    & Destroys resource and infrastructure \\ \\
(8) External forces on social actors                    & Major changes in political system 
   & Conflict, uncertainty, migration \\
   & Great migration, commodity prices, and regulation                              
   & Increased demand \\ \hline
\end{tabularx}
\caption{From \cite{anderies2004framework}. Links in socioecological systems.}
\label{anderies2004_table2}
\end{table*}

\begin{figure}[ht!]
      \centering
      \includegraphics[width = \linewidth]{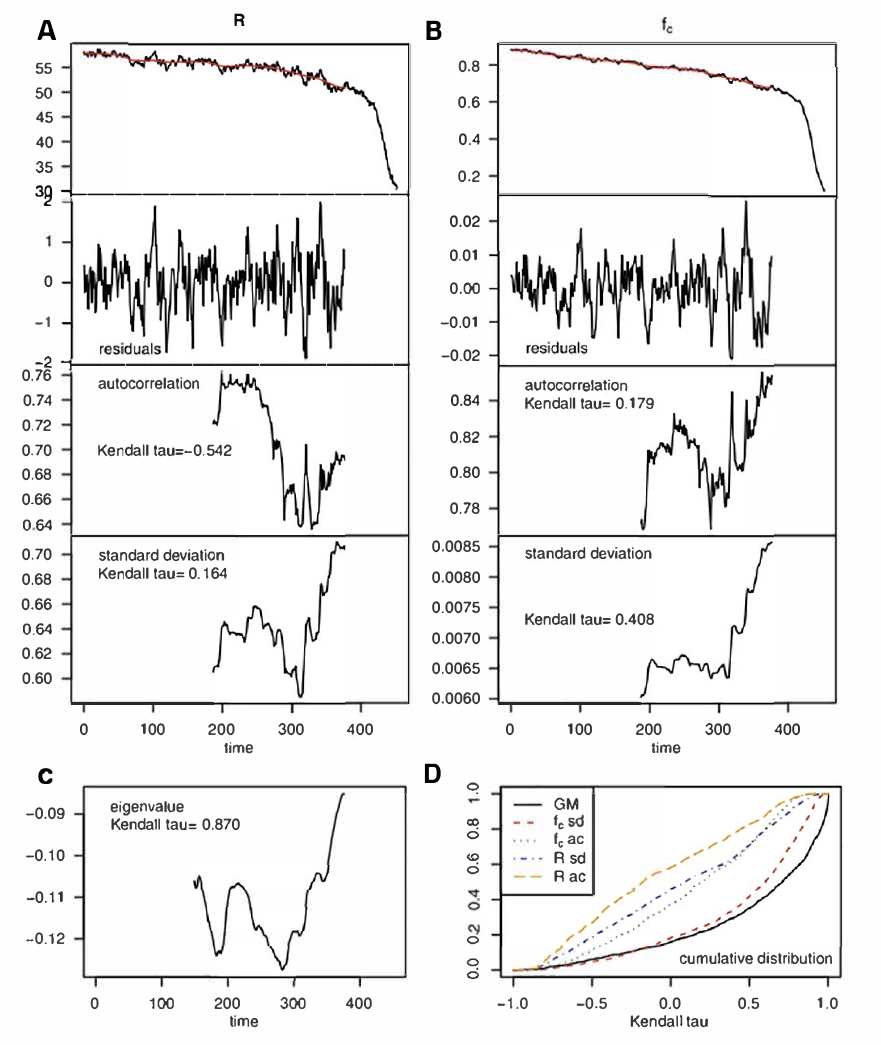} % figure 68
      \caption{Early warning signals. (A-B)Time series in the lead-up to a regime shift of the model (\textit{black line} (for parameters of the simulation, see text) with filtered fit (\textit{red line} only over the time range prior to the regime shift to be used in the following analysis), detrended fluctuations, and their autocorrelations and standard deviations. (C) Generalized model-based early warning signal preceding the regime shift. (D) Cumulative distribution of the Kendall $\tau$ statistic for different warning indicators (\textit{GM} = generalized model-based signal; \textit{ac} = autocorrelation; \textit{sd} = standard deviation) over 1,000 simulations of the regime shift with different noise realizations. All Kendall $\tau$ statistics shown were calculated over the time range 300 to 377.\\
\textit{Source:} The figure is from \cite{lade2013regime}.}
      \label{fig:lade2013_early}
    \end{figure}

Early warning signal approaches show potential for warning of socioecological regime shifts \cite{lade2013regime}. Such signals could be valuable in natural resource management and could be used to guide management responses to variable and changing resource levels or to changes in resource users. However, investigation of specific cases of socioecological regime shifts is required to ensure that, first, the required data are available for the studied cases and, second, that the resulting early warning signals are robust. The third and fundamental criterion through which an early warning signal should be evaluated in a specific case study is whether the signal provides a warning that can be noted early enough for the transition to be avoided. Fig.~\ref{fig:lade2013_early} shows the results of such analysis.

The theoretical framework built in Ref. \cite{suweis2014early} is used to demonstrate that rising variance-measure, evaluated, for example, using the maximum element of the covariance matrix of the network, is a valid leading indicator that a system is approaching instability. The authors show that this indicator's reliability and robustness depend more on the pattern of interactions within the network than on the network structure or the noise intensity. Mutualistic, scale-free, and small-world networks are less stable than are their antagonistic or random counterparts, but this leading indicator more reliably predicts instability for them than it does for their counterparts.

\subsubsection{Regime shifts in socioecological systems}

Regime shifts in socioecological systems are discussed in Ref. \cite{bauch2016early}. The authors observe that theoretical models of socioecological systems are receiving increasing attention. Hysteresis transitions in networks have been observed in coupled socioecological systems. The complex community structure of socioecological systems may create multiple small regime shifts rather than a single large regime shift. Human feedback in a socioecological system can be fundamental to shaping regime shifts. Resource collapse can be caused by surprising and counterintuitive changes that occur due to nonlinear feedback. Thus, failing to account for human feedback can lead to underestimation of the potential for regime shifts in ecological systems.

Four modeling approaches are explored in Ref. \cite{filatova2016regime} and applied to the study of regime shifts in coupled socioenvironmental systems. These approaches are statistical methods, models of system dynamics, models of equilibrium points, and agent-based modeling. A set of criteria has been established and used to (1) capture feedback between social and environmental systems, (2) represent the sources of regime shifts, (3) incorporate aspects of the complexity of the systems, and (4) deal with regime shift identification.

The role of social networks in socioecological regime shifts is investigated in Ref. \cite{sugiarto2015socioecological}. The authors also study the corresponding hysteresis caused by the local ostracism mechanism when various social and ecological parameters are applied. The results show that decreasing the degrees of the network nodes reduces the hysteresis effect and changes the tipping point. The numerical results and analytical estimations verify this conclusion. Interestingly, the hysteresis effect is more reliable in scale-free networks than in random networks, even when the two types of network have the same average degree.

A socioecological regime shift in a model of harvesters of a common-pool resource is studied in Ref. \cite{lade2013regime}. The authors find that such a shift may avoid overexploitation of the resource through social ostracism of noncomplying harvesters. The authors generalize their modeling to study the robustness of the regime shift to uncertainty when specific forms of the model components are used. Generalized modeling is introduced using a simple example of a single population X that can increase due to a gain process $G(X)$ and decrease due to a loss process $L(X)$. A generalized model in differential equation form for population X is 
  	\begin{equation}\label{populationX}
  	\frac{dX}{dt} = G(X) - L(X).
  	\end{equation}
The corresponding Jacobian Matrix $J$ of Eq. \ref{populationX} consists of a single element, which is also the eigenvalue $\lambda$  	
  	\begin{equation}
  	J = \lambda = G^{'*} - L^{'*}.
  	\end{equation}

Additional parameters related to scale, elasticity, and ratios are introduced to relate the model to real-world scenarios. By analyzing the Jacobian Matrix $J$ and referring to contextual knowledge, one could identify and evaluate possible bifurcations.

    \begin{figure}[ht!]
      \centering
      \includegraphics[width = \linewidth]{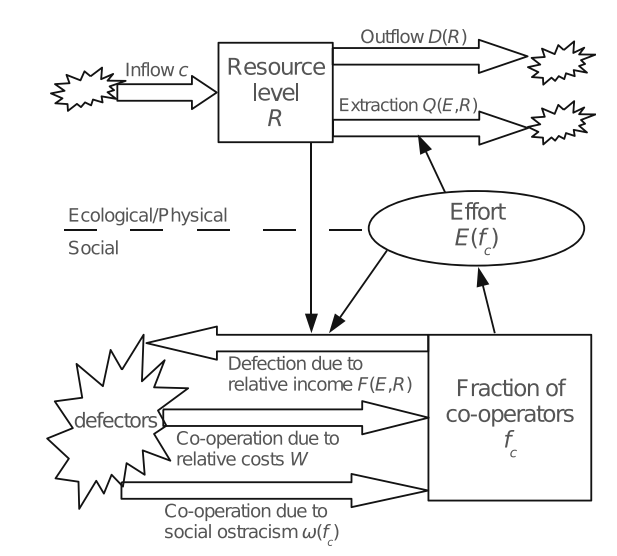} % figure 69
      \caption{Diagram of the generalized model. Double-line arrows represent flows, single-line arrows show influences, rectangles represent state variables,
       ovals show intermediate quantities, and explosion symbols represent the sources and sinks of flows.\\
\textit{Source:} Figure from \cite{lade2013regime}.}
      \label{fig:lade2013_demo}
    \end{figure}
    
Figure~\ref{fig:lade2013_demo} shows the actual model presented in Ref. \cite{lade2013regime}. The corresponding generalized model can be defined as follows:
    
    \begin{equation}
    \frac{dR}{dt} = c - D(R) - Q(E(f_c), R),
    \end{equation}
where $c$ is the resource inflow or growth rate (a parameter that is independent of the current resource level), $D(R)$ is the natural resource outflow rate or mortality, and $Q(E, R)$ rate of is the resource extraction. Here, $E(f_c)$ is the total effort exerted by the harvesters; it decreases as the proportion of cooperators increases. The changes in the fraction of cooperators are represented by the following equation:
    
    \begin{equation}
    \frac{d f_c}{dt} = f_c(1 - f_c)(-F(E(f_c), R) + W + \omega(f_c)).
    \end{equation}

    \begin{figure}[ht!]
      \centering
      \includegraphics[width = 0.8\linewidth]{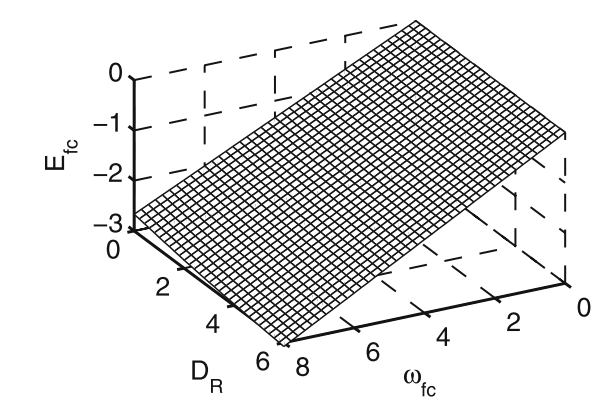} % figure 70
      \caption{ Surface of fold bifurcations for ranges of generalized parameter matching. \\
\textit{Source:} Figure from \cite{lade2013regime}.}
      \label{fig:lade2013_surface}
    \end{figure}

    \begin{figure}[ht!]
      \centering
      \includegraphics[width=\linewidth]{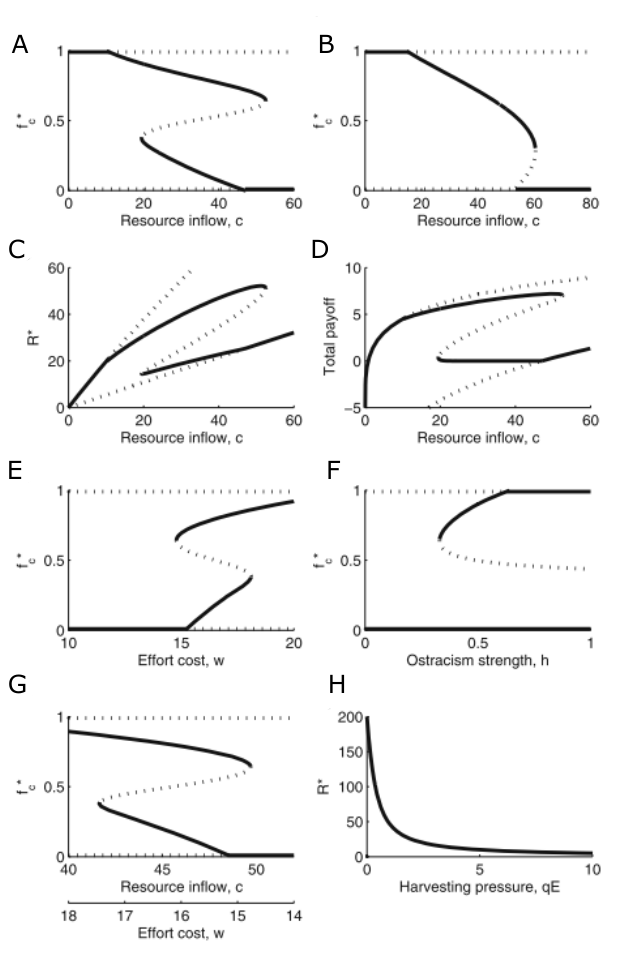} % figure 71
      \caption{Bifurcation diagrams of simulation models. \\
\textit{Source:} Figure from \cite{lade2013regime}.}
      \label{fig:lade2013_params}
    \end{figure}
    
Figure~\ref{fig:lade2013_surface} plots the surfaces of fold bifurcations for ranges of generalized parameters. Figure~\ref{fig:lade2013_params} shows bifurcation diagrams produced using simulation models on paired parameters. Two developments identified in Ref. \cite{scheffer2018quantifying} enable users to build a framework for understanding systemic resilience. The first is the establishment of dynamic indicators of resilience, i.e., early warning indicators. The second is an increasingly ubiquitous collection of dynamic time series data for humans and livestock; collection of these data was made possible by the rapid rise in technologies for automated recording through wearable electronics. The authors also suggest that humans and the livestock they consume may be seen as a complex network that has a strong impact on ecology. The dynamic indicators for this network can be estimated directly from the interactive dynamics of its nodes. Such an approach may help advance social and animal dynamic models from reductionism to a systemic paradigm.

The stability of our planet's ecology is directly impacted by the growing human population and the eating habits of that population. As pointed out in Ref. \cite{kanervarole}, one of the threats is increasing meat consumption, which requires high energy input compared to alternatives. Kanerva \cite{kanervarole} discusses how societies can voluntarily lower meat consumption.

\subsubsection{Difficulties in studies of resilience in social networks}
Integrated research efforts on sustainability, driven by the urgency of addressing climate change threats, are analyzed in Ref. \cite{olsson2015resilience}. The authors begin with a detailed review of the core concepts and principles in resilience theory that could lead to disciplinary tensions between the social and natural sciences. The authors then point out some of the difficulties that currently weaken the feasibility of studying resilience in social systems. The most important of these difficulties are the following:

\begin{itemize}
    \item Ontological differences exist (e.g., social scientists tend to study resilience on the individual level rather than on the community level that is considered in ecological studies). Researchers are reluctant to use the system as an ontological description of society. 
    
    \item The boundaries between disciplines have already been constructed, although the use of institutional lenses following traditional social science studies could help with this issue.
    
    \item Socioecological systems exhibit thresholds that, when exceeded, result in changes in system feedback that lead to changes in function and structure. Thus, the use of the ball and the undulating surface as an analogy in the visualization of tipping points is problematic when modeling social phenomena.
    
    \item Although self-organization is becoming an overriding organizational principle in complex system theory, in which resilience theory is rooted, its mechanism is dubious in the social system setting.
    
    \item The similarities between resilience theory and abandoned theories of functionalism that are now being replaced by theoretically stronger theories for understanding society raise doubts among social scientists about the applicability of resilience theory to their domain.
  
  \end{itemize}
  
In addition to the inherent complexity introduced by the nature of social system problems, scientists cannot bypass the difficulty of coupling multiple systems when studying a single social system resilience phenomenon.

The interdependence of regime shifts in ecosystems is studied in Ref. \cite{rocha2018cascading}. The authors construct a weighted directed network of 30 regime shifts from a database of over 300 case studies obtained from a literature review of more than 1000 scientific papers. The links in this network represent causal relationships (relationships that have cascading effects or simply share standard drivers) between each link endpoint. The analysis shows that the cascading effect accounts for $\sim 45\%$ of all regime shift couplings analyzed, implying structural dependence. The key lesson from this study is that regime shifts can be interconnected and that they should not be studied in isolation under the assumption that they are independent of each other. This lesson is valid for resilience studies in general and indicates that, to achieve precise modeling, we should always include all necessary contexts when studying social system resilience.

\section{Failures in critical infrastructure (CI) systems}\label{Engineering}

Critical infrastructures (CIs) are the backbone of human society and play an essential role in transporting materials and services between distant locations. These critical infrastructures span multiple scales of space and time and provide various flow services. Examples include power grids, gas and petroleum systems, telecommunications systems, banking and financial systems, and transportation, water distribution, and emergency management systems. Power grids, one example of critical infrastructure, often break down due to unexpected large-scale blackouts. The Internet, which symbolizes the beginning of the information era, has become one of the fastest-growing critical infrastructures. Many other critical infrastructures are embedded in various industrial sectors. Due to their crucial role in society, partial or complete dysfunction of critical infrastructure leads to large-scale damage. Thus, similar to resilience in ecology (Chapter \ref{Ecology}), biology (Chapter \ref{Biology}), and society (Chapter \ref{Social}), resilience is also vital to CIs.

While serving as the lifeline or backbone of the entire society, critical infrastructures are exposed to various degrees of risk of the occurrence of perturbations or malicious attacks. On the one hand, under specific scenarios, minor faults in a localized region may accumulate and propagate, causing a domino effect that leads to lethal cascading failures. For example, a tripped line caused a cascading failure in a power grid in North America in 2003, resulting in damage estimated at approximately 10 billion dollars. According to the North American Electrical Reliability Council (NERC)\cite{amin2000toward}, power outages affect nearly 700,000 customers annually. During the occurrence of the most torrential rain in 61 years on July 21, 2012, a city-wide traffic breakdown occurred in Beijing, causing economic losses of 11.6 billion yuan. According to statistics, traffic congestion costs Americans \$166 billion annually \footnote{https://www.truckinginfo.com/338932/traffic-congestion-costs-americans-166-billion-annually}. The eruption of the Icelandic volcano in 2010 led to the cancellation of at least 60\% of daily European airline flights, resulting in a breakdown in airline networks \footnote{The Sydney Morning Herald., http://goo.gl/gvy9Kg, (2010) (date of access: 04/04/2014).}. In addition to extreme weather, malicious cyberattacks on the Internet have been occurring ever more frequently \cite{genge2015system} and increasingly threaten to paralyze entire server systems.

On the other hand, critical infrastructures are usually complex and contain numerous nonlinearly coupled components that may adapt and learn from changing environments. Consequently, as a typical complex adaptive system, a CI system is more than the sum of its parts~\cite{rinaldi2001identifying}. The complexity of CIs lies in their emergent properties that arise from interactions and synergy over large-scale space and time. Therefore, CI systems can adapt and absorb unexpected disturbances while retaining their basic functionalities. 
Given the existence of unexpected circumstances and interdependence, it becomes increasingly urgent to understand and improve the resilience of critical infrastructures. Resilience makes possible adaptation, absorption, and recovery from various perturbations, faults, attacks, and environmental changes. The study of system resilience stems from the study of the self-recovery of ecosystems~\cite{holling1973resilience} (see also Chapter \ref{Ecology}) and biological systems \cite{veraart2012recovery} (see also Chapter \ref{Biology}). An ecosystem may automatically restore itself after it is altered by species invasion or environmental changes; similarly, a cellular network may automatically recover by changing the expression levels of specific genes. Since C. Holling and his colleagues introduced the concept of resilience of ecosystems~\cite{holling1973resilience}, resilience theory has been generalized not only to biology \cite{dai2012generic} but also to climate \cite{dakos2008slowing}, economics \cite{headey2015opinion}, and social systems \cite{barrett2014toward} (also see chapter \ref{Social}).
This suggests a new possible direction for system reliability management. Engineering designed to increase the resilience of critical infrastructures and other complex engineering systems has recently become widely known to the research community \cite{hollnagel2006resilience}.

This section summarizes the structure properties of critical infrastructure systems (Sec. \ref{5spc}) and the various failure models in CIs (Sec. \ref{5fmoci}), including the recent progress about interdependent networks. Due to the nature of infrastructure resilience, we also review the resilience engineering of CIs (Sec.\ref{5reoc}), including evaluation, prediction, and adaptive and control.

\subsection{Structural properties of CIs}\label{5spc}

Research on system resilience relies on our understanding of failures; a system's resilience is related to but differs from the system's reliability and vulnerability. Traditional reliability analysis generally follows a reductionist concept in which it is assumed that system failures occur due to failure of certain combinations of sets of components \cite{kastenberg2005assessing}. According to the analysis, failures of different parts of the system have no correlation or only a weak correlation with each other; thus, the analysis misses the possible strong dependency between components. However, components are not isolated but are interconnected with one another, forming a complex network. A network can be represented by a graph composed of nodes and links that connect the nodes. The road intersections represent nodes, and roads connecting two intersections are links in transportation. In the power grid, generators or transmitters constitute nodes, and transmission lines are links. On the Internet, routers are nodes, and cables are links. In this way, CI system failures can be described and seen as emergent behaviors that can be understood as the collective result of interactions among components at different hierarchical levels. A series of works on technical complex networks, including CIs, analyzes these networks from the point of view of complexity science \cite{kauffman1993origins, albert2002statistical}. Critical infrastructure networks usually contain numerous components that are coupled through nonlinear dynamical interactions. Their structural properties, which are discussed below, include spatiality, the small-world property, modularity, the rich-club property, and other properties, which determine the systems' resilience~\cite{riffonneau2011optimal}.

\subsubsection{Spatial aspects of CIs}

Although many complex systems, including social and biological systems, exhibit common topological properties, the complexities of CI networks differ from those of other complex networks. For example, the degree distributions are scale-free for some networks, such as the gene regulatory networks in biology. Many CI networks are embedded in space with certain constraints, resulting in degree distributions that are usually not scale-free~\cite{albert2004structural}. A good example of a CI network that has such constraints is road networks \cite{li2015percolation}. Some complex networks are found through a renormalization procedure measured in fractal dimensions to have self-similar structures \cite{song2005self}. CI networks usually have geometric constraints and are embedded in two- or three-dimensional (3D) space~\cite{wang2019local}. We call these networks, some of which are 2D or 3D lattices, ``spatial networks'' \cite{barthelemy2014spatial}. Network dimensionality is fundamental since it determines how the system is connected at different scales and has important implications for system function and failure. For a $d$-dimensional regular lattice with $N$ nodes, the average shortest path $\langle l\rangle$ scales as $\langle l\rangle \sim N^{1/d}$. Spatial networks usually have a few long-range connections that bridge distances between distant areas to balance optimal transportation efficiency and cost. Some examples are city highways in road networks and high-voltage transmission lines in power grids. Considering the power-law distribution of link length, spatial networks may have dimensions higher than those that are embedded in space \cite{daqing2011dimension}, suggesting the possible existence of hyperbolic space \cite{boguna2010sustaining}. For example, in a mobile phone communication network, the probability $P(r)$ of having a friend at distance $r$ decays as $P(r)\sim r^{-\delta}$, where $\delta=2$ ~\cite{daqing2011dimension}, and in the global airline network, the probability that an airport has a link to an airport at distance $r$ decays with the exponent $\delta=3$~\cite{bianconi2009assessing}. Note that the dimension characterizes spatially embedded networks with $0<\delta<\infty$. Dimension plays a central role in characterizing the structure of a network and determining the network's dynamical properties and its behavior near a critical point.

\subsubsection{Small-world connectivity of CIs}

Topologically, small-world networks usually have small distances and high clustering coefficients. Some examples of CIs in this category are road networks \cite{dong2020measuring}, railway networks \cite{sen2003small}, the worldwide maritime transportation network \cite{hu2009empirical}, and power grids \cite{watts1998collective}. The small-world phenomenon was first discovered in a letter-delivering experiment \cite{travers1977experimental} in which the distance between any two people in the world was found to be 5.2 on average. A network model developed in 1998~\cite{watts1998collective} explains the formation of a small-world network through random rewiring of a fraction $p$ of the links in a regular lattice. For small $p$, systems show new properties of short global path length and high local clustering coefficients, characteristics that rarely appear in the original regular lattices. According to their statistical properties, small-world networks can be classified into different types that have different scale constraints \cite{amaral2000classes}. Latora et al. \cite{latora2001efficient} introduced the concept of network efficiency as a measure of how efficiently communication networks exchange information and showed that small-world networks are globally and locally efficient. Kleinberg has also demonstrated that one can find short chains effectively with pure local information in small-world networks \cite{kleinberg2000navigation}.

Dynamically, CI networks also have small-world properties \cite{zeng2019switch}. For instance, in urban traffic networks, two different modes of critical percolation behavior appear alternately in the same network topology under different traffic dynamics. During rush hours, the system shows properties associated with critical percolation of a 2D lattice. During nonrush hours or days off, it shows percolation characteristics similar to those of small-world networks. Due to the presence or absence of highway roads in a metropolitan city, the system behaves as s long-range connection in the classical small-world network model \cite{watts1998collective}.

Based on evidence regarding both the topology and the dynamics of transportation networks, resilience is not purely a structural problem. Indeed, consider traffic in the transportation system, which can become gridlocked even if a natural disaster leaves the underlying roads, bridges, and railways relatively unscathed. There is mounting recognition that to address the resilience of networked CI systems, we must measure and control their dynamic states.

\subsubsection{Modularity of CIs}

Modular networks possess densely connected groups of nodes with weak connections between these nodes. CI networks are found to have modular structures. For instance, due to geographical constraints and geopolitical considerations, air transportation networks have typical modular structures \cite{guimera2005worldwide}. The modular structure of a system can be extracted by maximizing the quality function known as ``modularity'' over the possible network partitions \cite{newman2006modularity}. Higher modularity indicates the presence of denser intramodular and sparser intermodular connections. Intermodular links are essential for system robustness, while increasing modularity may decrease robustness against cascading failures \cite{babaei2011cascading}.

\subsubsection{Rich-club aspects of CIs}
The rich-club property of a complex network suggests the presence of dense connections between nodes with high centralities (including degree or betweenness); these connections form the whole complex system's core structure. An analytical expression that can be used to measure and analyze the rich-club phenomena in various networks, including CIs, is provided in~\cite{colizza2006detecting}. At the autonomous system (AS) level of Internet topology, the core tiers between nodes show the rich-club property \cite{zhou2004rich}. International trade and financial networks in high-income countries tend to form groups that may allow financial crises to spread rapidly among them \cite{schiavo2010international}. The rich-club phenomenon examines the group tendency of prominent elements at the top of the network hierarchy to control most of a system's resources \cite{opsahl2008prominence}. In addition to their effects on network topology, weighted rich-club networks are a nontrivial generalization, and their metrics and corresponding integrated detection methods are critical \cite{alstott2014unifying}. The international financial system can be represented by a weighted graph in which the nodes are countries and the links are debtor-creditor relationships between countries \cite{chinazzi2013post}. While performing a critical role in networks, rich-club nodes can also be targets of intentional attacks \cite{cohen2001breakdown}.

\subsection{Failure models of critical infrastructures}\label{5fmoci}
Critical infrastructures are vulnerable to two significant environmental perturbations: extreme weather (e.g., hurricanes, earthquakes, and tsunamis) and intentional attacks (``9.11'' attacks and Bali bombings). CI failures that occur as a result of these perturbations have complexity at both the macroscopic level (rare events) and the microscopic level (cascading failures). 
Historical statistics show that most CI failures are unusual events with low probabilities of occurrence and catastrophic consequences \cite{duenas2009cascading}. This poses challenges related to preparing effectively for the possibility of CI failures. For example, analysis of the empirical data \cite{carreras2003blackout} shows that blackouts have a scale-free distribution of failure size. That is, the possibility of massive outages is far greater than the possibility that would be estimated based on traditional statistical analysis. This intimates that self-organized criticality~\cite{bak1987self} in which the operating pressure of the system continues to increase due to the demand for increasing improvement in system efficiency, may be involved in such failures.
Under these conditions, underlying risks accumulate. The presence of two competing forces drives the system into a state of criticality. When perturbation enters the system, failure size shows a power-law distribution. In the long term, competition between efficient system operation and reliability is also observed, and this competition may generate an unstable balance \cite{carreras2013validating} that may produce massive blackouts in an unprecedented way. An example that illustrates this idea is the sandpile model \cite{dhar1990self}, in which increasing system pressure (addition of sand) can generate the scale-free distribution of sand cascades.
Criticality is believed to differ in different engineering systems as a result of tuning and optimization. Unlike many natural networks, CIs have significant system targets that are usually optimized based on considerations related to efficiency and cost. This ``highly optimized tolerance'' (HOT)'' requires a sophisticated topological configuration and can generate high system reliability \cite{moritz2005wildfires}. The HOT model can also produce a power-law distribution of failure size through various mechanisms.

\subsubsection{Cascading failure model}
Given the statistical properties of macroscopic CI failures, understanding and modeling the microscopic behavior of CI systems is critical. Failure of a CI causes substantial damage to the system itself and to other interdependent systems. CIs that transport materials and services are exposed to various types of perturbations, including component faults, extreme weather, dramatic changes in demand, and malicious attacks. When one or a few components fail, damage to the failing nodes causes transport to go through other routes and generates extra loads on these routes. These additional loads may induce overload on more sites and cause them to be paralyzed, an effect that is called ``cascading overloads''~\cite{motter2002cascade}. A positive feedback process of overloads then begins and will continue to amplify the damaging effect. This process, which poses the main threat to the CI's regular operation, does not stop until the system reaches a new steady state. Knowing the cascading failure propagation behavior of a system enables the evaluation of the underlying risk and the development of efficient mitigation strategies. Unlike visible spreading via contacts in most network dynamics, cascading failures due to overloads usually propagate through hidden paths to seemingly unexpected locations. Tremendous efforts have been made to isolate and localize the faults based on the assumption that cascading failures are short-range correlated. However, the frequency of massive blackouts in the United States was reported not to have decreased between 1984 and 2006, despite the enormous investments that were made in system reliability \cite{hines2008trends}. Based on real-world failure data, city traffic jams and faults in the power grid are spatially long-range correlated, decaying slowly with distance~\cite{daqing2014spatial}. The long-range correlations between failures explain why some existing mitigation efforts are inefficient and suggest a new paradigm that could be used to attempt to reduce this risk.

\begin{SCfigure*}
  \centering
  \includegraphics[width=0.6\textwidth]{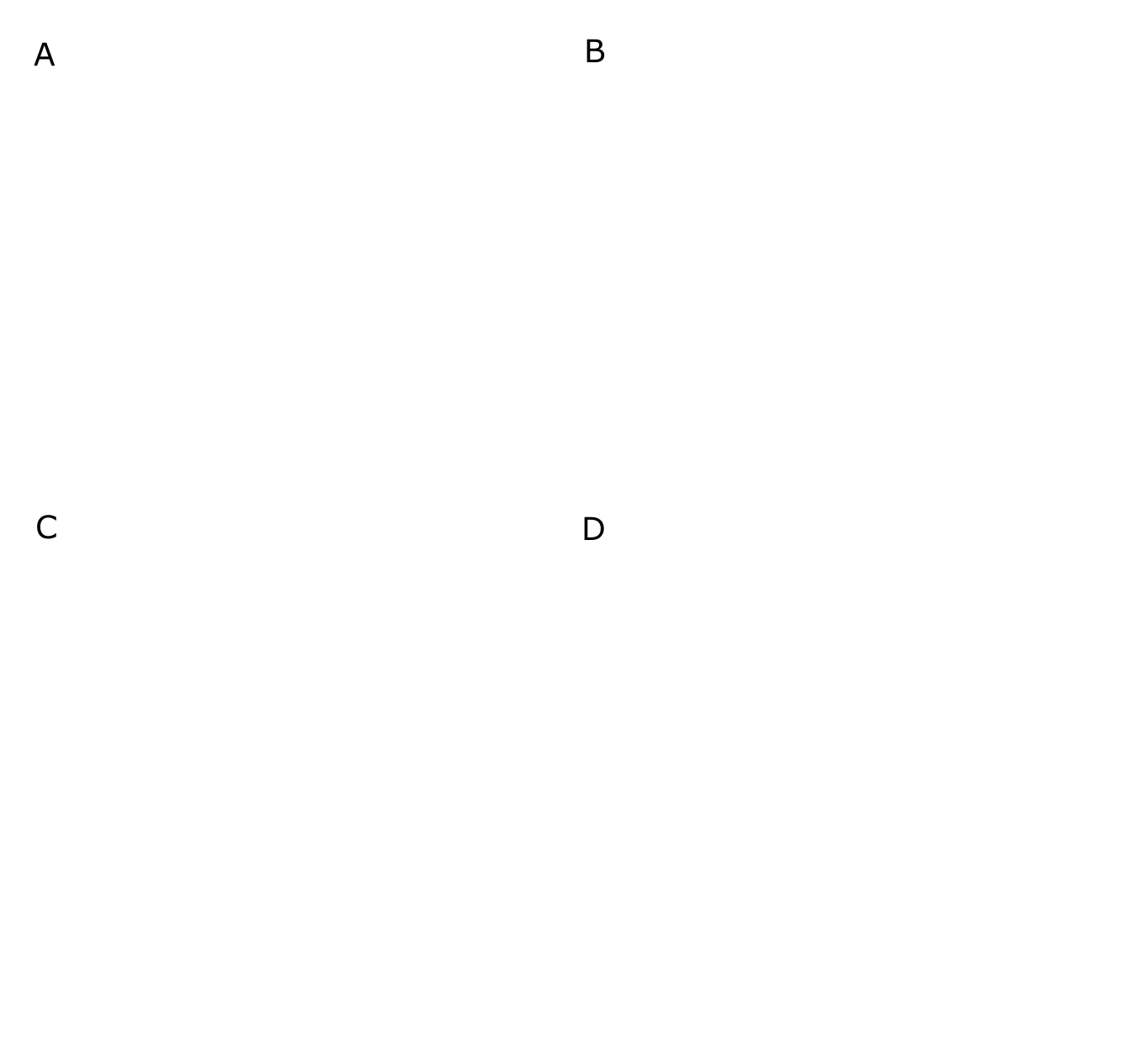} % figure 72
  \caption{Effect of a localized attack on a system with dependencies.
     (A) Propagation of local damage in a system of two interdependent diluted lattices with spatially constrained dependency links between the lattices (only one lattice is shown here). The hole on the right is above the critical size and spreads throughout the system, while the hole on the left is below the critical size and remains essentially the same size. (B) A localized circular failure of radius $r^c_h$ in a lattice with dependency links of length up to $r$. Outside the hole, the survival probability of a node increases with distance $\rho$ from the edge. The parameter $\rho_c$ denotes the distance from the edge of the hole at which the occupation probability is equal to the percolation threshold of a lattice without dependencies $p_c \approx 0.5927^{36}$. (C) Phase diagram of a lattice with dependencies or two interdependent lattices. Depending on the average degree $\langle k\rangle$ and on the dependency length $r$, the system is stable, unstable or metastable. The circles illustrate the increase (when $\langle k\rangle$ increases) in the critical attack size ($r^c_h$) that leads to system collapse in the metastable region. (D) As the size of the system increases, the minimal number of nodes needed to cause the system to collapse increases linearly for random attacks but remains constant ($\approx 300$) for localized attacks. This figure was obtained for a system of interdependent lattices diluted to $\langle k\rangle \approx 2.9$ and $r = 15$ (in the metastable phase-see c), with 1000 runs for each data point. Cited from \cite{berezin2015localized}.}
  \label{figure1}
\end{SCfigure*}

In addition to the long-range spatial correlation of CI failures, the spatiotemporal spreading of failures is also unique and worthy of study. One traffic jam in a local area of a city may cause subsequent jams nearby or a few kilometers away. These common phenomena differ in a fundamental way from events modeled based on the assumptions made in contagion models, including the SIS and SIR models, in the complex network sciences.
 Spatiotemporal propagation of overload failures follows a wave-like pattern~\cite{zhao2016spatio} and occurs at a rate that is independent of the network's spatial structure. This propagation speed demonstrates not only that CI failures are infectious to neighboring sites but also that they spread to sites at a given characteristic distance (measured by their constant velocity) \cite{zhao2016spatio}. It has been confirmed by realistic blackout data analysis that the spreading of failures is nonlocal with respect to both topological distance and geographical distance \cite{yang2017small}. Theoretically, these nonlocal spreading features of cascading overloads could be modeled by adding dependency links to original network structures \cite{berezin2015localized}, as shown in Fig. \ref{figure1}. Although much progress has been made in recent years thanks to the availability of big data, predicting the occurrence of propagating cascade failures in CI networks is still challenging yet pressing to solve. Greater fundamental understanding and additional modeling efforts are required to quantify cascading failures in CI and thereby make it possible to design and improve system reliability and resilience.

A meaningful way in which to understand the propagation of cascading failures is based on overload models, including the Motter-Lai model \cite{motter2002cascade}, the CA models \cite{eisenblatter1998jamming} for transportation and the CASCADE models for power grids \cite{dobson2004branching}. In these models, flow dynamics have different features, and corresponding cascading failures display specific characteristics. The overloads may cause partial or complete breakdown of the networks when intentional attacks \cite{motter2002cascade} or random perturbations \cite{daqing2014spatial} occur. In extreme cases, failure of the single node with the largest load is sufficient to destroy the entire system or a substantial part of it \cite{motter2002cascade}. While the Motter-Lai model assumes that the overloaded node is permanently removed, Crucitti et al. \cite{crucitti2004model} proposed a dynamical model with link efficiency update rules. In this model, the overloaded nodes are unsteady but may recover their functionality, mimicking the Internet's congestion dynamics. They also found that attacking the node with the largest load may decrease the efficiency of the network to the level at which it collapses. Simonsen et al. \cite{simonsen2008transient} proposed another type of dynamical cascading failure model and found that flow dynamics with transient oscillations or overshooting may cause more damage than is observed in static models. Load redistribution during cascading failures is not always deterministic. Thus, Lehmann et al. \cite{lehmann2010stochastic} proposed a stochastic model that was analytically solved using generalized branching processes. The CASCADE model, which is at a higher abstraction level and can theoretically be solved using the Galton-Watson branching process \cite{kim2010approximating}, was developed mainly to model cascading failures on the power grid \cite{dobson2004branching}.

Load redistribution and network structure are essential for cascading failures to occur. Wang et al. \cite{wang2008attack} studied cascading failures in scale-free networks based on the local load redistribution rule. They found that the system reaches its most resilient level under specific redistribution parameters. Based on the global load redistribution rule in the Motter-Lai model, Zhao et al. \cite{zhao2004attack} formulated the cascading process as a phase transition process. In this model, the cascading failure causes the entire network to completely break down below the transition point. Xia et al. \cite{xia2010cascading} studied cascading failures in small-world networks and found that heterogeneous betweenness is a critical factor for network resilience against cascading failures. Wang et al. \cite{wang2004cascading} studied cascading failures in coupled map lattices. They found that cascading failures occur much more easily in small-world and scale-free coupled map lattices than in globally coupled map lattices. While most of these models initiate cascading overloads from given nodes with certain structural features, the percolation framework has also been used to study the critical condition of the giant component \cite{daqing2014spatial}.

\subsubsection{Critical transitions}
In recent decades, the primary applications of cascading overload models have been CI networks. For instance, Wang et al. \cite{wang2009cascade} applied such models to the US power grid and found, surprisingly, that attacking the nodes with the lowest loads produces more damage than attacking the nodes with the highest loads. Menck et al. \cite{menck2014dead} explored the stability mechanism of the power grid and found that the presence of local structures of dead ends and dead trees considerably diminished system stability. For North American power grids, the failure of a single substation could lead to up to a 25\% loss of transmission efficiency due to an overload cascade \cite{kinney2005modeling}. The Italian power grid \cite{crucitti2004topological} and the EU grid \cite{Asztalos2014Cascading} have also been studied using cascading failure models. Blackouts are now mostly modeled as cascading overloads in a series of outage lines with consideration of AC or DC flow dynamics \cite{cetinay2017comparing}. Because cascading overloads typically occur over short time scales, the OPA model~\cite{carreras2002dynamics} has been developed to incorporate the long-term range of the power grid when the network is continuously upgraded to meet increasing load demand. The OPA model was validated using data from the Western Electricity Coordinating Council (WECC) electrical transmission system \cite{carreras2013validating}. To further understand realistic cascading failures, a simulation model that combines power networks with protection systems was proposed \cite{song2015dynamic}. Compared with a simple DC-power-flow-based quasi-steady-state model, this model generates similar results for the early stages of cascading but substantially different results for later stages. When supply and demand features, including increasing demand and renewable sources, are considered, cascading failures could become the first-order transition in the large system size limit \cite{pahwa2014abruptness}. The threat of cascading failures in critical information structures has also been demonstrated \cite{ren2018stochastic}. The model used in the above work incorporates two uncertain conditions regarding the particular next failure node and the time required before the next node state transition occurs.

Next, we review work on critical transitions in transportation infrastructure. Based on real-world data, Treiterer et al. \cite{treiterer1974hysteresis} identified the existence of ``phantom traffic jams''. These traffic jams occur spontaneously with no apparent cause in the absence of accidents or bottlenecks. Such phantom traffic jams can be traced back to lane changes or flow dynamics \cite{kerner1994structure}. Furthermore, large perturbations are found to propagate against the direction of vehicle flow \cite{edie1958traffic,mika1969dual}. Traffic jam formation and propagation in high-dimensional networks are more complicated than in 1D highways. Equilibrium in traffic is a central concept that determines the flow assignment and the resulting distribution. In weighted networks \cite{wu2007cascading}, cascading failures show three types of dynamic behavior: slow, fast, and stationary. For urban traffic, based on the percolation approach, a new framework for studying urban traffic has been proposed \cite{li2015percolation}. When roads become congested and are considered ``effectively removed'', the emergence and formation of urban-scale congestion could be viewed as a percolation process. The critical point of this percolation process could be used to evaluate the urban transportation system's resilience. Thus, the dynamic organization of the transportation network could be analyzed at multiple scales, showing the spatiotemporal imbalance between traffic demand and supply.

\subsubsection{Interdependent networks}
Critical infrastructure systems depend on each other, creating more vulnerabilities than does a single isolated system \cite{rinaldi2001identifying}. For example, the 1998 failure of the Galaxy 4 telecommunications satellite caused the breakdown of most pagers and disrupted a broad spectrum of other CI systems, including systems related to the financial sector and emergency systems. The frequency of failure propagation between different CI systems is becoming higher due to the increasing embedding of information systems into existing CI systems to form so-called ``cyber-physical systems''. Moreover, the interdependence of non-information CI systems is also increasing, given the essential service each CI provides. For example, blackouts in the power grid can cause breakdown of transportation or water distribution networks due to the indispensable role of power in the operation of these systems. There are different types of interdependencies in CI, including physical dependency, cyber dependency, logical dependency, and others \cite{rinaldi2001identifying}. The interdependence relationships can be tight or loose depending on the functional processes involved.

Current models for interdependent CIs focus on mapping hidden functional dependency, a type of dependency that is distinct from the connectivity links in a single network. In 2010, Buldyrev et al. \cite{buldyrev2010catastrophic} developed a framework for analyzing the robustness of interdependent networks subject to cascading failures. The failure of even a small number of elements within a single network may trigger a catastrophic cascade of failures that destroy global connectivity. In a fully interdependent case, each node in a network depends on a functioning node in other networks, and vice versa. The system shows a first-order discontinuous phase transition that is dramatically different from the second-order continuous phase transition found in isolated networks. This phenomenon is caused by the presence of two types of links: connectivity links within each network and dependence links between networks. Connectivity links enable the network to perform its function, and dependence links illustrate that the function of a given node in one network depends crucially on nodes in other networks. The addition of dependency links is found to change a system's robustness significantly. An interdependent network with a  high density of dependency links disintegrates in the form of a first-order phase transition, whereas an interdependent network with a low density of dependency links falls apart through a second-order transition \cite{parshani2011critical}. Later, this framework was generalized to handle the situation of $n$ interdependent networks; the above work revealed that percolation in a single network is a limiting case of the general situation of $n$ interdependent networks \cite{gao2011robustness,gao2012networks}.

As is the case for CI networks embedded in 2D space with cost constraints, once spatially embedded, interdependent networks become extremely vulnerable \cite{bashan2013extreme}. In contrast to nonembedded networks, interdependent networks have no critical fraction of dependency links, and any small fraction of dependency links leads to an abrupt collapse. There may be cascading overloads inside each network and dependency failures between different networks~\cite{zhang2013robustness} with the consideration of flow dynamics. These combined cascading failures can generate more damage than can failures in classical interdependent systems. Complex interdependence has also been modeled as a cyber network overlaying a physical network \cite{yagan2012optimal}. An optimum interlink allocation strategy to protect against random attacks is proposed when the topology of every single network is unknown. Interdependent network models make it possible to design complex systems with more overall robustness and to develop new mitigation methods.

Despite the existence of models for cascading failures in interdependent CI, it remains challenging to evaluate and mitigate cascading failures and to eventually control their occurrence and propagation. In the current state of the art, there are a few measures that can be adopted to manage CI failures. Prevention and planning are two of the most common approaches to avoiding catastrophic failures; they make it possible to improve the resilience of CIs by allocating resources effectively to various parts and stages of the CIs \cite{scaparra2008bilevel}. Nevertheless, preventing all CI failures by planning and enumerating all possible failure scenarios is challenging. Another standard method is crisis management \cite{boin2007preparing}, which emphasizes top-down responses to catastrophes. Crisis management must organize mitigation resources in a timely manner, especially in the ``aftershock'' stage. While these two methods focus on the two end stages of cascading failures, a more systematic CI management framework is required; the development of such a framework is the central task for improving system resilience for critical infrastructures.

\subsection{Resilience engineering of CIs}\label{5reoc}

Resilience concepts have been increasingly applied in a growing number of areas. While reliability \cite{conrad2006critical} emphasizes the capability of a system to continue to function under perturbations, resilience requires the system to bounce back rapidly and strongly after suffering serious damage. Resilience is also more than protection \cite{yusta2011methodologies}; in the context of a critical infrastructure, resilience should include recovery from the degradation of system functions. As a new concept for critical infrastructure, resilience has attracted much attention and has been defined in various contexts \cite{woods2015four, dong2019robust}. Resilience emphasizes adaptation, absorption, and recovery from the failure environment. The realization of this comprehensive ability can be divided into three stages: evaluation, prediction, and adaptation.

\subsubsection{Evaluation}

\begin{figure}
  \centering
  \includegraphics[width=0.48\textwidth]{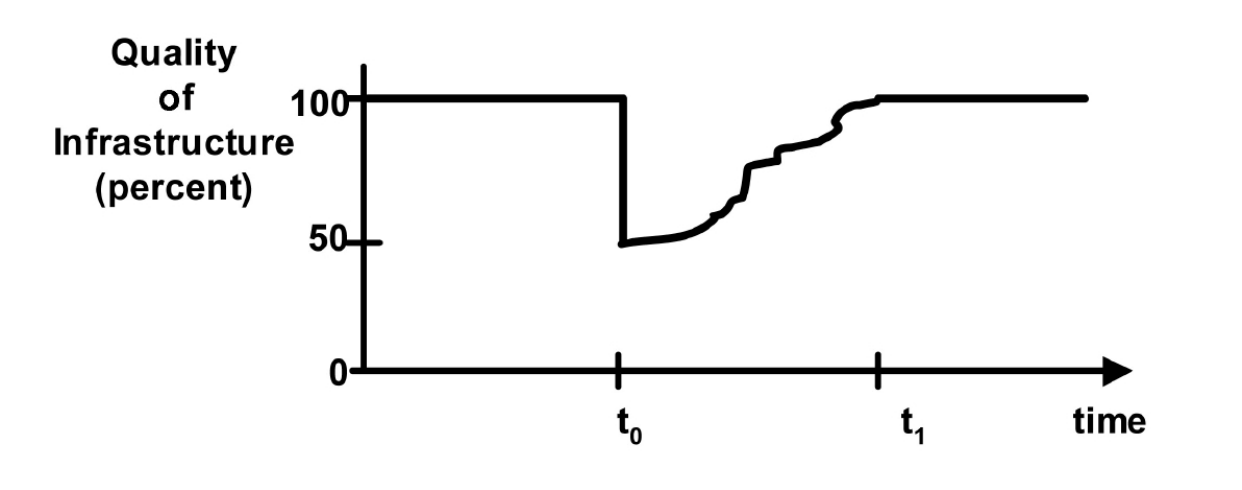} % figure 73
  \caption{Conceptual definition of resilience. \\
        \textit{Source:} The figure is from \cite{bruneau2003framework}.}
  \label{figure2}
\end{figure}

For a critical infrastructure, the first step is to evaluate the system's resilience. Earlier work on evaluating a system's resilience was performed by the earthquake community, for whom ``resilience'' means measuring how the system performs when earthquakes and extreme weather occur. Loss of system performance (the performance of a single CI or that of comprehensive urban systems) in the scenario of an earthquake is compared to the benchmark with consideration of robustness and rapidity \cite{bruneau2003framework}, as shown in Fig. \ref{figure2}. The loss of resilience can be defined as

\begin{equation}
  RL=\int_{t_{0}}^{t_{1}}[1-Q(t)]dt,
\end{equation}
\begin{quote}
  $RL$: resilience loss, \\
  $Q(t)$: service quality of the community (limited to the range from 0\% to 100\%). 
\end{quote}

\noindent
This work suggests that resilience can be conceptualized using the four interrelated dimensions shown in Fig. \ref{figure3}: technical, organizational, social, and economic. Technical resilience refers to the performance of physical systems when faced with earthquake threats. Organizational resilience concerns the ability of organizations to respond to disaster emergencies while continuing to perform their core functions. Social resilience (as discussed in Chapter \ref{Social}) concerns the system's ability to alleviate the negative effects of service losses caused by the social aspect. Economic resilience minimizes the direct and indirect economic losses due to earthquakes. Zobel \cite{zobel2011representing} proposed a similar definition of resilience based on the ``resilience triangle'' (Fig. \ref{figure4}); the proposed definition has then been extended to scenarios of partial recovery from multiple disruptive events \cite{zobel2014characterizing}:

\begin{figure} [!ht]
  \centering
  \includegraphics[width=0.48\textwidth]{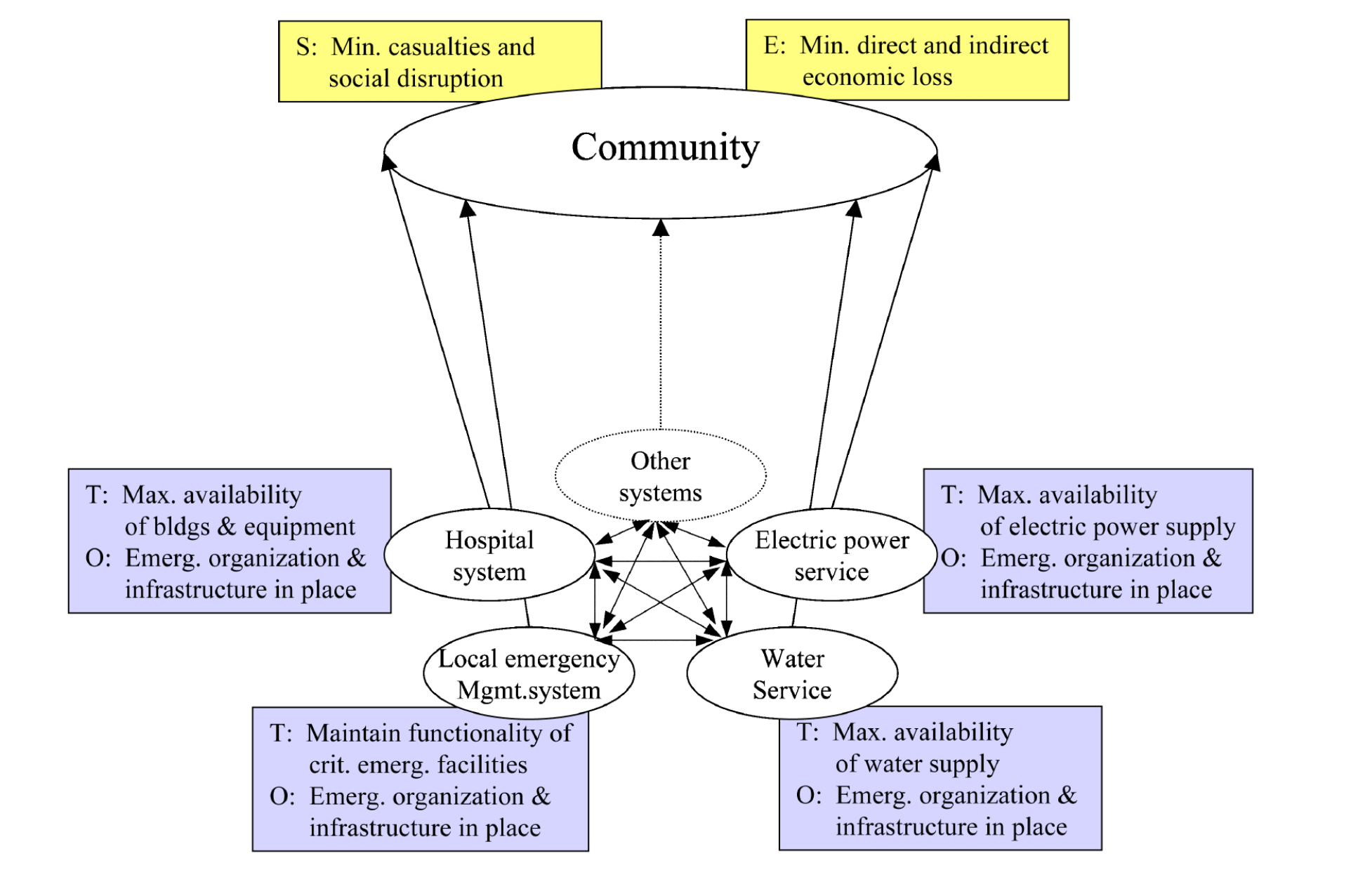} % figure 74
  \caption{A framework for assessing seismic resilience. \\
        \textit{Source:} The figure is from \cite{bruneau2003framework}.}
  \label{figure3}
\end{figure}

\begin{figure} [!ht]
  \centering
  \includegraphics[width=0.48\textwidth]{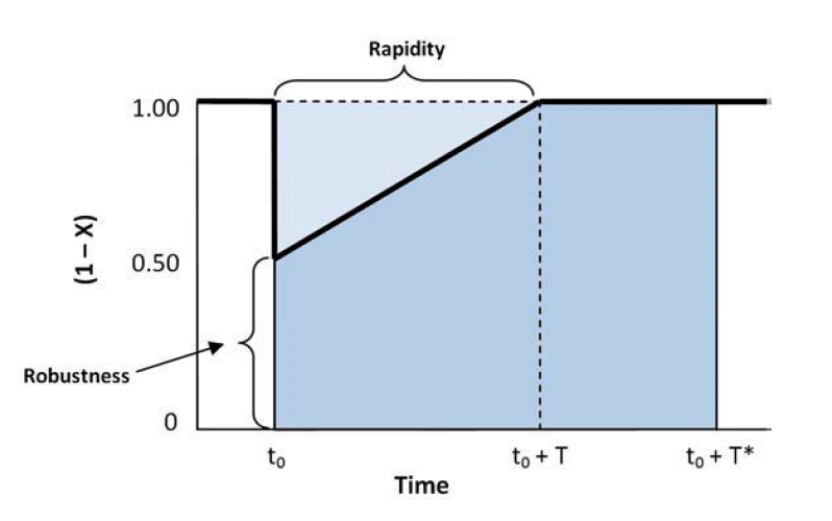} % figure 75
  \caption{A definition of resilience. \\
        \textit{Source:} The figure is from \cite{zobel2011representing}.}
  \label{figure4}
\end{figure}

\begin{equation}
  R(X,T)=\frac{T^*-XT/2}{T^*}=1-\frac{XT}{2T^*},
\end{equation}
\begin{quote}
$R(X,T)$: predicted resilience function, \\
$X$: the initial loss value, \\
$T$: the recovery time, and\\
$T^*$: a strict upper bound on the set of possible values for $T$.
\end{quote}

\noindent
{\bf Mathematical models for evaluation.}
Henry et al. \cite{henry2012generic} developed a time-dependent resilience metric and used it to measure the resistance of a system to disruption during different stages. The metric, which includes reliability, vulnerability and recoverability, is written as

\begin{equation}
  R_{\phi}(t_r|e^j)=\frac{\phi(t_r|e^j)-\phi(t_d|e^j)}{\phi(t_0)-\phi(t_d|e^j)},
\end{equation}
\begin{quote}
  $R_{\phi}(t_r|e^j)$: the value of resilience; indicates the proportion of delivery function that has been recovered from its disrupted state, \\
  $\phi()$: delivery function (or the so-called figure-of-merit), \\
  $e^j$: disruptive event, \\
  $t_r$: the time set for evaluating the current system resilience, \\ 
  $t_d$: the time at which the system transits to its final disrupted state, and \\
  $t_0$: the original time. 
\end{quote}

\noindent
Francis et al. \cite{francis2014metric} proposed a dynamical resilience metric that explicitly incorporates recovery speed. They also considered a two-dimensional system that is capable of absorbing perturbations, possesses adaptive capacity, and is able to recover in the postdisaster stage, as follows:

\begin{equation}
  \rho_i = S_p\frac{F_r}{F_o}\frac{F_d}{F_o}, \\
\end{equation}
\begin{equation}
  S_p =
    \begin{cases}
    (t_{\delta}/t_r^*)exp[-a(t_r-t_r^*)] & \text{for $t_r >= t_r^*$},\\
    (t_{\delta}/t_r^*) & \text{otherwise},
    \end{cases}
\end{equation}
\begin{quote}
  $\rho_i$: resilience factor, \\
  $S_p$: speed recovery factor, \\
  $F_r$: the system's performance at a new stable level after recovery efforts have been exhausted, \\
  $F_d$: the system's performance level immediately following the disruption,\\
  $F_o$: the performance level of the original stable system, \\
  $t_{\delta}$: slack time, \\
  $t_r$: the time to final recovery (i.e., new equilibrium state), \\ 
  $t_r^*$: the time needed to complete initial recovery actions, and \\
  $a$: a parameter that controls for decay in resilience attributable to the time required to reach a new equilibrium.
\end{quote}

\noindent
In addition to the deterministic definition of resilience, a probabilistic metric is considered in the assessment of resilience. Chang et al. \cite{chang2004measuring} defined probabilistic resilience based on the probability that the loss in performance of the initial system following a disruption will be less than the maximum acceptable performance loss and that the time to full recovery will be less than the maximum acceptable disruption time. This differs from the deterministic metric, which is mainly for disasters with complete information. It can be written as

\begin{equation}
  R = P(A|i)=P(r_0 < r^* \text{ and } t_1 < t^*),
\end{equation}
\begin{quote}
  $R$: system resilience, \\
  $P(A|i)$: (resilience defined as) the probability that the system of interest will meet predefined performance standards $A$ in a scenario seismic event of magnitude $i$, \\
  $r_0$: the initial loss, \\
  $r*$: the maximum acceptable loss, \\
  $t_1$: the time to full recovery, and \\
  $t*$: the maximum acceptable disruption time.
\end{quote}

\noindent
Ouyang et al. \cite{ouyang2012three} proposed a resilience metric based on the expectation of a system's performance when the system is exposed to perturbations. The metric is

\begin{equation}
\begin{split}
  AR &= E\left[ \frac{\int_{0}^{T}P(t)dt}{\int_{0}^{T}TP(t)dt} \right] \\
  &= E\left[ \frac{\int_{0}^{T}TP(t)dt-\sum_{n=1}^{N(T)}AIA_n(t_n)}{\int_{0}^{T}TP(t)dt} \right],
  \end{split}
\end{equation}
\begin{quote}
  $AR$: a time-dependent expected annual resilience metric, \\
  $E[]$: the expected value, \\
  $T$: a time interval of one year ($T$=1 year=365 days), \\
  $P(t)$: the actual performance curve, \\
  $TP(t)$: the target performance curve, \\
  $N(t)$: the total number of event occurrences during $T$, \\ 
  $t_n$: the time of occurrence of the {$n$}th event, and \\ 
  $AIA_n(t_n)$: the area between the real performance curve and the targeted performance curve (i.e., the impact area) for the occurrence of the {$n$}th event at time $t_n$.
\end{quote}

\noindent
Youn et al. \cite{youn2011resilience} considered resilience as the sum of the passive survival rate (reliability) and the proactive survival rate (restoration); this differs from most definitions, which integrate a system's performance during the entire disruption process. In Youn''s work, resilience was defined as

\begin{equation}
  \Phi(resilience) = R(reliability) + \rho(restoration),
\end{equation}
\begin{quote}
  $\Phi(resilience)$: the conceptual definition of engineering resilience,\\
  $R(reliability)$: the rate of passive survival, and \\
  $\rho(restoration)$: the rate of proactive survival.
\end{quote}

\noindent
Ayyub \cite{ayyub2014systems} incorporated the effect of system aging in the resilience assessment; the model included different failure profiles of brittle, ductile, and graceful failures, as shown in the following definition:

\begin{equation}
  R_e = \frac{T_i + F{\Delta}T_f+R{\Delta}T_r}{T_i + {\Delta}T_f+{\Delta}T_r},
\end{equation}
\begin{quote}
  $R_e$: system resilience, \\
  $T_i$: time to incident, \\ 
  $F$: failure profile measuring the system's robustness and redundancy, \\ 
  $R$: recovery profile measuring the system's resourcefulness and the speed at which it returns to its original performance level,\\
  ${\Delta}T_f$: failure duration, and \\
  ${\Delta}T_r$: recovery duration.
\end{quote}

\noindent
{\bf Conceptual models for evaluation.}
Ouyang et al. \cite{ouyang2012three} proposed a multistage framework for analyzing infrastructure resilience. The three stages include disaster prevention during normal operation, cascading failures resulting from the initial failure, and a recovery stage. Taking the power transmission grid in Harris County, Texas, USA, as a case study, the authors compare an original power grid model with several resilience models and validate the effectiveness of these models under conditions of random hazards and hurricane hazards. Cai et al. \cite{cai2018availability} proposed an availability-based engineering metric for the measurement of resilience that depends mainly on engineering system structure and maintenance resources. System resilience is evaluated by the developed method based on a dynamic Bayesian network. Renschler et al. \cite{renschler2010framework} defined resilience in seven dimensions (the so-called PEOPLES dimensions: P, Population and demographics; E, Environmental/Ecosystem; O, Organized governmental services; P, Physical infrastructure; L, Lifestyle and community competence; Eco, Economic development; and S, Social-cultural capital). Based on Renschler's definition, Vincenzo et al. \cite{arcidiacono2012community} analyzed the resilience of transportation systems to extreme events. Considering the interdependencies among systems, categories, and dimensions, any plan for recovery should be evaluated based on its ability to maximize the resilience index. Comes et al. \cite{comes2014measuring} evaluated the resilience of power grids and that of the New York subway system and compared them with infrastructures of the same types in different regions.

\begin{figure}[!ht]
  \centering
  \includegraphics[width=0.48\textwidth]{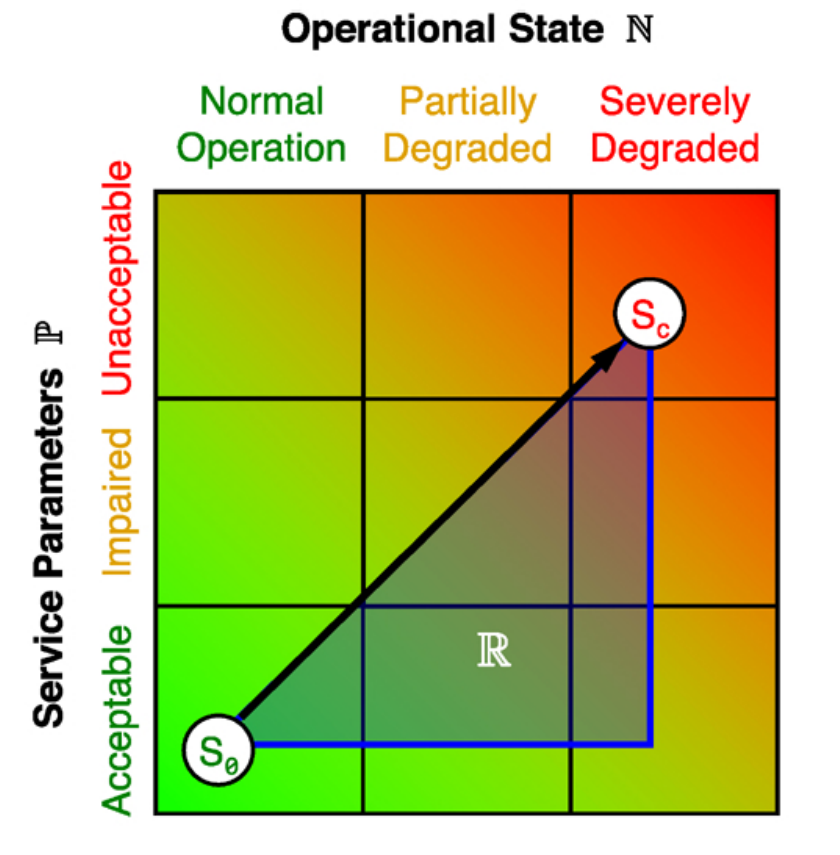} % figure 76
  \caption{Resilience measured in state space.\\
        \textit{Source:} The figure is from \cite{sterbenz2013evaluation}. }
  \label{figure5}
\end{figure}

For analysis of the resilience of transportation networks, Ip et al. \cite{ip2011resilience} proposed a network resilience metric expressed as the weighted sum of node resilience. The resilience of each node is evaluated based on the weighted average number of reliable passageways to other nodes in the network. Using this resilience metric, transportation networks are optimized with a structure optimization model. For information CI networks, resilience is defined similarly, but it also includes concepts developed earlier based on survivability \cite{sterbenz2013evaluation}. Sterbenz et al. \cite{sterbenz2013evaluation} defined resilience at any given layer as a (negative) change in service corresponding to a (negative) change in the operating conditions. Network operational space is divided into regular operation, partially degraded, and severely degraded regions, while the service space is divided into acceptable, impaired, and unacceptable regions. The resilience  of a particular scenario at a particular layer boundary is then represented as the area under the resilience trajectory, as shown in Fig. \ref{figure5}. Fang et al. \cite{fang2016resilience} proposed two metrics of optimal repair time and resilience reduction and used them to evaluate the importance of individual components. These two metrics can determine the repair priority of failed components 
and thereby make it possible to achieve higher resilience and minimize potential losses if component repair is delayed.

\begin{SCfigure*}
  \centering
  \includegraphics[width=0.74\textwidth]{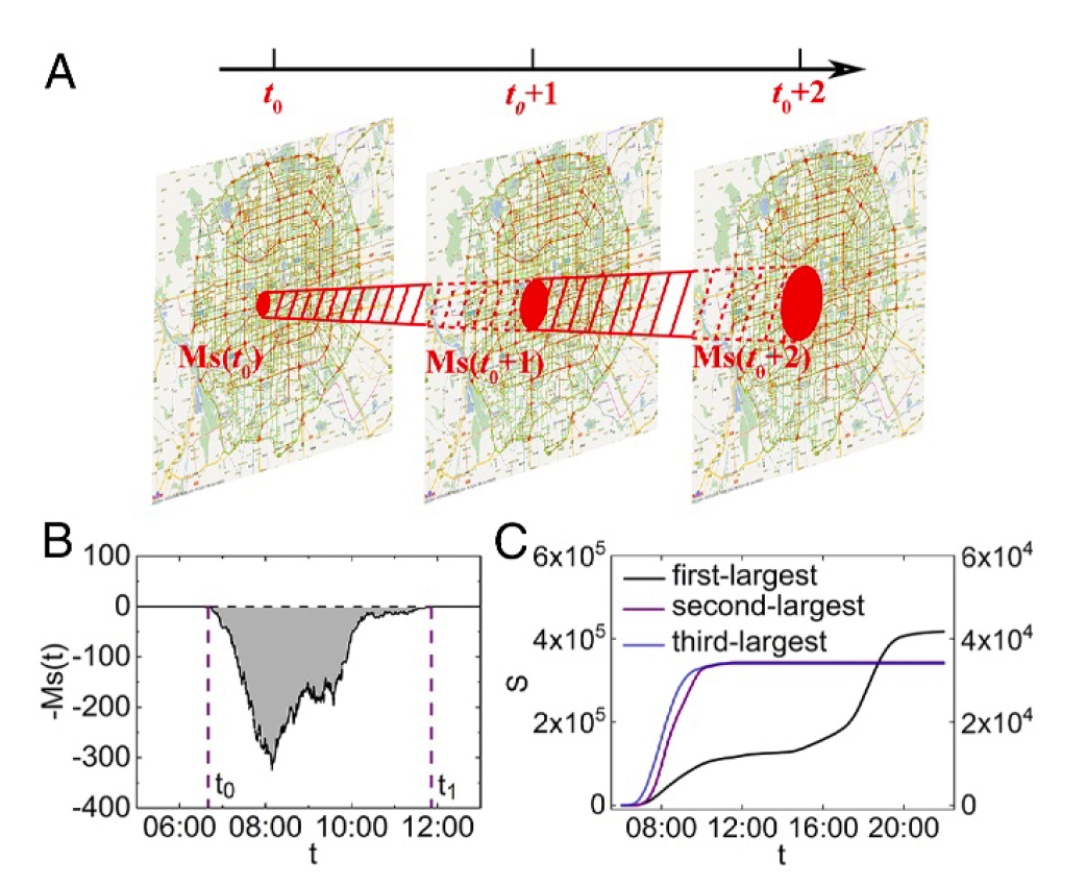} % figure 77
\caption{
Traffic resilience defined based on spatiotemporal jammed clusters. (A) Illustration of the evolution of a jammed cluster in a city. The links shown in red are considered congested. All red links in the shadow belong to the same jammed cluster. (B) Cross-sectional area $M_s(t)$ of the second-largest jammed cluster that occurred on October 26, 2015, in Beijing. Since resilience is reduced during the jam, we plot the negative of $M_s(t)$ as a function of time. Traffic resilience is represented by the gray area. The gray area is the size of the spatiotemporal jammed cluster (S) shown in red in A. The time span between $t_0$ and $t_1$ represents the recovery time of the system ($T=t_1-t_0 + 1$). (C) Cluster sizes of the first-, second-, and third-largest jammed clusters on October 26, 2015 in Beijing as a function of time (the sizes of the second- and third-largest clusters can be read using the right-axis scale).
    Cited from \cite{zhang2019scale}.}
  \label{figure6}
\end{SCfigure*}

\noindent
{\bf Spatiotemporal network models for evaluation.}
Most of the resilience metrics that have been used to date to describe the variation in a system's performance during perturbations are dimensionless. However, critical infrastructures are typically spatiotemporal systems, and as such their failure occurs in both space and time. This spatiotemporal failure behavior and corresponding system response make it necessary to use a more sophisticated framework to evaluate and analyze a CI system's resilience. In \cite{zhang2019scale}, Zhang et al. proposed a spatiotemporal definition that reflects the resilience feature of CI systems, as shown in Fig. \ref{figure6}. The definition is expressed as
\begin{equation}
  S = \int_{t_0}^{t_1}M_s(t)dt,
\end{equation}
\begin{quote}
  $S$: spatiotemporal resilience loss, and \\
  $M_s(t)$: size of the failure cluster at a snapshot of the temporal layer $t$.
\end{quote}

\noindent
This resilience quantification integrates the spatial spreading of traffic congestion during its lifespan. More surprisingly, with the new spatiotemporal definition of system resilience, it is found that system resilience follows a stable scale-free distribution. Such a distribution is usually the result of self-organized criticality. This result contributes to the longstanding discussion of whether resilience is an intrinsic property of the system \cite{francis2014metric} in that it shows that traffic resilience has similar scaling laws for different days and different cities. This scale-free distribution is stable across different working days in Beijing and Shenzhen cities, and it depends on only a few macroscopic parameters, including system dimensions and demand. In addition to system resilience, the time required to recover from jamming failures is also found to follow a scale-free distribution, but that distribution has a different exponent. Note that on the temporal scale, the scaling relation between the lifetime of the traffic jam and system size is found in a 1D lattice cellular automaton model \cite{nagatani1995self}. On the spatial scale, spatial correlations in traffic flow fluctuations are also found to follow a power-law decay model \cite{daqing2014spatial}. With these spatial and temporal scales, a combined spatiotemporal scaling of traffic jams that allows one to view the jams in a stereo way is possible. Resilience that is independent of typical efficiencies \cite{ganin2017resilience} may be explained by stable resilience scaling.

\noindent
{\bf Reliability of networks.} System reliability \cite{billinton1992reliability} refers to fundamental quantities related to system resilience. Reliability engineering focuses on the ability of a system or component to function under the stated conditions for a specified period \cite{geraci1991ieee}. In addition to the statistical measurement of failures, reliability engineering analyzes failure mechanisms \cite{stojadinovic1983failure}, predicts possible causes of failure, and attempts to increase the likelihood that the system will function successfully during its entire lifetime. In other words, reliability engineering requires inherently proactive management of the whole system over its entire lifetime, including design, testing, operation, maintenance, and other activities \cite{barnard20083}. Reliability engineering was first applied to electronic equipment because of its high failure rate, while reliability engineering of complex systems has developed in a different way. For example, fault tree analysis, a classical and efficient method for assessment of system reliability, could help identify the root cause of system failure.
The emergence of system failures in complex systems usually does not reflect the sum of independent component failures. This poses challenges when one attempts to use fault tree analysis and other traditional methods of reliability analysis. Classical reliability methods assume the absence of or weak coupling between components, protocols, and failures. One possible solution to this problem is to combine reliability engineering with network science \cite{zio2007complexity}; such an approach has been proven valid for the analysis of complex systems \cite{li2015network}.

Existing reliability studies of network systems focus on connectivity. Connectivity reliability measures the probability of the existence of connection paths between network nodes. Connectivity reliability can be further classified as two-terminal, $k$-terminal and all-terminal problems \cite{hardy2007k,ramirez2005monte,ramirez2008all}. A network is considered structurally reliable if there is at least one path between the required terminals. Natural connectivity with acute discrimination for different networks is proposed and is derived from the graph spectrum when considering redundancy \cite{jun2010natural}. In addition to the reliability metric based on probability theory, another metric, referred to as belief reliability, has been proposed \cite{zhang2018belief}; it can measure aleatory uncertainty and epistemic uncertainty. Connectivity reliability is usually considered in scenarios such as earthquakes \cite{dong2018post}, floods \cite{dong2020integrated}, and other emergency situations \cite{mostafizi2019agent} in which the state of roads is binary without considering travel dynamics. In addition to the basic connectivity function, reliability also refers to a certain service level from the user demand side. Travel time reliability, which is the probability that a trip can be completed within a given period of time, is then developed to fill this gap \cite{asakura1991road}. Asakura \cite{asakura1999reliability} studied reliability measures for travel between an origin and a destination (OD) in a degraded road network in which links had been damaged by natural disasters. While these reliability indices have their own advantages, measuring reliability more comprehensively remains challenging. Chen et al. \cite{chen1999capacity} defined travel capacity reliability as the probability that a network can meet a certain level of travel demand at the required service level. Considering the fluctuation in demand, Shao et al. \cite{shao2006reliability} studied traffic reliability using a stochastic user-equilibrium-traffic-assignment model solved by a proposed heuristic solution algorithm. To explore the spatiotemporal dimension of travel reliability, Li et al. \cite{li2015percolation} proposed a new model for the measurement of urban traffic operational reliability that features a dynamic traffic cluster in a system composed only of high-velocity roads. This model describes how well the network traffic above a certain service level covers the city from the point of view of network operators rather than from the point of view of a single user.

\noindent
{\bf Vulnerability in networks.}
Since the introduction of the concept of resilience in the study of critical infrastructure, resilience engineering has been related to safety management \cite{righi2015systematic}, a factor that is also a focus in risk analysis. Given the unprecedented risks and the possibly catastrophic damage resulting from them, risk analysis \cite{moteff2005risk} is also one of the critical methods used to understand system resilience. Critical infrastructures are often threatened by natural hazards, and risk assessment for those situations is related to the vulnerabilities exposed to these disasters \cite{guikema2009natural}. Hazard assessment mainly studies the disaster itself, while vulnerability focuses on the exposure and economic losses suffered by critical infrastructures due to failures. Interdependence is also another source of system vulnerability \cite{ouyang2009methodological}. Concepts of uncertainty regarding risk usually involve one of two approaches \cite{der2009aleatory}. The first approach considers risk an inherent property of the system that can be measured by the severity of the harmful consequences and their probability of occurrence \cite{haimes2006definition}. The other approach considers risk from the Bayesian perspective \cite{apeland2002quantifying} and focuses on developing an epistemic framework. Overall, risks arise from the inherent uncertainties in a system itself and from our limited knowledge about the system. These uncertainties regarding unexpected and adverse consequences could pose various threats to CI systems. While these uncertainties can hardly be fully evaluated in risk assessment, resilience may provide a complementary method for risk management and may help improve the system's ability to absorb or adapt to these uncertainties.

In addition to uncertainty evaluation, another critical step in risk analysis is the exploration of the vulnerable parts of the system. This step focuses on system recovery. Vulnerability is usually defined through the degree of exposure to hazards and losses due to the risks that are present \cite{douglas2007physical}. For CI networks, another kind of feature, spatial properties \cite{bashan2013extreme}, is recognized as another source of vulnerability. Various indicators can be used to measure the vulnerability of a link as the service provided by a CI degrades to a level at which there is a possibility of its being closed. For road networks, several link importance indices and site exposure indices have been proposed \cite{jenelius2006importance} based on the variation in generalized travel cost when these links are disrupted. For power grids, identifying the most vulnerable locations in the grid has been studied from the perspective of network survivability analysis, taking into consideration the spatial correlation among outages  \cite{bernstein2014power}. Water distribution networks are critical for city life and may be the targets of intentional attacks. The vulnerability of a water distribution network depends on its hydraulic components and its network topology \cite{shuang2014node}. Chang et al. \cite{chang2001measuring} proposed measurements for postdisaster transportation performance. Tuncel et al. \cite{tuncel2010risk} proposed risk management for supply chain networks based on the failure mode, effects, the criticality analysis (FMECA) technique, and Petri net simulation.

In some cases, critical infrastructures are subject to multiple hazards. While some related studies \cite{pearce1984stochastic, ghosn2005load} have focused on analyzing the probability of occurrence of joint hazards, other risk assessment methods for this scenario have also been proposed \cite{deco2011risk}. For example, considering the multiple hazards of earthquakes and hurricanes, Kameshwar et al. \cite{kameshwar2014multi} proposed a parameterized fragility-based multihazard risk-assessment procedure for highway bridges.

Vulnerability in transportation is defined \cite{berdica2002introduction} as ``a susceptibility to incidents that can result in considerable reductions in road network serviceability''. Compared with reliability, vulnerability usually refers to more harmful consequences that have a lower probability of occurring \cite{mattsson2015vulnerability}. Based on an integrated equilibrium model for a large-scale transportation system, Nicholson et al. studied the critical components based on their socioeconomic impacts \cite{nicholson1997degradable}.

In network analysis, vulnerability is related to nodes or links that possess specific topological features. For example, the most vulnerable part of a scale-free network is usually considered to have the largest degree (i.e., number of links) \cite{albert1999internet}. Other topological features are also found to be related to vulnerability. Weak relationships among the communities in a modular network can cause extensive damage when vulnerability is removed. Critical links in the airline network are explored using memetic algorithms \cite{du2017identifying} and are found to differ from the links that possess the highest topological importance. Basoz \cite{basoz1997risk} proposed a risk assessment method consisting of three parts: (i) retrofitting of critical infrastructures as a means of predisaster mitigation; (ii) predisaster emergency response planning; and (iii) emergency response operations. He measured the importance of the components of the system by network analysis and decision analysis. Chang et al. \cite{chang2002disaster} proposed a vulnerability metric based on business sector, size, and building occupancy tenure to provide a predictor of business loss. Hong et al. \cite{hong2015vulnerability} performed an analysis of the vulnerability of the Chinese railway system to flood risks and provided an effective maintenance strategy considering link vulnerability and burden.

Critical infrastructures are usually modeled as spatially embedded networks, and vulnerability analysis performed in this work has produced exciting and surprising results. The resilience of critical infrastructures to natural disasters, including earthquakes and landslides, could be modeled as localized attacks on spatial networks. Berezin et al. \cite{berezin2015localized} proposed a general theoretical model for localized damage to a spatially embedded network with functional dependency. They found that localized damage can cause substantially more harm to spatial networks than can equivalent amounts of random damage. Localized damage will generate cascading failure when the attack size exceeds a critical value that is independent of the system size (i.e., a zero fraction). An empirical study \cite{yang2017small} confirmed this modeling result. Using North American power grid data collected between 2008 and 2013, the authors found that significant cascading failures are usually associated with concurrent events that occur close to each other in the vulnerable set. Localized damage models have also been extended to interdependent infrastructures in analyses of terrorist attacks \cite{wu2016modeling} and in vulnerability analysis \cite{ouyang2016critical}.

\subsubsection{Prediction}

The prediction of the spatiotemporal propagation of cascading failures could determine the timing and amount of mitigation resource allocation in corresponding real-time resilience management. At the same time, uncertainties regarding the emergence and proliferation of cascading overloads bring fundamental unpredictability to this process. Most studies of the resilience of CIs and other engineering systems assume a single equilibrium and focus on the system's ability to ``bounce back'' to its original state after perturbations occur. 
Due to the clearly demonstrated complexity of dynamical systems, systems may in practice have multiple states, and this can be studied in the context of the theory of multistate systems 
\cite{lisnianski2003multi}. Multistate systems are now applied to evaluate and optimize system reliability \cite{yeh2006k}. While analysis of a multistate system requires the availability of probability data for all the system components, this is rarely possible in real-world cases due to limited budgets and limitations on the amount of time available for observation. Thus, fuzzy set theory is combined with the traditional multistate system to achieve a realistic estimation at an acceptable computational cost. Ding et al. extended multistate system analysis through the development of a fuzzy universal generating function \cite{ding2008afuzzy} in which performance rates and corresponding probabilities are handled with fuzzy values.

Multiple equilibrium states exist for ecological and climate systems, allowing the system to shift from one state to another at its tipping point when it experiences a given disturbance \cite{scheffer2010complex}. Critical infrastructure states can also evolve into different states during daily operations or emergencies. Moving back to the original state may require reaching a different tipping point; this is referred to as the hysteresis property \cite{chakrabarti1999dynamic} and is widely observed in various physical systems. Among critical infrastructures, the hysteresis of traffic dynamics is mainly studied. Traffic hysteresis in freeway traffic was first observed in real-world data in 1974 \cite{treiterer1974hysteresis}. A traffic jam can spontaneously appear during a phase transition without obvious reasons and may be accompanied by a hysteresis phenomenon \cite{treiterer1975investigation}. In traffic studies, fundamental relationships among velocity, flow, and density or occupancy have been observed for a long time \cite{godfrey1969mechanism}. The so-called fundamental diagram is widely accepted and has been used in various scenarios, e.g., in modeling the flow-density relation. The fundamental diagram can be derived in various ways \cite{zhang1999mathematical}: (1) from statistical modeling; (2) from car-following, and (3) from fluid analogies. However, this diagram is challenged by the effects of traffic hysteresis; for example, the relationship between density and speed shows ``loop'' behavior.

\begin{figure*}
  \centering
  \includegraphics[width=0.8\textwidth]{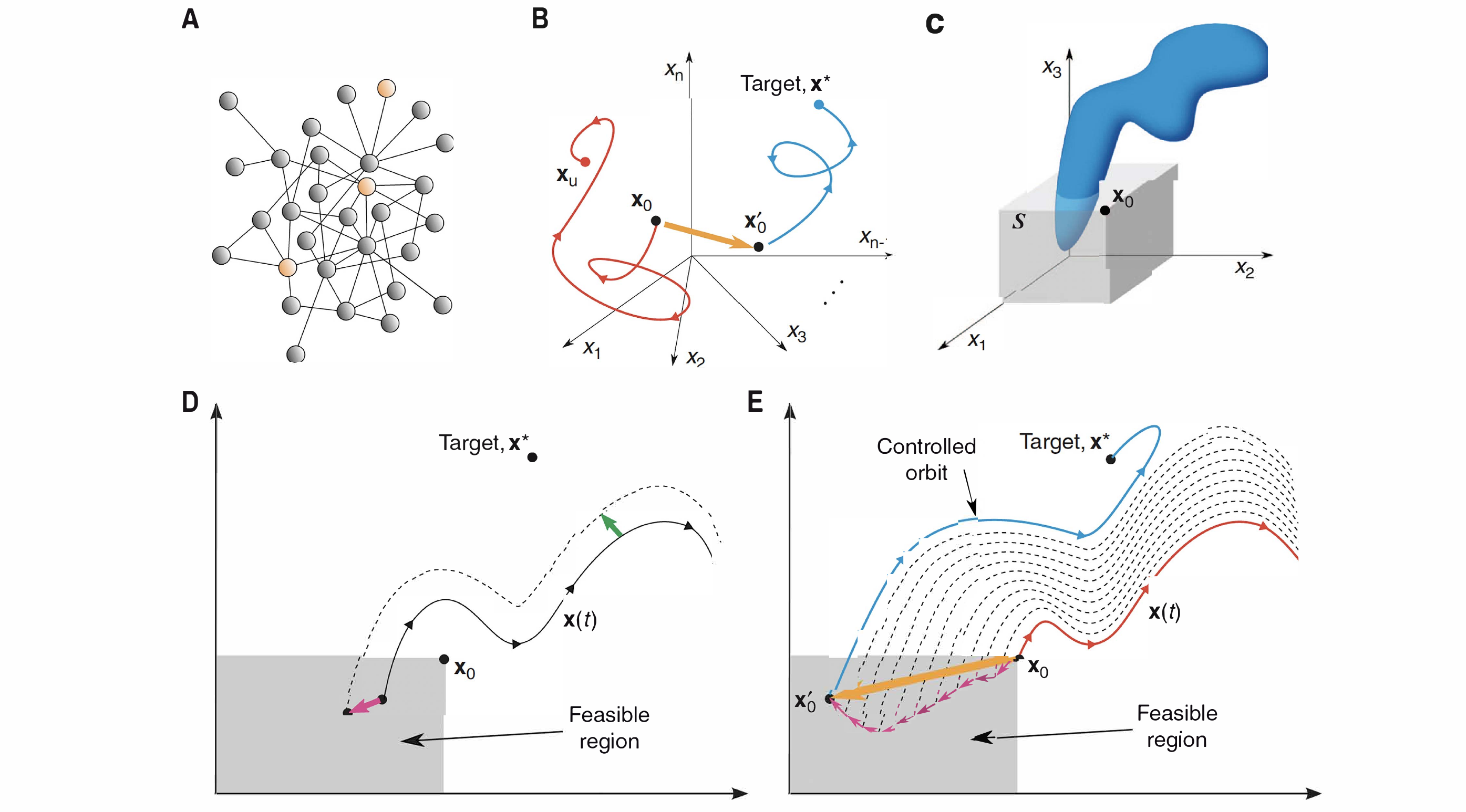} % figure 78
  \caption{Bringing a network to its desired state through perturbation of the initial state. (A) The goal is to drive the network to a desired state by perturbing nodes in a control set (a set consisting of one or more nodes that are accessible to compensatory perturbations). (B) State space portrait. In the absence of control, the network at an initial state $x_0$ evolves to an undesirable equilibrium $x_u$ in the $n$-dimensional state space (red curve). Perturbing the initial state (orange arrow) causes the network to reach a new state that evolves to the desired target state $x^*$ (blue curve). (C) In this example, the network is controllable if and only if the corresponding slice of the target's basin of attraction (blue volume) intersects the region of eligible perturbations (gray volume). (D) A perturbation of a given initial condition (magenta arrow) results in a perturbation of its orbit (green arrow) at the point of closest approach to the target. (E) This process generates orbits that are increasingly closer to the target (dashed curves) and is repeated until a perturbed state $x^\prime_0$ that evolves to the target is identified.
  \\
  \textit{Source:} The figure is from \cite{cornelius2013realistic}.
  }
\label{figure7}
\end{figure*}

Various theoretical models have been proposed to explain the phenomenon of hysteresis. It is conjectured \cite{newell1965instability} that hysteresis is generated by asymmetric behavior of drivers of acceleration and deceleration. Zhang \cite{zhang1999mathematical} proposed a mathematical theory in which acceleration, deceleration, and equilibrium flow are distinguished and used it to model the hysteresis phenomenon. The empirical results are well predicted by the proposed approach. Chen et al. \cite{chen2012microscopic} also suggested that traffic hysteresis occurs when drivers' reactions to traffic oscillations are not symmetric. Through the incorporation of velocity-dependent randomization, Barlovic et al. \cite{barlovic1998metastable} demonstrated metastable states through slow-to-start behavior using the Nagel-Schreckenberg model. Without using  the 1D lattice that was used in most of the above models, Hu et al. \cite{hu2007phase} found hysteresis phenomena related to information traffic in a scale-free network. They observed a hysteresis loop of two branches in the fundamental diagram: the upper branch is obtained by adding packets in the free-flow state, while the lower branch is obtained by removing packages from the jammed state. These modeling results suggest the existence of multiple states of traffic dynamics.

Systems may undergo a regime shift involving an abrupt transition when they evolve                                                                                                                                                                                                                                                                                                                                                                                                                                                                                                                                                                                                                                                                                                                                                                                                                                                                                                                                                                                                                                                                                                                                                                                                                                                                                                                                                                                                                                                                                                                                                                                                                                                                                                                                                                                                                                                                                                                                                                                                                                                                                                                                                                                                                                                                                                                                                                                                                                                                                                                                                                                                                                                                                                                                                                                                                                                                                                                                                                                                                                                                                                                                                                                                                                                                                                                                                                                                                                                                                                                                                                                                                                                                                                                                                                                                                                                                                                                                                                                                                                                                                                                                                                                                                                                                                                                                                                                                                                                                                                                                                                                                                                                                                                                                                                                                                                                                                                                                                                                                                                                                                                                                                                                                                                                                                                                                                                                                                                                                                                                                                                                                                                                                                                                                                                                                                                                                                                                                                                                                                                                                                                                                                                                                                                                                                                                                                                                                                                                                                                                                                                                                                                                                                                                                                                                                                                                                                                                                                                                                                                                                                                                                                                                                                                                                                                                                                                                                                                                                                                                                                                                                                                                                                                                                                                                                                                                                                                                                                                                                                                                                                                                                                                                                                                                                                                                                                                                                                                                                                                                                                                                                                                                                                                                                                                                                                                                                                                                                                                                                                                                                                                                                                                                                                                                                                                                                                                                                                                                                                                                                                                                                                                                                                                                                                                                                                                                                                                                                                                                                                                                                                                                                                                                                                                                                                                                                                                                                                                                                                                                                                                                                                                                                                                                                                                                                                                                                                                                                                                                                                                                                                                                                                                                                                                                                                                                                                                                                                                                                                                                                                                                                                                                                                                                                                                                                                                                                                                                                                                                                                                                                                                                                                                                                                                                                                                                                                                                                                                                                                                                                                                                                                                                                                                                                                                                                                                                                                                                                                                                                                                                                                                                                                                                                                                                                                                                                                                                                                                                                                                                                                                                                                                                                                                                                                                                                                                                                                                                                                                                                                                                                                                                                                                                                                                                                                                                                                                                                                                                                                                                                                                                                                                                                                                                                                                                                                                                                                                                                                                                                                                                                                                                                                                                                                                                                                                                                                                                                                                                                                                                                                                                                                                                                                                                                                                                                                                                                                                                                                                                                                                                                                                                                                                                                                                                                                                                                                                                                                                                                                                                                                                                                                                                                                                                                                                                                                                                                                                                                                                                                                                                                                                                                                                                                                                                                                                                                                                                                                                                                                                                                                                                                                                                                                                                                                                                                                                                                                                                                                                                                                                                                                                                                                                                                                                                                                                                                                                                                                                                                                                                                                                                                                                                                                                                                                                                                                                                                                                                                                                                                                                                                                                                                                                                                                                                                                                                                                                                                                                                                                                                                                                                                                                                                                                                                                                                                                                                                                                                                                                                                                                                                                                                                                                                                                                                                                                                                                                                                                                                                                                                                                                                                                                                                                                                                                                                                                                                                                                                                                                                                                                                                                                                                                                                                                                                                                                                                                                                                                                                                                                                                                                                                                                                                                                                                                                                                                                                                                                                                                                                                                                                                                                                                                                                                                                                                                                                                                                                                                                                                                                                                                                                                                                                                                                                                                                                                                                                                                                                                                                                                                                                                                                                                                                                                                                                                                                                                                                                                                                                                                                                                                                                                                                                                                                                                                                                                                                                                                                                                                                                                                                                                                                                                                                                                                                                                                                                                                                                                                                                                                                                                                                                                                                                                                                                                                                                                                                                                                                                                                                                                                                                                                                                                                                                                                                                                                                                                                                                                                                                                                                                                                                                                                                                                                                                                                                                                                                                                                                                                                                                                                                                                                                                                                                                                                                                                                                                                                                                                                                                                                                                                                                                                                                                                                                                                                                                                                                                                                                                                                                                                                                                                                                                                                                                                                                                                                                                                                                                                                                                                                                                                                                                                                                                                                                                                                                                                                                                                                                                                                                                                                                                                                                                                                                                                                                                                                                                                                                                                                                                                                                                                                                                                                                                                                                                                                                                                                                                                                                                                                                                                                                                                                                                                                                                                                                                                                                                                                                                                                                                                                                                                                                                                                                                                                                                                                                                                                                                                                                                                                                                                                                                                                                                                                                                                                                                                                                                                                                                                                                                                                                                                                                                                                                                                                                                                                                                                                                                                                                                                                                                                                                                                                                                                                                                                                                                                                                                                                                                                                                                                                                                                                                                                                                                                                                                                                                                                                                                                                                                                                                                                                                                                                                                                                                                                                                                                                                                                                                                                                                                                                                                                                                                                                                                                                                                                                                                                                                                                                                                                                                                                                                                                                                                                                                                                                                                                                                                                                                                                                                                                                                                                                                                                                                                                                                                                                                                                                                                                                                                                                                                                                                                                                                                                                                                                                                                                                                                                                                                                                                                                                                                                                                                                                                                                                                                                                                                                                                                                                                                                                                                                                                                                                                                                                                                                                                                                                                                                                                                                                                                                                                                                                                                                                                                                                                                                                                                                                                                                                                                                                                                                                                                                                                                                                                                                                                                                                                                                                                                                                                                                                                                                                                                                                                                                                                                                                                                                                                                                                                                                                                                                                                                                                                                                                                                                                                                                                                                                                                                                                                                                                                                                                                                                                                                                                                                                                                                                                                                                                                                                                                                                                                                                                                                                                                                                                                                                                                                                                                                                                                                                                                                                                                                                                                                                                                                                                                                                                                                                                                                                                                                                                                                                                                                                                                                                                                                                                                                                                                                                                                                                                                                                                                                                                                                                                                                                                                                                                                                                                                                                                                                                                                                                                                                                                                                                                                                                                                                                                                                                                                                                                                                                                                                                                                                                                                                                                                                                                                                                                                                                                                                                                                                                                                                                                                                                                                                                                                                                                                                                                                                                                                                                                                                                                                                                                                                                                                                                                                                                                                                                                                                                                                                                                                                                                                                                                                                                                                                                                                                                                                                                                                                                                                                                                                                                                                                                                                                                                                                                                                                                                                                                                                                                                              from one metastable state to another. This regime shift is usually highly unpredictable, posing a significant challenge for predicting the behavior of complex systems \cite{scheffer2012anticipating}. The tipping point often marks the critical threshold of a regime shift. When a system approaches its tipping point, a positive feedback mechanism pushing the system to another state becomes much more substantial. This leads to a domino effect that produces cascading failures when critical infrastructure robustness or reliability is considered. Homogeneous systems with dense connections may have strong local resilience yet be globally fragile \cite{scheffer2012anticipating}.

Heterogeneous modular systems may also have the capacity to adapt to gradual changes in the environment. 
Knowing the relationship between system resilience and structure can help predict the behavior of social-technological infrastructures more accurately. Prediction of closeness to the tipping point is an urgent and critical need for avoiding catastrophic collapse and improving system resilience. One classical solution suggests looking for ``critical slowing down'' during which a system's recovery to the original state becomes slower as the system approaches the vicinity of the tipping point \cite{van2007slow,veraart2012recovery,ives1995measuring,carpenter2006rising,kleinen2003potential,livina2007modified}. When systems display intense fluctuation, an alternative indicator is developed to infer the shape of the system state landscape \cite{livina2010potential,hirota2011global}. The change in the potential system landscape also reflects the shifts among different attractors. Although these indicators have been applied successfully in a variety of disciplines, the development of valid applicable indicators is far from complete. As the availability of high-resolution spatiotemporal data increases, more indicators could be explored; spatial information is sometimes more informative and robust than temporal indicators, especially for critical infrastructures. For example, early-warning signals could be designed based on percolation methods for traffic dynamics \cite{zeng2020multiple}.

\subsubsection{Adaptation and control}

The ultimate goal of system resilience is improvement of the system's ability to absorb and adapt to unexpected risks and faults. With big data and the development of control technology, CI systems have become increasingly automated and intelligent. One significant difference between CI systems and other engineering systems is that human behaviors and decisions are deeply involved in every inch of CI systems. For example, traffic congestion has become an ``urban disease'' that impedes urban development. Its primary ``pathogenesis'' may be an imbalance between the supply of and the demand for spatiotemporal transportation resources, manifested by the mismatch between the rapid growth of short-term traffic demand and the slow improvement in traffic supply capacity. Urban traffic has the following features: increasing in the short term, being differentially distributed in the whole space, and jamming quickly in local areas. In a system with limited transportation resources, congestion can become challenging to dissipate and may propagate in space and time, resulting in possible regional dysfunctions in urban traffic. In extreme weather, important events or other emergencies, city-scale congestion and significant traffic capacity degradation may occur, leading to calls for system adaptation and recovery capability.

\begin{figure}
  \centering
  \includegraphics[width=0.95\linewidth]{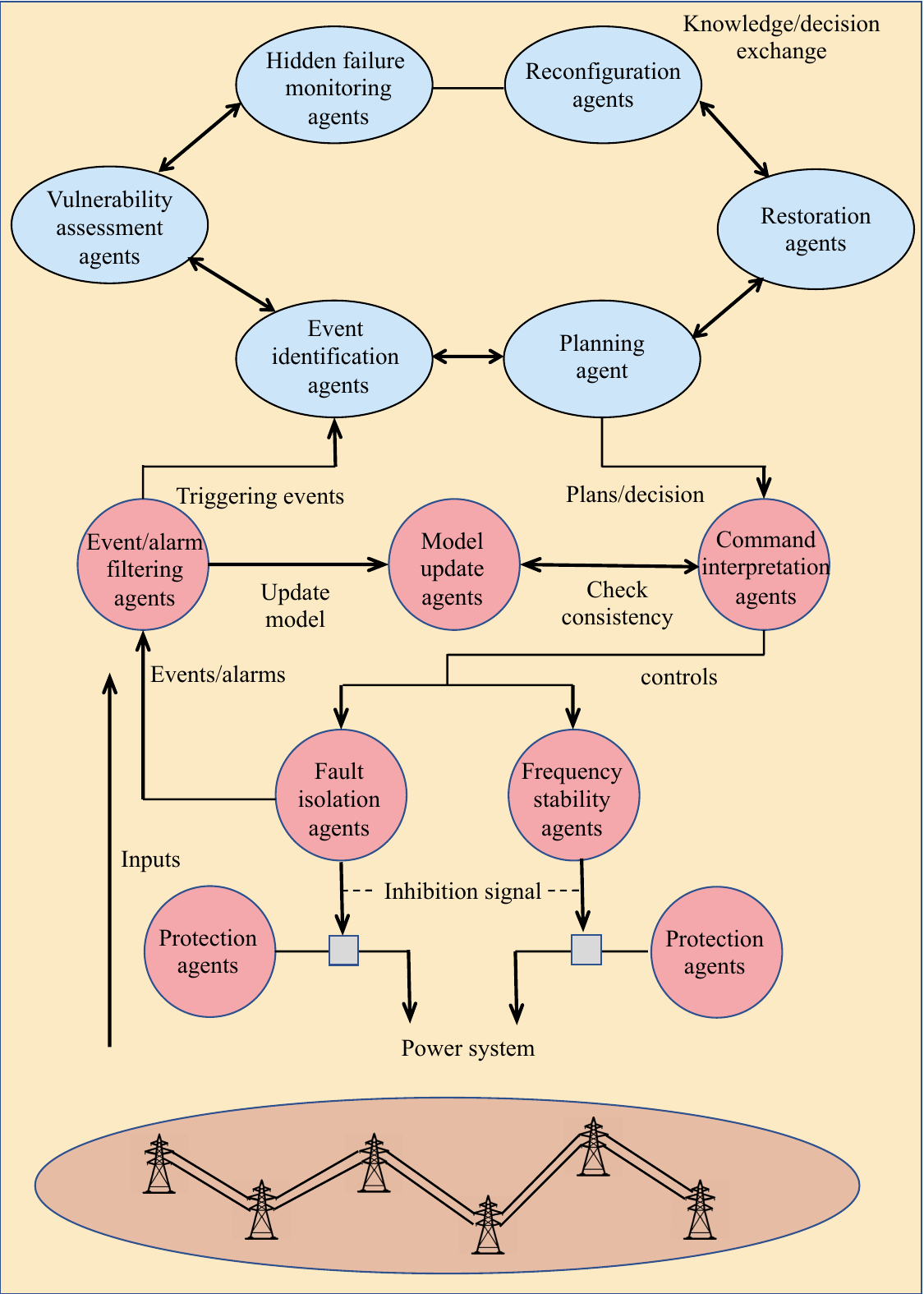} % figure 79
  \caption{
  A multiagent system design. This design organizes agents into three layers. The reactive layer (bottom) consists of agents that perform reprogrammed self-healing actions that must be initiated immediately. Reactive agents, the goal of which is autonomous and fast control, are in every local subsystem. The agents in the coordination (middle) layer include heuristic knowledge that can be used to identify which triggering events in the reactive layer are urgent, important, or resource-consuming. The agents in the coordination layer, whose goal is consistency, also update the system's real-world model and check whether the plans (or commands) from the deliberative (top) layer represent the system's current status. If the plans do not match the real-world model, the agents in the middle layer trigger the deliberative layer to modify the plans. The deliberative layer consists of cognitive agents that have goals and explicit plans that allow them to achieve their goals. The goals of the agents in this layer are dependability, robustness, and self-healing.
  \textit{Source:} The figure is from \cite{amin2000toward}.
  }
  \label{figure8}
\end{figure}

The adaptation and recovery of CIs could be achieved through the ultimate control of complex systems. Recent studies of controllability have attracted much attention to the control of complex networks \cite{liu2011controllability,gao2014target}. These studies offer theoretical tools that can be used to identify a subset of driver nodes that can control the whole system through proper control strategies. Liu et al. \cite{liu2011controllability} developed conceptual tools for studying the controllability of an arbitrary, complex, directed network through a set of driver nodes that exert time-dependent control of the system's entire dynamics. Within this framework of network controllability, nodes and links could be ranked based on their importance in controlling the whole network. For example, connections could be classified as critical, redundant, or ordinary based on their influence on the controllability of the network \cite{liu2011controllability}.

From the system control point of view, the realization of system resilience that allows recovery to a healthy state depends on whether we can tune the system to the desired attractor in terms of control theory. For linear systems, this could be achieved using a minimal control energy strategy. However, most of the critical infrastructures are not linear systems. There are no general solutions for controlling nonlinear systems, especially those with numerous components and complex interactions. A variety of approaches have been proposed that take into account the specific background of the problem under consideration. Urban traffic systems are found to follow the dynamic pattern of the macroscopic fundamental diagram (MFD), allowing the controller to perform state control accordingly. For example, optimal perimeter control for two regions through model predictive control with the controller at the border between the two regions is proposed \cite{geroliminis2012optimal}. For power grids, Cornelius et al. proposed bringing the network to its desired state through perturbation of the initial state to the attractor basin of the desired state \cite{cornelius2013realistic}, as shown in Fig. \ref{figure7}.

A self-healing framework for the realization of system adaptation for critical infrastructures has been proposed \cite{amin2000toward}. This framework is described in Fig. \ref{figure8}. It is based on intelligent agents that are distributed at different levels over the whole system. These agents make decisions according to their environments and interact with each other. Liu et al. \cite{liu2014modeling} proposed a self-healing model of protection against cascading overloads in which agents decide the restoration timing and resources. Optimal restoration timing is found to enable the system to move away from the edge of collapse. Lin et al. \cite{lin2016restorative} proposed a self-healing transmission network reconfiguration algorithm that considers electrical betweenness. Quattrociocchi et al. \cite{quattrociocchi2014self} achieved self-healing capability using distributed communication protocols. They studied the effects of redundancy on healing performance in the presence of various connectivity patterns. Gallos et al. \cite{gallos2015simple} proposed a local-information-based self-healing algorithm by adding after-damage links that are as short as possible, considering the fraction of failed neighbors. Shang \cite{shang2015impact} studied the impact of self-healing capability on network robustness in a network percolation framework.

\section{Conclusions and future perspectives}\label{Conclusion}

\subsection{Conclusions}
Resilience has long been recognized as a defining property of many dynamical complex systems \cite{holling1973resilience}. Moreover, network resilience has already become a new emerging subfield in network science \cite{gao2016universal}. This report reviews several primary theoretical tools and empirical studies that analyze the resilience of complex networked systems. To quantify a complex system's resilience, we first need to uncover its dynamics, which are usually modeled as ordinary differential equations. We could detect the attractors in low-dimensional systems by calculating the fixed points of the equations \cite{scheffer1993alternative, mumby2007thresholds} and use the quasi-potential landscape to show the trajectories of transitions between different attractors \cite{waddington2014strategy, fang2018cell}. For high-dimensional networks, in addition to the traditional linearization method \cite{koopman1931hamiltonian, korda2018linear}, the dimension-reduction method based on mean-field theory has been shown to be a powerful tool for predicting tipping points \cite{gao2016universal, jiang2018predicting}. We could also use changes in network structure as early warning signals of regime shifts \cite{chen2012detecting, yang2018dynamic}.

We have incorporated the primary theoretical tools discussed above into a discussion of their applications and have extensively reviewed state-of-the-art progress while analyzing the resilience of ecological, biological, social, and infrastructure systems. A series of topics, including alternative stable states (also called multiple stable states), regime shifts, feedback loops, and early warning signals, have been addressed in mathematical models and empirical studies. We have also emphasized the impact of the network on resilience. We point out that network robustness and stability are distinct from network resilience; robustness focuses on network topology, stability is a property of systems that show fluctuation around a specific fixed point, and resilience is a broader concept within which the ability to withstand perturbations, recovery, and adaptation are the three essential elements. The collapse of ecosystems and global warming threaten the well-being of entire societies or even the global population. Yet, they often result in communities split into two groups, one supporting remedies, the other denying the need for them, leading to the strong political polarization of societies~\cite{macy2021polarization}.

\subsection{Future perspectives}
Given the rapid advances in the resilience of complex networks, we have discussed the results of both theoretical and empirical studies that will likely remain with us for many years to come. In any emerging field, there is a large body of relevant questions and upcoming developments that may further increase the relevance of the discussed frameworks. Below, we highlight several research topics that are currently being addressed to achieve a comprehensive understanding of networked system resilience.

\subsubsection{Uncovering network dynamics from data on system evolution}
As we discussed in Chapter~\ref{Robustness}, network dynamics is a prerequisite for resilience analysis. Due to the complex and often unknown mechanisms that operate within complex networks, it is difficult to determine the appropriate equations, and only rarely are sets of parameters and mechanistic models available for complex systems. Thanks to the accumulation of massive data, capturing the detailed node-level dynamics of biological, social, and technological systems has become possible, and doing so could help us begin to understand the inner mechanisms of complex systems. Developing tools for discovering network dynamic models directly from data \cite{barzel2015constructing, yuan2019data} or simply from network topology \cite{harush2017dynamic}, even in the absence of a mechanistic model, has also become popular. Previous studies have shown that a complex system's response to perturbations is driven by a small number of universal mechanisms \cite{barzel2013universality}. The authors of the cited study also proposed a method for inferring the microscopic dynamics of a complex system from observations of its response to external perturbations \cite{barzel2015constructing}, allowing us to construct an effective dynamic model of the system. By separating the contributions of network topology from those of network dynamics, Barzel et al. \cite{harush2017dynamic, hens2019spatiotemporal} developed general tools that could be used to translate a network's topology into predictions of its observed propagation patterns.

Despite these excellent efforts, tools for uncovering the dynamics of specific real-world systems are still lacking. First, these studies assume that all nodes in a given network follow the same dynamical pattern, whereas nodes in a real network may exhibit very different dynamic behavior. For example, in a protein interaction network, two proteins may be connected due to their simple physical binding/separation or because they participate in the same metabolic reaction that produces other compounds \cite{rual2005towards}. Second, the dynamic models presented in these studies are all ordinary differential equations, and time is the argument. Other variables, such as spatial distance, could be vital driving forces in dynamic changes. Hence, a more general model represented by partial differential equations may be needed. Finally, each component's time scale may vary over a vast range from seconds to months, a fact that brings significant challenges to dynamic modeling. A feasible way in which to consider all these complexities is to combine prior knowledge about the real-world system with data-driven dynamic model discovery methods.

\subsubsection{Empirical studies of resilience}
Despite the extensive studies of network resilience and related topics such as alternative stable states, regime shifts, and feedback loops, there are far more theoretical studies of resilience than empirical studies. For example, compared to the plethora of mathematical models demonstrating the existence of alternative stable states\cite{scheffer2001catastrophic}, there are only a few empirical studies, especially experimental studies conducted in the real world rather than in laboratories. The reason for this is that most real networks are too complex and interconnected to permit real-world experimentation. There are always gaps between theoretical models and natural systems; for example, the universal dimension-reduction method for predicting the resilience of complex networks is based on the assumption that no negative interactions occur in networks \cite{gao2016universal}.

The remaining challenges in this field include bridging the gaps between theoretical models and real networks and applying resilience theory to various real-world systems such as traffic networks, power grids, human mobility networks~\cite{zhang2018spatiotemporal}, and ecological and biological systems. For these real-world systems, it is crucial that we develop strategies to improve their resilience and prevent collapse. Currently, massive data on these systems have been obtained. Predicting the tipping point in these systems directly from data rather than by using models remains an open problem. In addition, previous studies treat structure and dynamics independently as a way of investigating how structural perturbations affect the system's resilience. However, real-world systems contain ``adaptive cycle'' \cite{holling2002resilience} through which they adjust their dynamics to compensate for structural perturbations. It remains a challenge to determine the interplay between structure and dynamics.

\subsubsection{Resilience of networks of networks}
It is increasingly clear that networks in the real world are not isolated but are interdependent with one another. Examples are interdependent critical infrastructures such as electric power networks that provide power for the pumping and control systems of a water network. In turn, the water network provides water for cooling and emissions reduction in the power network, a fuel network provides fuel for the generators in the electric power network, and the electric power network provides the power needed to pump the oil used in the fuel network \cite{gao2015recent}. Sets of networks may be coupled to form multilayer networks \cite{boccaletti2014structure, pan2020phase} or networks of networks \cite{liu2019multiple}. As discussed in Chapter~\ref{Robustness}, studies that focus mainly on structural integrity and robustness have already been conducted \cite{buldyrev2010catastrophic, radicchi2013abrupt}. In contrast, resilience is related to network dynamics, and coupled or interdependent networks may be characterized by different time scales and structural patterns, making the analysis of the network resilience extremely difficult. Early attempts have focused on specific dimension-reduction methods for ecological mutualistic networks \cite{jiang2018predicting} or have taken purely numerical approaches \cite{boccaletti2014structure}. Despite these efforts, developing a general framework for analyzing the resilience of networks of networks remains a challenge and can trigger a series of future studies.

\subsubsection{Resilience of networks for which there is incomplete information}
Most real networks are incomplete due to their complexity and due to a lack of complete knowledge about the original systems. For example, there may be lack of knowledge about hidden relationships in social networks \cite{cai2005mining}, unmeasured connections in biological networks \cite{menche2015uncovering}, and adaptively changing interactions between species in ecological networks \cite{guimaraes2011evolution}. It is essential that we develop tools that can be used to infer the original network structure and dynamics when the available information is incomplete. The application of a branch of network science called link prediction, which aims to uncover such hidden relationships \cite{menche2015uncovering, zeng2018prediction}, could provide a step towards this goal. It would still be difficult to achieve this goal entirely. The more significant challenge is that real complete networks are usually too large to handle. For example, the social network Facebook attracted more than 2.5 billion users in 2019. Hence, an inverse problem has attracted much attention. That is, given a vast real network, how can we derive a representative sample \cite{leskovec2006sampling}? Graph sampling is a technique that can be used to create a small but representative sample (an incomplete network) by choosing a subset of nodes and links from the original network\cite{li2015random}. Several methods of graph sampling, including random walk sampling \cite{gjoka2010walking}, breadth-first sampling \cite{wilson2009user}, and scaling-down sampling \cite{leskovec2006sampling}, have been proposed and have been shown to be able to keep the network small in scale while capturing specific properties of the original network.

Unfortunately, all current studies in which graph sampling is used consider mainly the topological properties of networks and ignore their nonlinear dynamics. A comprehensive analytical framework is needed to predict the original system's true dynamic behavior using an incomplete network. This framework must include an unbiased graph sampling method that preserves in the sampled nodes the dynamics experienced by the nodes in the original network. This could facilitate other related studies on large-scale dynamical networks, such as controllability analysis \cite{liu2016control, liu2016controllability} and studies of coevolution spreading dynamics \cite{wang2019coevolution}. Based on such a framework, future studies on how to predict the dynamics and resilience of incomplete networks of networks could be another essential branch of network resilience.

\subsubsection{Controlling network resilience }
A reflection of our ultimate understanding of a complex system is our ability to control its behavior \cite{liu2016control}. This requires an accurate network topology model, an understanding of the system's dynamics, and powerful tools for analyzing the network's resilience. Early related studies on controlling complex networks have focused on the controllability of linear and nonlinear systems \cite{liu2011controllability, whalen2015observability} or on controlling collective behavior in complex networks \cite{delellis2010fully}. The latter studies ignored the strength of interactions among dynamical components, making them unsuitable for controlling network resilience. As discussed in Chapter. \ref{Ecology}, Jiang et al. \cite{jiang2019harnessing} investigated how to manage or control tipping points in real-world complex and nonlinear dynamical networks in ecology by altering the way in which species extinction occurs, changing it from massive extinction of all species to the gradual extinction of individual species. In this way, the occurrence of global extinction is substantially delayed. The goal, which is to avoid reaching the tipping point, is realized by merely fixing the abundance of some species.

As we discussed, many relevant systems exhibit a hysteresis phenomenon in which they remain in the dysfunctional state despite having reconstructed their damaged topology. To address this challenge, Sanhedrai et al. \cite{sanhedrai2020reviving} develop a two-step recovery scheme. The first step in this scheme is topological reconstruction of the system to a point at which it can be revived; in the second step, dynamic interventions designed to restore the system's lost functionality are made. By applying this scheme to a range of nonlinear network dynamics, they identify a complex system's recoverable phase, a state in which the system can be restored by a microscopic intervention, i.e., by controlling only a single node. Recently, Ma et al.\cite{ma2020universality} developed a new theory that explains noise-induced resilience restoration through a combination of nucleation theory and a dimension reduction approach\cite{gao2016universal}. By applying the new theory to four types of lattice-based ecosystems, two cluster modes that exhibit different transition patterns are distinguished based on system size and noise strength.

Controlling network resilience is extremely difficult but deserves our effort because it is critical for infrastructural networks. There is a practical and immediate need to control the dynamics of information flow and those of social and organizational networks and an ongoing need to understand the control principles of adversary networks. We could move towards such a goal by developing a theory for quantifying the safe operating space \cite{scheffer2015creating} and proposing strategies that can be used to enhance network resilience and recoverability in complex networks and multilayer networks. Strategies for accomplishing this could include intervention in node state, addition/diminution of links, rewiring, and weight changes. Resilience control and optimization represent a new direction in network science and offer new avenues for controlling the dynamic behavior of real-world systems.

%\end{comment}

{\bf Acknowledgements.} We acknowledge the support of the National Natural Science Foundation of China
(Grants no. 62172170, 71822101, 71890973/71890970, 71771009, and 72071011), the US National Science Foundation under Grant No. 2047488, and the Rensselaer-IBM AI Research Collaboration. B.K.S. was partially supported by DARPA under contracts W911NF-17-C-0099 and HR001121C0165.

\bibliographystyle{unsrt}
\bibliography{Ref_Resilience_PR.bib}

\begin{thebibliography}{100}

\bibitem{venegas2005self}
Jose~G Venegas, Tilo Winkler, Guido Musch, Marcos F~Vidal Melo, Dominick
  Layfield, Nora Tgavalekos, Alan~J Fischman, Ronald~J Callahan, Giacomo
  Bellani, and R~Scott Harris.
\newblock Self-organized patchiness in asthma as a prelude to catastrophic
  shifts.
\newblock {\em Nature}, 434(7034):777, 2005.

\bibitem{perrings1998resilience}
Charles Perrings.
\newblock Resilience in the dynamics of economy-environment systems.
\newblock {\em Environmental and Resource Economics}, 11(3-4):503--520, 1998.

\bibitem{may1977thresholds}
Robert~M May.
\newblock Thresholds and breakpoints in ecosystems with a multiplicity of
  stable states.
\newblock {\em Nature}, 269(5628):471, 1977.

\bibitem{lyapunov1992general}
Aleksandr~Mikhailovich Lyapunov.
\newblock The general problem of the stability of motion.
\newblock {\em International journal of control}, 55(3):531--534, 1992.

\bibitem{gao2016universal}
Jianxi Gao, Baruch Barzel, and Albert-L{\'a}szl{\'o} Barab{\'a}si.
\newblock Universal resilience patterns in complex networks.
\newblock {\em Nature}, 530(7590):307, 2016.

\bibitem{chang2011systematic}
Rui Chang, Robert Shoemaker, and Wei Wang.
\newblock Systematic search for recipes to generate induced pluripotent stem
  cells.
\newblock {\em PLoS computational biology}, 7(12):e1002300, 2011.

\bibitem{devi2020locust}
Sharmila Devi.
\newblock Locust swarms in east africa could be “a catastrophe”.
\newblock {\em The Lancet}, 395(10224):547, 2020.

\bibitem{Resilienceshift}
\url{https://www.resilienceshift.org/bushfires-resilience/}.

\bibitem{chen2012smart}
Pin-Yu Chen, Shin-Ming Cheng, and Kwang-Cheng Chen.
\newblock Smart attacks in smart grid communication networks.
\newblock {\em IEEE Communications Magazine}, 50(8):24--29, 2012.

\bibitem{isaac2010security}
Jes{\'u}s~T{\'e}llez Isaac, Sherali Zeadally, and Jos{\'e}~Sierra Camara.
\newblock Security attacks and solutions for vehicular ad hoc networks.
\newblock {\em IET communications}, 4(7):894--903, 2010.

\bibitem{mumby2007thresholds}
Peter~J Mumby, Alan Hastings, and Helen~J Edwards.
\newblock Thresholds and the resilience of caribbean coral reefs.
\newblock {\em Nature}, 450(7166):98, 2007.

\bibitem{van2012response}
EH~van Nes, M~Holmgren, M~Hirota, and M~Scheffer.
\newblock Response to comment on global resilience of tropical forest and
  savanna to critical transitions.
\newblock {\em Science}, 336(6081):541--541, 2012.

\bibitem{veraart2012recovery}
Annelies~J Veraart, Elisabeth~J Faassen, Vasilis Dakos, Egbert~H van Nes,
  Miquel L{\"u}rling, and Marten Scheffer.
\newblock Recovery rates reflect distance to a tipping point in a living
  system.
\newblock {\em Nature}, 481(7381):357, 2012.

\bibitem{dai2013slower}
Lei Dai, Kirill~S Korolev, and Jeff Gore.
\newblock Slower recovery in space before collapse of connected populations.
\newblock {\em Nature}, 496(7445):355, 2013.

\bibitem{centola2018experimental}
Damon Centola, Joshua Becker, Devon Brackbill, and Andrea Baronchelli.
\newblock Experimental evidence for tipping points in social convention.
\newblock {\em Science}, 360(6393):1116--1119, 2018.

\bibitem{prakash2012threshold}
B~Aditya Prakash, Deepayan Chakrabarti, Nicholas~C Valler, Michalis Faloutsos,
  and Christos Faloutsos.
\newblock Threshold conditions for arbitrary cascade models on arbitrary
  networks.
\newblock {\em Knowledge and information systems}, 33(3):549--575, 2012.

\bibitem{legido2020high}
Helena Legido-Quigley, Nima Asgari, Yik~Ying Teo, Gabriel~M Leung, Hitoshi
  Oshitani, Keiji Fukuda, Alex~R Cook, Li~Yang Hsu, Kenji Shibuya, and David
  Heymann.
\newblock Are high-performing health systems resilient against the covid-19
  epidemic?
\newblock {\em The Lancet}, 395(10227):848--850, 2020.

\bibitem{ivanov2020viability}
Dmitry Ivanov and Alexandre Dolgui.
\newblock Viability of intertwined supply networks: extending the supply chain
  resilience angles towards survivability. a position paper motivated by
  covid-19 outbreak.
\newblock {\em International Journal of Production Research},
  58(10):2904--2915, 2020.

\bibitem{helbing2013globally}
Dirk Helbing.
\newblock Globally networked risks and how to respond.
\newblock {\em Nature}, 497(7447):51--59, 2013.

\bibitem{rocha2018cascading}
Juan~C Rocha, Garry Peterson, {\"O}rjan Bodin, and Simon Levin.
\newblock Cascading regime shifts within and across scales.
\newblock {\em Science}, 362(6421):1379--1383, 2018.

\bibitem{steffen2011australian}
Will Steffen, John Sims, James Walcott, and Greg Laughlin.
\newblock Australian agriculture: coping with dangerous climate change.
\newblock {\em Regional Environmental Change}, 11(1):205--214, 2011.

\bibitem{rohr2014structural}
Rudolf~P Rohr, Serguei Saavedra, and Jordi Bascompte.
\newblock On the structural stability of mutualistic systems.
\newblock {\em Science}, 345(6195):1253497, 2014.

\bibitem{ives2007stability}
Anthony~R Ives and Stephen~R Carpenter.
\newblock Stability and diversity of ecosystems.
\newblock {\em science}, 317(5834):58--62, 2007.

\bibitem{allesina2012stability}
Stefano Allesina and Si~Tang.
\newblock Stability criteria for complex ecosystems.
\newblock {\em Nature}, 483(7388):205, 2012.

\bibitem{motiian2017few}
Saeid Motiian, Quinn Jones, Seyed Iranmanesh, and Gianfranco Doretto.
\newblock Few-shot adversarial domain adaptation.
\newblock In {\em Advances in Neural Information Processing Systems}, pages
  6670--6680, 2017.

\bibitem{chen2015net2net}
Tianqi Chen, Ian Goodfellow, and Jonathon Shlens.
\newblock Net2net: Accelerating learning via knowledge transfer.
\newblock {\em arXiv preprint arXiv:1511.05641}, 2015.

\bibitem{dong2019robust}
Shangjia Dong, Haizhong Wang, Ali Mostafavi, and Jianxi Gao.
\newblock Robust component: a robustness measure that incorporates access to
  critical facilities under disruptions.
\newblock {\em Journal of the Royal Society Interface}, 16(157):20190149, 2019.

\bibitem{holling2002resilience}
Crawford~S Holling and Lance~H Gunderson.
\newblock Resilience and adaptive cycles.
\newblock {\em In: Panarchy: Understanding Transformations in Human and Natural
  Systems, 25-62}, 2002.

\bibitem{AustraliaBushfires}
\url{https://interestingengineering.com/6-images-from-the-australia-bushfires-that-.show-the-resilience-of-nature}.

\bibitem{bhamra2011resilience}
Ran Bhamra, Samir Dani, and Kevin Burnard.
\newblock Resilience: the concept, a literature review and future directions.
\newblock {\em International Journal of Production Research},
  49(18):5375--5393, 2011.

\bibitem{fisher2015more}
Len Fisher.
\newblock More than 70 ways to show resilience.
\newblock {\em Nature}, 518(7537):35--35, 2015.

\bibitem{holling1973resilience}
Crawford~S Holling.
\newblock Resilience and stability of ecological systems.
\newblock {\em Annual review of ecology and systematics}, 4(1):1--23, 1973.

\bibitem{haimes2009complex}
Yacov~Y Haimes.
\newblock On the complex definition of risk: A systems-based approach.
\newblock {\em Risk analysis}, 29(12):1647--1654, 2009.

\bibitem{bruneau2003framework}
Michel Bruneau, Stephanie~E Chang, Ronald~T Eguchi, George~C Lee, Thomas~D
  O'Rourke, Andrei~M Reinhorn, Masanobu Shinozuka, Kathleen Tierney, William~A
  Wallace, and Detlof Von~Winterfeldt.
\newblock A framework to quantitatively assess and enhance the seismic
  resilience of communities.
\newblock {\em Earthquake spectra}, 19(4):733--752, 2003.

\bibitem{ahn2015reflections}
Joonhong Ahn, Cathryn Carson, Mikael Jensen, Kohta Juraku, Shinya Nagasaki, and
  Satoru Tanaka.
\newblock {\em Reflections on the Fukushima Daiichi nuclear accident}.
\newblock Springer, 2015.

\bibitem{hirota2011global}
Marina Hirota, Milena Holmgren, Egbert~H Van~Nes, and Marten Scheffer.
\newblock Global resilience of tropical forest and savanna to critical
  transitions.
\newblock {\em Science}, 334(6053):232--235, 2011.

\bibitem{dai2012generic}
Lei Dai, Daan Vorselen, Kirill~S Korolev, and Jeff Gore.
\newblock Generic indicators for loss of resilience before a tipping point
  leading to population collapse.
\newblock {\em Science}, 336(6085):1175--1177, 2012.

\bibitem{holling1996engineering}
Crawford~Stanley Holling.
\newblock Engineering resilience versus ecological resilience.
\newblock {\em Engineering within ecological constraints}, 31(1996):32, 1996.

\bibitem{buldyrev2010catastrophic}
Sergey~V Buldyrev, Roni Parshani, Gerald Paul, H~Eugene Stanley, and Shlomo
  Havlin.
\newblock Catastrophic cascade of failures in interdependent networks.
\newblock {\em Nature}, 464(7291):1025, 2010.

\bibitem{scheffer2001catastrophic}
Marten Scheffer, Steve Carpenter, Jonathan~A Foley, Carl Folke, and Brian
  Walker.
\newblock Catastrophic shifts in ecosystems.
\newblock {\em Nature}, 413(6856):591, 2001.

\bibitem{scheffer2009early}
Marten Scheffer, Jordi Bascompte, William~A Brock, Victor Brovkin, Stephen~R
  Carpenter, Vasilis Dakos, Hermann Held, Egbert~H Van~Nes, Max Rietkerk, and
  George Sugihara.
\newblock Early-warning signals for critical transitions.
\newblock {\em Nature}, 461(7260):53, 2009.

\bibitem{barabasi1999emergence}
Albert-L{\'a}szl{\'o} Barab{\'a}si and R{\'e}ka Albert.
\newblock Emergence of scaling in random networks.
\newblock {\em science}, 286(5439):509--512, 1999.

\bibitem{cohen2010complex}
Reuven Cohen and Shlomo Havlin.
\newblock {\em Complex networks: structure, robustness and function}.
\newblock Cambridge university press, 2010.

\bibitem{zhou2019bayesian}
Wei Zhou, Omid Ardakanian, Hai-Tao Zhang, and Ye~Yuan.
\newblock Bayesian learning-based harmonic state estimation in distribution
  systems with smart meter and dpmu data.
\newblock {\em IEEE Transactions on Smart Grid}, 11(1):832--845, 2019.

\bibitem{teichmann2004gene}
Sarah~A Teichmann and M~Madan Babu.
\newblock Gene regulatory network growth by duplication.
\newblock {\em Nature genetics}, 36(5):492--496, 2004.

\bibitem{liu2014detection}
Xueming Liu and Linqiang Pan.
\newblock Detection of driver metabolites in the human liver metabolic network
  using structural controllability analysis.
\newblock {\em BMC systems biology}, 8(1):51, 2014.

\bibitem{liu2016control}
Yang-Yu Liu and Albert-L{\'a}szl{\'o} Barab{\'a}si.
\newblock Control principles of complex systems.
\newblock {\em Reviews of Modern Physics}, 88(3):035006, 2016.

\bibitem{wang2016data}
Wen-Xu Wang, Ying-Cheng Lai, and Celso Grebogi.
\newblock Data based identification and prediction of nonlinear and complex
  dynamical systems.
\newblock {\em Physics Reports}, 644:1--76, 2016.

\bibitem{barzel2015constructing}
Baruch Barzel, Yang-Yu Liu, and Albert-L{\'a}szl{\'o} Barab{\'a}si.
\newblock Constructing minimal models for complex system dynamics.
\newblock {\em Nature communications}, 6:7186, 2015.

\bibitem{korda2018linear}
Milan Korda and Igor Mezi{\'c}.
\newblock Linear predictors for nonlinear dynamical systems: Koopman operator
  meets model predictive control.
\newblock {\em Automatica}, 93:149--160, 2018.

\bibitem{budivsic2012applied}
Marko Budi{\v{s}}i{\'c}, Ryan Mohr, and Igor Mezi{\'c}.
\newblock Applied koopmanism.
\newblock {\em Chaos: An Interdisciplinary Journal of Nonlinear Science},
  22(4):047510, 2012.

\bibitem{williams2015data}
Matthew~O Williams, Ioannis~G Kevrekidis, and Clarence~W Rowley.
\newblock A data--driven approximation of the koopman operator: Extending
  dynamic mode decomposition.
\newblock {\em Journal of Nonlinear Science}, 25(6):1307--1346, 2015.

\bibitem{koopman1931hamiltonian}
Bernard~O Koopman.
\newblock Hamiltonian systems and transformation in hilbert space.
\newblock {\em Proceedings of the National Academy of Sciences},
  17(5):315--318, 1931.

\bibitem{kohda1994explicit}
Tohru KOHDA and Akio Tsuneda.
\newblock Explicit evaluations of correlation functions of chebyshev binary and
  bit sequences based on perron--frobenius operator.
\newblock {\em IEICE Transactions on Fundamentals of Electronics,
  Communications and Computer Sciences}, 77(11):1794--1800, 1994.

\bibitem{mauroy2013isostables}
Alexandre Mauroy, Igor Mezi{\'c}, and Jeff Moehlis.
\newblock Isostables, isochrons, and koopman spectrum for the action--angle
  representation of stable fixed point dynamics.
\newblock {\em Physica D: Nonlinear Phenomena}, 261:19--30, 2013.

\bibitem{tu2017collapse}
Chengyi Tu, Jacopo Grilli, Friedrich Schuessler, and Samir Suweis.
\newblock Collapse of resilience patterns in generalized lotka-volterra
  dynamics and beyond.
\newblock {\em Physical Review E}, 95(6):062307, 2017.

\bibitem{jiang2018predicting}
Junjie Jiang, Zi-Gang Huang, Thomas~P Seager, Wei Lin, Celso Grebogi, Alan
  Hastings, and Ying-Cheng Lai.
\newblock Predicting tipping points in mutualistic networks through dimension
  reduction.
\newblock {\em Proceedings of the National Academy of Sciences}, page
  201714958, 2018.

\bibitem{laurence2019spectral}
Edward Laurence, Nicolas Doyon, Louis~J Dub{\'e}, and Patrick Desrosiers.
\newblock Spectral dimension reduction of complex dynamical networks.
\newblock {\em Physical Review X}, 9(1):011042, 2019.

\bibitem{gao2017complex}
Zhong-Ke Gao, Michael Small, and Juergen Kurths.
\newblock Complex network analysis of time series.
\newblock {\em EPL (Europhysics Letters)}, 116(5):50001, 2017.

\bibitem{beisner2003alternative}
Beatrix~E Beisner, Daniel~T Haydon, and Kim Cuddington.
\newblock Alternative stable states in ecology.
\newblock {\em Frontiers in Ecology and the Environment}, 1(7):376--382, 2003.

\bibitem{ghaffarizadeh2014multistable}
Ahmadreza Ghaffarizadeh, Nicholas~S Flann, and Gregory~J Podgorski.
\newblock Multistable switches and their role in cellular differentiation
  networks.
\newblock {\em BMC bioinformatics}, 15(7):S7, 2014.

\bibitem{bauch2016early}
Chris~T Bauch, Ram Sigdel, Joe Pharaon, and Madhur Anand.
\newblock Early warning signals of regime shifts in coupled human--environment
  systems.
\newblock {\em Proceedings of the National Academy of Sciences},
  113(51):14560--14567, 2016.

\bibitem{fraccascia2018resilience}
Luca Fraccascia, Ilaria Giannoccaro, and Vito Albino.
\newblock Resilience of complex systems: state of the art and directions for
  future research.
\newblock {\em Complexity}, 2018, 2018.

\bibitem{fletcher2013psychological}
David Fletcher and Mustafa Sarkar.
\newblock Psychological resilience: A review and critique of definitions,
  concepts, and theory.
\newblock {\em European psychologist}, 18(1):12, 2013.

\bibitem{kupers2014turbulence}
Roland Kupers.
\newblock {\em Turbulence: a corporate perspective on collaborating for
  resilience}.
\newblock Amsterdam University Press, 2014.

\bibitem{yang2019hybrid}
Fut~Kuo Yang, Aleksander Cholewinski, Li~Yu, Geoffrey Rivers, and Boxin Zhao.
\newblock A hybrid material that reversibly switches between two stable solid
  states.
\newblock {\em Nature Materials}, 18(8):874, 2019.

\bibitem{capano2017resilience}
Giliberto Capano and Jun~Jie Woo.
\newblock Resilience and robustness in policy design: A critical appraisal.
\newblock {\em Policy Sciences}, 50(3):399--426, 2017.

\bibitem{winson193227}
CG~Winson.
\newblock Report on a method for measuring the resilience of wool.
\newblock {\em Journal of the Textile Institute Transactions},
  23(12):T386--T393, 1932.

\bibitem{murphy1974coping}
Lois~B Murphy et~al.
\newblock Coping, vulnerability, and resilience in childhood.
\newblock {\em Coping and adaptation}, pages 69--100, 1974.

\bibitem{urruty2016stability}
Nicolas Urruty, Delphine Tailliez-Lefebvre, and Christian Huyghe.
\newblock Stability, robustness, vulnerability and resilience of agricultural
  systems. a review.
\newblock {\em Agronomy for sustainable development}, 36(1):15, 2016.

\bibitem{meredith2018applying}
Hannah~R Meredith, Virgile Andreani, Helena~R Ma, Allison~J Lopatkin, Anna~J
  Lee, Deverick~J Anderson, Gregory Batt, and Lingchong You.
\newblock Applying ecological resistance and resilience to dissect bacterial
  antibiotic responses.
\newblock {\em Science advances}, 4(12):eaau1873, 2018.

\bibitem{parsons2017social}
Talcott Parsons, Edward~A Shils, and Neil~J Smelser.
\newblock The social system.
\newblock In {\em Toward a general theory of action}, pages 190--233.
  Routledge, 2017.

\bibitem{hollnagel2006resilience}
Erik Hollnagel, David~D Woods, and Nancy Leveson.
\newblock {\em Resilience engineering: Concepts and precepts}.
\newblock Ashgate Publishing, Ltd., 2006.

\bibitem{holling2004complex}
Crawford~S Holling.
\newblock From complex regions to complex worlds.
\newblock {\em Ecology and society}, 9(1), 2004.

\bibitem{standish2014resilience}
Rachel~J Standish, Richard~J Hobbs, Margaret~M Mayfield, Brandon~T Bestelmeyer,
  Katherine~N Suding, Loretta~L Battaglia, Valerie Eviner, Christine~V Hawkes,
  Vicky~M Temperton, Viki~A Cramer, et~al.
\newblock Resilience in ecology: Abstraction, distraction, or where the action
  is?
\newblock {\em Biological Conservation}, 177:43--51, 2014.

\bibitem{nelson2004oscillations}
DE~Nelson, AEC Ihekwaba, M~Elliott, JR~Johnson, CA~Gibney, BE~Foreman,
  G~Nelson, V~See, CA~Horton, DG~Spiller, et~al.
\newblock Oscillations in nf-$\kappa$b signaling control the dynamics of gene
  expression.
\newblock {\em Science}, 306(5696):704--708, 2004.

\bibitem{volkov2003neutral}
Igor Volkov, Jayanth~R Banavar, Stephen~P Hubbell, and Amos Maritan.
\newblock Neutral theory and relative species abundance in ecology.
\newblock {\em Nature}, 424(6952):1035, 2003.

\bibitem{kitsak2010identification}
Maksim Kitsak, Lazaros~K Gallos, Shlomo Havlin, Fredrik Liljeros, Lev Muchnik,
  H~Eugene Stanley, and Hern{\'a}n~A Makse.
\newblock Identification of influential spreaders in complex networks.
\newblock {\em Nature physics}, 6(11):888, 2010.

\bibitem{newman2003structure}
Mark~EJ Newman.
\newblock The structure and function of complex networks.
\newblock {\em SIAM review}, 45(2):167--256, 2003.

\bibitem{pan2020phase}
Liming Pan, Dan Yang, Wei Wang, Shimin Cai, Tao Zhou, and Ying-Cheng Lai.
\newblock Phase diagrams of interacting spreading dynamics in complex networks.
\newblock {\em Physical Review Research}, 2(2):023233, 2020.

\bibitem{ma2018randomly}
Huanfei Ma, Siyang Leng, Kazuyuki Aihara, Wei Lin, and Luonan Chen.
\newblock Randomly distributed embedding making short-term high-dimensional
  data predictable.
\newblock {\em Proceedings of the National Academy of Sciences},
  115(43):E9994--E10002, 2018.

\bibitem{tanaka2015dynamical}
Gouhei Tanaka, Kai Morino, and Kazuyuki Aihara.
\newblock Dynamical robustness of complex biological networks.
\newblock In {\em Mathematical Approaches to Biological Systems}, pages 29--53.
  Springer, 2015.

\bibitem{alon2006introduction}
Uri Alon.
\newblock {\em An introduction to systems biology: design principles of
  biological circuits}.
\newblock CRC press, 2006.

\bibitem{boguna2013nature}
Marian Bogun{\'a}, Claudio Castellano, and Romualdo Pastor-Satorras.
\newblock Nature of the epidemic threshold for the
  susceptible-infected-susceptible dynamics in networks.
\newblock {\em Physical review letters}, 111(6):068701, 2013.

\bibitem{barzel2013universality}
Baruch Barzel and Albert-L{\'a}szl{\'o} Barab{\'a}si.
\newblock Universality in network dynamics.
\newblock {\em Nature physics}, 9(10):673, 2013.

\bibitem{duan2019universal}
Dongli Duan, Changchun Lv, Shubin Si, Zhen Wang, Daqing Li, Jianxi Gao, Shlomo
  Havlin, H~Eugene Stanley, and Stefano Boccaletti.
\newblock Universal behavior of cascading failures in interdependent networks.
\newblock {\em Proceedings of the National Academy of Sciences},
  116(45):22452--22457, 2019.

\bibitem{takens1981detecting}
Floris Takens.
\newblock Detecting strange attractors in turbulence.
\newblock In {\em Dynamical systems and turbulence, Warwick 1980}, pages
  366--381. Springer, 1981.

\bibitem{wang2011predicting}
Wen-Xu Wang, Rui Yang, Ying-Cheng Lai, Vassilios Kovanis, and Celso Grebogi.
\newblock Predicting catastrophes in nonlinear dynamical systems by compressive
  sensing.
\newblock {\em Physical Review Letters}, 106(15):154101, 2011.

\bibitem{tilman2001forecasting}
David Tilman, Joseph Fargione, Brian Wolff, Carla D'antonio, Andrew Dobson,
  Robert Howarth, David Schindler, William~H Schlesinger, Daniel Simberloff,
  and Deborah Swackhamer.
\newblock Forecasting agriculturally driven global environmental change.
\newblock {\em science}, 292(5515):281--284, 2001.

\bibitem{mumby2013evidence}
Peter~J Mumby, Robert~S Steneck, and Alan Hastings.
\newblock Evidence for and against the existence of alternate attractors on
  coral reefs.
\newblock {\em Oikos}, 122(4):481--491, 2013.

\bibitem{moffett2015multiple}
Kevan Moffett, William Nardin, Sonia Silvestri, Chen Wang, and Stijn Temmerman.
\newblock Multiple stable states and catastrophic shifts in coastal wetlands:
  Progress, challenges, and opportunities in validating theory using remote
  sensing and other methods.
\newblock {\em Remote Sensing}, 7(8):10184--10226, 2015.

\bibitem{dudgeon2010phase}
Steve~R Dudgeon, Richard~B Aronson, John~F Bruno, and William~F Precht.
\newblock Phase shifts and stable states on coral reefs.
\newblock {\em Marine Ecology Progress Series}, 413:201--216, 2010.

\bibitem{scheffer2003catastrophic}
Marten Scheffer and Stephen~R Carpenter.
\newblock Catastrophic regime shifts in ecosystems: linking theory to
  observation.
\newblock {\em Trends in ecology \& evolution}, 18(12):648--656, 2003.

\bibitem{folke2004regime}
Carl Folke, Steve Carpenter, Brian Walker, Marten Scheffer, Thomas Elmqvist,
  Lance Gunderson, and Crawford~Stanley Holling.
\newblock Regime shifts, resilience, and biodiversity in ecosystem management.
\newblock {\em Annu. Rev. Ecol. Evol. Syst.}, 35:557--581, 2004.

\bibitem{crandall1971bifurcation}
Michael~G Crandall and Paul~H Rabinowitz.
\newblock Bifurcation from simple eigenvalues.
\newblock {\em Journal of Functional Analysis}, 8(2):321--340, 1971.

\bibitem{zeeman1979catastrophe}
Erik~Christopher Zeeman.
\newblock Catastrophe theory.
\newblock In {\em Structural Stability in Physics}, pages 12--22. Springer,
  1979.

\bibitem{carpenter2001alternate}
Stephen~R Carpenter.
\newblock Alternate states of ecosystems: evidence and some implications.
\newblock {\em Ecology: achievement and challenge}, pages 357--383, 2001.

\bibitem{broock1996test}
William~A Broock, Jos{\'e}~Alexandre Scheinkman, W~Davis Dechert, and Blake
  LeBaron.
\newblock A test for independence based on the correlation dimension.
\newblock {\em Econometric reviews}, 15(3):197--235, 1996.

\bibitem{carpenter1997dystrophy}
Stephen~R Carpenter and Michael~L Pace.
\newblock Dystrophy and eutrophy in lake ecosystems: implications of
  fluctuating inputs.
\newblock {\em Oikos}, pages 3--14, 1997.

\bibitem{sutherland1974multiple}
John~P Sutherland.
\newblock Multiple stable points in natural communities.
\newblock {\em The American Naturalist}, 108(964):859--873, 1974.

\bibitem{holmgren2001nino}
Milena Holmgren, Marten Scheffer, Exequiel Ezcurra, Julio~R Guti{\'e}rrez, and
  Godefridus~MJ Mohren.
\newblock El ni{\~n}o effects on the dynamics of terrestrial ecosystems.
\newblock {\em Trends in Ecology \& Evolution}, 16(2):89--94, 2001.

\bibitem{scheffer2007regime}
Marten Scheffer and Erik Jeppesen.
\newblock Regime shifts in shallow lakes.
\newblock {\em Ecosystems}, 10(1):1--3, 2007.

\bibitem{wissel1984universal}
C~Wissel.
\newblock A universal law of the characteristic return time near thresholds.
\newblock {\em Oecologia}, 65(1):101--107, 1984.

\bibitem{van2007slow}
Egbert~H Van~Nes and Marten Scheffer.
\newblock Slow recovery from perturbations as a generic indicator of a nearby
  catastrophic shift.
\newblock {\em The American Naturalist}, 169(6):738--747, 2007.

\bibitem{ovaskainen2002transient}
Otso Ovaskainen and Ilkka Hanski.
\newblock Transient dynamics in metapopulation response to perturbation.
\newblock {\em Theoretical population biology}, 61(3):285--295, 2002.

\bibitem{ives1995measuring}
Anthony~R Ives.
\newblock Measuring resilience in stochastic systems.
\newblock {\em Ecological Monographs}, 65(2):217--233, 1995.

\bibitem{dakos2008slowing}
Vasilis Dakos, Marten Scheffer, Egbert~H van Nes, Victor Brovkin, Vladimir
  Petoukhov, and Hermann Held.
\newblock Slowing down as an early warning signal for abrupt climate change.
\newblock {\em Proceedings of the National Academy of Sciences},
  105(38):14308--14312, 2008.

\bibitem{lenton2008using}
Timothy~M Lenton, Richard~J Myerscough, Robert Marsh, Valerie~N Livina,
  Andrew~R Price, Simon~J Cox, and GENIE team.
\newblock Using genie to study a tipping point in the climate system.
\newblock {\em Philosophical Transactions of the Royal Society A: Mathematical,
  Physical and Engineering Sciences}, 367(1890):871--884, 2008.

\bibitem{berglund2002metastability}
Nils Berglund and Barbara Gentz.
\newblock Metastability in simple climate models: pathwise analysis of slowly
  driven langevin equations.
\newblock {\em Stochastics and Dynamics}, 2(03):327--356, 2002.

\bibitem{scheffer1993alternative}
Marten Scheffer, SH~Hosper, ML~Meijer, Brian Moss, and Erik Jeppesen.
\newblock Alternative equilibria in shallow lakes.
\newblock {\em Trends in ecology \& evolution}, 8(8):275--279, 1993.

\bibitem{xiong2003positive}
Wen Xiong and James~E Ferrell~Jr.
\newblock A positive-feedback-based bistable `memory module' that governs a
  cell fate decision.
\newblock {\em Nature}, 426(6965):460, 2003.

\bibitem{jiang2020true}
Chunheng Jiang, Jianxi Gao, and Malik Magdon-Ismail.
\newblock True nonlinear dynamics from incomplete networks.
\newblock In {\em Proceedings of the AAAI Conference on Artificial
  Intelligence}, volume~34, pages 131--138, 2020.

\bibitem{jiang2020inferring}
Chunheng Jiang, Jianxi Gao, and Malik Magdon-Ismail.
\newblock Inferring degrees from incomplete networks and nonlinear dynamics.
\newblock In Christian Bessiere, editor, {\em Proceedings of the Twenty-Ninth
  International Joint Conference on Artificial Intelligence, {IJCAI-20}}, pages
  3307--3313. International Joint Conferences on Artificial Intelligence
  Organization, 7 2020.
\newblock Main track.

\bibitem{zhang2020resilience}
Yongtao Zhang, Cunqi Shao, Shibo He, and Jianxi Gao.
\newblock Resilience centrality in complex networks.
\newblock {\em Physical Review E}, 101(2):022304, 2020.

\bibitem{dai2015relation}
Lei Dai, Kirill~S Korolev, and Jeff Gore.
\newblock Relation between stability and resilience determines the performance
  of early warning signals under different environmental drivers.
\newblock {\em Proceedings of the National Academy of Sciences},
  112(32):10056--10061, 2015.

\bibitem{shade2012fundamentals}
Ashley Shade, Hannes Peter, Steven~D Allison, Didier Baho, Merc{\`e} Berga,
  Helmut B{\"u}rgmann, David~H Huber, Silke Langenheder, Jay~T Lennon,
  Jennifer~BH Martiny, et~al.
\newblock Fundamentals of microbial community resistance and resilience.
\newblock {\em Frontiers in microbiology}, 3:417, 2012.

\bibitem{callaway2000network}
Duncan~S Callaway, Mark~EJ Newman, Steven~H Strogatz, and Duncan~J Watts.
\newblock Network robustness and fragility: Percolation on random graphs.
\newblock {\em Physical review letters}, 85(25):5468, 2000.

\bibitem{albert2000error}
R{\'e}ka Albert, Hawoong Jeong, and Albert-L{\'a}szl{\'o} Barab{\'a}si.
\newblock Error and attack tolerance of complex networks.
\newblock {\em nature}, 406(6794):378--382, 2000.

\bibitem{may1972limit}
Robert~M May.
\newblock Limit cycles in predator-prey communities.
\newblock {\em Science}, 177(4052):900--902, 1972.

\bibitem{may1972will}
Robert~M May.
\newblock Will a large complex system be stable?
\newblock {\em Nature}, 238(5364):413--414, 1972.

\bibitem{may1971stability}
Robert~M May.
\newblock Stability in multispecies community models.
\newblock {\em Mathematical Biosciences}, 12(1-2):59--79, 1971.

\bibitem{turner1993revised}
Monica~G Turner, William~H Romme, Robert~H Gardner, Robert~V O'Neill, and
  Timothy~K Kratz.
\newblock A revised concept of landscape equilibrium: disturbance and stability
  on scaled landscapes.
\newblock {\em Landscape Ecology}, 8(3):213--227, 1993.

\bibitem{donohue2016navigating}
Ian Donohue, Helmut Hillebrand, Jos{\'e}~M Montoya, Owen~L Petchey, Stuart~L
  Pimm, Mike~S Fowler, Kevin Healy, Andrew~L Jackson, Miguel Lurgi, Deirdre
  McClean, et~al.
\newblock Navigating the complexity of ecological stability.
\newblock {\em Ecology Letters}, 19(9):1172--1185, 2016.

\bibitem{watt1968computer}
Kenneth~EF Watt.
\newblock A computer approach to analysis of data on weather, population
  fluctuations, and disease.
\newblock {\em Biometerology. Oregon St. Univ. Press, Corvallis}, 1968.

\bibitem{butler2018stability}
Stacey Butler and James~P O’Dwyer.
\newblock Stability criteria for complex microbial communities.
\newblock {\em Nature communications}, 9(1):2970, 2018.

\bibitem{coyte2015ecology}
Katharine~Z Coyte, Jonas Schluter, and Kevin~R Foster.
\newblock The ecology of the microbiome: networks, competition, and stability.
\newblock {\em Science}, 350(6261):663--666, 2015.

\bibitem{kitano2004biological}
Hiroaki Kitano.
\newblock Biological robustness.
\newblock {\em Nature Reviews Genetics}, 5(11):826--837, 2004.

\bibitem{liu2015vulnerability}
Xueming Liu, Hao Peng, and Jianxi Gao.
\newblock Vulnerability and controllability of networks of networks.
\newblock {\em Chaos, Solitons \& Fractals}, 80:125--138, 2015.

\bibitem{gao2015recent}
Jianxi Gao, Xueming Liu, Daqing Li, and Shlomo Havlin.
\newblock Recent progress on the resilience of complex networks.
\newblock {\em Energies}, 8(10):12187--12210, 2015.

\bibitem{barabasi2016network}
Albert-L{\'a}szl{\'o} Barab{\'a}si et~al.
\newblock {\em Network science}.
\newblock Cambridge university press, 2016.

\bibitem{liu2017controllability}
Xueming Liu, Linqiang Pan, H~Eugene Stanley, and Jianxi Gao.
\newblock Controllability of giant connected components in a directed network.
\newblock {\em Physical Review E}, 95(4):042318, 2017.

\bibitem{cohen2000resilience}
Reuven Cohen, Keren Erez, Daniel Ben-Avraham, and Shlomo Havlin.
\newblock Resilience of the internet to random breakdowns.
\newblock {\em Physical review letters}, 85(21):4626, 2000.

\bibitem{hu2011percolation}
Yanqing Hu, Baruch Ksherim, Reuven Cohen, and Shlomo Havlin.
\newblock Percolation in interdependent and interconnected networks: Abrupt
  change from second-to first-order transitions.
\newblock {\em Physical Review E}, 84(6):066116, 2011.

\bibitem{radicchi2013abrupt}
Filippo Radicchi and Alex Arenas.
\newblock Abrupt transition in the structural formation of interconnected
  networks.
\newblock {\em Nature Physics}, 9(11):717--720, 2013.

\bibitem{baxter2012avalanche}
GJ~Baxter, SN~Dorogovtsev, AV~Goltsev, and JFF Mendes.
\newblock Avalanche collapse of interdependent networks.
\newblock {\em Physical Review Letters}, 109(24):248701, 2012.

\bibitem{gao2012networks}
Jianxi Gao, Sergey~V Buldyrev, H~Eugene Stanley, and Shlomo Havlin.
\newblock Networks formed from interdependent networks.
\newblock {\em Nature Physics}, 8(1):40, 2012.

\bibitem{parshani2010interdependent}
Roni Parshani, Sergey~V Buldyrev, and Shlomo Havlin.
\newblock Interdependent networks: Reducing the coupling strength leads to a
  change from a first to second order percolation transition.
\newblock {\em Physical review letters}, 105(4):048701, 2010.

\bibitem{gao2011robustness}
Jianxi Gao, Sergey~V Buldyrev, Shlomo Havlin, and H~Eugene Stanley.
\newblock Robustness of a network of networks.
\newblock {\em Physical Review Letters}, 107(19):195701, 2011.

\bibitem{liu2016breakdown}
Xueming Liu, H~Eugene Stanley, and Jianxi Gao.
\newblock Breakdown of interdependent directed networks.
\newblock {\em Proceedings of the National Academy of Sciences},
  113(5):1138--1143, 2016.

\bibitem{huang2011robustness}
Xuqing Huang, Jianxi Gao, Sergey~V Buldyrev, Shlomo Havlin, and H~Eugene
  Stanley.
\newblock Robustness of interdependent networks under targeted attack.
\newblock {\em Physical Review E}, 83(6):065101, 2011.

\bibitem{xu2021breakdown}
W~Xu, L~Pan, and X~Liu.
\newblock Breakdown in interdependent directed networks under targeted attacks.
\newblock {\em EPL (Europhysics Letters)}, 133(6):68004, 2021.

\bibitem{dong2012percolation}
Gaogao Dong, Jianxi Gao, Lixin Tian, Ruijin Du, and Yinghuan He.
\newblock Percolation of partially interdependent networks under targeted
  attack.
\newblock {\em Physical Review E}, 85(1):016112, 2012.

\bibitem{dong2013robustness}
Gaogao Dong, Jianxi Gao, Ruijin Du, Lixin Tian, H~Eugene Stanley, and Shlomo
  Havlin.
\newblock Robustness of network of networks under targeted attack.
\newblock {\em Physical Review E}, 87(5):052804, 2013.

\bibitem{bi2014explosive}
Hongjie Bi, Xin Hu, Xiyun Zhang, Yong Zou, Zonghua Liu, and Shuguang Guan.
\newblock Explosive oscillation death in coupled stuart-landau oscillators.
\newblock {\em EPL (Europhysics Letters)}, 108(5):50003, 2014.

\bibitem{jeong2000large}
Hawoong Jeong, B{\'a}lint Tombor, R{\'e}ka Albert, Zoltan~N Oltvai, and A-L
  Barab{\'a}si.
\newblock The large-scale organization of metabolic networks.
\newblock {\em Nature}, 407(6804):651, 2000.

\bibitem{knowlton1992thresholds}
Nancy Knowlton.
\newblock Thresholds and multiple stable states in coral reef community
  dynamics.
\newblock {\em American Zoologist}, 32(6):674--682, 1992.

\bibitem{done1992phase}
Terry~J Done.
\newblock Phase shifts in coral reef communities and their ecological
  significance.
\newblock {\em Hydrobiologia}, 247(1-3):121--132, 1992.

\bibitem{petraitis2004detection}
Peter~S Petraitis and Steve~R Dudgeon.
\newblock Detection of alternative stable states in marine communities.
\newblock {\em Journal of Experimental Marine Biology and Ecology},
  300(1-2):343--371, 2004.

\bibitem{connell1983evidence}
Joseph~H Connell and Wayne~P Sousa.
\newblock On the evidence needed to judge ecological stability or persistence.
\newblock {\em The American Naturalist}, 121(6):789--824, 1983.

\bibitem{sinclair1995serengeti}
Anthony Ronald~Entrican Sinclair and Peter Arcese.
\newblock {\em Serengeti II: dynamics, management, and conservation of an
  ecosystem}, volume~2.
\newblock University of Chicago Press, 1995.

\bibitem{gunderson2000ecological}
Lance~H Gunderson.
\newblock Ecological resilience—in theory and application.
\newblock {\em Annual review of ecology and systematics}, 31(1):425--439, 2000.

\bibitem{laycock1991stable}
William~A Laycock.
\newblock Stable states and thresholds of range condition on north american
  rangelands: a viewpoint.
\newblock {\em Rangeland Ecology \& Management/Journal of Range Management
  Archives}, 44(5):427--433, 1991.

\bibitem{wilson1992positive}
J~Bastow Wilson and Andrew~DQ Agnew.
\newblock Positive-feedback switches in plant communities.
\newblock In {\em Advances in ecological research}, volume~23, pages 263--336.
  Elsevier, 1992.

\bibitem{lewontin1969meaning}
Richard~C Lewontin.
\newblock The meaning of stability.
\newblock In {\em Brookhaven symposia in biology}, volume~22, pages 13--24,
  1969.

\bibitem{law1993alternative}
Richard Law and R~Daniel Morton.
\newblock Alternative permanent states of ecological communities.
\newblock {\em Ecology}, 74(5):1347--1361, 1993.

\bibitem{jorgensen2014encyclopedia}
Sven~Erik Jorgensen and Brian Fath.
\newblock {\em Encyclopedia of ecology}.
\newblock Newnes, 2014.

\bibitem{blackwood2012effect}
Julie~C Blackwood, Alan Hastings, and Peter~J Mumby.
\newblock The effect of fishing on hysteresis in caribbean coral reefs.
\newblock {\em Theoretical ecology}, 5(1):105--114, 2012.

\bibitem{scheffer1990multiplicity}
Marten Scheffer.
\newblock Multiplicity of stable states in freshwater systems.
\newblock In {\em Biomanipulation tool for water management}, pages 475--486.
  Springer, 1990.

\bibitem{heck1971statistical}
Henry~d'A Heck.
\newblock Statistical theory of cooperative binding to proteins. hill equation
  and the binding potential.
\newblock {\em Journal of the American Chemical Society}, 93(1):23--29, 1971.

\bibitem{vasilakopoulos2015resilience}
Paraskevas Vasilakopoulos and C~Tara Marshall.
\newblock Resilience and tipping points of an exploited fish population over
  six decades.
\newblock {\em Global Change Biology}, 21(5):1834--1847, 2015.

\bibitem{scheffer1997ecology}
Marten Scheffer.
\newblock {\em Ecology of shallow lakes}, volume~22.
\newblock Springer Science \& Business Media, 1997.

\bibitem{janssen2014alternative}
Annette~BG Janssen, Sven Teurlincx, Shuqing An, Jan~H Janse, Hans~W Paerl, and
  Wolf~M Mooij.
\newblock Alternative stable states in large shallow lakes?
\newblock {\em Journal of Great Lakes Research}, 40(4):813--826, 2014.

\bibitem{janssen2015research}
Annette~BG Janssen, Sven Teurlincx, and Wolf~M Mooij.
\newblock Research summary: Alternative stable states in large shallow lakes.
\newblock {\em Lake Scientist}, 2015.

\bibitem{hughes1994catastrophes}
Terence~P Hughes.
\newblock Catastrophes, phase shifts, and large-scale degradation of a
  caribbean coral reef.
\newblock {\em Science}, 265(5178):1547--1551, 1994.

\bibitem{heffernan2008wetlands}
James~B Heffernan.
\newblock Wetlands as an alternative stable state in desert streams.
\newblock {\em Ecology}, 89(5):1261--1271, 2008.

\bibitem{CoralReefGlossary}
\url{http://oceantippingpoints.org/our-work/glossary}.

\bibitem{marani2013vegetation}
Marco Marani, Cristina Da~Lio, and Andrea D’Alpaos.
\newblock Vegetation engineers marsh morphology through multiple competing
  stable states.
\newblock {\em Proceedings of the National Academy of Sciences},
  110(9):3259--3263, 2013.

\bibitem{wang2013does}
Chen Wang and Stijn Temmerman.
\newblock Does biogeomorphic feedback lead to abrupt shifts between alternative
  landscape states: An empirical study on intertidal flats and marshes.
\newblock {\em Journal of Geophysical Research: Earth Surface},
  118(1):229--240, 2013.

\bibitem{wang2016biogeomorphic}
Chen Wang, Qiao Wang, Dieter Meire, Wandong Ma, Chuanqing Wu, Zhen Meng, Johan
  Van~de Koppel, Peter Troch, Ronny Verhoeven, Tom De~Mulder, et~al.
\newblock Biogeomorphic feedback between plant growth and flooding causes
  alternative stable states in an experimental floodplain.
\newblock {\em Advances in water resources}, 93:223--235, 2016.

\bibitem{cahoon2003mass}
Donald~R Cahoon, Philippe Hensel, John Rybczyk, Karen~L McKee, C~Edward
  Proffitt, and Brian~C Perez.
\newblock Mass tree mortality leads to mangrove peat collapse at bay islands,
  honduras after hurricane mitch.
\newblock {\em Journal of ecology}, 91(6):1093--1105, 2003.

\bibitem{carniello2014sediment}
L~Carniello, S~Silvestri, M~Marani, A~D'Alpaos, V~Volpe, and A~Defina.
\newblock Sediment dynamics in shallow tidal basins: In situ observations,
  satellite retrievals, and numerical modeling in the venice lagoon.
\newblock {\em Journal of Geophysical Research: Earth Surface},
  119(4):802--815, 2014.

\bibitem{carr2010stability}
J~Carr, P~D'odorico, K~McGlathery, and PL~Wiberg.
\newblock Stability and bistability of seagrass ecosystems in shallow coastal
  lagoons: Role of feedbacks with sediment resuspension and light attenuation.
\newblock {\em Journal of Geophysical Research: Biogeosciences}, 115(G3), 2010.

\bibitem{walker1997resilience}
Brian~H Walker, Jenny~L Langridge, and F~McFarlane.
\newblock Resilience of an australian savanna grassland to selective and
  non-selective perturbations.
\newblock {\em Australian Journal of Ecology}, 22(2):125--135, 1997.

\bibitem{ludwig1996landscape}
John Ludwig, D~Tongway, K~Hodgkinson, D~Freudenberger, and J~Noble.
\newblock {\em Landscape ecology, function and management: principles from
  Australia's rangelands}.
\newblock Csiro Publishing, 1996.

\bibitem{dublin1990elephants}
Holly~T Dublin, Alan~RE Sinclair, and J~McGlade.
\newblock Elephants and fire as causes of multiple stable states in the
  serengeti-mara woodlands.
\newblock {\em The Journal of Animal Ecology}, pages 1147--1164, 1990.

\bibitem{barnosky2012approaching}
Anthony~D Barnosky, Elizabeth~A Hadly, Jordi Bascompte, Eric~L Berlow, James~H
  Brown, Mikael Fortelius, Wayne~M Getz, John Harte, Alan Hastings, Pablo~A
  Marquet, et~al.
\newblock Approaching a state shift in earth’s biosphere.
\newblock {\em Nature}, 486(7401):52, 2012.

\bibitem{steneck2013ecosystem}
Robert~S Steneck, Amanda Leland, Douglas~C McNaught, and John Vavrinec.
\newblock Ecosystem flips, locks, and feedbacks: the lasting effects of
  fisheries on maine's kelp forest ecosystem.
\newblock {\em Bulletin of Marine Science}, 89(1):31--55, 2013.

\bibitem{graham2015predicting}
Nicholas~AJ Graham, Simon Jennings, M~Aaron MacNeil, David Mouillot, and
  Shaun~K Wilson.
\newblock Predicting climate-driven regime shifts versus rebound potential in
  coral reefs.
\newblock {\em Nature}, 518(7537):94, 2015.

\bibitem{wernberg2016climate}
Thomas Wernberg, Scott Bennett, Russell~C Babcock, Thibaut De~Bettignies,
  Katherine Cure, Martial Depczynski, Francois Dufois, Jane Fromont,
  Christopher~J Fulton, Renae~K Hovey, et~al.
\newblock Climate-driven regime shift of a temperate marine ecosystem.
\newblock {\em Science}, 353(6295):169--172, 2016.

\bibitem{wernberg2010decreasing}
Thomas Wernberg, Mads~S Thomsen, Fernando Tuya, Gary~A Kendrick, Peter~A
  Staehr, and Benjamin~D Toohey.
\newblock Decreasing resilience of kelp beds along a latitudinal temperature
  gradient: potential implications for a warmer future.
\newblock {\em Ecology letters}, 13(6):685--694, 2010.

\bibitem{wernberg2013extreme}
Thomas Wernberg, Dan~A Smale, Fernando Tuya, Mads~S Thomsen, Timothy~J
  Langlois, Thibaut De~Bettignies, Scott Bennett, and Cecile~S Rousseaux.
\newblock An extreme climatic event alters marine ecosystem structure in a
  global biodiversity hotspot.
\newblock {\em Nature Climate Change}, 3(1):78, 2013.

\bibitem{dayton1984catastrophic}
Paul~K Dayton and Mia~J Tegner.
\newblock Catastrophic storms, el ni{\~n}o, and patch stability in a southern
  california kelp community.
\newblock {\em Science}, 224(4646):283--285, 1984.

\bibitem{martinez2003recovery}
Enrique~A Mart{\'\i}nez, Leyla C{\'a}rdenas, and Raquel Pinto.
\newblock Recovery and genetic diversity of the intertidal kelp lessonia
  nigrescens (phaeophyceae) 20 years after el nino 1982/831.
\newblock {\em Journal of Phycology}, 39(3):504--508, 2003.

\bibitem{bennett2015tropical}
Scott Bennett, Thomas Wernberg, Euan~S Harvey, Julia Santana-Garcon, and
  Benjamin~J Saunders.
\newblock Tropical herbivores provide resilience to a climate-mediated phase
  shift on temperate reefs.
\newblock {\em Ecology letters}, 18(7):714--723, 2015.

\bibitem{burrows2014geographical}
Michael~T Burrows, David~S Schoeman, Anthony~J Richardson, Jorge~Garcia
  Molinos, Ary Hoffmann, Lauren~B Buckley, Pippa~J Moore, Christopher~J Brown,
  John~F Bruno, Carlos~M Duarte, et~al.
\newblock Geographical limits to species-range shifts are suggested by climate
  velocity.
\newblock {\em Nature}, 507(7493):492, 2014.

\bibitem{bennett2016great}
Scott Bennett, Thomas Wernberg, Sean~D Connell, Alistair~J Hobday, Craig~R
  Johnson, and Elvira~S Poloczanska.
\newblock The ‘great southern reef’: social, ecological and economic value
  of australia’s neglected kelp forests.
\newblock {\em Marine and Freshwater Research}, 67(1):47--56, 2016.

\bibitem{rietkerk1997alternate}
Max Rietkerk and Johan van~de Koppel.
\newblock Alternate stable states and threshold effects in semi-arid grazing
  systems.
\newblock {\em Oikos}, pages 69--76, 1997.

\bibitem{rietkerk2004self}
Max Rietkerk, Stefan~C Dekker, Peter~C De~Ruiter, and Johan van~de Koppel.
\newblock Self-organized patchiness and catastrophic shifts in ecosystems.
\newblock {\em Science}, 305(5692):1926--1929, 2004.

\bibitem{klausmeier1999regular}
Christopher~A Klausmeier.
\newblock Regular and irregular patterns in semiarid vegetation.
\newblock {\em Science}, 284(5421):1826--1828, 1999.

\bibitem{lejeune2002localized}
Olivier Lejeune, Mustapha Tlidi, and Pierre Couteron.
\newblock Localized vegetation patches: a self-organized response to resource
  scarcity.
\newblock {\em Physical Review E}, 66(1):010901, 2002.

\bibitem{rietkerk2004putative}
M~Rietkerk, SC~Dekker, MJ~Wassen, AWM Verkroost, and MFP Bierkens.
\newblock A putative mechanism for bog patterning.
\newblock {\em The American Naturalist}, 163(5):699--708, 2004.

\bibitem{getzin2016discovery}
Stephan Getzin, Hezi Yizhaq, Bronwyn Bell, Todd~E Erickson, Anthony~C Postle,
  Itzhak Katra, Omer Tzuk, Yuval~R Zelnik, Kerstin Wiegand, Thorsten Wiegand,
  et~al.
\newblock Discovery of fairy circles in australia supports self-organization
  theory.
\newblock {\em Proceedings of the National Academy of Sciences},
  113(13):3551--3556, 2016.

\bibitem{carpenter2003regime}
Stephen~R Carpenter and Otto Kinne.
\newblock {\em Regime shifts in lake ecosystems: pattern and variation},
  volume~15.
\newblock International Ecology Institute Oldendorf/Luhe, Germany, 2003.

\bibitem{liu2015systems}
Jianguo Liu, Harold Mooney, Vanessa Hull, Steven~J Davis, Joanne Gaskell,
  Thomas Hertel, Jane Lubchenco, Karen~C Seto, Peter Gleick, Claire Kremen,
  et~al.
\newblock Systems integration for global sustainability.
\newblock {\em Science}, 347(6225):1258832, 2015.

\bibitem{vonlanthen2012eutrophication}
Pascal Vonlanthen, David Bittner, Alan~G Hudson, Kyle~A Young, Ruedi
  M{\"u}ller, B{\"a}nz Lundsgaard-Hansen, Denis Roy, Sacha Di~Piazza,
  Carlo~Rodolfo Largiad{\`e}r, and Ole Seehausen.
\newblock Eutrophication causes speciation reversal in whitefish adaptive
  radiations.
\newblock {\em Nature}, 482(7385):357, 2012.

\bibitem{diaz2008spreading}
Robert~J Diaz and Rutger Rosenberg.
\newblock Spreading dead zones and consequences for marine ecosystems.
\newblock {\em science}, 321(5891):926--929, 2008.

\bibitem{altieri2017tropical}
Andrew~H Altieri, Seamus~B Harrison, Janina Seemann, Rachel Collin, Robert~J
  Diaz, and Nancy Knowlton.
\newblock Tropical dead zones and mass mortalities on coral reefs.
\newblock {\em Proceedings of the National Academy of Sciences},
  114(14):3660--3665, 2017.

\bibitem{biggs2018regime}
Reinette Biggs, Garry Peterson, and Juan Rocha.
\newblock The regime shifts database: a framework for analyzing regime shifts
  in social-ecological systems.
\newblock {\em Ecology and Society}, 23(3), 2018.

\bibitem{rocha2015regime}
Juan~Carlos Rocha, Garry~D Peterson, and Reinette Biggs.
\newblock Regime shifts in the anthropocene: drivers, risks, and resilience.
\newblock {\em PLoS One}, 10(8):e0134639, 2015.

\bibitem{nelson2006anthropogenic}
Gerald~C Nelson, A~Dobermann, N~Nakicenovic, and BC~O'Neill.
\newblock Anthropogenic drivers of ecosystem change: an overview.
\newblock {\em Ecology and Society}, 11(2), 2006.

\bibitem{lane2008emergence}
David~C Lane.
\newblock The emergence and use of diagramming in system dynamics: a critical
  account.
\newblock {\em Systems Research and Behavioral Science: The Official Journal of
  the International Federation for Systems Research}, 25(1):3--23, 2008.

\bibitem{scheffer2012anticipating}
Marten Scheffer, Stephen~R Carpenter, Timothy~M Lenton, Jordi Bascompte,
  William Brock, Vasilis Dakos, Johan Van~de Koppel, Ingrid~A Van~de Leemput,
  Simon~A Levin, Egbert~H Van~Nes, et~al.
\newblock Anticipating critical transitions.
\newblock {\em science}, 338(6105):344--348, 2012.

\bibitem{scheffer2018seeing}
Marten Scheffer and Egbert~H Van~Nes.
\newblock Seeing a global web of connected systems.
\newblock {\em Science}, 362(6421):1357--1357, 2018.

\bibitem{kronke2019dynamics}
Jonathan Kr{\"o}nke, Nico Wunderling, Ricarda Winkelmann, Arie Staal, Benedikt
  Stumpf, Obbe~A Tuinenburg, and Jonathan~F Donges.
\newblock Dynamics of tipping cascades on complex networks.
\newblock {\em arXiv preprint arXiv:1905.05476}, 2019.

\bibitem{gaucherel2017potential}
Cedric Gaucherel and Vincent Moron.
\newblock Potential stabilizing points to mitigate tipping point interactions
  in earth's climate.
\newblock {\em International Journal of Climatology}, 37(1):399--408, 2017.

\bibitem{drake2010early}
John~M Drake and Blaine~D Griffen.
\newblock Early warning signals of extinction in deteriorating environments.
\newblock {\em Nature}, 467(7314):456, 2010.

\bibitem{barnosky2011has}
Anthony~D Barnosky, Nicholas Matzke, Susumu Tomiya, Guinevere~OU Wogan, Brian
  Swartz, Tiago~B Quental, Charles Marshall, Jenny~L McGuire, Emily~L Lindsey,
  Kaitlin~C Maguire, et~al.
\newblock Has the earth’s sixth mass extinction already arrived?
\newblock {\em Nature}, 471(7336):51, 2011.

\bibitem{marshall2006explaining}
Charles~R Marshall.
\newblock Explaining the cambrian “explosion” of animals.
\newblock {\em Annu. Rev. Earth Planet. Sci.}, 34:355--384, 2006.

\bibitem{hoek2008last}
Wim~Z Hoek.
\newblock The last glacial-interglacial transition.
\newblock {\em Episodes}, 31(2):226--229, 2008.

\bibitem{koch2006late}
Paul~L Koch and Anthony~D Barnosky.
\newblock Late quaternary extinctions: State of the debate.
\newblock {\em Annu. Rev. Ecol. Evol. Syst}, 37:215--50, 2006.

\bibitem{graham1996spatial}
Russell~W Graham, Ernest~L Lundelius, Mary~Ann Graham, Erich~K Schroeder,
  Rickard~S Toomey, Elaine Anderson, Anthony~D Barnosky, James~A Burns,
  Charles~S Churcher, Donald~K Grayson, et~al.
\newblock Spatial response of mammals to late quaternary environmental
  fluctuations.
\newblock {\em Science}, 272(5268):1601--1606, 1996.

\bibitem{barnosky2008megafauna}
Anthony~D Barnosky.
\newblock Megafauna biomass tradeoff as a driver of quaternary and future
  extinctions.
\newblock {\em Proceedings of the National Academy of Sciences}, 105(Supplement
  1):11543--11548, 2008.

\bibitem{steffen2011anthropocene}
Will Steffen, {\AA}sa Persson, Lisa Deutsch, Jan Zalasiewicz, Mark Williams,
  Katherine Richardson, Carole Crumley, Paul Crutzen, Carl Folke, Line Gordon,
  et~al.
\newblock The anthropocene: From global change to planetary stewardship.
\newblock {\em Ambio}, 40(7):739, 2011.

\bibitem{mcdaniel2002increased}
Carl~N McDaniel and David~N Borton.
\newblock Increased human energy use causes biological diversity loss and
  undermines prospects for sustainability.
\newblock {\em Bioscience}, 52(10):929--936, 2002.

\bibitem{lenton2011early}
Timothy~M Lenton.
\newblock Early warning of climate tipping points.
\newblock {\em Nature climate change}, 1(4):201, 2011.

\bibitem{hastings2010regime}
Alan Hastings and Derin~B Wysham.
\newblock Regime shifts in ecological systems can occur with no warning.
\newblock {\em Ecology letters}, 13(4):464--472, 2010.

\bibitem{schroeder2009fractals}
Manfred Schroeder.
\newblock {\em Fractals, chaos, power laws: Minutes from an infinite paradise}.
\newblock Courier Corporation, 2009.

\bibitem{guttal2008changing}
Vishwesha Guttal and Ciriyam Jayaprakash.
\newblock Changing skewness: an early warning signal of regime shifts in
  ecosystems.
\newblock {\em Ecology letters}, 11(5):450--460, 2008.

\bibitem{kuznetsov2013elements}
Yuri~A Kuznetsov.
\newblock {\em Elements of applied bifurcation theory}, volume 112.
\newblock Springer Science \& Business Media, 2013.

\bibitem{strogatz2018nonlinear}
Steven~H Strogatz.
\newblock {\em Nonlinear dynamics and chaos: with applications to physics,
  biology, chemistry, and engineering}.
\newblock CRC Press, 2018.

\bibitem{chisholm2009critical}
Ryan~A Chisholm and Elise Filotas.
\newblock Critical slowing down as an indicator of transitions in two-species
  models.
\newblock {\em Journal of theoretical biology}, 257(1):142--149, 2009.

\bibitem{vandermeer1999basin}
John Vandermeer and Peter Yodzis.
\newblock Basin boundary collision as a model of discontinuous change in
  ecosystems.
\newblock {\em Ecology}, 80(6):1817--1827, 1999.

\bibitem{rinaldi2000geometric}
Sergio Rinaldi and Marten Scheffer.
\newblock Geometric analysis of ecological models with slow and fast processes.
\newblock {\em Ecosystems}, 3(6):507--521, 2000.

\bibitem{leung2000bifurcation}
HK~Leung.
\newblock Bifurcation of synchronization as a nonequilibrium phase transition.
\newblock {\em Physica A: Statistical Mechanics and its Applications},
  281(1-4):311--317, 2000.

\bibitem{hanski1998metapopulation}
Ilkka Hanski.
\newblock Metapopulation dynamics.
\newblock {\em Nature}, 396(6706):41, 1998.

\bibitem{scheffer2003slow}
Marten Scheffer, Frances Westley, and William Brock.
\newblock Slow response of societies to new problems: causes and costs.
\newblock {\em Ecosystems}, 6(5):493--502, 2003.

\bibitem{kefi2007spatial}
Sonia K{\'e}fi, Max Rietkerk, Concepci{\'o}n~L Alados, Yolanda Pueyo,
  Vasilios~P Papanastasis, Ahmed ElAich, and Peter~C De~Ruiter.
\newblock Spatial vegetation patterns and imminent desertification in
  mediterranean arid ecosystems.
\newblock {\em Nature}, 449(7159):213, 2007.

\bibitem{turing1990chemical}
Alan~Mathison Turing.
\newblock The chemical basis of morphogenesis.
\newblock {\em Bulletin of mathematical biology}, 52(1-2):153--197, 1990.

\bibitem{gardner2003long}
Toby~A Gardner, Isabelle~M C{\^o}t{\'e}, Jennifer~A Gill, Alastair Grant, and
  Andrew~R Watkinson.
\newblock Long-term region-wide declines in caribbean corals.
\newblock {\em Science}, 301(5635):958--960, 2003.

\bibitem{moore2018predicting}
John~C Moore.
\newblock Predicting tipping points in complex environmental systems.
\newblock {\em Proceedings of the National Academy of Sciences},
  115(4):635--636, 2018.

\bibitem{jiang2019harnessing}
Junjie Jiang, Alan Hastings, and Ying-Cheng Lai.
\newblock Harnessing tipping points in complex ecological networks.
\newblock {\em Journal of the Royal Society Interface}, 16(158):20190345, 2019.

\bibitem{hui2006carrying}
Cang Hui.
\newblock Carrying capacity, population equilibrium, and environment's maximal
  load.
\newblock {\em Ecological Modelling}, 192(1-2):317--320, 2006.

\bibitem{courchamp1999inverse}
Franck Courchamp, Tim Clutton-Brock, and Bryan Grenfell.
\newblock Inverse density dependence and the allee effect.
\newblock {\em Trends in ecology \& evolution}, 14(10):405--410, 1999.

\bibitem{rahmstorf2002ocean}
Stefan Rahmstorf.
\newblock Ocean circulation and climate during the past 120,000 years.
\newblock {\em Nature}, 419(6903):207, 2002.

\bibitem{nishikawa2014controlling}
Takashi Nishikawa and Edward Ott.
\newblock Controlling systems that drift through a tipping point.
\newblock {\em Chaos: An Interdisciplinary Journal of Nonlinear Science},
  24(3):033107, 2014.

\bibitem{vidiella2018exploiting}
Blai Vidiella, Josep Sardany{\'e}s, and Ricard Sol{\'e}.
\newblock Exploiting delayed transitions to sustain semiarid ecosystems after
  catastrophic shifts.
\newblock {\em Journal of The Royal Society Interface}, 15(143):20180083, 2018.

\bibitem{pace2017reversal}
Michael~L Pace, Ryan~D Batt, Cal~D Buelo, Stephen~R Carpenter, Jonathan~J Cole,
  Jason~T Kurtzweil, and Grace~M Wilkinson.
\newblock Reversal of a cyanobacterial bloom in response to early warnings.
\newblock {\em Proceedings of the National Academy of Sciences},
  114(2):352--357, 2017.

\bibitem{biggs2009turning}
Reinette Biggs, Stephen~R Carpenter, and William~A Brock.
\newblock Turning back from the brink: detecting an impending regime shift in
  time to avert it.
\newblock {\em Proceedings of the National academy of Sciences},
  106(3):826--831, 2009.

\bibitem{dupont2003structure}
Yoko~L Dupont, Dennis~M Hansen, and Jens~M Olesen.
\newblock Structure of a plant--flower-visitor network in the high-altitude
  sub-alpine desert of tenerife, canary islands.
\newblock {\em Ecography}, 26(3):301--310, 2003.

\bibitem{jiang2019irrelevance}
Junjie Jiang and Ying-Cheng Lai.
\newblock Irrelevance of linear controllability to nonlinear dynamical
  networks.
\newblock {\em Nature communications}, 10(1):1--10, 2019.

\bibitem{chechetka2017materially}
Svetlana~A Chechetka, Yue Yu, Masayoshi Tange, and Eijiro Miyako.
\newblock Materially engineered artificial pollinators.
\newblock {\em Chem}, 2(2):224--239, 2017.

\bibitem{rundlof2015seed}
Maj Rundl{\"o}f, Georg~KS Andersson, Riccardo Bommarco, Ingemar Fries, Veronica
  Hederstr{\"o}m, Lina Herbertsson, Ove Jonsson, Bj{\"o}rn~K Klatt, Thorsten~R
  Pedersen, Johanna Yourstone, et~al.
\newblock Seed coating with a neonicotinoid insecticide negatively affects wild
  bees.
\newblock {\em Nature}, 521(7550):77, 2015.

\bibitem{graham2013managing}
Nicholas~AJ Graham, David~R Bellwood, Joshua~E Cinner, Terry~P Hughes, Albert~V
  Norstr{\"o}m, and Magnus Nystr{\"o}m.
\newblock Managing resilience to reverse phase shifts in coral reefs.
\newblock {\em Frontiers in Ecology and the Environment}, 11(10):541--548,
  2013.

\bibitem{macneil2015recovery}
M~Aaron MacNeil, Nicholas~AJ Graham, Joshua~E Cinner, Shaun~K Wilson, Ivor~D
  Williams, Joseph Maina, Steven Newman, Alan~M Friedlander, Stacy Jupiter,
  Nicholas~VC Polunin, et~al.
\newblock Recovery potential of the world's coral reef fishes.
\newblock {\em Nature}, 520(7547):341, 2015.

\bibitem{chung2019building}
Anne~Elizabeth Chung, Lisa~M Wedding, Alison~L Green, Alan~M Friedlander, Grace
  Goldberg, Amber Meadows, and Mark~A Hixon.
\newblock Building coral reef resilience through spatial herbivore management.
\newblock {\em Frontiers in Marine Science}, 6:98, 2019.

\bibitem{sanhedrai2022reviving}
Hillel Sanhedrai, Jianxi Gao, Amir Bashan, Moshe Schwartz, Shlomo Havlin, and
  Baruch Barzel.
\newblock Reviving a failed network through microscopic interventions.
\newblock {\em Nature Physics}, pages 1--12, 2022.

\bibitem{ma2021universality}
Cheng Ma, Gyorgy Korniss, Boleslaw~K Szymanski, and Jianxi Gao.
\newblock Universality of noise-induced resilience restoration in
  spatially-extended ecological systems.
\newblock {\em Communications Physics}, 4(1):1--12, 2021.

\bibitem{schultz2015adaptive}
Lisen Schultz, Carl Folke, Henrik {\"O}sterblom, and Per Olsson.
\newblock Adaptive governance, ecosystem management, and natural capital.
\newblock {\em Proceedings of the National Academy of Sciences},
  112(24):7369--7374, 2015.

\bibitem{peterson1998ecological}
Garry Peterson, Craig~R Allen, and Crawford~Stanley Holling.
\newblock Ecological resilience, biodiversity, and scale.
\newblock {\em Ecosystems}, 1(1):6--18, 1998.

\bibitem{martin2011regional}
Ron Martin.
\newblock Regional economic resilience, hysteresis and recessionary shocks.
\newblock {\em Journal of economic geography}, 12(1):1--32, 2011.

\bibitem{song2015integrating}
Hyun-Seob Song, Ryan~S Renslow, Jim~K Fredrickson, and Stephen~R Lindemann.
\newblock Integrating ecological and engineering concepts of resilience in
  microbial communities.
\newblock {\em Frontiers in microbiology}, 6:1298, 2015.

\bibitem{zobel2011representing}
Christopher~W Zobel.
\newblock Representing perceived tradeoffs in defining disaster resilience.
\newblock {\em Decision Support Systems}, 50(2):394--403, 2011.

\bibitem{rose2007economic}
Adam Rose.
\newblock Economic resilience to natural and man-made disasters:
  Multidisciplinary origins and contextual dimensions.
\newblock {\em Environmental Hazards}, 7(4):383--398, 2007.

\bibitem{ouyang2015resilience}
Min Ouyang and Zhenghua Wang.
\newblock Resilience assessment of interdependent infrastructure systems: With
  a focus on joint restoration modeling and analysis.
\newblock {\em Reliability Engineering \& System Safety}, 141:74--82, 2015.

\bibitem{smith2008brief}
Bruce~W Smith, Jeanne Dalen, Kathryn Wiggins, Erin Tooley, Paulette
  Christopher, and Jennifer Bernard.
\newblock The brief resilience scale: assessing the ability to bounce back.
\newblock {\em International journal of behavioral medicine}, 15(3):194--200,
  2008.

\bibitem{shade2011resistance}
Ashley Shade, Jordan~S Read, David~G Welkie, Timothy~K Kratz, Chin~H Wu, and
  Katherine~D McMahon.
\newblock Resistance, resilience and recovery: aquatic bacterial dynamics after
  water column disturbance.
\newblock {\em Environmental microbiology}, 13(10):2752--2767, 2011.

\bibitem{golubovich2014safety}
Juliya Golubovich, Chu-Hsiang Chang, and Erin~M Eatough.
\newblock Safety climate, hardiness, and musculoskeletal complaints: A mediated
  moderation model.
\newblock {\em Applied Ergonomics}, 45(3):757--766, 2014.

\bibitem{chaumot2012molecular}
Arnaud Chaumot, Jean-Luc Da~Lage, Oscar Maestro, David Martin, Thomas Iwema,
  Frederic Brunet, Xavier Belles, Vincent Laudet, and Fran{\c{c}}ois Bonneton.
\newblock Molecular adaptation and resilience of the insect’s nuclear
  receptor usp.
\newblock {\em BMC evolutionary biology}, 12(1):199, 2012.

\bibitem{paniagua2013farmers}
Angel Paniagua.
\newblock Farmers in remote rural areas: The worth of permanence in the place.
\newblock {\em Land Use Policy}, 35:1--7, 2013.

\bibitem{waide1976engineering}
JACK~B Waide and JACKSON~R Webster.
\newblock Engineering systems analysis: applicability to ecosystems.
\newblock {\em Systems analysis and simulation in ecology}, 4:329--371, 1976.

\bibitem{doring2015resilience}
Thomas~F D{\"o}ring, Anja Vieweger, Marco Pautasso, Mette Vaarst, Maria~R
  Finckh, and Martin~S Wolfe.
\newblock Resilience as a universal criterion of health.
\newblock {\em Journal of the Science of Food and Agriculture}, 95(3):455--465,
  2015.

\bibitem{kirouac2009cell}
Daniel~C Kirouac, Gerard~J Madlambayan, Mei Yu, Edward~A Sykes, Caryn Ito, and
  Peter~W Zandstra.
\newblock Cell--cell interaction networks regulate blood stem and progenitor
  cell fate.
\newblock {\em Molecular systems biology}, 5(1), 2009.

\bibitem{huttlin2017architecture}
Edward~L Huttlin, Raphael~J Bruckner, Joao~A Paulo, Joe~R Cannon, Lily Ting,
  Kurt Baltier, Greg Colby, Fana Gebreab, Melanie~P Gygi, Hannah Parzen, et~al.
\newblock Architecture of the human interactome defines protein communities and
  disease networks.
\newblock {\em Nature}, 545(7655):505--509, 2017.

\bibitem{laurent1999multistability}
Michel Laurent and Nicolas Kellershohn.
\newblock Multistability: a major means of differentiation and evolution in
  biological systems.
\newblock {\em Trends in biochemical sciences}, 24(11):418--422, 1999.

\bibitem{yu2016physical}
Chong Yu and Jin Wang.
\newblock A physical mechanism and global quantification of breast cancer.
\newblock {\em PloS one}, 11(7), 2016.

\bibitem{gardner2000construction}
Timothy~S Gardner, Charles~R Cantor, and James~J Collins.
\newblock Construction of a genetic toggle switch in escherichia coli.
\newblock {\em Nature}, 403(6767):339, 2000.

\bibitem{angeli2004detection}
David Angeli, James~E Ferrell, and Eduardo~D Sontag.
\newblock Detection of multistability, bifurcations, and hysteresis in a large
  class of biological positive-feedback systems.
\newblock {\em Proceedings of the National Academy of Sciences},
  101(7):1822--1827, 2004.

\bibitem{wilhelm2009smallest}
Thomas Wilhelm.
\newblock The smallest chemical reaction system with bistability.
\newblock {\em BMC systems biology}, 3(1):90, 2009.

\bibitem{ferrell2002self}
James~E Ferrell~Jr.
\newblock Self-perpetuating states in signal transduction: positive feedback,
  double-negative feedback and bistability.
\newblock {\em Current opinion in cell biology}, 14(2):140--148, 2002.

\bibitem{liu2014identifying}
Rui Liu, Xiangtian Yu, Xiaoping Liu, Dong Xu, Kazuyuki Aihara, and Luonan Chen.
\newblock Identifying critical transitions of complex diseases based on a
  single sample.
\newblock {\em Bioinformatics}, 30(11):1579--1586, 2014.

\bibitem{wang2009bistable}
Lei Wang, Brandon~L Walker, Stephen Iannaccone, Devang Bhatt, Patrick~J
  Kennedy, and T~Tse William.
\newblock Bistable switches control memory and plasticity in cellular
  differentiation.
\newblock {\em Proceedings of the National Academy of Sciences},
  106(16):6638--6643, 2009.

\bibitem{bhattacharya2010bistable}
Sudin Bhattacharya, Rory~B Conolly, Norbert~E Kaminski, Russell~S Thomas,
  Melvin~E Andersen, and Qiang Zhang.
\newblock A bistable switch underlying b-cell differentiation and its
  disruption by the environmental contaminant 2, 3, 7,
  8-tetrachlorodibenzo-p-dioxin.
\newblock {\em Toxicological sciences}, 115(1):51--65, 2010.

\bibitem{monod1961general}
Jacques Monod and Fran{\c{c}}ois Jacob.
\newblock General conclusions: teleonomic mechanisms in cellular metabolism,
  growth, and differentiation.
\newblock In {\em Cold Spring Harbor symposia on quantitative biology},
  volume~26, pages 389--401. Cold Spring Harbor Laboratory Press, 1961.

\bibitem{lebar2014bistable}
Tina Lebar, Urban Bezeljak, Anja Golob, Miha Jerala, Lucija Kadunc,
  Bo{\v{s}}tjan Pir{\v{s}}, Martin Stra{\v{z}}ar, Du{\v{s}}an Vu{\v{c}}ko,
  Uro{\v{s}} Zupan{\v{c}}i{\v{c}}, Mojca Ben{\v{c}}ina, et~al.
\newblock A bistable genetic switch based on designable dna-binding domains.
\newblock {\em Nature communications}, 5(1):1--13, 2014.

\bibitem{sabouri2008antagonism}
Mohsen Sabouri-Ghomi, Andrea Ciliberto, Sandip Kar, Bela Novak, and John~J
  Tyson.
\newblock Antagonism and bistability in protein interaction networks.
\newblock {\em Journal of theoretical biology}, 250(1):209--218, 2008.

\bibitem{ferrell1996tripping}
James~E Ferrell.
\newblock Tripping the switch fantastic: how a protein kinase cascade can
  convert graded inputs into switch-like outputs.
\newblock {\em Trends in biochemical sciences}, 21(12):460--466, 1996.

\bibitem{markevich2004signaling}
Nick~I Markevich, Jan~B Hoek, and Boris~N Kholodenko.
\newblock Signaling switches and bistability arising from multisite
  phosphorylation in protein kinase cascades.
\newblock {\em The Journal of cell biology}, 164(3):353--359, 2004.

\bibitem{pomerening2003building}
Joseph~R Pomerening, Eduardo~D Sontag, and James~E Ferrell.
\newblock Building a cell cycle oscillator: hysteresis and bistability in the
  activation of cdc2.
\newblock {\em Nature cell biology}, 5(4):346--351, 2003.

\bibitem{ferrell2011modeling}
James~E Ferrell~Jr, Tony Yu-Chen Tsai, and Qiong Yang.
\newblock Modeling the cell cycle: why do certain circuits oscillate?
\newblock {\em Cell}, 144(6):874--885, 2011.

\bibitem{huang2006bistable}
Guanghua Huang, Huafeng Wang, Song Chou, Xinyi Nie, Jiangye Chen, and Haoping
  Liu.
\newblock Bistable expression of wor1, a master regulator of white--opaque
  switching in candida albicans.
\newblock {\em Proceedings of the National Academy of Sciences},
  103(34):12813--12818, 2006.

\bibitem{warren2005chemical}
Patrick~B Warren and Pieter~Rein ten Wolde.
\newblock Chemical models of genetic toggle switches.
\newblock {\em The Journal of Physical Chemistry B}, 109(14):6812--6823, 2005.

\bibitem{schultz2008extinction}
Daniel Schultz, Aleksandra~M Walczak, Jos{\'e}~N Onuchic, and Peter~G Wolynes.
\newblock Extinction and resurrection in gene networks.
\newblock {\em Proceedings of the National Academy of Sciences},
  105(49):19165--19170, 2008.

\bibitem{lai2016understanding}
Xin Lai, Olaf Wolkenhauer, and Julio Vera.
\newblock Understanding microrna-mediated gene regulatory networks through
  mathematical modelling.
\newblock {\em Nucleic acids research}, 44(13):6019--6035, 2016.

\bibitem{schroter2015fgf}
Christian Schr{\"o}ter, Pau Ru{\'e}, Jonathan~Peter Mackenzie, and
  Alfonso~Martinez Arias.
\newblock Fgf/mapk signaling sets the switching threshold of a bistable circuit
  controlling cell fate decisions in embryonic stem cells.
\newblock {\em Development}, 142(24):4205--4216, 2015.

\bibitem{bednarz2014revisiting}
Michael Bednarz, Jennifer~A Halliday, Christophe Herman, and Ido Golding.
\newblock Revisiting bistability in the lysis/lysogeny circuit of bacteriophage
  lambda.
\newblock {\em PloS one}, 9(6):e100876, 2014.

\bibitem{kramer2004engineered}
Beat~P Kramer, Alessandro~Usseglio Viretta, Marie Daoud-El~Baba, Dominique
  Aubel, Wilfried Weber, and Martin Fussenegger.
\newblock An engineered epigenetic transgene switch in mammalian cells.
\newblock {\em Nature biotechnology}, 22(7):867, 2004.

\bibitem{lai2012modeling}
Xin Lai, Olaf Wolkenhauer, and Julio Vera.
\newblock Modeling mirna regulation in cancer signaling systems: mir-34a
  regulation of the p53/sirt1 signaling module.
\newblock In {\em Computational Modeling of Signaling Networks}, pages 87--108.
  Springer, 2012.

\bibitem{martinez2018bistable}
Rosa Martinez-Corral, Jintao Liu, G{\"u}rol~M S{\"u}el, and Jordi
  Garcia-Ojalvo.
\newblock Bistable emergence of oscillations in growing bacillus subtilis
  biofilms.
\newblock {\em Proceedings of the National Academy of Sciences},
  115(36):E8333--E8340, 2018.

\bibitem{bala2018bistability}
Saminu~Iliyasu Bala and Nura Mohammad~Rabiu Ahmad.
\newblock Bistability analysis in mathematical model of m-phase control in
  xenopus oocyte extracts.
\newblock {\em Computational and Applied Mathematics}, 37(3):2667--2692, 2018.

\bibitem{fang2018cell}
Xiaona Fang, Qiong Liu, Christopher Bohrer, Zach Hensel, Wei Han, Jin Wang, and
  Jie Xiao.
\newblock Cell fate potentials and switching kinetics uncovered in a classic
  bistable genetic switch.
\newblock {\em Nature communications}, 9(1):1--9, 2018.

\bibitem{pomerening2005systems}
Joseph~R Pomerening, Sun~Young Kim, and James~E Ferrell~Jr.
\newblock Systems-level dissection of the cell-cycle oscillator: bypassing
  positive feedback produces damped oscillations.
\newblock {\em Cell}, 122(4):565--578, 2005.

\bibitem{coudreuse2010driving}
Damien Coudreuse and Paul Nurse.
\newblock Driving the cell cycle with a minimal cdk control network.
\newblock {\em Nature}, 468(7327):1074--1079, 2010.

\bibitem{goldbeter2002computational}
Albert Goldbeter.
\newblock Computational approaches to cellular rhythms.
\newblock {\em Nature}, 420(6912):238--245, 2002.

\bibitem{novak1993numerical}
Bela Novak and John~J Tyson.
\newblock Numerical analysis of a comprehensive model of m-phase control in
  xenopus oocyte extracts and intact embryos.
\newblock {\em Journal of cell science}, 106(4):1153--1168, 1993.

\bibitem{doree2002cdc2}
Marcel Dor{\'e}e and Tim Hunt.
\newblock From cdc2 to cdk1: when did the cell cycle kinase join its cyclin
  partner?
\newblock {\em Journal of cell science}, 115(12):2461--2464, 2002.

\bibitem{rata2018two}
Scott Rata, Maria F Suarez~Peredo Rodriguez, Stephy Joseph, Nisha Peter,
  Fabio~Echegaray Iturra, Fengwei Yang, Anotida Madzvamuse, Jan~G Ruppert,
  Kumiko Samejima, Melpomeni Platani, et~al.
\newblock Two interlinked bistable switches govern mitotic control in mammalian
  cells.
\newblock {\em Current biology}, 28(23):3824--3832, 2018.

\bibitem{oikonomou2010frequency}
Catherine Oikonomou and Frederick~R Cross.
\newblock Frequency control of cell cycle oscillators.
\newblock {\em Current opinion in genetics \& development}, 20(6):605--612,
  2010.

\bibitem{abrieu2001interplay}
Ariane Abrieu, Marcel Dor{\'e}e, and Daniel Fisher.
\newblock The interplay between cyclin-b-cdc2 kinase (mpf) and map kinase
  during maturation of oocytes.
\newblock {\em Journal of cell science}, 114(2):257--267, 2001.

\bibitem{gotoh1995initiation}
Yukiko Gotoh, Norihisa Masuyama, Karen Dell, Kyoko Shirakabe, and Eisuke
  Nishida.
\newblock Initiation of xenopus oocyte maturation by activation of the
  mitogen-activated protein kinase cascade.
\newblock {\em Journal of Biological Chemistry}, 270(43):25898--25904, 1995.

\bibitem{ferrell1998biochemical}
James~E Ferrell and Eric~M Machleder.
\newblock The biochemical basis of an all-or-none cell fate switch in xenopus
  oocytes.
\newblock {\em Science}, 280(5365):895--898, 1998.

\bibitem{st2008determination}
Fran{\c{c}}ois St-Pierre and Drew Endy.
\newblock Determination of cell fate selection during phage lambda infection.
\newblock {\em Proceedings of the National Academy of Sciences},
  105(52):20705--20710, 2008.

\bibitem{trinh2017cell}
Jimmy~T Trinh, Tam{\'a}s Sz{\'e}kely, Qiuyan Shao, G{\'a}bor Bal{\'a}zsi, and
  Lanying Zeng.
\newblock Cell fate decisions emerge as phages cooperate or compete inside
  their host.
\newblock {\em Nature communications}, 8(1):1--13, 2017.

\bibitem{ozbudak2004multistability}
Ertugrul~M Ozbudak, Mukund Thattai, Han~N Lim, Boris~I Shraiman, and Alexander
  Van~Oudenaarden.
\newblock Multistability in the lactose utilization network of escherichia
  coli.
\newblock {\em Nature}, 427(6976):737, 2004.

\bibitem{cai2007dedifferentiation}
SA~Cai, Xiaobing Fu, and Zhiyong Sheng.
\newblock Dedifferentiation: a new approach in stem cell research.
\newblock {\em Bioscience}, 57(8):655--662, 2007.

\bibitem{ahrends2014controlling}
Robert Ahrends, Asuka Ota, Kyle~M Kovary, Takamasa Kudo, Byung~Ouk Park, and
  Mary~N Teruel.
\newblock Controlling low rates of cell differentiation through noise and
  ultrahigh feedback.
\newblock {\em Science}, 344(6190):1384--1389, 2014.

\bibitem{chang2006multistable}
Hannah~H Chang, Philmo~Y Oh, Donald~E Ingber, and Sui Huang.
\newblock Multistable and multistep dynamics in neutrophil differentiation.
\newblock {\em BMC cell biology}, 7(1):11, 2006.

\bibitem{goldbeter2007sharp}
Albert Goldbeter, Didier Gonze, and Olivier Pourqui{\'e}.
\newblock Sharp developmental thresholds defined through bistability by
  antagonistic gradients of retinoic acid and fgf signaling.
\newblock {\em Developmental dynamics: an official publication of the American
  Association of Anatomists}, 236(6):1495--1508, 2007.

\bibitem{tian2004bistability}
Tianhai Tian and Kevin Burrage.
\newblock Bistability and switching in the lysis/lysogeny genetic regulatory
  network of bacteriophage $\lambda$.
\newblock {\em Journal of Theoretical Biology}, 227(2):229--237, 2004.

\bibitem{pedraza2005noise}
Juan~M Pedraza and Alexander van Oudenaarden.
\newblock Noise propagation in gene networks.
\newblock {\em Science}, 307(5717):1965--1969, 2005.

\bibitem{raser2005noise}
Jonathan~M Raser and Erin~K O'Shea.
\newblock Noise in gene expression: origins, consequences, and control.
\newblock {\em Science}, 309(5743):2010--2013, 2005.

\bibitem{vujovic2019notch}
Filip Vujovic, Neil Hunter, and Ramin~M Farahani.
\newblock Notch pathway: a bistable inducer of biological noise?
\newblock {\em Cell Communication and Signaling}, 17(1):1--13, 2019.

\bibitem{mitosch2017noisy}
Karin Mitosch, Georg Rieckh, and Tobias Bollenbach.
\newblock Noisy response to antibiotic stress predicts subsequent single-cell
  survival in an acidic environment.
\newblock {\em Cell systems}, 4(4):393--403, 2017.

\bibitem{balazsi2011cellular}
G{\'a}bor Bal{\'a}zsi, Alexander van Oudenaarden, and James~J Collins.
\newblock Cellular decision making and biological noise: from microbes to
  mammals.
\newblock {\em Cell}, 144(6):910--925, 2011.

\bibitem{garcia2016transcription}
Tatiana Garc{\'\i}a-Muse and Andr{\'e}s Aguilera.
\newblock Transcription--replication conflicts: how they occur and how they are
  resolved.
\newblock {\em Nature reviews Molecular cell biology}, 17(9):553, 2016.

\bibitem{skourti2014double}
Konstantina Skourti-Stathaki and Nicholas~J Proudfoot.
\newblock A double-edged sword: R loops as threats to genome integrity and
  powerful regulators of gene expression.
\newblock {\em Genes \& development}, 28(13):1384--1396, 2014.

\bibitem{van1992stochastic}
Nicolaas~Godfried Van~Kampen.
\newblock {\em Stochastic processes in physics and chemistry}, volume~1.
\newblock Elsevier, 1992.

\bibitem{pajaro2019transient}
Manuel P{\'a}jaro, Irene Otero-Muras, Carlos V{\'a}zquez, and Antonio~A Alonso.
\newblock Transient hysteresis and inherent stochasticity in gene regulatory
  networks.
\newblock {\em Nature communications}, 10(1):1--7, 2019.

\bibitem{lestas2010fundamental}
Ioannis Lestas, Glenn Vinnicombe, and Johan Paulsson.
\newblock Fundamental limits on the suppression of molecular fluctuations.
\newblock {\em Nature}, 467(7312):174--178, 2010.

\bibitem{paulsson2004summing}
Johan Paulsson.
\newblock Summing up the noise in gene networks.
\newblock {\em Nature}, 427(6973):415--418, 2004.

\bibitem{siciliano2013mirnas}
Velia Siciliano, Immacolata Garzilli, Chiara Fracassi, Stefania Criscuolo,
  Simona Ventre, and Diego Di~Bernardo.
\newblock Mirnas confer phenotypic robustness to gene networks by suppressing
  biological noise.
\newblock {\em Nature communications}, 4(1):1--7, 2013.

\bibitem{barkai2000circadian}
Naama Barkai and Stanislas Leibler.
\newblock Circadian clocks limited by noise.
\newblock {\em Nature}, 403(6767):267--268, 2000.

\bibitem{acar2005enhancement}
Murat Acar, Attila Becskei, and Alexander van Oudenaarden.
\newblock Enhancement of cellular memory by reducing stochastic transitions.
\newblock {\em Nature}, 435(7039):228--232, 2005.

\bibitem{young2001time}
Michael~W Young and Steve~A Kay.
\newblock Time zones: a comparative genetics of circadian clocks.
\newblock {\em Nature Reviews Genetics}, 2(9):702--715, 2001.

\bibitem{gonze2006circadian}
Didier Gonze and Albert Goldbeter.
\newblock Circadian rhythms and molecular noise.
\newblock {\em Chaos: An Interdisciplinary Journal of Nonlinear Science},
  16(2):026110, 2006.

\bibitem{gupta2013transcriptional}
Chinmaya Gupta, Jos{\'e}~Manuel L{\'o}pez, William Ott, Kre{\v{s}}imir
  Josi{\'c}, and Matthew~R Bennett.
\newblock Transcriptional delay stabilizes bistable gene networks.
\newblock {\em Physical review letters}, 111(5):058104, 2013.

\bibitem{josic2011stochastic}
Kre{\v{s}}imir Josi{\'c}, Jos{\'e}~Manuel L{\'o}pez, William Ott, LieJune
  Shiau, and Matthew~R Bennett.
\newblock Stochastic delay accelerates signaling in gene networks.
\newblock {\em PLoS computational biology}, 7(11):e1002264, 2011.

\bibitem{li2009effects}
Gene-Wei Li, Otto~G Berg, and Johan Elf.
\newblock Effects of macromolecular crowding and dna looping on gene regulation
  kinetics.
\newblock {\em Nature Physics}, 5(4):294, 2009.

\bibitem{carvalho2019antibiotic}
Gabriel Carvalho, Christiane Forestier, and Jean-Denis Mathias.
\newblock Antibiotic resilience: a necessary concept to complement antibiotic
  resistance?
\newblock {\em Proceedings of the Royal Society B}, 286(1916):20192408, 2019.

\bibitem{d2006effects}
Daniela D'Amato, Maria~Rosaria Corbo, Matteo Alessandro~Del Nobile, and Milena
  Sinigaglia.
\newblock Effects of temperature, ammonium and glucose concentrations on yeast
  growth in a model wine system.
\newblock {\em International journal of food science \& technology},
  41(10):1152--1157, 2006.

\bibitem{koschwanez2011sucrose}
John~H Koschwanez, Kevin~R Foster, and Andrew~W Murray.
\newblock Sucrose utilization in budding yeast as a model for the origin of
  undifferentiated multicellularity.
\newblock {\em PLoS biology}, 9(8), 2011.

\bibitem{celiker2013cellular}
Hasan Celiker and Jeff Gore.
\newblock Cellular cooperation: insights from microbes.
\newblock {\em Trends in cell biology}, 23(1):9--15, 2013.

\bibitem{hope2017experimental}
Elyse~A Hope, Clara~J Amorosi, Aaron~W Miller, Kolena Dang, Caiti~Smukowski
  Heil, and Maitreya~J Dunham.
\newblock Experimental evolution reveals favored adaptive routes to cell
  aggregation in yeast.
\newblock {\em Genetics}, 206(2):1153--1167, 2017.

\bibitem{schuergers2016cyanobacteria}
Nils Schuergers, Tchern Lenn, Ronald Kampmann, Markus~V Meissner, Tiago
  Esteves, Maja Temerinac-Ott, Jan~G Korvink, Alan~R Lowe, Conrad~W Mullineaux,
  and Annegret Wilde.
\newblock Cyanobacteria use micro-optics to sense light direction.
\newblock {\em Elife}, 5:e12620, 2016.

\bibitem{montgomery2014regulation}
Beronda~L Montgomery.
\newblock The regulation of light sensing and light-harvesting impacts the use
  of cyanobacteria as biotechnology platforms.
\newblock {\em Frontiers in bioengineering and biotechnology}, 2:22, 2014.

\bibitem{luimstra2018blue}
Veerle~M Luimstra, J~Merijn Schuurmans, Antonie~M Verschoor, Klaas~J
  Hellingwerf, Jef Huisman, and Hans~CP Matthijs.
\newblock Blue light reduces photosynthetic efficiency of cyanobacteria through
  an imbalance between photosystems i and ii.
\newblock {\em Photosynthesis research}, 138(2):177--189, 2018.

\bibitem{gerla2011photoinhibition}
Daan~J Gerla, Wolf~M Mooij, and Jef Huisman.
\newblock Photoinhibition and the assembly of light-limited phytoplankton
  communities.
\newblock {\em Oikos}, 120(3):359--368, 2011.

\bibitem{wiltbank2019diverse}
Lisa~B Wiltbank and David~M Kehoe.
\newblock Diverse light responses of cyanobacteria mediated by phytochrome
  superfamily photoreceptors.
\newblock {\em Nature Reviews Microbiology}, 17(1):37--50, 2019.

\bibitem{photoinhibition}
Transition to extinction due to photo-inhibition.
\newblock \url{http://www.early-warning-signals.org/?page_id=565}.

\bibitem{zilliges2008extracellular}
Yvonne Zilliges, Jan-Christoph Kehr, Stefan Mikkat, Christiane Bouchier,
  Nicole~Tandeau de~Marsac, Thomas Boerner, and Elke Dittmann.
\newblock An extracellular glycoprotein is implicated in cell-cell contacts in
  the toxic cyanobacterium microcystis aeruginosa pcc 7806.
\newblock {\em Journal of bacteriology}, 190(8):2871--2879, 2008.

\bibitem{mitosch2014bacterial}
Karin Mitosch and Tobias Bollenbach.
\newblock Bacterial responses to antibiotics and their combinations.
\newblock {\em Environmental microbiology reports}, 6(6):545--557, 2014.

\bibitem{meredith2015collective}
Hannah~R Meredith, Jaydeep~K Srimani, Anna~J Lee, Allison~J Lopatkin, and
  Lingchong You.
\newblock Collective antibiotic tolerance: mechanisms, dynamics and
  intervention.
\newblock {\em Nature chemical biology}, 11(3):182, 2015.

\bibitem{brock2009non}
Amy Brock, Hannah Chang, and Sui Huang.
\newblock Non-genetic heterogeneity—a mutation-independent driving force for
  the somatic evolution of tumours.
\newblock {\em Nature Reviews Genetics}, 10(5):336, 2009.

\bibitem{li2014landscape}
Chunhe Li and Jin Wang.
\newblock Landscape and flux reveal a new global view and physical
  quantification of mammalian cell cycle.
\newblock {\em Proceedings of the National Academy of Sciences},
  111(39):14130--14135, 2014.

\bibitem{zhou2012quasi}
Joseph~Xu Zhou, MDS Aliyu, Erik Aurell, and Sui Huang.
\newblock Quasi-potential landscape in complex multi-stable systems.
\newblock {\em Journal of the Royal Society Interface}, 9(77):3539--3553, 2012.

\bibitem{gardiner1985handbook}
Crispin~W Gardiner et~al.
\newblock {\em Handbook of stochastic methods}, volume~3.
\newblock springer Berlin, 1985.

\bibitem{wang2008potential}
Jin Wang, Li~Xu, and Erkang Wang.
\newblock Potential landscape and flux framework of nonequilibrium networks:
  Robustness, dissipation, and coherence of biochemical oscillations.
\newblock {\em Proceedings of the National Academy of Sciences},
  105(34):12271--12276, 2008.

\bibitem{zhang2018exploring}
Kun Zhang and Jin Wang.
\newblock Exploring the underlying mechanisms of the xenopus laevis embryonic
  cell cycle.
\newblock {\em The Journal of Physical Chemistry B}, 122(21):5487--5499, 2018.

\bibitem{wang2010potential}
Jin Wang, Li~Xu, Erkang Wang, and Sui Huang.
\newblock The potential landscape of genetic circuits imposes the arrow of time
  in stem cell differentiation.
\newblock {\em Biophysical journal}, 99(1):29--39, 2010.

\bibitem{waddington2014strategy}
Conrad~Hal Waddington.
\newblock {\em The strategy of the genes}.
\newblock Routledge, 2014.

\bibitem{ferrell2012bistability}
James~E Ferrell~Jr.
\newblock Bistability, bifurcations, and waddington's epigenetic landscape.
\newblock {\em Current biology}, 22(11):R458--R466, 2012.

\bibitem{wang2011quantifying}
Jin Wang, Kun Zhang, Li~Xu, and Erkang Wang.
\newblock Quantifying the waddington landscape and biological paths for
  development and differentiation.
\newblock {\em Proceedings of the National Academy of Sciences},
  108(20):8257--8262, 2011.

\bibitem{huang2007bifurcation}
Sui Huang, Yan-Ping Guo, Gillian May, and Tariq Enver.
\newblock Bifurcation dynamics in lineage-commitment in bipotent progenitor
  cells.
\newblock {\em Developmental biology}, 305(2):695--713, 2007.

\bibitem{li2013quantifying}
Chunhe Li and Jin Wang.
\newblock Quantifying cell fate decisions for differentiation and reprogramming
  of a human stem cell network: landscape and biological paths.
\newblock {\em PLoS computational biology}, 9(8), 2013.

\bibitem{yu2019landscape}
Chong Yu, Qiong Liu, Cong Chen, Jian Yu, and Jin Wang.
\newblock Landscape perspectives of tumor, emt, and development.
\newblock {\em Physical biology}, 16(5):051003, 2019.

\bibitem{armitage1954age}
Peter Armitage and Richard Doll.
\newblock The age distribution of cancer and a multi-stage theory of
  carcinogenesis.
\newblock {\em British journal of cancer}, 8(1):1, 1954.

\bibitem{korolev2014turning}
Kirill~S Korolev, Joao~B Xavier, and Jeff Gore.
\newblock Turning ecology and evolution against cancer.
\newblock {\em Nature Reviews Cancer}, 14(5):371--380, 2014.

\bibitem{wenbo2017uncovering}
Li~Wenbo and Jin Wang.
\newblock Uncovering the underlying mechanism of cancer tumorigenesis and
  development under an immune microenvironment from global quantification of
  the landscape.
\newblock {\em Journal of The Royal Society Interface}, 14(131):20170105, 2017.

\bibitem{stark2007oscillations}
Jaroslav Stark, Cliburn Chan, and Andrew~JT George.
\newblock Oscillations in the immune system.
\newblock {\em Immunological reviews}, 216(1):213--231, 2007.

\bibitem{yu2017epigenetic}
Bingfei Yu, Kai Zhang, J~Justin Milner, Clara Toma, Runqiang Chen, James~P
  Scott-Browne, Renata~M Pereira, Shane Crotty, John~T Chang, Matthew~E Pipkin,
  et~al.
\newblock Epigenetic landscapes reveal transcription factors that regulate cd8+
  t cell differentiation.
\newblock {\em Nature immunology}, 18(5):573, 2017.

\bibitem{hutchings2004marine}
Jeffrey~A Hutchings and John~D Reynolds.
\newblock Marine fish population collapses: consequences for recovery and
  extinction risk.
\newblock {\em BioScience}, 54(4):297--309, 2004.

\bibitem{guttal2009spatial}
Vishwesha Guttal and Ciriyam Jayaprakash.
\newblock Spatial variance and spatial skewness: leading indicators of regime
  shifts in spatial ecological systems.
\newblock {\em Theoretical Ecology}, 2(1):3--12, 2009.

\bibitem{rindi2017direct}
Luca Rindi, Martina Dal~Bello, Lei Dai, Jeff Gore, and Lisandro
  Benedetti-Cecchi.
\newblock Direct observation of increasing recovery length before collapse of a
  marine benthic ecosystem.
\newblock {\em Nature ecology \& evolution}, 1(6):1--7, 2017.

\bibitem{clements2018indicators}
Christopher~F Clements and Arpat Ozgul.
\newblock Indicators of transitions in biological systems.
\newblock {\em Ecology letters}, 21(6):905--919, 2018.

\bibitem{clements2016including}
Christopher~F Clements and Arpat Ozgul.
\newblock Including trait-based early warning signals helps predict population
  collapse.
\newblock {\em Nature communications}, 7:10984, 2016.

\bibitem{berghof2019body}
Tom Berghof, Henk Bovenhuis, and Han Mulder.
\newblock Body weight deviations as indicator for resilience in layer chickens.
\newblock {\em Frontiers in Genetics}, 10:1216, 2019.

\bibitem{dakos2019ecosystem}
Vasilis Dakos, Blake Matthews, Andrew~P Hendry, Jonathan Levine, Nicolas
  Loeuille, Jon Norberg, Patrik Nosil, Marten Scheffer, and Luc De~Meester.
\newblock Ecosystem tipping points in an evolving world.
\newblock {\em Nature ecology \& evolution}, page~1, 2019.

\bibitem{trefois2015critical}
Christophe Trefois, Paul~MA Antony, Jorge Goncalves, Alexander Skupin, and Rudi
  Balling.
\newblock Critical transitions in chronic disease: transferring concepts from
  ecology to systems medicine.
\newblock {\em Current opinion in biotechnology}, 34:48--55, 2015.

\bibitem{balling2019diagnosing}
Rudi Balling, Jorge Goncalves, Stefano Magni, Laurent Mombaerts, Alice Oldano,
  and Alexander Skupin.
\newblock From diagnosing diseases to predicting diseases.
\newblock In {\em Curious2018}, pages 95--103. Springer, 2019.

\bibitem{rikkert2016slowing}
Olde Rikkert, GM~Marcel, Vasilis Dakos, Timothy~G Buchman, Rob~de Boer, Leon
  Glass, Ang{\'e}lique~OJ Cramer, Simon Levin, Egbert van Nes, George Sugihara,
  et~al.
\newblock Slowing down of recovery as generic risk marker for acute severity
  transitions in chronic diseases.
\newblock {\em Critical care medicine}, 44(3):601--606, 2016.

\bibitem{namazi2016signal}
Hamidreza Namazi, Vladimir~V Kulish, Jamal Hussaini, Jalal Hussaini, Ali
  Delaviz, Fatemeh Delaviz, Shaghayegh Habibi, and Sara Ramezanpoor.
\newblock A signal processing based analysis and prediction of seizure onset in
  patients with epilepsy.
\newblock {\em Oncotarget}, 7(1):342, 2016.

\bibitem{holmes2019attitudes}
Emily Holmes, Siobhan Bourke, and Catrin Plumpton.
\newblock Attitudes towards epilepsy in the uk population: Results from a 2018
  national survey.
\newblock {\em Seizure}, 65:12--19, 2019.

\bibitem{meisel2012scaling}
Christian Meisel and Christian Kuehn.
\newblock Scaling effects and spatio-temporal multilevel dynamics in epileptic
  seizures.
\newblock {\em PLoS One}, 7(2), 2012.

\bibitem{mormann2016seizure}
Florian Mormann and Ralph~G Andrzejak.
\newblock Seizure prediction: making mileage on the long and winding road.
\newblock {\em Brain}, 139(6):1625--1627, 2016.

\bibitem{wilkat2019no}
Theresa Wilkat, Thorsten Rings, and Klaus Lehnertz.
\newblock No evidence for critical slowing down prior to human epileptic
  seizures.
\newblock {\em Chaos: An Interdisciplinary Journal of Nonlinear Science},
  29(9):091104, 2019.

\bibitem{kramer2012human}
Mark~A Kramer, Wilson Truccolo, Uri~T Eden, Kyle~Q Lepage, Leigh~R Hochberg,
  Emad~N Eskandar, Joseph~R Madsen, Jong~W Lee, Atul Maheshwari, Eric Halgren,
  et~al.
\newblock Human seizures self-terminate across spatial scales via a critical
  transition.
\newblock {\em Proceedings of the National Academy of Sciences},
  109(51):21116--21121, 2012.

\bibitem{borsboom2013network}
Denny Borsboom and Ang{\'e}lique~OJ Cramer.
\newblock Network analysis: an integrative approach to the structure of
  psychopathology.
\newblock {\em Annual review of clinical psychology}, 9:91--121, 2013.

\bibitem{bringmann2013network}
Laura~F Bringmann, Nathalie Vissers, Marieke Wichers, Nicole Geschwind, Peter
  Kuppens, Frenk Peeters, Denny Borsboom, and Francis Tuerlinckx.
\newblock A network approach to psychopathology: new insights into clinical
  longitudinal data.
\newblock {\em PloS one}, 8(4):e60188, 2013.

\bibitem{van2014critical}
Ingrid~A van~de Leemput, Marieke Wichers, Ang{\'e}lique~OJ Cramer, Denny
  Borsboom, Francis Tuerlinckx, Peter Kuppens, Egbert~H van Nes, Wolfgang
  Viechtbauer, Erik~J Giltay, Steven~H Aggen, et~al.
\newblock Critical slowing down as early warning for the onset and termination
  of depression.
\newblock {\em Proceedings of the National Academy of Sciences}, 111(1):87--92,
  2014.

\bibitem{wichers2016critical}
Marieke Wichers, Peter~C Groot, ESM Psychosystems, EWS Group, et~al.
\newblock Critical slowing down as a personalized early warning signal for
  depression.
\newblock {\em Psychotherapy and psychosomatics}, 85(2):114--116, 2016.

\bibitem{quail2015predicting}
Thomas Quail, Alvin Shrier, and Leon Glass.
\newblock Predicting the onset of period-doubling bifurcations in noisy cardiac
  systems.
\newblock {\em Proceedings of the National Academy of Sciences},
  112(30):9358--9363, 2015.

\bibitem{hsieh2014changing}
Nan-Hung Hsieh, Yi-Hsien Cheng, and Chung-Min Liao.
\newblock Changing variance and skewness as leading indicators for detecting
  ozone exposure-associated lung function decrement.
\newblock {\em Stochastic environmental research and risk assessment},
  28(8):2205--2216, 2014.

\bibitem{tambuyzer2014interleukin}
Tim Tambuyzer, Tine De~Waele, Koen Chiers, Daniel Berckmans, Bruno~M Goddeeris,
  and Jean-Marie Aerts.
\newblock Interleukin-6 dynamics as a basis for an early-warning monitor for
  sepsis and inflammation in individual pigs.
\newblock {\em Research in veterinary science}, 96(3):460--463, 2014.

\bibitem{chen2012detecting}
Luonan Chen, Rui Liu, Zhi-Ping Liu, Meiyi Li, and Kazuyuki Aihara.
\newblock Detecting early-warning signals for sudden deterioration of complex
  diseases by dynamical network biomarkers.
\newblock {\em Scientific reports}, 2:342, 2012.

\bibitem{stuart2003gene}
Joshua~M Stuart, Eran Segal, Daphne Koller, and Stuart~K Kim.
\newblock A gene-coexpression network for global discovery of conserved genetic
  modules.
\newblock {\em science}, 302(5643):249--255, 2003.

\bibitem{jeong2014overexpression}
Hae~Min Jeong, Mi~Jeong Kwon, and Young~Kee Shin.
\newblock Overexpression of cancer-associated genes via epigenetic derepression
  mechanisms in gynecologic cancer.
\newblock {\em Frontiers in oncology}, 4:12, 2014.

\bibitem{zhu2019identification}
Sha Zhu, Lili Jiang, Liuyan Wang, Lingli Wang, Cong Zhang, Yu~Ma, and Tao
  Huang.
\newblock Identification of key genes and specific pathways potentially
  involved in androgen-independent, mitoxantrone-resistant prostate cancer.
\newblock {\em Cancer management and research}, 11:419, 2019.

\bibitem{liu2013dynamical}
Rui Liu, Kazuyuki Aihara, and Luonan Chen.
\newblock Dynamical network biomarkers for identifying critical transitions and
  their driving networks of biologic processes.
\newblock {\em Quantitative Biology}, 1(2):105--114, 2013.

\bibitem{liu2019single}
Rui Liu, Pei Chen, and Luonan Chen.
\newblock Single-sample landscape entropy reveals the imminent phase transition
  during disease progression.
\newblock {\em Bioinformatics}, 2019.

\bibitem{atkinson1999nod}
Mark~A Atkinson and Edward~H Leiter.
\newblock The nod mouse model of type 1 diabetes: as good as it gets?
\newblock {\em Nature medicine}, 5(6):601, 1999.

\bibitem{hayden2002islet}
Melvin~R Hayden.
\newblock Islet amyloid, metabolic syndrome, and the natural progressive
  history of type 2 diabetes mellitus.
\newblock {\em Jop}, 3(5):126--38, 2002.

\bibitem{li2013detecting}
Meiyi Li, Tao Zeng, Rui Liu, and Luonan Chen.
\newblock Detecting tissue-specific early warning signals for complex diseases
  based on dynamical network biomarkers: study of type 2 diabetes by
  cross-tissue analysis.
\newblock {\em Briefings in bioinformatics}, 15(2):229--243, 2013.

\bibitem{liu2013detecting}
Xiaoping Liu, Rui Liu, Xing-Ming Zhao, and Luonan Chen.
\newblock Detecting early-warning signals of type 1 diabetes and its leading
  biomolecular networks by dynamical network biomarkers.
\newblock {\em BMC medical genomics}, 6(2):S8, 2013.

\bibitem{zeng2014deciphering}
Tao Zeng, Chuan-chao Zhang, Wanwei Zhang, Rui Liu, Juan Liu, and Luonan Chen.
\newblock Deciphering early development of complex diseases by progressive
  module network.
\newblock {\em Methods}, 67(3):334--343, 2014.

\bibitem{chen2019detecting}
Pei Chen, Ely Chen, Luonan Chen, Xianghong~Jasmine Zhou, and Rui Liu.
\newblock Detecting early-warning signals of influenza outbreak based on
  dynamic network marker.
\newblock {\em Journal of cellular and molecular medicine}, 23(1):395--404,
  2019.

\bibitem{liu2018hunt}
Rui Liu, Jinzeng Wang, Masao Ukai, Ki~Sewon, Pei Chen, Yutaka Suzuki, Haiyun
  Wang, Kazuyuki Aihara, Mariko Okada-Hatakeyama, and Luonan Chen.
\newblock Hunt for the tipping point during endocrine resistance process in
  breast cancer by dynamic network biomarkers.
\newblock {\em Journal of molecular cell biology}, 2018.

\bibitem{liu2017quantifying}
Xiaoping Liu, Xiao Chang, Rui Liu, Xiangtian Yu, Luonan Chen, and Kazuyuki
  Aihara.
\newblock Quantifying critical states of complex diseases using single-sample
  dynamic network biomarkers.
\newblock {\em PLoS computational biology}, 13(7):e1005633, 2017.

\bibitem{yang2018dynamic}
Biwei Yang, Meiyi Li, Wenqing Tang, Weixin Liu, Si~Zhang, Luonan Chen, and
  Jinglin Xia.
\newblock Dynamic network biomarker indicates pulmonary metastasis at the
  tipping point of hepatocellular carcinoma.
\newblock {\em Nature communications}, 9(1):678, 2018.

\bibitem{mojtahedi2016cell}
Mitra Mojtahedi, Alexander Skupin, Joseph Zhou, Ivan~G Casta{\~n}o, Rebecca~YY
  Leong-Quong, Hannah Chang, Kalliopi Trachana, Alessandro Giuliani, and Sui
  Huang.
\newblock Cell fate decision as high-dimensional critical state transition.
\newblock {\em PLoS biology}, 14(12):e2000640, 2016.

\bibitem{buganim2012single}
Yosef Buganim, Dina~A Faddah, Albert~W Cheng, Elena Itskovich, Styliani
  Markoulaki, Kibibi Ganz, Sandy~L Klemm, Alexander van Oudenaarden, and Rudolf
  Jaenisch.
\newblock Single-cell expression analyses during cellular reprogramming reveal
  an early stochastic and a late hierarchic phase.
\newblock {\em Cell}, 150(6):1209--1222, 2012.

\bibitem{richard2016single}
Ang{\'e}lique Richard, Lo{\"\i}s Boullu, Ulysse Herbach, Arnaud Bonnafoux,
  Val{\'e}rie Morin, Elodie Vallin, Anissa Guillemin, Nan~Papili Gao, Rudiyanto
  Gunawan, J{\'e}r{\'e}mie Cosette, et~al.
\newblock Single-cell-based analysis highlights a surge in cell-to-cell
  molecular variability preceding irreversible commitment in a differentiation
  process.
\newblock {\em PLoS biology}, 14(12):e1002585, 2016.

\bibitem{lesterhuis2017dynamic}
W~Joost Lesterhuis, Anthony Bosco, Michael~J Millward, Michael Small, Anna~K
  Nowak, and Richard~A Lake.
\newblock Dynamic versus static biomarkers in cancer immune checkpoint
  blockade: unravelling complexity.
\newblock {\em Nature Reviews Drug Discovery}, 16(4):264, 2017.

\bibitem{craciun2006understanding}
Gheorghe Craciun, Yangzhong Tang, and Martin Feinberg.
\newblock Understanding bistability in complex enzyme-driven reaction networks.
\newblock {\em Proceedings of the National Academy of Sciences},
  103(23):8697--8702, 2006.

\bibitem{Heidler2014relationship}
Richard Heidler, Marcus. Gamper, Andreas Herz, and Florian E.
\newblock Relationship patterns in the 19th century: The friendship network in
  a german boys’ school class from 1880 to 1881 revisited.
\newblock {\em Sociometry}, 27:1--13, 2014.

\bibitem{Moreno1938statistics}
Jacob~L. Moreno and Jennings. H.H.
\newblock Statistics of social configurations.
\newblock {\em Sociometry}, 1:342--373, 1938.

\bibitem{Freeman2004development}
Linton~C Freeman.
\newblock {\em The The Development of Social Network Analysis: A Study in the
  Sociology of Science}.
\newblock Empirical Press, Vancuver BC, 2004.

\bibitem{Moreno1932application}
Jacob~L Moreno.
\newblock {\em Application of the group method to classification}.
\newblock National Committee on Prisons and Prison Labor, New York, NY, 2014.

\bibitem{Newman1999scaling}
M.~E.~J. Newman and D.~J Watts.
\newblock Scaling and percolation in the small-world network model.
\newblock {\em Phys. A Stat. Mech. Its Appl}, 310:7332--7342, Dec 1999.

\bibitem{peng2015collective}
Hao Peng, Dandan Zhao, Xueming Liu, and Jianxi Gao.
\newblock Collective motion in a network of self-propelled agent systems.
\newblock {\em PloS one}, 10(12), 2015.

\bibitem{lazer2009life}
David Lazer, D~Brewer, N~Christakis, J~Fowler, and G~King.
\newblock Life in the network: the coming age of computational social.
\newblock {\em Science}, 323(5915):721--723, 2009.

\bibitem{magis2010community}
Kristen Magis.
\newblock Community resilience: An indicator of social sustainability.
\newblock {\em Society and Natural Resources}, 23(5):401--416, 2010.

\bibitem{aldrich2015social}
Daniel~P Aldrich and Michelle~A Meyer.
\newblock Social capital and community resilience.
\newblock {\em American behavioral scientist}, 59(2):254--269, 2015.

\bibitem{cutter2014geographies}
Susan~L Cutter, Kevin~D Ash, and Christopher~T Emrich.
\newblock The geographies of community disaster resilience.
\newblock {\em Global environmental change}, 29:65--77, 2014.

\bibitem{nardini2008s}
Cecilia Nardini, Bal{\'a}zs Kozma, and Alain Barrat.
\newblock Who’s talking first? consensus or lack thereof in coevolving
  opinion formation models.
\newblock {\em Physical review letters}, 100(15):158701, 2008.

\bibitem{xie2011social}
Jierui Xie, Sameet Sreenivasan, Gyorgy Korniss, Weituo Zhang, Chjan Lim, and
  Boleslaw~K Szymanski.
\newblock Social consensus through the influence of committed minorities.
\newblock {\em Physical Review E}, 84(1):011130, 2011.

\bibitem{granovetter1978threshold}
Mark Granovetter.
\newblock Threshold models of collective behavior.
\newblock {\em American journal of sociology}, 83(6):1420--1443, 1978.

\bibitem{komareji2013resilience}
Mohammad Komareji and Roland Bouffanais.
\newblock Resilience and controllability of dynamic collective behaviors.
\newblock {\em PLoS one}, 8(12), 2013.

\bibitem{lu2009naming}
Qiming Lu, Gyorgy Korniss, and Boleslaw~K Szymanski.
\newblock The naming game in social networks: community formation and consensus
  engineering.
\newblock {\em Journal of Economic Interaction and Coordination}, 4(2):221,
  2009.

\bibitem{xie2012evolution}
Jierui Xie, Jeffrey Emenheiser, Matthew Kirby, Sameet Sreenivasan, Boleslaw~K
  Szymanski, and Gyorgy Korniss.
\newblock Evolution of opinions on social networks in the presence of competing
  committed groups.
\newblock {\em PLoS One}, 7(3), 2012.

\bibitem{pickering2016analysis}
William Pickering, Boleslaw~K Szymanski, and Chjan Lim.
\newblock Analysis of the high-dimensional naming game with committed
  minorities.
\newblock {\em Physical Review E}, 93(5):052311, 2016.

\bibitem{mobilia2007role}
Mauro Mobilia, Anna Petersen, and Sidney Redner.
\newblock On the role of zealotry in the voter model.
\newblock {\em Journal of Statistical Mechanics: Theory and Experiment},
  2007(08):P08029, 2007.

\bibitem{singh2016competing}
Pramesh Singh, Sameet Sreenivasan, Boleslaw~K Szymanski, and Gyorgy Korniss.
\newblock Competing effects of social balance and influence.
\newblock {\em Physical Review E}, 93(4):042306, 2016.

\bibitem{bryan2005fostering}
Julia Bryan.
\newblock Fostering educational resilience and achievement in urban schools
  through school-family-community partnerships.
\newblock {\em Professional School Counseling}, pages 219--227, 2005.

\bibitem{almedom2005social}
Astier~M Almedom.
\newblock Social capital and mental health: An interdisciplinary review of
  primary evidence.
\newblock {\em Social science \& medicine}, 61(5):943--964, 2005.

\bibitem{eakin2017opinion}
Hallie Eakin, Luis~A Boj{\'o}rquez-Tapia, Marco~A Janssen, Matei Georgescu,
  David Manuel-Navarrete, Enrique~R Vivoni, Ana~E Escalante, Andres
  Baeza-Castro, M~Mazari-Hiriart, and Amy~M Lerner.
\newblock Opinion: urban resilience efforts must consider social and political
  forces.
\newblock {\em Proceedings of the National Academy of Sciences},
  114(2):186--189, 2017.

\bibitem{christakis2013social}
Nicholas~A Christakis and James~H Fowler.
\newblock Social contagion theory: examining dynamic social networks and human
  behavior.
\newblock {\em Statistics in medicine}, 32(4):556--577, 2013.

\bibitem{dynes2005community}
Russell~R Dynes.
\newblock Community social capital as the primary basis for resilience.
\newblock 2005.

\bibitem{baronchelli2008depth}
Andrea Baronchelli, Vittorio Loreto, and Luc Steels.
\newblock In-depth analysis of the naming game dynamics: the homogeneous mixing
  case.
\newblock {\em International Journal of Modern Physics C}, 19(05):785--812,
  2008.

\bibitem{meng2019event}
Haofei Meng, Hai-Tao Zhang, Zhen Wang, and Guanrong Chen.
\newblock Event-triggered control for semi-global robust consensus of a class
  of nonlinear uncertain multi-agent systems.
\newblock {\em IEEE Transactions on Automatic Control}, 2019.

\bibitem{liu2019collective}
Bin Liu, Zhiyong Chen, Hai-Tao Zhang, Xudong Wang, Tao Geng, Housheng Su, and
  Jin Zhao.
\newblock Collective dynamics and control for multiple unmanned surface
  vessels.
\newblock {\em IEEE Transactions on Control Systems Technology},
  28(6):2540--2547, 2019.

\bibitem{gao2014naming}
Yuan Gao, Guanrong Chen, and Rosa~HM Chan.
\newblock Naming game on networks: let everyone be both speaker and hearer.
\newblock {\em Scientific reports}, 4:6149, 2014.

\bibitem{zhang2011social}
W~Zhang, C~Lim, Sameet Sreenivasan, Jierui Xie, Boleslaw~K Szymanski, and
  Gyorgy Korniss.
\newblock Social influencing and associated random walk models: Asymptotic
  consensus times on the complete graph.
\newblock {\em Chaos: An Interdisciplinary Journal of Nonlinear Science},
  21(2):025115, 2011.

\bibitem{zhang2012analytic}
Weituo Zhang, Chjan Lim, and Boleslaw~K Szymanski.
\newblock Analytic treatment of tipping points for social consensus in large
  random networks.
\newblock {\em Physical Review E}, 86(6):061134, 2012.

\bibitem{zhang2014opinion}
Weituo Zhang, Chjan~C Lim, Gyorgy Korniss, and Boleslaw~K Szymanski.
\newblock Opinion dynamics and influencing on random geometric graphs.
\newblock {\em Scientific reports}, 4(1):1--9, 2014.

\bibitem{doyle2017effects}
Casey Doyle, Boleslaw~K Szymanski, and Gyorgy Korniss.
\newblock Effects of communication burstiness on consensus formation and
  tipping points in social dynamics.
\newblock {\em Physical Review E}, 95(6):062303, 2017.

\bibitem{niu2017impact}
Xiang Niu, Casey Doyle, Gyorgy Korniss, and Boleslaw~K Szymanski.
\newblock The impact of variable commitment in the naming game on consensus
  formation.
\newblock {\em Scientific reports}, 7(1):1--11, 2017.

\bibitem{thompson2014propensity}
Andrew~M Thompson, Boleslaw~K Szymanski, and Chjan~C Lim.
\newblock Propensity and stickiness in the naming game: Tipping fractions of
  minorities.
\newblock {\em Physical review E}, 90(4):042809, 2014.

\bibitem{marvel2012encouraging}
Seth~A Marvel, Hyunsuk Hong, Anna Papush, and Steven~H Strogatz.
\newblock Encouraging moderation: clues from a simple model of ideological
  conflict.
\newblock {\em Physical review letters}, 109(11):118702, 2012.

\bibitem{couzin2011uninformed}
Iain~D Couzin, Christos~C Ioannou, G{\"u}ven Demirel, Thilo Gross, Colin~J
  Torney, Andrew Hartnett, Larissa Conradt, Simon~A Levin, and Naomi~E Leonard.
\newblock Uninformed individuals promote democratic consensus in animal groups.
\newblock {\em science}, 334(6062):1578--1580, 2011.

\bibitem{doyle2016social}
Casey Doyle, Sameet Sreenivasan, Boleslaw~K Szymanski, and Gyorgy Korniss.
\newblock Social consensus and tipping points with opinion inertia.
\newblock {\em Physica A: Statistical Mechanics and its Applications},
  443:316--323, 2016.

\bibitem{holme2006nonequilibrium}
Petter Holme and Mark~EJ Newman.
\newblock Nonequilibrium phase transition in the coevolution of networks and
  opinions.
\newblock {\em Physical Review E}, 74(5):056108, 2006.

\bibitem{singh2013threshold}
Pramesh Singh, Sameet Sreenivasan, Boleslaw~K Szymanski, and Gyorgy Korniss.
\newblock Threshold-limited spreading in social networks with multiple
  initiators.
\newblock {\em Scientific reports}, 3:2330, 2013.

\bibitem{Jankowski2017Balancing}
Jaros{\l}aw Jankowski, Piotr Br{\'o}dka, Przemys{\l}aw Kazienko, Boleslaw~K
  Szymanski, Rados{\l}aw Michalski, and Tomasz Kajdanowicz.
\newblock Balancing speed and coverage by sequential seeding in complex
  networks.
\newblock {\em Scientific reports}, 7(1):1--11, 2017.

\bibitem{Jankowski2018Probing}
Jaros{\l}aw Jankowski, Boleslaw~K Szymanski, Przemys{\l}aw Kazienko,
  Rados{\l}aw Michalski, and Piotr Br{\'o}dka.
\newblock Probing limits of information spread with sequential seeding.
\newblock {\em Scientific reports}, 8(1):1--9, 2018.

\bibitem{karampourniotis2015impact}
Panagiotis~D Karampourniotis, Sameet Sreenivasan, Boleslaw~K Szymanski, and
  Gyorgy Korniss.
\newblock The impact of heterogeneous thresholds on social contagion with
  multiple initiators.
\newblock {\em PloS one}, 10(11), 2015.

\bibitem{Sood2005voter}
V.~Sood and S~Redner.
\newblock Voter model on heterogeneous graphs.
\newblock {\em Phys. Rev. Lett}, 94:178701, May 2005.

\bibitem{Axelrod1997dissemination}
R~Axelrod.
\newblock The dissemination of culture: A model with local convergence and
  global polarization.
\newblock {\em The Journal of Conflict Resolution}, 41:203, 1997.

\bibitem{singh2012accelerating}
Pramesh Singh, Sameet Sreenivasan, Boleslaw~K Szymanski, and Gyorgy Korniss.
\newblock Accelerating consensus on coevolving networks: The effect of
  committed individuals.
\newblock {\em Physical Review E}, 85(4):046104, 2012.

\bibitem{beekman2001phase}
Madeleine Beekman, David~JT Sumpter, and Francis~LW Ratnieks.
\newblock Phase transition between disordered and ordered foraging in pharaoh's
  ants.
\newblock {\em Proceedings of the National Academy of Sciences},
  98(17):9703--9706, 2001.

\bibitem{toffin2009shape}
Etienne Toffin, David Di~Paolo, Alexandre Campo, Claire Detrain, and Jean-Louis
  Deneubourg.
\newblock Shape transition during nest digging in ants.
\newblock {\em Proceedings of the National Academy of Sciences},
  106(44):18616--18620, 2009.

\bibitem{doering2018social}
Grant~Navid Doering, Inon Scharf, Holly~V Moeller, and Jonathan~N Pruitt.
\newblock Social tipping points in animal societies in response to heat stress.
\newblock {\em Nature ecology \& evolution}, 2(8):1298--1305, 2018.

\bibitem{middleton2016resilience}
Eliza~JT Middleton and Tanya Latty.
\newblock Resilience in social insect infrastructure systems.
\newblock {\em Journal of The Royal Society Interface}, 13(116):20151022, 2016.

\bibitem{loengarov2008phase}
Andreas Loengarov and Valery Tereshko.
\newblock Phase transitions and bistability in honeybee foraging dynamics.
\newblock {\em Artificial life}, 14(1):111--120, 2008.

\bibitem{wood2019evolving}
Connor Wood, Robert~NL Fitt, and Lesley~T Lancaster.
\newblock Evolving social dynamics prime thermal tolerance during a poleward
  range shift.
\newblock {\em Biological Journal of the Linnean Society}, 126(3):574--586,
  2019.

\bibitem{phillips2004maximum}
Steven~J Phillips, Miroslav Dud{\'\i}k, and Robert~E Schapire.
\newblock A maximum entropy approach to species distribution modeling.
\newblock In {\em Proceedings of the twenty-first international conference on
  Machine learning}, page~83, 2004.

\bibitem{pruitt2018social}
Jonathan~N Pruitt, Andrew Berdahl, Christina Riehl, Noa Pinter-Wollman, Holly~V
  Moeller, Elizabeth~G Pringle, Lucy~M Aplin, Elva~JH Robinson, Jacopo Grilli,
  Pamela Yeh, et~al.
\newblock Social tipping points in animal societies.
\newblock {\em Proceedings of the Royal Society B: Biological Sciences},
  285(1887):20181282, 2018.

\bibitem{stokols2013enhancing}
Daniel Stokols, Raul~Perez Lejano, and John Hipp.
\newblock Enhancing the resilience of human--environment systems: A social
  ecological perspective.
\newblock {\em Ecology and Society}, 18(1), 2013.

\bibitem{folke2006resilience}
Carl Folke.
\newblock Resilience: The emergence of a perspective for social--ecological
  systems analyses.
\newblock {\em Global environmental change}, 16(3):253--267, 2006.

\bibitem{olsson2015resilience}
Lennart Olsson, Anne Jerneck, Henrik Thoren, Johannes Persson, and David
  O’Byrne.
\newblock Why resilience is unappealing to social science: Theoretical and
  empirical investigations of the scientific use of resilience.
\newblock {\em Science advances}, 1(4):e1400217, 2015.

\bibitem{dutting2010building}
Gisela D{\"u}tting and David Sogge.
\newblock Building safety nets in the global politic: Ngo collaboration for
  solidarity and sustainability.
\newblock {\em Development}, 53(3):350--355, 2010.

\bibitem{gittell2006relationships}
Jody~Hoffer Gittell, Kim Cameron, Sandy Lim, and Victor Rivas.
\newblock Relationships, layoffs, and organizational resilience: Airline
  industry responses to september 11.
\newblock {\em The Journal of Applied Behavioral Science}, 42(3):300--329,
  2006.

\bibitem{marshall2007resource}
Nadine~A Marshall, D~Mark Fenton, Paul~A Marshall, and Steven~G Sutton.
\newblock How resource dependency can influence social resilience within a
  primary resource industry.
\newblock {\em Rural Sociology}, 72(3):359--390, 2007.

\bibitem{giannone2011market}
Domenico Giannone, Michele Lenza, and Lucrezia Reichlin.
\newblock Market freedom and the global recession.
\newblock {\em IMF Economic Review}, 59(1):111--135, 2011.

\bibitem{ostrom2009general}
Elinor Ostrom.
\newblock A general framework for analyzing sustainability of social-ecological
  systems.
\newblock {\em Science}, 325(5939):419--422, 2009.

\bibitem{walker2004resilience}
Brian Walker, Crawford~S Holling, Stephen~R Carpenter, and Ann Kinzig.
\newblock Resilience, adaptability and transformability in social--ecological
  systems.
\newblock {\em Ecology and society}, 9(2), 2004.

\bibitem{anderies2004framework}
John~M Anderies, Marco~A Janssen, and Elinor Ostrom.
\newblock A framework to analyze the robustness of social-ecological systems
  from an institutional perspective.
\newblock {\em Ecology and society}, 9(1), 2004.

\bibitem{lade2013regime}
Steven~J Lade, Alessandro Tavoni, Simon~A Levin, and Maja Schl{\"u}ter.
\newblock Regime shifts in a social-ecological system.
\newblock {\em Theoretical ecology}, 6(3):359--372, 2013.

\bibitem{suweis2014early}
Samir Suweis and Paolo D'Odorico.
\newblock Early warning signs in social-ecological networks.
\newblock {\em PloS one}, 9(7), 2014.

\bibitem{filatova2016regime}
Tatiana Filatova, J~Gary Polhill, and Stijn Van~Ewijk.
\newblock Regime shifts in coupled socio-environmental systems: review of
  modelling challenges and approaches.
\newblock {\em Environmental modelling \& software}, 75:333--347, 2016.

\bibitem{sugiarto2015socioecological}
Hendrik~Santoso Sugiarto, Ning~Ning Chung, Choy~Heng Lai, and Lock~Yue Chew.
\newblock Socioecological regime shifts in the setting of complex social
  interactions.
\newblock {\em Physical Review E}, 91(6):062804, 2015.

\bibitem{scheffer2018quantifying}
Marten Scheffer, J~Elizabeth Bolhuis, Denny Borsboom, Timothy~G Buchman,
  Sanne~MW Gijzel, Dave Goulson, Jan~E Kammenga, Bas Kemp, Ingrid~A van~de
  Leemput, Simon Levin, et~al.
\newblock Quantifying resilience of humans and other animals.
\newblock {\em Proceedings of the National Academy of Sciences},
  115(47):11883--11890, 2018.

\bibitem{kanervarole}
Minna~Marjatta Kanerva.
\newblock {\em The role of discourses in a transformation of social practices
  towards sustainability : The case of meat eating related practices}.
\newblock PhD thesis, University of Bremen, Germany, 2019.

\bibitem{amin2000toward}
Massoud Amin.
\newblock Toward self-healing infrastructure systems.
\newblock {\em Computer}, (8):44--53, 2000.

\bibitem{genge2015system}
B{\'e}la Genge, Istv{\'a}n Kiss, and Piroska Haller.
\newblock A system dynamics approach for assessing the impact of cyber attacks
  on critical infrastructures.
\newblock {\em International Journal of Critical Infrastructure Protection},
  10:3--17, 2015.

\bibitem{rinaldi2001identifying}
Steven~M Rinaldi, James~P Peerenboom, and Terrence~K Kelly.
\newblock Identifying, understanding, and analyzing critical infrastructure
  interdependencies.
\newblock {\em IEEE control systems magazine}, 21(6):11--25, 2001.

\bibitem{headey2015opinion}
Derek Headey and Christopher~B Barrett.
\newblock Opinion: Measuring development resilience in the world’s poorest
  countries.
\newblock {\em Proceedings of the National Academy of Sciences},
  112(37):11423--11425, 2015.

\bibitem{barrett2014toward}
Christopher~B Barrett and Mark~A Constas.
\newblock Toward a theory of resilience for international development
  applications.
\newblock {\em Proceedings of the National Academy of Sciences},
  111(40):14625--14630, 2014.

\bibitem{kastenberg2005assessing}
WE~Kastenberg.
\newblock Assessing and managing the security of complex systems: Shifting the
  rams paradigm.
\newblock {\em System analysis for a more secure world: Application of system
  analysis and RAMS to security of complex systems}, 2005.

\bibitem{kauffman1993origins}
Stuart~A Kauffman.
\newblock {\em The origins of order: Self-organization and selection in
  evolution}.
\newblock OUP USA, 1993.

\bibitem{albert2002statistical}
R{\'e}ka Albert and Albert-L{\'a}szl{\'o} Barab{\'a}si.
\newblock Statistical mechanics of complex networks.
\newblock {\em Reviews of modern physics}, 74(1):47, 2002.

\bibitem{riffonneau2011optimal}
Yann Riffonneau, Seddik Bacha, Franck Barruel, and Stephane Ploix.
\newblock Optimal power flow management for grid connected pv systems with
  batteries.
\newblock {\em IEEE Transactions on sustainable energy}, 2(3):309--320, 2011.

\bibitem{albert2004structural}
R{\'e}ka Albert, Istv{\'a}n Albert, and Gary~L Nakarado.
\newblock Structural vulnerability of the {N}orth {A}merican power grid.
\newblock {\em Physical review E}, 69(2):025103, 2004.

\bibitem{li2015percolation}
Daqing Li, Bowen Fu, Yunpeng Wang, Guangquan Lu, Yehiel Berezin, H~Eugene
  Stanley, and Shlomo Havlin.
\newblock Percolation transition in dynamical traffic network with evolving
  critical bottlenecks.
\newblock {\em Proceedings of the National Academy of Sciences},
  112(3):669--672, 2015.

\bibitem{song2005self}
Chaoming Song, Shlomo Havlin, and Hernan~A Makse.
\newblock Self-similarity of complex networks.
\newblock {\em Nature}, 433(7024):392, 2005.

\bibitem{wang2019local}
Weiping Wang, Saini Yang, H~Eugene Stanley, and Jianxi Gao.
\newblock Local floods induce large-scale abrupt failures of road networks.
\newblock {\em Nature communications}, 10(1):2114, 2019.

\bibitem{barthelemy2014spatial}
Marc Barth{\'e}lemy.
\newblock {\em Spatial networks}.
\newblock Springer, 2014.

\bibitem{daqing2011dimension}
Li~Daqing, Kosmas Kosmidis, Armin Bunde, and Shlomo Havlin.
\newblock Dimension of spatially embedded networks.
\newblock {\em Nature Physics}, 7(6):481, 2011.

\bibitem{boguna2010sustaining}
Mari{\'a}n Bogun{\'a}, Fragkiskos Papadopoulos, and Dmitri Krioukov.
\newblock Sustaining the internet with hyperbolic mapping.
\newblock {\em Nature Communications}, 1:62, 2010.

\bibitem{bianconi2009assessing}
Ginestra Bianconi, Paolo Pin, and Matteo Marsili.
\newblock Assessing the relevance of node features for network structure.
\newblock {\em Proceedings of the National Academy of Sciences},
  106(28):11433--11438, 2009.

\bibitem{dong2020measuring}
Shangjia Dong, Alireza Mostafizi, Haizhong Wang, Jianxi Gao, and Xiaopeng Li.
\newblock Measuring the topological robustness of transportation networks to
  disaster-induced failures: A percolation approach.
\newblock {\em Journal of Infrastructure Systems}, 26(2):04020009, 2020.

\bibitem{sen2003small}
Parongama Sen, Subinay Dasgupta, Arnab Chatterjee, PA~Sreeram, G~Mukherjee, and
  SS~Manna.
\newblock Small-world properties of the indian railway network.
\newblock {\em Physical Review E}, 67(3):036106, 2003.

\bibitem{hu2009empirical}
Yihong Hu and Daoli Zhu.
\newblock Empirical analysis of the worldwide maritime transportation network.
\newblock {\em Physica A: Statistical Mechanics and its Applications},
  388(10):2061--2071, 2009.

\bibitem{watts1998collective}
Duncan~J Watts and Steven~H Strogatz.
\newblock Collective dynamics of ‘small-world’ networks.
\newblock {\em Nature}, 393(6684):440, 1998.

\bibitem{travers1977experimental}
Jeffrey Travers and Stanley Milgram.
\newblock An experimental study of the small world problem.
\newblock In {\em Social Networks}, pages 179--197. Elsevier, 1977.

\bibitem{amaral2000classes}
Lu{\i}s A~Nunes Amaral, Antonio Scala, Marc Barthelemy, and H~Eugene Stanley.
\newblock Classes of small-world networks.
\newblock {\em Proceedings of the National Academy of Sciences},
  97(21):11149--11152, 2000.

\bibitem{latora2001efficient}
Vito Latora and Massimo Marchiori.
\newblock Efficient behavior of small-world networks.
\newblock {\em Physical review letters}, 87(19):198701, 2001.

\bibitem{kleinberg2000navigation}
Jon~M Kleinberg.
\newblock Navigation in a small world.
\newblock {\em Nature}, 406(6798):845, 2000.

\bibitem{zeng2019switch}
Guanwen Zeng, Daqing Li, Shengmin Guo, Liang Gao, Ziyou Gao, H~Eugene Stanley,
  and Shlomo Havlin.
\newblock Switch between critical percolation modes in city traffic dynamics.
\newblock {\em Proceedings of the National Academy of Sciences}, 116(1):23--28,
  2019.

\bibitem{guimera2005worldwide}
Roger Guimera, Stefano Mossa, Adrian Turtschi, and LA~Nunes Amaral.
\newblock The worldwide air transportation network: Anomalous centrality,
  community structure, and cities' global roles.
\newblock {\em Proceedings of the National Academy of Sciences},
  102(22):7794--7799, 2005.

\bibitem{newman2006modularity}
Mark~EJ Newman.
\newblock Modularity and community structure in networks.
\newblock {\em Proceedings of the National Academy of Sciences},
  103(23):8577--8582, 2006.

\bibitem{babaei2011cascading}
Mahmoudreza Babaei, Hamed Ghassemieh, and Mahdi Jalili.
\newblock Cascading failure tolerance of modular small-world networks.
\newblock {\em IEEE Transactions on Circuits and Systems II: Express Briefs},
  58(8):527--531, 2011.

\bibitem{colizza2006detecting}
Vittoria Colizza, Alessandro Flammini, M~Angeles Serrano, and Alessandro
  Vespignani.
\newblock Detecting rich-club ordering in complex networks.
\newblock {\em Nature Physics}, 2(2):110, 2006.

\bibitem{zhou2004rich}
Shi Zhou and Ra{\'u}l~J Mondrag{\'o}n.
\newblock The rich-club phenomenon in the internet topology.
\newblock {\em IEEE Communications Letters}, 8(3):180--182, 2004.

\bibitem{schiavo2010international}
Stefano Schiavo, Javier Reyes, and Giorgio Fagiolo.
\newblock International trade and financial integration: a weighted network
  analysis.
\newblock {\em Quantitative Finance}, 10(4):389--399, 2010.

\bibitem{opsahl2008prominence}
Tore Opsahl, Vittoria Colizza, Pietro Panzarasa, and Jose~J Ramasco.
\newblock Prominence and control: the weighted rich-club effect.
\newblock {\em Physical review letters}, 101(16):168702, 2008.

\bibitem{alstott2014unifying}
Jeff Alstott, Pietro Panzarasa, Mikail Rubinov, Edward~T Bullmore, and Petra~E
  V{\'e}rtes.
\newblock A unifying framework for measuring weighted rich clubs.
\newblock {\em Scientific Reports}, 4:7258, 2014.

\bibitem{chinazzi2013post}
Matteo Chinazzi, Giorgio Fagiolo, Javier~A Reyes, and Stefano Schiavo.
\newblock Post-mortem examination of the international financial network.
\newblock {\em Journal of Economic Dynamics and Control}, 37(8):1692--1713,
  2013.

\bibitem{cohen2001breakdown}
Reuven Cohen, Keren Erez, Daniel Ben-Avraham, and Shlomo Havlin.
\newblock Breakdown of the internet under intentional attack.
\newblock {\em Physical review letters}, 86(16):3682, 2001.

\bibitem{duenas2009cascading}
Leonardo Due{\~n}as-Osorio and Srivishnu~Mohan Vemuru.
\newblock Cascading failures in complex infrastructure systems.
\newblock {\em Structural Safety}, 31(2):157--167, 2009.

\bibitem{carreras2003blackout}
Benjamin~A Carreras, Vickie~E Lynch, David~E Newman, and Ian Dobson.
\newblock Blackout mitigation assessment in power transmission systems.
\newblock In {\em 36th Annual Hawaii International Conference on System
  Sciences, 2003. Proceedings of the IEEE}, pages 10--pp. IEEE, 2003.

\bibitem{bak1987self}
Per Bak, Chao Tang, and Kurt Wiesenfeld.
\newblock Self-organized criticality: An explanation of the 1/f noise.
\newblock {\em Physical review letters}, 59(4):381, 1987.

\bibitem{carreras2013validating}
Benjamin~A Carreras, David~E Newman, Ian Dobson, and Naga~S Degala.
\newblock Validating opa with wecc data.
\newblock In {\em 2013 46th Hawaii International Conference on System
  Sciences}, pages 2197--2204. IEEE, 2013.

\bibitem{dhar1990self}
Deepak Dhar.
\newblock Self-organized critical state of sandpile automaton models.
\newblock {\em Physical Review Letters}, 64(14):1613, 1990.

\bibitem{moritz2005wildfires}
Max~A Moritz, Marco~E Morais, Lora~A Summerell, JM~Carlson, and John Doyle.
\newblock Wildfires, complexity, and highly optimized tolerance.
\newblock {\em Proceedings of the National Academy of Sciences},
  102(50):17912--17917, 2005.

\bibitem{motter2002cascade}
Adilson~E Motter and Ying-Cheng Lai.
\newblock Cascade-based attacks on complex networks.
\newblock {\em Physical Review E}, 66(6):065102, 2002.

\bibitem{hines2008trends}
Paul Hines, Jay Apt, and Sarosh Talukdar.
\newblock Trends in the history of large blackouts in the united states.
\newblock In {\em 2008 IEEE Power and Energy Society General Meeting-Conversion
  and Delivery of Electrical Energy in the 21st Century}, pages 1--8. IEEE,
  2008.

\bibitem{daqing2014spatial}
Li~Daqing, Jiang Yinan, Kang Rui, and Shlomo Havlin.
\newblock Spatial correlation analysis of cascading failures: congestions and
  blackouts.
\newblock {\em Scientific Reports}, 4:5381, 2014.

\bibitem{zhao2016spatio}
Jichang Zhao, Daqing Li, Hillel Sanhedrai, Reuven Cohen, and Shlomo Havlin.
\newblock Spatio-temporal propagation of cascading overload failures in
  spatially embedded networks.
\newblock {\em Nature Communications}, 7:10094, 2016.

\bibitem{yang2017small}
Yang Yang, Takashi Nishikawa, and Adilson~E Motter.
\newblock Small vulnerable sets determine large network cascades in power
  grids.
\newblock {\em Science}, 358(6365):eaan3184, 2017.

\bibitem{berezin2015localized}
Yehiel Berezin, Amir Bashan, Michael~M Danziger, Daqing Li, and Shlomo Havlin.
\newblock Localized attacks on spatially embedded networks with dependencies.
\newblock {\em Scientific Reports}, 5(8934), 2015.

\bibitem{eisenblatter1998jamming}
B~Eisenbl{\"a}tter, L~Santen, A~Schadschneider, and M~Schreckenberg.
\newblock Jamming transition in a cellular automaton model for traffic flow.
\newblock {\em Physical Review E}, 57(2):1309, 1998.

\bibitem{dobson2004branching}
Ian Dobson, Benjamin~A Carreras, and David~E Newman.
\newblock A branching process approximation to cascading load-dependent system
  failure.
\newblock In {\em 37th Annual Hawaii International Conference on System
  Sciences, 2004. Proceedings of the}, pages 10--pp. IEEE, 2004.

\bibitem{crucitti2004model}
Paolo Crucitti, Vito Latora, and Massimo Marchiori.
\newblock Model for cascading failures in complex networks.
\newblock {\em Physical Review E}, 69(4):045104, 2004.

\bibitem{simonsen2008transient}
Ingve Simonsen, Lubos Buzna, Karsten Peters, Stefan Bornholdt, and Dirk
  Helbing.
\newblock Transient dynamics increasing network vulnerability to cascading
  failures.
\newblock {\em Physical review letters}, 100(21):218701, 2008.

\bibitem{lehmann2010stochastic}
J{\"o}rg Lehmann and Jakob Bernasconi.
\newblock Stochastic load-redistribution model for cascading failure
  propagation.
\newblock {\em Physical Review E}, 81(3):031129, 2010.

\bibitem{kim2010approximating}
Janghoon Kim and Ian Dobson.
\newblock Approximating a loading-dependent cascading failure model with a
  branching process.
\newblock {\em IEEE Transactions on Reliability}, 59(4):691--699, 2010.

\bibitem{wang2008attack}
Jianwei Wang, Lili Rong, Liang Zhang, and Zhongzhi Zhang.
\newblock Attack vulnerability of scale-free networks due to cascading
  failures.
\newblock {\em Physica A: Statistical Mechanics and its Applications},
  387(26):6671--6678, 2008.

\bibitem{zhao2004attack}
Liang Zhao, Kwangho Park, and Ying-Cheng Lai.
\newblock Attack vulnerability of scale-free networks due to cascading
  breakdown.
\newblock {\em Physical review E}, 70(3):035101, 2004.

\bibitem{xia2010cascading}
Yongxiang Xia, Jin Fan, and David Hill.
\newblock Cascading failure in watts--strogatz small-world networks.
\newblock {\em Physica A: Statistical Mechanics and its Applications},
  389(6):1281--1285, 2010.

\bibitem{wang2004cascading}
Xiao~Fan Wang and Jian Xu.
\newblock Cascading failures in coupled map lattices.
\newblock {\em Physical Review E}, 70(5):056113, 2004.

\bibitem{wang2009cascade}
Jian-Wei Wang and Li-Li Rong.
\newblock Cascade-based attack vulnerability on the us power grid.
\newblock {\em Safety Science}, 47(10):1332--1336, 2009.

\bibitem{menck2014dead}
Peter~J Menck, Jobst Heitzig, J{\"u}rgen Kurths, and Hans~Joachim Schellnhuber.
\newblock How dead ends undermine power grid stability.
\newblock {\em Nature Communications}, 5:3969, 2014.

\bibitem{kinney2005modeling}
Ryan Kinney, Paolo Crucitti, Reka Albert, and Vito Latora.
\newblock Modeling cascading failures in the {N}orth {A}merican power grid.
\newblock {\em The European Physical Journal B-Condensed Matter and Complex
  Systems}, 46(1):101--107, 2005.

\bibitem{crucitti2004topological}
Paolo Crucitti, Vito Latora, and Massimo Marchiori.
\newblock A topological analysis of the italian electric power grid.
\newblock {\em Physica A: Statistical Mechanics and its Applications},
  338(1-2):92--97, 2004.

\bibitem{Asztalos2014Cascading}
Andrea Asztalos, Sameet Sreenivasan, Boleslaw~K Szymanski, and Gyorgy Korniss.
\newblock Cascading failures in spatially-embedded random networks.
\newblock {\em PLoS One}, 9(1), 2014.

\bibitem{cetinay2017comparing}
Hale Cetinay, Saleh Soltan, Fernando~A Kuipers, Gil Zussman, and Piet
  Van~Mieghem.
\newblock Comparing the effects of failures in power grids under the ac and dc
  power flow models.
\newblock {\em IEEE Transactions on Network Science and Engineering},
  5(4):301--312, 2017.

\bibitem{carreras2002dynamics}
Benjamin~A Carreras, Vickie~E Lynch, Ian Dobson, and David~E Newman.
\newblock Dynamics, criticality and self-organization in a model for blackouts
  in power transmission systems.
\newblock In {\em Proceedings of the 35th Annual Hawaii International
  Conference on System Sciences}, pages 9--pp. IEEE, 2002.

\bibitem{song2015dynamic}
Jiajia Song, Eduardo Cotilla-Sanchez, Goodarz Ghanavati, and Paul~DH Hines.
\newblock Dynamic modeling of cascading failure in power systems.
\newblock {\em IEEE Transactions on Power Systems}, 31(3):2085--2095, 2015.

\bibitem{pahwa2014abruptness}
Sakshi Pahwa, Caterina Scoglio, and Antonio Scala.
\newblock Abruptness of cascade failures in power grids.
\newblock {\em Scientific Reports}, 4:3694, 2014.

\bibitem{ren2018stochastic}
Wendi Ren, Jiajing Wu, Xi~Zhang, Rong Lai, and Liang Chen.
\newblock A stochastic model of cascading failure dynamics in communication
  networks.
\newblock {\em IEEE Transactions on Circuits and Systems II: Express Briefs},
  65(5):632--636, 2018.

\bibitem{treiterer1974hysteresis}
Joseph Treiterer and Jeffrey Myers.
\newblock The hysteresis phenomenon in traffic flow.
\newblock {\em Transportation and Traffic Theory}, 6:13--38, 1974.

\bibitem{kerner1994structure}
Boris~S Kerner and P~Konh{\"a}user.
\newblock Structure and parameters of clusters in traffic flow.
\newblock {\em Physical Review E}, 50(1):54, 1994.

\bibitem{edie1958traffic}
Leslie~C Edie and Robert~S Foote.
\newblock Traffic flow in tunnels.
\newblock In {\em Highway Research Board Proceedings}, volume~37, 1958.

\bibitem{mika1969dual}
HS~Mika, JB~Kreer, and LS~Yuan.
\newblock Dual mode behavior of freeway traffic.
\newblock {\em Highway Research Record}, 279:1--13, 1969.

\bibitem{wu2007cascading}
JJ~Wu, HJ~Sun, and ZY~Gao.
\newblock Cascading failures on weighted urban traffic equilibrium networks.
\newblock {\em Physica A: Statistical Mechanics and its Applications},
  386(1):407--413, 2007.

\bibitem{parshani2011critical}
Roni Parshani, Sergey~V Buldyrev, and Shlomo Havlin.
\newblock Critical effect of dependency groups on the function of networks.
\newblock {\em Proceedings of the National Academy of Sciences},
  108(3):1007--1010, 2011.

\bibitem{bashan2013extreme}
Amir Bashan, Yehiel Berezin, Sergey~V Buldyrev, and Shlomo Havlin.
\newblock The extreme vulnerability of interdependent spatially embedded
  networks.
\newblock {\em Nature Physics}, 9(10):667--672, 2013.

\bibitem{zhang2013robustness}
Peng Zhang, Baisong Cheng, Zhuang Zhao, Daqing Li, Guangquan Lu, Yunpeng Wang,
  and Jinghua Xiao.
\newblock The robustness of interdependent transportation networks under
  targeted attack.
\newblock {\em EPL (Europhysics Letters)}, 103(6):68005, 2013.

\bibitem{yagan2012optimal}
Osman Yagan, Dajun Qian, Junshan Zhang, and Douglas Cochran.
\newblock Optimal allocation of interconnecting links in cyber-physical
  systems: Interdependence, cascading failures, and robustness.
\newblock {\em IEEE Transactions on Parallel and Distributed Systems},
  23(9):1708--1720, 2012.

\bibitem{scaparra2008bilevel}
Maria~P Scaparra and Richard~L Church.
\newblock A bilevel mixed-integer program for critical infrastructure
  protection planning.
\newblock {\em Computers \& Operations Research}, 35(6):1905--1923, 2008.

\bibitem{boin2007preparing}
Arjen Boin and Allan McConnell.
\newblock Preparing for critical infrastructure breakdowns: the limits of
  crisis management and the need for resilience.
\newblock {\em Journal of Contingencies and Crisis Management}, 15(1):50--59,
  2007.

\bibitem{conrad2006critical}
Stephen~H Conrad, Rene~J LeClaire, Gerard~P O'Reilly, and Huseyin Uzunalioglu.
\newblock Critical national infrastructure reliability modeling and analysis.
\newblock {\em Bell Labs Technical Journal}, 11(3):57--71, 2006.

\bibitem{yusta2011methodologies}
Jose~M Yusta, Gabriel~J Correa, and Roberto Lacal-Ar{\'a}ntegui.
\newblock Methodologies and applications for critical infrastructure
  protection: State-of-the-art.
\newblock {\em Energy Policy}, 39(10):6100--6119, 2011.

\bibitem{woods2015four}
DD~Woods.
\newblock Four concepts for resilience and their implications for systems
  safety in the face of complexity.
\newblock {\em Reliab Eng Syst Saf}, 141:5--9, 2015.

\bibitem{zobel2014characterizing}
Christopher~W Zobel and Lara Khansa.
\newblock Characterizing multi-event disaster resilience.
\newblock {\em Computers \& Operations Research}, 42:83--94, 2014.

\bibitem{henry2012generic}
Devanandham Henry and Jose~Emmanuel Ramirez-Marquez.
\newblock Generic metrics and quantitative approaches for system resilience as
  a function of time.
\newblock {\em Reliability Engineering \& System Safety}, 99:114--122, 2012.

\bibitem{francis2014metric}
Royce Francis and Behailu Bekera.
\newblock A metric and frameworks for resilience analysis of engineered and
  infrastructure systems.
\newblock {\em Reliability Engineering \& System Safety}, 121:90--103, 2014.

\bibitem{chang2004measuring}
Stephanie~E Chang and Masanobu Shinozuka.
\newblock Measuring improvements in the disaster resilience of communities.
\newblock {\em Earthquake Spectra}, 20(3):739--755, 2004.

\bibitem{ouyang2012three}
Min Ouyang, Leonardo Due{\~n}as-Osorio, and Xing Min.
\newblock A three-stage resilience analysis framework for urban infrastructure
  systems.
\newblock {\em Structural Safety}, 36:23--31, 2012.

\bibitem{youn2011resilience}
Byeng~D Youn, Chao Hu, and Pingfeng Wang.
\newblock Resilience-driven system design of complex engineered systems.
\newblock {\em Journal of Mechanical Design}, 133(10):101011, 2011.

\bibitem{ayyub2014systems}
Bilal~M Ayyub.
\newblock Systems resilience for multihazard environments: Definition, metrics,
  and valuation for decision making.
\newblock {\em Risk Analysis}, 34(2):340--355, 2014.

\bibitem{cai2018availability}
Baoping Cai, Min Xie, Yonghong Liu, Yiliu Liu, and Qiang Feng.
\newblock Availability-based engineering resilience metric and its
  corresponding evaluation methodology.
\newblock {\em Reliability Engineering \& System Safety}, 172:216--224, 2018.

\bibitem{renschler2010framework}
Chris~S Renschler, Amy~E Frazier, Lucy~A Arendt, Gian~Paolo Cimellaro, Andrei~M
  Reinhorn, and Michel Bruneau.
\newblock {\em A framework for defining and measuring resilience at the
  community scale: The PEOPLES resilience framework}.
\newblock MCEER Buffalo, 2010.

\bibitem{arcidiacono2012community}
Vincenzo Arcidiacono, Gian~Paolo Cimellaro, AM~Reinhorn, and M~Bruneau.
\newblock Community resilience evaluation including interdependencies.
\newblock In {\em 15th world conference on earthquake engineering (15WCEE)},
  pages 24--28, 2012.

\bibitem{comes2014measuring}
Tina Comes and Bartel Van~de Walle.
\newblock Measuring disaster resilience: The impact of hurricane sandy on
  critical infrastructure systems.
\newblock {\em ISCRAM}, 11:195--204, 2014.

\bibitem{ip2011resilience}
Wai~Hung Ip and Dingwei Wang.
\newblock Resilience and friability of transportation networks: evaluation,
  analysis and optimization.
\newblock {\em IEEE Systems Journal}, 5(2):189--198, 2011.

\bibitem{sterbenz2013evaluation}
James~PG Sterbenz, Egemen~K {\c{C}}etinkaya, Mahmood~A Hameed, Abdul Jabbar,
  Shi Qian, and Justin~P Rohrer.
\newblock Evaluation of network resilience, survivability, and disruption
  tolerance: analysis, topology generation, simulation, and experimentation.
\newblock {\em Telecommunication systems}, 52(2):705--736, 2013.

\bibitem{fang2016resilience}
Yi-Ping Fang, Nicola Pedroni, and Enrico Zio.
\newblock Resilience-based component importance measures for critical
  infrastructure network systems.
\newblock {\em IEEE Transactions on Reliability}, 65(2):502--512, 2016.

\bibitem{zhang2019scale}
Limiao Zhang, Guanwen Zeng, Daqing Li, Hai-Jun Huang, H~Eugene Stanley, and
  Shlomo Havlin.
\newblock Scale-free resilience of real traffic jams.
\newblock {\em Proceedings of the National Academy of Sciences},
  116(18):8673--8678, 2019.

\bibitem{nagatani1995self}
Takashi Nagatani.
\newblock Self-organized criticality and scaling in lifetime of traffic jams.
\newblock {\em Journal of the Physical Society of Japan}, 64(1):31--34, 1995.

\bibitem{ganin2017resilience}
Alexander~A Ganin, Maksim Kitsak, Dayton Marchese, Jeffrey~M Keisler, Thomas
  Seager, and Igor Linkov.
\newblock Resilience and efficiency in transportation networks.
\newblock {\em Science Advances}, 3(12):e1701079, 2017.

\bibitem{billinton1992reliability}
Roy Billinton and Ronald~Norman Allan.
\newblock {\em Reliability evaluation of engineering systems}.
\newblock Springer, 1992.

\bibitem{geraci1991ieee}
Anne Geraci, Freny Katki, Louise McMonegal, Bennett Meyer, John Lane, Paul
  Wilson, Jane Radatz, Mary Yee, Hugh Porteous, and Fredrick Springsteel.
\newblock {\em IEEE standard computer dictionary: Compilation of IEEE standard
  computer glossaries}.
\newblock IEEE Press, 1991.

\bibitem{stojadinovic1983failure}
ND~Stojadinovi{\'c}.
\newblock Failure physics of integrated circuits—a review.
\newblock {\em Microelectronics Reliability}, 23(4):609--707, 1983.

\bibitem{barnard20083}
RWA Barnard.
\newblock 3.2. 2 what is wrong with reliability engineering?
\newblock In {\em INCOSE International Symposium}, volume~18, pages 357--365.
  Wiley Online Library, 2008.

\bibitem{zio2007complexity}
Enrico Zio.
\newblock From complexity science to reliability efficiency: a new way of
  looking at complex network systems and critical infrastructures.
\newblock {\em International Journal of Critical Infrastructures},
  3(3-4):488--508, 2007.

\bibitem{li2015network}
Daqing Li, Qiong Zhang, Enrico Zio, Shlomo Havlin, and Rui Kang.
\newblock Network reliability analysis based on percolation theory.
\newblock {\em Reliability Engineering \& System Safety}, 142:556--562, 2015.

\bibitem{hardy2007k}
Gary Hardy, Corinne Lucet, and Nikolaos Limnios.
\newblock K-terminal network reliability measures with binary decision
  diagrams.
\newblock {\em IEEE Transactions on Reliability}, 56(3):506--515, 2007.

\bibitem{ramirez2005monte}
Jose~E Ramirez-Marquez and David~W Coit.
\newblock A monte-carlo simulation approach for approximating multi-state
  two-terminal reliability.
\newblock {\em Reliability Engineering \& System Safety}, 87(2):253--264, 2005.

\bibitem{ramirez2008all}
Jos{\'e}~Emmanuel Ramirez-Marquez and Claudio~M Rocco.
\newblock All-terminal network reliability optimization via probabilistic
  solution discovery.
\newblock {\em Reliability Engineering \& System Safety}, 93(11):1689--1697,
  2008.

\bibitem{jun2010natural}
WU~Jun, Mauricio Barahona, Tan Yue-Jin, and Deng Hong-Zhong.
\newblock Natural connectivity of complex networks.
\newblock {\em Chinese Physics Letters}, 27(7):078902, 2010.

\bibitem{zhang2018belief}
Qingyuan Zhang, Rui Kang, and Meilin Wen.
\newblock Belief reliability for uncertain random systems.
\newblock {\em IEEE Transactions on Fuzzy Systems}, 26(6):3605--3614, 2018.

\bibitem{dong2018post}
Shangjia Dong, Alireza Mostafizi, Haizhong Wang, and Peter Bosa.
\newblock Post-disaster mobility in disrupted transportation network: Case
  study of portland, oregon.
\newblock {\em Journal of Urban Planning and Development}, 2018.

\bibitem{dong2020integrated}
Shangjia Dong, Amir Esmalian, Hamed Farahmand, and Ali Mostafavi.
\newblock An integrated physical-social analysis of disrupted access to
  critical facilities and community service-loss tolerance in urban flooding.
\newblock {\em Computers, Environment and Urban Systems}, 80:101443, 2020.

\bibitem{mostafizi2019agent}
Alireza Mostafizi, Haizhong Wang, Dan Cox, and Shangjia Dong.
\newblock An agent-based vertical evacuation model for a near-field tsunami:
  Choice behavior, logical shelter locations, and life safety.
\newblock {\em International journal of disaster risk reduction}, 34:467--479,
  2019.

\bibitem{asakura1991road}
Yasuo Asakura and Masuo Kashiwadani.
\newblock Road network reliability caused by daily fluctuation of traffic flow.
\newblock In {\em PTRC Summer Annual Meeting, 19th, 1991, University of Sussex,
  United Kingdom}, 1991.

\bibitem{asakura1999reliability}
Yasuo Asakura.
\newblock Reliability measures of an origin and destination pair in a
  deteriorated road network with variable flows.
\newblock In {\em Transportation Networks: Recent Methodological Advances.
  Selected Proceedings of the 4th EURO Transportation MeetingAssociation of
  European Operational Research Societies}, 1999.

\bibitem{chen1999capacity}
Anthony Chen, Hai Yang, Hong~K Lo, and Wilson~H Tang.
\newblock A capacity related reliability for transportation networks.
\newblock {\em Journal of Advanced Transportation}, 33(2):183--200, 1999.

\bibitem{shao2006reliability}
Hu~Shao, William~HK Lam, and Mei~Lam Tam.
\newblock A reliability-based stochastic traffic assignment model for network
  with multiple user classes under uncertainty in demand.
\newblock {\em Networks and Spatial Economics}, 6(3-4):173--204, 2006.

\bibitem{righi2015systematic}
Angela~Weber Righi, Tarcisio~Abreu Saurin, and Priscila Wachs.
\newblock A systematic literature review of resilience engineering: Research
  areas and a research agenda proposal.
\newblock {\em Reliability Engineering \& System Safety}, 141:142--152, 2015.

\bibitem{moteff2005risk}
John Moteff.
\newblock Risk management and critical infrastructure protection: Assessing,
  integrating, and managing threats, vulnerabilities and consequences.
\newblock Library of Congress Washington DC Congressional Research Service,
  2005.

\bibitem{guikema2009natural}
Seth~D Guikema.
\newblock Natural disaster risk analysis for critical infrastructure systems:
  An approach based on statistical learning theory.
\newblock {\em Reliability Engineering \& System Safety}, 94(4):855--860, 2009.

\bibitem{ouyang2009methodological}
Min Ouyang, Liu Hong, Zi-Jun Mao, Ming-Hui Yu, and Fei Qi.
\newblock A methodological approach to analyze vulnerability of interdependent
  infrastructures.
\newblock {\em Simulation Modelling Practice and Theory}, 17(5):817--828, 2009.

\bibitem{der2009aleatory}
Armen Der~Kiureghian and Ove Ditlevsen.
\newblock Aleatory or epistemic? does it matter?
\newblock {\em Structural Safety}, 31(2):105--112, 2009.

\bibitem{haimes2006definition}
Yacov~Y Haimes.
\newblock On the definition of vulnerabilities in measuring risks to
  infrastructures.
\newblock {\em Risk Analysis: An International Journal}, 26(2):293--296, 2006.

\bibitem{apeland2002quantifying}
S~Apeland, Terje Aven, and Thomas Nilsen.
\newblock Quantifying uncertainty under a predictive, epistemic approach to
  risk analysis.
\newblock {\em Reliability Engineering \& System Safety}, 75(1):93--102, 2002.

\bibitem{douglas2007physical}
John Douglas.
\newblock Physical vulnerability modelling in natural hazard risk assessment.
\newblock {\em Natural Hazards and Earth System Science}, 7(2):283--288, 2007.

\bibitem{jenelius2006importance}
Erik Jenelius, Tom Petersen, and Lars-G{\"o}ran Mattsson.
\newblock Importance and exposure in road network vulnerability analysis.
\newblock {\em Transportation Research Part A: Policy and Practice},
  40(7):537--560, 2006.

\bibitem{bernstein2014power}
Andrey Bernstein, Daniel Bienstock, David Hay, Meric Uzunoglu, and Gil Zussman.
\newblock Power grid vulnerability to geographically correlated
  failures—analysis and control implications.
\newblock In {\em IEEE INFOCOM 2014-IEEE Conference on Computer
  Communications}, pages 2634--2642. IEEE, 2014.

\bibitem{shuang2014node}
Qing Shuang, Mingyuan Zhang, and Yongbo Yuan.
\newblock Node vulnerability of water distribution networks under cascading
  failures.
\newblock {\em Reliability Engineering \& System Safety}, 124:132--141, 2014.

\bibitem{chang2001measuring}
Stephanie~E Chang and Nobuoto Nojima.
\newblock Measuring post-disaster transportation system performance: the 1995
  kobe earthquake in comparative perspective.
\newblock {\em Transportation Research Part A: Policy and Practice},
  35(6):475--494, 2001.

\bibitem{tuncel2010risk}
Gonca Tuncel and G{\"u}lg{\"u}n Alpan.
\newblock Risk assessment and management for supply chain networks: A case
  study.
\newblock {\em Computers in Industry}, 61(3):250--259, 2010.

\bibitem{pearce1984stochastic}
Howard~T Pearce and YK~Wen.
\newblock Stochastic combination of load effects.
\newblock {\em Journal of Structural Engineering}, 110(7):1613--1629, 1984.

\bibitem{ghosn2005load}
Michel Ghosn.
\newblock Load combination factors for extreme events.
\newblock {\em Transportation Research Record: Journal of the Transportation
  Research Board}, (CD 11-S), 2005.

\bibitem{deco2011risk}
Alberto Dec{\`o} and Dan~M Frangopol.
\newblock Risk assessment of highway bridges under multiple hazards.
\newblock {\em Journal of Risk Research}, 14(9):1057--1089, 2011.

\bibitem{kameshwar2014multi}
Sabarethinam Kameshwar and Jamie~E Padgett.
\newblock Multi-hazard risk assessment of highway bridges subjected to
  earthquake and hurricane hazards.
\newblock {\em Engineering Structures}, 78:154--166, 2014.

\bibitem{berdica2002introduction}
Katja Berdica.
\newblock An introduction to road vulnerability: what has been done, is done
  and should be done.
\newblock {\em Transport Policy}, 9(2):117--127, 2002.

\bibitem{mattsson2015vulnerability}
Lars-G{\"o}ran Mattsson and Erik Jenelius.
\newblock Vulnerability and resilience of transport systems--a discussion of
  recent research.
\newblock {\em Transportation Research Part A: Policy and Practice}, 81:16--34,
  2015.

\bibitem{nicholson1997degradable}
Alan Nicholson and Zhen-Ping Du.
\newblock Degradable transportation systems: an integrated equilibrium model.
\newblock {\em Transportation Research Part B: Methodological}, 31(3):209--223,
  1997.

\bibitem{albert1999internet}
R{\'e}ka Albert, Hawoong Jeong, and Albert-L{\'a}szl{\'o} Barab{\'a}si.
\newblock Internet: Diameter of the world-wide web.
\newblock {\em Nature}, 401(6749):130, 1999.

\bibitem{du2017identifying}
Wenbo Du, Boyuan Liang, Gang Yan, Oriol Lordan, and Xianbin Cao.
\newblock Identifying vital edges in chinese air route network via memetic
  algorithm.
\newblock {\em Chinese Journal of Aeronautics}, 30(1):330--336, 2017.

\bibitem{basoz1997risk}
NI~Basoz.
\newblock Risk assessment for highway transportation systems.
\newblock 1997.

\bibitem{chang2002disaster}
Stephanie~E Chang and Anthony Falit-Baiamonte.
\newblock Disaster vulnerability of businesses in the 2001 nisqually
  earthquake.
\newblock {\em Global Environmental Change Part B: Environmental Hazards},
  4(2):59--71, 2002.

\bibitem{hong2015vulnerability}
Liu Hong, Min Ouyang, Srinivas Peeta, Xiaozheng He, and Yongze Yan.
\newblock Vulnerability assessment and mitigation for the chinese railway
  system under floods.
\newblock {\em Reliability Engineering \& System Safety}, 137:58--68, 2015.

\bibitem{wu2016modeling}
Baichao Wu, Aiping Tang, and Jie Wu.
\newblock Modeling cascading failures in interdependent infrastructures under
  terrorist attacks.
\newblock {\em Reliability Engineering \& System Safety}, 147:1--8, 2016.

\bibitem{ouyang2016critical}
Min Ouyang.
\newblock Critical location identification and vulnerability analysis of
  interdependent infrastructure systems under spatially localized attacks.
\newblock {\em Reliability Engineering \& System Safety}, 154:106--116, 2016.

\bibitem{lisnianski2003multi}
Anatoly Lisnianski and Gregory Levitin.
\newblock {\em Multi-state system reliability: assessment, optimization and
  applications}, volume~6.
\newblock World Scientific Publishing Company, 2003.

\bibitem{yeh2006k}
Wei-Chang Yeh.
\newblock The k-out-of-n acyclic multistate-node networks reliability
  evaluation using the universal generating function method.
\newblock {\em Reliability Engineering \& System Safety}, 91(7):800--808, 2006.

\bibitem{ding2008afuzzy}
Yi~Ding and Anatoly Lisnianski.
\newblock Fuzzy universal generating functions for multi-state system
  reliability assessment.
\newblock {\em Fuzzy Sets and Systems}, 159(3):307--324, 2008.

\bibitem{scheffer2010complex}
Marten Scheffer.
\newblock Complex systems: foreseeing tipping points.
\newblock {\em Nature}, 467(7314):411, 2010.

\bibitem{chakrabarti1999dynamic}
Bikas~K Chakrabarti and Muktish Acharyya.
\newblock Dynamic transitions and hysteresis.
\newblock {\em Reviews of Modern Physics}, 71(3):847, 1999.

\bibitem{treiterer1975investigation}
Joseph Treiterer.
\newblock Investigation of traffic dynamics by aerial photogrammetry
  techniques.
\newblock Technical report, 1975.

\bibitem{godfrey1969mechanism}
JW~Godfrey.
\newblock The mechanism of a road network.
\newblock {\em Traffic Engineering \& Control}, 8(8), 1969.

\bibitem{zhang1999mathematical}
H~Michael Zhang.
\newblock A mathematical theory of traffic hysteresis.
\newblock {\em Transportation Research Part B: Methodological}, 33(1):1--23,
  1999.

\bibitem{cornelius2013realistic}
Sean~P Cornelius, William~L Kath, and Adilson~E Motter.
\newblock Realistic control of network dynamics.
\newblock {\em Nature Communications}, 4:1942, 2013.

\bibitem{newell1965instability}
Gordon~Frank Newell.
\newblock Instability in dense highway traffic: A review.
\newblock 1965.

\bibitem{chen2012microscopic}
Danjue Chen, Jorge~A Laval, Soyoung Ahn, and Zuduo Zheng.
\newblock Microscopic traffic hysteresis in traffic oscillations: A behavioral
  perspective.
\newblock {\em Transportation Research Part B: Methodological},
  46(10):1440--1453, 2012.

\bibitem{barlovic1998metastable}
Robert Barlovic, Ludger Santen, Andreas Schadschneider, and Michael
  Schreckenberg.
\newblock Metastable states in cellular automata for traffic flow.
\newblock {\em The European Physical Journal B-Condensed Matter and Complex
  Systems}, 5(3):793--800, 1998.

\bibitem{hu2007phase}
Mao-Bin Hu, Wen-Xu Wang, Rui Jiang, Qing-Song Wu, and Yong-Hong Wu.
\newblock Phase transition and hysteresis in scale-free network traffic.
\newblock {\em Physical Review E}, 75(3):036102, 2007.

\bibitem{carpenter2006rising}
Stephen~R Carpenter and William~A Brock.
\newblock Rising variance: a leading indicator of ecological transition.
\newblock {\em Ecology letters}, 9(3):311--318, 2006.

\bibitem{kleinen2003potential}
Thomas Kleinen, Hermann Held, and Gerhard Petschel-Held.
\newblock The potential role of spectral properties in detecting thresholds in
  the earth system: application to the thermohaline circulation.
\newblock {\em Ocean Dynamics}, 53(2):53--63, 2003.

\bibitem{livina2007modified}
Valerie~N Livina and Timothy~M Lenton.
\newblock A modified method for detecting incipient bifurcations in a dynamical
  system.
\newblock {\em Geophysical Research Letters}, 34(3), 2007.

\bibitem{livina2010potential}
Valerie~N Livina, F~Kwasniok, and Timothy~M Lenton.
\newblock Potential analysis reveals changing number of climate states during
  the last 60 kyr.
\newblock {\em Climate of the Past}, 6(1):77--82, 2010.

\bibitem{zeng2020multiple}
Guanwen Zeng, Jianxi Gao, Louis Shekhtman, Shengmin Guo, Weifeng Lv, Jianjun
  Wu, Hao Liu, Orr Levy, Daqing Li, Ziyou Gao, et~al.
\newblock Multiple metastable network states in urban traffic.
\newblock {\em Proceedings of the National Academy of Sciences},
  117(30):17528--17534, 2020.

\bibitem{liu2011controllability}
Yang-Yu Liu, Jean-Jacques Slotine, and Albert-L{\'a}szl{\'o} Barab{\'a}si.
\newblock Controllability of complex networks.
\newblock {\em nature}, 473(7346):167--173, 2011.

\bibitem{gao2014target}
Jianxi Gao, Yang-Yu Liu, Raissa~M D'souza, and Albert-L{\'a}szl{\'o}
  Barab{\'a}si.
\newblock Target control of complex networks.
\newblock {\em Nature Communications}, 5:5415, 2014.

\bibitem{geroliminis2012optimal}
Nikolas Geroliminis, Jack Haddad, and Mohsen Ramezani.
\newblock Optimal perimeter control for two urban regions with macroscopic
  fundamental diagrams: A model predictive approach.
\newblock {\em IEEE Transactions on Intelligent Transportation Systems},
  14(1):348--359, 2012.

\bibitem{liu2014modeling}
Chaoran Liu, Daqing Li, Bowen Fu, Shunkun Yang, Yunpeng Wang, and Guangquan Lu.
\newblock Modeling of self-healing against cascading overload failures in
  complex networks.
\newblock {\em EPL (Europhysics Letters)}, 107(6):68003, 2014.

\bibitem{lin2016restorative}
Zhenzhi Lin, Fushuan Wen, and Yusheng Xue.
\newblock A restorative self-healing algorithm for transmission systems based
  on complex network theory.
\newblock {\em IEEE Transactions on Smart Grid}, 7(4):2154--2162, 2016.

\bibitem{quattrociocchi2014self}
Walter Quattrociocchi, Guido Caldarelli, and Antonio Scala.
\newblock Self-healing networks: redundancy and structure.
\newblock {\em Plos One}, 9(2):e87986, 2014.

\bibitem{gallos2015simple}
Lazaros~K Gallos and Nina~H Fefferman.
\newblock Simple and efficient self-healing strategy for damaged complex
  networks.
\newblock {\em Physical Review E}, 92(5):052806, 2015.

\bibitem{shang2015impact}
Yilun Shang.
\newblock Impact of self-healing capability on network robustness.
\newblock {\em Physical Review E}, 91(4):042804, 2015.

\bibitem{macy2021polarization}
Michael~W Macy, Manqing Ma, Daniel~R Tabin, Jianxi Gao, and Boleslaw~K
  Szymanski.
\newblock Polarization and tipping points.
\newblock {\em Proceedings of the National Academy of Sciences}, 118(50), 2021.

\bibitem{yuan2019data}
Ye~Yuan, Xiuchuan Tang, Wei Zhou, Wei Pan, Xiuting Li, Hai-Tao Zhang, Han Ding,
  and Jorge Goncalves.
\newblock Data driven discovery of cyber physical systems.
\newblock {\em Nature communications}, 10(1):1--9, 2019.

\bibitem{harush2017dynamic}
Uzi Harush and Baruch Barzel.
\newblock Dynamic patterns of information flow in complex networks.
\newblock {\em Nature communications}, 8(1):1--11, 2017.

\bibitem{hens2019spatiotemporal}
Chittaranjan Hens, Uzi Harush, Simi Haber, Reuven Cohen, and Baruch Barzel.
\newblock Spatiotemporal signal propagation in complex networks.
\newblock {\em Nature Physics}, 15(4):403--412, 2019.

\bibitem{rual2005towards}
Jean-Fran{\c{c}}ois Rual, Kavitha Venkatesan, Tong Hao, Tomoko
  Hirozane-Kishikawa, Am{\'e}lie Dricot, Ning Li, Gabriel~F Berriz, Francis~D
  Gibbons, Matija Dreze, Nono Ayivi-Guedehoussou, et~al.
\newblock Towards a proteome-scale map of the human protein--protein
  interaction network.
\newblock {\em Nature}, 437(7062):1173--1178, 2005.

\bibitem{zhang2018spatiotemporal}
Hai-Tao Zhang, Tao Zhu, Dongfei Fu, Bowen Xu, Xiao-Pu Han, and Duxin Chen.
\newblock Spatiotemporal property and predictability of large-scale human
  mobility.
\newblock {\em Physica A: Statistical Mechanics and its Applications},
  495:40--48, 2018.

\bibitem{boccaletti2014structure}
Stefano Boccaletti, Ginestra Bianconi, Regino Criado, Charo~I Del~Genio,
  Jes{\'u}s G{\'o}mez-Gardenes, Miguel Romance, Irene Sendina-Nadal, Zhen Wang,
  and Massimiliano Zanin.
\newblock The structure and dynamics of multilayer networks.
\newblock {\em Physics Reports}, 544(1):1--122, 2014.

\bibitem{liu2019multiple}
Xueming Liu, Linqiang Pan, H~Eugene Stanley, and Jianxi Gao.
\newblock Multiple phase transitions in networks of directed networks.
\newblock {\em Physical Review E}, 99(1):012312, 2019.

\bibitem{cai2005mining}
Deng Cai, Zheng Shao, Xiaofei He, Xifeng Yan, and Jiawei Han.
\newblock Mining hidden community in heterogeneous social networks.
\newblock In {\em Proceedings of the 3rd international workshop on Link
  discovery}, pages 58--65, 2005.

\bibitem{menche2015uncovering}
J{\"o}rg Menche, Amitabh Sharma, Maksim Kitsak, Susan~Dina Ghiassian, Marc
  Vidal, Joseph Loscalzo, and Albert-L{\'a}szl{\'o} Barab{\'a}si.
\newblock Uncovering disease-disease relationships through the incomplete
  interactome.
\newblock {\em Science}, 347(6224):1257601, 2015.

\bibitem{guimaraes2011evolution}
Paulo~R Guimar{\~a}es~Jr, Pedro Jordano, and John~N Thompson.
\newblock Evolution and coevolution in mutualistic networks.
\newblock {\em Ecology letters}, 14(9):877--885, 2011.

\bibitem{zeng2018prediction}
Xiangxiang Zeng, Li~Liu, Linyuan L{\"u}, and Quan Zou.
\newblock Prediction of potential disease-associated micrornas using structural
  perturbation method.
\newblock {\em Bioinformatics}, 34(14):2425--2432, 2018.

\bibitem{leskovec2006sampling}
Jure Leskovec and Christos Faloutsos.
\newblock Sampling from large graphs.
\newblock In {\em Proceedings of the 12th ACM SIGKDD international conference
  on Knowledge discovery and data mining}, pages 631--636, 2006.

\bibitem{li2015random}
Rong-Hua Li, Jeffrey~Xu Yu, Lu~Qin, Rui Mao, and Tan Jin.
\newblock On random walk based graph sampling.
\newblock In {\em 2015 IEEE 31st International Conference on Data Engineering},
  pages 927--938. IEEE, 2015.

\bibitem{gjoka2010walking}
Minas Gjoka, Maciej Kurant, Carter~T Butts, and Athina Markopoulou.
\newblock Walking in facebook: A case study of unbiased sampling of osns.
\newblock In {\em 2010 Proceedings IEEE Infocom}, pages 1--9. Ieee, 2010.

\bibitem{wilson2009user}
Christo Wilson, Bryce Boe, Alessandra Sala, Krishna~PN Puttaswamy, and Ben~Y
  Zhao.
\newblock User interactions in social networks and their implications.
\newblock In {\em Proceedings of the 4th ACM European conference on Computer
  systems}, pages 205--218, 2009.

\bibitem{liu2016controllability}
Xueming Liu and Linqiang Pan.
\newblock Controllability of the better chosen partial networks.
\newblock {\em Physica A: Statistical Mechanics and its Applications},
  456:120--127, 2016.

\bibitem{wang2019coevolution}
Wei Wang, Quan-Hui Liu, Junhao Liang, Yanqing Hu, and Tao Zhou.
\newblock Coevolution spreading in complex networks.
\newblock {\em Physics Reports}, 2019.

\bibitem{whalen2015observability}
Andrew~J Whalen, Sean~N Brennan, Timothy~D Sauer, and Steven~J Schiff.
\newblock Observability and controllability of nonlinear networks: The role of
  symmetry.
\newblock {\em Physical Review X}, 5(1):011005, 2015.

\bibitem{delellis2010fully}
Pietro DeLellis, Mario di~Bernardo, and Luiz Felipe~R Turci.
\newblock Fully adaptive pinning control of complex networks.
\newblock In {\em Proceedings of 2010 IEEE international symposium on circuits
  and systems}, pages 685--688. IEEE, 2010.

\bibitem{sanhedrai2020reviving}
Hillel Sanhedrai, Jianxi Gao, Moshe Schwartz, Shlomo Havlin, and Baruch Barzel.
\newblock Reviving a failed network via microscopic interventions.
\newblock {\em arXiv preprint arXiv:2011.14919}, 2020.

\bibitem{ma2020universality}
Cheng Ma, Gyorgy Korniss, Boleslaw~K Szymanski, and Jianxi Gao.
\newblock Universality of noise-induced resilience restoration of ecological
  systems.
\newblock {\em arXiv preprint arXiv:2011.11808}, 2020.

\bibitem{scheffer2015creating}
M~Scheffer, S~Barrett, SR~Carpenter, C~Folke, Andy~J Green, M~Holmgren,
  TP~Hughes, S~Kosten, IA~Van~de Leemput, DC~Nepstad, et~al.
\newblock Creating a safe operating space for iconic ecosystems.
\newblock {\em Science}, 347(6228):1317--1319, 2015.

\end{thebibliography}
%\bibliography{bib_main_bks}

\end{document}